\documentclass[12pt]{report}
\usepackage[utf8]{inputenc}
\usepackage[margin=2.5cm]{geometry}
\usepackage{setspace}
\usepackage{graphicx}
\usepackage[toc,page]{appendix}
\graphicspath{ {Images/} }
\usepackage[colorlinks=true, linkcolor  = blue,urlcolor=blue,citecolor=violet]{hyperref}
\usepackage{xcolor}
\usepackage{amsmath, amssymb}
\usepackage{float}
\usepackage{physics}
\usepackage{subfigure}
\usepackage{cite}
\usepackage[font=small,labelfont=bf]{caption}
\usepackage[bottom]{footmisc}

\usepackage{caption}
\usepackage{scalerel}
\usepackage{amsthm,latexsym,amsxtra,graphicx,appendix,epstopdf,feynmf,hyperref,setspace,fix-cm}
\usepackage[normalem]{ulem}
\usepackage{makeidx}
\usepackage{fullpage}
\usepackage{amsmath}
\usepackage{amssymb}
\usepackage{setspace}
\usepackage{bbm}
\usepackage{dsfont}
\usepackage{epsfig}
\usepackage[font=footnotesize,labelfont=bf,justification=centerlast,width=.94\textwidth]{caption}
\usepackage{cite}
 \usepackage{multirow}
\usepackage{array,booktabs}
\usepackage{subfigure}
\usepackage{mathtools}
\usepackage{color}
\usepackage[makeroom]{cancel}
\usepackage{tikz}
\usepackage{pgfplots}
\usetikzlibrary{intersections, pgfplots.fillbetween}
\usepackage{tensor}
\usepackage{simplewick}
\usepackage{frcursive}
\usepackage{pgfornament,multicol}
\usepackage{caption}
\usepackage{hyperref}

\usepackage[skip=0.2cm, indent=0pt]{parskip}

\hypersetup{
    bookmarks=true,%
    colorlinks,%
    citecolor=blue,%
    filecolor=blue,%
    linkcolor=blue,%
    urlcolor=blue
}
\usepackage{float}
\usepackage{slashed}
\usepackage{tikz}
\usetikzlibrary{decorations.pathmorphing}

\def\r{\mathfrak{r}}
\def\frakr{\mathfrak{r}}
\def\bomega{{\boldsymbol{\omega}}}

\newcommand{\bi}{\begin{itemize}}
\newcommand{\ei}{\end{itemize}}
\newcommand{\bea}{\begin{eqnarray}}
\newcommand{\eea}{\end{eqnarray}}
\newcommand{\be}{\begin{equation}}
\newcommand{\ee}{\end{equation}}

\def\XXint#1#2#3{{\setbox0=\hbox{$#1{#2#3}{\int}$}
     \vcenter{\hbox{$#2#3$}}\kern-.5\wd0}}

\def\={\, = \,}

\numberwithin{equation}{section}
\allowdisplaybreaks 

\author{Chawakorn Maneerat}
\date{April 2026}

\begin{document}
\begin{titlepage}
    \begin{center}

        \Huge
        \textsc{General Relativity with\\ Finite Boundaries}

        \vspace{2.5cm}

        \LARGE
        \textsc{Chawakorn Maneerat}

        \vspace{1.5cm}

        \large
        \textsc{A Thesis}\\[0.4em]
        \textsc{Presented for the Degree}\\[0.4em]
        \textsc{of Doctor of Philosophy}\\[0.4em]
        \textsc{in the Subject of}\\[0.4em]
        \textsc{Theoretical Physics}

        \vspace{1.5em}

        \textsc{Advisors: Damián A. Galante, Dionysios Anninos}

        \vspace{1.5em}

        \includegraphics[width=0.3\textwidth]{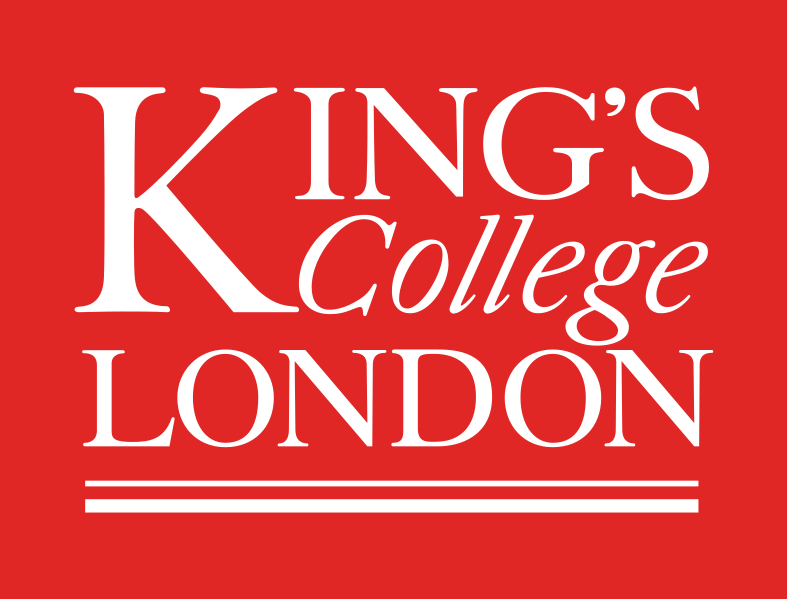}\\

        \vspace{1.5em}

        Department of Mathematics\\[0.3em]
        King's College London\\[0.3em]
        United Kingdom\\[0.3em]
        July 2026

    \end{center}
\end{titlepage}

\chapter*{Abstract}
In this thesis, we consider general relativity on a manifold with a timelike
boundary of finite size, across all signs of the cosmological constant
$\Lambda$. We impose conformal boundary conditions, whereby the conformal
class of the induced metric and the trace of the extrinsic curvature $K$ are
fixed at the boundary. In Lorentzian signature,
we analyse the linearised Einstein equations about Minkowski, de Sitter (dS), and
anti-de Sitter (AdS) space for a variety of boundaries, employing the
Kodama-Ishibashi formalism adapted to manifolds with a boundary. In
Euclidean signature, we study the thermodynamics of black hole and
cosmological horizons enclosed by the boundary, computing the leading
semiclassical approximation of the gravitational path integral, for which
the boundary data define a conformal canonical ensemble of fixed conformal
temperature and $K$. For $\Lambda = 0$, a spherical boundary of constant $K$
admits modes growing exponentially in time, which can be controlled by
varying $K$ and are absent for a flat boundary with vanishing extrinsic
curvature. The thermodynamics of a black hole enclosed by the boundary
retains the Bekenstein-Hawking form of the entropy. For $\Lambda > 0$, we
present the thermodynamic structure of dS in three and four
spacetime dimensions. In dS$_3$, the spacetime is thermally stable
for all $K$, while in dS$_4$ the cosmic patch, the region
enclosing the de Sitter horizon, is thermally stable for sufficiently large
$K$, in contrast to the Dirichlet problem for which its specific heat is
always negative. Dynamically, we uncover novel linearised modes about the
cosmic patch, which are linked to the quasinormal modes of the static patch
in the worldline limit, and to the shear and sound modes of a fluid
dynamical system as the boundary approaches the dS horizon. For
$\Lambda < 0$, global AdS$_4$ with a finite boundary admits, in addition to
the usual normal modes, exponentially growing modes at large angular
momentum, which recede to increasingly large angular momentum as $K\ell \to
3$ and the boundary approaches the asymptotic boundary. For planar AdS$_4$,
the analogous novel modes exhibit no complex frequencies. In Euclidean
signature, a coordinate-dependent $K(x^m)$ plays the role of a source for
the Weyl factor $\boldsymbol{\omega}(x^m)$, and we speculate on its
holographic interpretation near the AdS$_4$ asymptotic boundary.
\chapter*{Acknowledgements}
At the beginning of my PhD, I was told that doing a PhD is similar to running a marathon: one should not sprint over a short distance, but rather stay consistent over a long period. As finishing this thesis perhaps marks the last kilometre of my marathon, I would like to say a few things to the people who have made this marathon a meaningful journey.

First, I would like to thank everyone I have met along the way, including those who have been running right next to me: all the amazing PhD peers and postdocs at King's College London. You have not only stimulated my scientific curiosity, but also shown me how much fun it is to talk science with someone.

I would also like to thank my viva examiners, Toby Wiseman and Michael Anderson, for their thoughtful questions and valuable feedback.

Next, I would like to thank the rest of my wonderful collaborators: Weam Abou Hamdan, Raúl Arias, Silvia Georgescu, and Andrew Svesko. Working alongside you and learning from each of you has been a true pleasure.

Now, the marathon coaches. I would like to thank my supervisors, Damián Galante and Dionysios Anninos. I am not sure if I have expressed this enough, but I am truly grateful to have you as my supervisors. There were times when I felt tired from this long journey, but the people who reignited my excitement for physics, every single time, were always you two. Discussing physics in your offices, or on the District line, will forever be an unforgettable memory to me.

Now, the person who has been my mental support throughout the last four years, Anna Srikitkul. Thank you for cheering me on from the sidelines, even remotely. Our little movie nights and chats have always been what I look forward to when I want to take a break from physics.

Finally, I would like to dedicate this thesis to my parents, Pongsak Maneerat and Wilai Phocharachoke. Long before this marathon began, you were the ones who taught me how to take my first steps. Thank you for supporting me every step of the way.

\onehalfspacing
\tableofcontents
\singlespacing

\chapter{Introduction}\label{chap: intro}
%

\section{Motivation}
\label{sec:motivation}

A foundational question in quantum gravity concerns its gauge-invariant observables. The problem persists even at large distances, where quantum fluctuations are suppressed and the theory is organised according to the effective field theory framework (EFT) into a derivative expansion whose leading term is classical general relativity.

General relativity is a theory of dynamical spacetime, with diffeomorphism invariance as its gauge ambiguity. Any computation requires a choice of coordinates, while physical predictions must not depend on it. As in non-gravitational theories, one is often interested in phenomena confined to a quasi-local region, since experimental apparatus is finite in extent. When spacetime itself fluctuates, however, constructing local observables in a diffeomorphism-invariant way is far from obvious. One avatar of the problem is the absence of a natural reference frame. Among the proposals in the literature is the relational approach \cite{geheniau1956invariants,Komar:1958ymq,Bergmann:1961wa,Tambornino:2011vg}, which defines observables as physical quantities measured relative to one another. It has proved useful perturbatively around a given background. Away from one, analytic control has proven elusive.

The problem is much better understood when spacetime exhibits asymptotic structure. For $\Lambda < 0$, the Einstein equations permit a series expansion of the metric near spatial infinity, the Fefferman--Graham expansion \cite{fefferman1985conformal}, whose freely specifiable data comprise a conformal equivalence class of codimension-one metrics together with a transverse, traceless, rank-two symmetric tensor. A family of spacetimes with a fixed asymptotic metric is called asymptotically anti-de Sitter (AdS). Fixing the asymptotic data endows such spacetimes with a rigid, non-gravitating timelike boundary at infinity, a natural reference frame relative to which observables can be measured. Boundary correlation functions, defined by an appropriate limit of the bulk correlators, thereby provide unambiguous observables for the gravitational physics in the interior, and form a basic building block of the gravitational side of the AdS/CFT correspondence \cite{Maldacena:1997re,Witten:1998qj,deHaro:2000vlm,Anderson:2004yi}. For $\Lambda = 0$, null infinity plays an analogous role to a certain extent \cite{Bondi:1962px,Sachs:1962wk}, with the gravitational $S$-matrix furnishing the unambiguous observables.

\textbf{Spacetime without asymptotia.} This strategy is unavailable when spacetime possesses no asymptotia. Expanding universes are phenomenologically relevant examples. The simplest toy model is the maximally symmetric solution with $\Lambda>0$, de Sitter space (dS), see \cite{Spradlin:2001pw,Anninos:2012qw,Galante:2023uyf} for reviews. The global structure of four-dimensional dS ($\text{dS}_4$) is described by the metric
\begin{equation}
\label{eq:globaldS}
\mathrm{d} s^2 = -\mathrm{d}\tau^2 + \ell^2 \cosh^2\frac{\tau}{\ell}\, \mathrm{d}\Omega_3^2\,, \qquad \ell \equiv \sqrt{\frac{3}{\Lambda}}\,.
\end{equation}
The Cauchy surface at constant time is compact, so there is no spatial infinity. The static patch of $\text{dS}_4$, namely the region causally accessible to an observer within it, is instead described by
\begin{equation}
\label{eq:staticpatch}
\mathrm{d} s^2 = -\left(1-\frac{r^2}{\ell^2}\right)\mathrm{d} t^2 + \frac{\mathrm{d} r^2}{1-\frac{r^2}{\ell^2}} + r^2\, \mathrm{d}\Omega_2^2\,, \qquad r \in (0,\ell)\,.
\end{equation}
The inertial observer sits at $r=0$, enclosed by the cosmological horizon at $r=\ell$. Any gravitational phenomenon behind the horizon is forever out of causal contact, so the physically relevant portion of spacetime is always finite. Moreover, the cosmological horizon appears to carry an entropy \cite{Gibbons:1977mu}, much as a black hole horizon does. Understanding what this entropy counts is an important step towards a quantum theory of an expanding universe like our own, and a well-defined notion of observable in the static patch perhaps plays an important role \cite{Anninos:2011af,anninos2018infrared, Witten:2023qsv}.

The approach followed in this thesis is to append to spacetime a finite timelike boundary $\Gamma$ and impose boundary conditions on it. These freeze part of the boundary data, rendering $\Gamma$ a non-gravitating reference frame relative to which observables can be constructed unambiguously. Physically, one may think of $\Gamma$ as a place, \emph{a laboratory}, where observers reside. In dS spacetime, with $\Gamma$ inside the static patch, one may go further and regard $\Gamma$ as \emph{an observatory}, collecting data from gravitational phenomena away from it and reconstructing the large-scale structure of spacetime.

\textbf{Gravity in a box.} Gravity with a timelike boundary is not a new subject. In numerical relativity, simulating spacetime with finite computational resources requires a formulation in a finite region, see \cite{Sarbach:2012pr}. Of interest in this thesis are questions involving quantum gravity. Along this line, perhaps the first systematic use of general relativity on a manifold with boundary is the work of Gibbons-Hawking-York \cite{yorkbdy,Gibbons:1976ue}, which identified the boundary term supplementing the Einstein--Hilbert action so that the variational principle is well defined upon fixing the induced boundary metric, so-called Dirichlet boundary conditions. This later played a central role in the quasi-local thermodynamics of black holes in the Euclidean gravitational path integral \cite{York:1986it} and in the quasi-local stress tensor of Brown and York \cite{Brown:1992br}. A boundary placed infinitesimally away from a black hole horizon revealed the resemblance of horizon dynamics to fluid dynamics, known as the membrane paradigm \cite{Damour:1978cg, znajek1978electric,Price:1986yy,Parikh:1997ma}. Anti-de Sitter space with a finite Dirichlet boundary has been related to the $T\bar T$-deformation of conformal field theory \cite{Smirnov:2016lqw, Cavaglia:2016oda,McGough:2016lol,Hartman:2018tkw}, in an attempt to understand the emergence of the radial direction in AdS/CFT. Two-dimensional dilaton-gravity with a finite boundary has led to new insights into near-extremal black holes at low temperature \cite{Maldacena:2016upp}.

In this thesis, we take a step back and ask whether general relativity with a finite boundary is mathematically consistent. The minimal requirement is perhaps that it admits a \emph{well-posed} initial boundary value formulation, involving a mathematically sensible choice of boundary conditions. Section \ref{sec:mathbackground} provides a brief overview of initial (boundary) value problems of general PDEs and their well-posedness. Section \ref{sec:ibvpgravity} discusses the initial (boundary) value formulation of general relativity and reports on the well-posedness of Dirichlet and conformal boundary conditions.

\section{Mathematical background}
\label{sec:mathbackground}

As a preliminary, we review the structure of partial differential equations (PDEs) on manifolds with boundary, with an emphasis on well-posedness, see \cite{Evans2010}. Let $\mathcal{M}$ be a manifold with boundary $\partial\mathcal{M}$, and let $u$ be a field on $\mathcal{M}$ subject to a second-order differential equation $Lu=0$. The central question is what data must be prescribed on $\partial\mathcal{M}$ to determine $u$ in the bulk, in a mathematically controlled sense.

The non-linear PDEs of interest are quasi-linear and second order,
\begin{equation}
\label{eq:quasilinear}
L u \equiv a^{\mu\nu}(u,\partial u, x)\, \partial_\mu \partial_\nu u - F(u,\partial u, x) = 0\,,
\end{equation}
where $\partial_\mu \equiv \frac{\partial}{\partial x^\mu}$. The operator $L$ is linear in the second-derivative term, while $a^{\mu\nu}$ and $F$ may depend arbitrarily on $u$, $\partial_\mu u$, and $x^\mu$. The extension to several coupled degrees of freedom is immediate, with $a^{\mu\nu}$ and $F$ acquiring additional field indices.

The classification of \eqref{eq:quasilinear} rests on its short-distance behaviour. Near a point $x$ the coefficients vary slowly, so $L$ is approximated by a linear operator with frozen coefficients, i.e.\ with $a^{\mu\nu}$ and $F$ evaluated at $x$ and treated as constant. Replacing $\partial_\mu \to i\, p_\mu$ in the highest-derivative term yields the leading symbol of $L$,
\begin{equation}
\label{eq:symbol}
\sigma(p) \equiv a^{\mu\nu} p_\mu p_\nu\,,
\end{equation}
with $p_\mu$ an arbitrary constant one-form. The type of $L$ is fixed by the algebraic properties of $\sigma(p)$, regarded as a quadratic form on the space of $p_\mu$. If $a^{\mu\nu}$ is non-degenerate and $\sigma(p)$ has fixed sign for every $p_\mu \neq 0$, then $L$ is elliptic. If instead $\sigma(p)$ has Lorentzian signature $(-,+,\dots,+)$, splitting $p_\mu$-space into timelike ($\sigma(p)<0$), spacelike ($\sigma(p)>0$) and null ($\sigma(p)=0$) directions, then $L$ is hyperbolic. The Klein-Gordon operator on Minkowski space, $L = \eta^{\mu\nu}\partial_\mu \partial_\nu$, provides a basic example. Its leading symbol,
\begin{equation}
\label{eq:KGsymbol}
\sigma(p) = \eta^{\mu\nu} p_\mu p_\nu\,,
\end{equation}
has Lorentzian signature, so $L$ is hyperbolic, whereas its Euclidean continuation has definite signature and is elliptic.

This classification dictates what data may consistently be specified on $\partial\mathcal{M}$ so as to determine a unique bulk solution. The wave equation on Minkowski space, $\partial^\mu \partial_\mu u = 0$, is hyperbolic, and its solution is fixed by Cauchy data, the initial profile together with its normal derivative, prescribed on a spacelike surface. Poisson's equation, by contrast, is elliptic, and its solution is fixed instead by the potential prescribed on the bounding surface.

More physically, treating \eqref{eq:quasilinear} as the equation of motion of a closed classical system, free of external influence, one seeks to control the solution through data specified on the boundary. The solution should be entirely determined by this data, and an infinitesimal variation of the data should induce a correspondingly small variation of the solution. These requirements are formalised, following Hadamard, in the notion of \emph{well-posedness}. Equation \eqref{eq:quasilinear} together with a prescription of data on $\partial\mathcal{M}$ is well-posed provided it admits
\begin{itemize}
\item existence of a solution,
\item uniqueness of the solution,
\item and continuous dependence of the solution on the data.
\end{itemize}
Existence and uniqueness together allow one to parametrise solutions of \eqref{eq:quasilinear} by the prescribed data. Continuous dependence requires a specification of the norm on both the space of data and the space of solutions, with respect to which nearby data yield nearby solutions.\footnote{Ill-posed problems in the Hadamard sense indeed occur in nature \cite{EnglGroetsch1987}. Classic examples include the backward heat equation, where exponential damping of short-wavelength modes makes the time-reversed reconstruction unstable, and the Cauchy problem for the Laplace equation, relevant to electrocardiography and geophysical prospecting. Such problems are typically addressed through regularisation \cite{Tikhonov:1977}.}

We now turn to the distinct formulations of \eqref{eq:quasilinear} as an evolution problem for data specified on $\partial\mathcal{M}$.

\paragraph{Initial value problem.} The first class arises when $L$ is hyperbolic and $\Sigma \equiv \partial\mathcal{M}$ is a spacelike surface with respect to $\sigma(p)$, that is, one-forms normal to $\Sigma$ are timelike. One may then treat $Lu=0$ as evolving the initial data prescribed on $\Sigma$,
\begin{equation}
\label{eq:IVPdata}
u\big|_\Sigma = f\,, \qquad n^\mu \partial_\mu u \big|_\Sigma = g\,,
\end{equation}
where $n^\mu$ is the unit timelike normal to $\Sigma$ and $f,g$ are prescribed functions on $\Sigma$. This is the initial value problem (IVP).

\paragraph{Boundary value problem.} The second class arises when $L$ is elliptic. Since $\sigma(p)$ is definite, there is no causal structure, and one solves $Lu=0$ by imposing boundary conditions on $\partial\mathcal{M}$ comprising half as many data as in the IVP case. The choice of boundary data is not unique. The standard choices are
\begin{align}
u\big|_{\partial\mathcal{M}} &= f  &&\text{(Dirichlet)}\,, \label{eq:Dirichlet}\\
n^\mu \partial_\mu u\big|_{\partial\mathcal{M}} &= g  &&\text{(Neumann)}\,. \label{eq:Neumann}
\end{align}
One may also impose a mixed (Robin) boundary condition, consisting of a (non-)linear combination of Dirichlet and Neumann data.

\paragraph{Initial boundary value problem.} The last class arises when $L$ is hyperbolic and $\partial\mathcal{M}$ decomposes into a spacelike component $\Sigma$ and a timelike component $\Gamma$, i.e.\ $\partial\mathcal{M} = \Sigma \cup \Gamma$. The standard treatment imposes initial conditions \eqref{eq:IVPdata} on $\Sigma$ and boundary conditions \eqref{eq:Dirichlet} on $\Gamma$, supplemented by a compatibility condition relating the initial and boundary data at the corner $C \equiv \Sigma \cap \Gamma$.

\subsection{Initial (boundary) value formulation of wave equation}
\label{sec:waveequation}

As a concrete illustration, consider the wave equation on two-dimensional Minkowski space,
\begin{equation}
\label{eq:waveeq}
-\partial_t^2 \phi + \partial_x^2 \phi = 0\,,
\end{equation}
whose general solution is a superposition of left- and right-moving plane waves,
\begin{equation}
\label{eq:planewaves}
\phi(t,x) = \phi_-(x-t) + \phi_+(x+t)\,,
\end{equation}
with $\phi_\pm$ arbitrary functions of a single variable. This closed form makes \eqref{eq:waveeq} a good toy model for the formulations above and their well-posedness on a manifold with boundary.

\paragraph{Well-posed initial value problem.} In the absence of a timelike boundary, \eqref{eq:waveeq} admits a well-posed IVP formulation. The Cauchy data are prescribed on the $t=0$ slice $\Sigma$,
\begin{equation}
\label{eq:waveIVPdata}
\phi(0,x) = f(x)\,, \qquad \partial_t \phi(0,x) = g(x)\,,
\end{equation}
for arbitrary smooth $f(x)$ and $g(x)$. Substituting \eqref{eq:planewaves} into \eqref{eq:waveIVPdata}, the unique solution is given by d'Alembert's formula,
\begin{equation}
\label{eq:dalembert}
\phi(t,x) = \frac{f(x-t)+f(x+t)}{2} + \frac{1}{2}\int_{x-t}^{x+t} \mathrm{d} s \, g(s)\,.
\end{equation}
Existence holds since the construction places no obstruction on $(f,g)$, and uniqueness follows since the difference of any two solutions with identical initial data vanishes. Continuous dependence on $(f,g)$ can be established with respect to various commonly used norms, e.g.\ the Sobolev norm, for which we refer to \cite{Evans2010} for more details.

\paragraph{Well-posed initial boundary value problem.} We now supplement the Cauchy surface at $t=0$ with a timelike boundary at $x=0$, restricting the physical region to $x>0$, $t>0$. This problem admits an IBVP formulation with a Dirichlet boundary condition. On the half-line $x\geq 0$ we impose
\begin{equation}
\label{eq:IBVPinitial}
\phi(0,x) = f(x)\,, \qquad \partial_t\phi(0,x) = g(x)\,, \qquad x \geq 0\,,
\end{equation}
and at $x=0$ the Dirichlet condition
\begin{equation}
\label{eq:IBVPdirichlet}
\phi(t,0) = h(t)\,,
\end{equation}
with $f$, $g$, and $h$ arbitrary. Substituting \eqref{eq:planewaves} into the initial and boundary conditions yields
\begin{equation}
\label{eq:IBVPsolution}
\phi(t,x) =
\begin{cases}
\displaystyle \frac{f(x-t)+f(x+t)}{2} + \frac{1}{2}\int_{x-t}^{x+t}\mathrm{d} s\, g(s)\,, & t \leq x\,,\\[3ex]
\displaystyle \frac{-f(t-x)+f(x+t)}{2} + \frac{1}{2}\int_{t-x}^{x+t}\mathrm{d} s\, g(s) + h(t-x)\,, & 0 \leq x < t\,.
\end{cases}
\end{equation}
On the region $t \leq x$ the solution coincides with the IVP result, while the Dirichlet data $h(t)$ propagate only into $0 \leq x < t$, as expected from causality. As before, the absence of obstructions to $(f,g)$ and $h$ ensures existence, while uniqueness is immediate from \eqref{eq:IBVPsolution}.

Evaluating the solution at the corner $t=x=0$, we find a compatibility condition relating $(f,g)$ to $h$,
\begin{equation}
\label{eq:compatibility}
\partial_x^{2n} f(0) = \partial_t^{2n} h(0)\,, \qquad \partial_x^{2n} g(0) = \partial_t^{2n+1} h(0)\,, \qquad n = 0,1,2,\dots\,.
\end{equation}
This infinite tower of conditions is closely tied to the smoothness of $\phi(t,x)$ across the null ray emanating from the corner. Truncating at finite $n$ yields a shock-wave-type solution, with a discontinuity in a sufficiently high derivative of $\phi$.

\paragraph{Ill-posed initial boundary value problem.} It is interesting to also consider a situation where the problem is ill-posed. We again impose \eqref{eq:IBVPinitial} on the half-line $x \geq 0$, but at $x = 0$ we now impose the boundary condition
\begin{equation}
\label{eq:badBC}
\partial_t \phi(t,0) + \partial_x \phi(t,0) = \tilde h(t)\,,
\end{equation}
with the boundary data $\tilde h(t)$ an arbitrary function. The general solution is
\begin{equation}
\label{eq:badsolution}
\phi(t,x) =
\begin{cases}
\displaystyle \frac{f(x-t)+f(x+t)}{2} + \frac{1}{2}\int_{x-t}^{x+t}\mathrm{d} s\, g(s)\,, & t \leq x\,,\\[3ex]
\displaystyle \frac{f(x-t)+f(x+t)}{2} + \frac{1}{2}\int_{x-t}^{x+t}\mathrm{d} s\, g(s) + a(x-t)\,, & 0 \leq x < t\,,
\end{cases}
\end{equation}
where $a$ is an arbitrary function with compact support on the negative axis. The initial data are moreover subject to a constraint following from \eqref{eq:badBC},
\begin{equation}
\label{eq:badconstraint}
\partial_x f(x) + g(x) = \tilde h(x)\,, \qquad x \geq 0\,.
\end{equation}
Both existence and uniqueness now fail. The zero mode $a(x-t)$ carries vanishing initial and boundary data, so infinitely many distinct solutions obey identical conditions. Conversely, the constraint \eqref{eq:badconstraint} restricts the admissible data, so that for generic data no solution exists.

\paragraph{Loss of continuous dependence.} A more delicate failure mode arises when solutions exist and are unique, yet do not depend continuously on the data. The classic example, due to Hadamard, is the elliptic continuation of \eqref{eq:waveeq}, namely the Laplace equation $\partial_t^2\phi + \partial_x^2\phi = 0$, treated (against the guidance of its symbol) as an evolution problem in $t$ with Cauchy data at $t=0$. Consider the family of data labelled by $k > 0$,
\begin{equation}
\label{eq:hadamarddata}
\phi(0,x) = 0\,, \qquad \partial_t \phi(0,x) = e^{-\sqrt{k}}\sin (k x)\,,
\end{equation}
for which the corresponding solutions read
\begin{equation}
\label{eq:hadamardsolution}
\phi_k(t,x) = \frac{e^{-\sqrt{k}}}{k}\, \sinh(k t) \sin(k x)\,.
\end{equation}
For each $k$ the solution exists, and standard uniqueness theorems for analytic data guarantee it is the only one. Nevertheless, as $k \to \infty$ the data \eqref{eq:hadamarddata} tend to zero uniformly, together with all of their derivatives, while the solution \eqref{eq:hadamardsolution} at any fixed $t>0$ grows without bound, owing to the factor $\sinh(kt) \sim e^{kt}$. An arbitrarily small deformation of trivial initial data thus produces an arbitrarily large response, and the map from data to solutions fails to be continuous in any norm built from finitely many derivatives.\footnote{In fact, this  example of loss of continuous dependence also suggests a possible obstruction to global-in-time existence. Specifically, one can construct a sequence of finite sums of modes whose initial data converge to analytic data, while the corresponding solutions become unbounded at sufficiently late times. This raises the possibility that the limiting data admit a local analytic solution that cannot be continued to arbitrarily late times, and hence it is worth revisiting a global-in-time existence in this setting.} The amplification $e^{kt}$ grows with the wavenumber $k$, so the shorter the wavelength, the more violent the response. 

\section{Initial (boundary) value formulation of gravity}
\label{sec:ibvpgravity}

Equipped with these PDE tools, we turn to general relativity on manifolds with boundary, focusing on its BVP and IBVP formulations and their well-posedness, see \cite{Wald:1984rg} for the discussion on IVP formulation of general relativity. For simplicity we restrict throughout to four-dimensional pure Einstein gravity without cosmological constant. The generalisation to higher dimensions or non-vanishing $\Lambda$ is straightforward and alters none of the structural features we discuss.

The governing PDEs are the Einstein field equations
\begin{equation}
\label{eq:einstein}
R_{\mu\nu} - \frac{1}{2} g_{\mu\nu} R = 0\,,
\end{equation}
in four dimensions a system of ten coupled non-linear second-order PDEs for the ten metric components. In Lorentzian signature, \eqref{eq:einstein} is of second-order hyperbolic type. Two features nevertheless make its IVP, and the related formulations, non-trivial.

First, four of the ten equations are constraints rather than evolution equations. The naive initial data, the initial metric profile together with its time derivative, cannot be prescribed freely. This stands in tension with existence.

Second, the six remaining equations do not suffice to fix the metric completely. This under-determination reflects diffeomorphism invariance, with infinitely many distinct-looking solutions describing the same geometry. This stands in tension with uniqueness.

\paragraph{An aside on Maxwell theory.} Both features are already visible in a simpler, linear setting, namely Maxwell theory formulated in terms of the gauge field $A_\mu$. In the absence of charged matter the equations of motion are
\begin{equation}
\label{eq:maxwell}
\partial^\mu F_{\mu\nu} = 0\,, \qquad F_{\mu\nu} = \partial_\mu A_\nu - \partial_\nu A_\mu\,,
\end{equation}
which, expanded out, read
\begin{equation}
\label{eq:maxwellexpanded}
\partial^\mu \partial_\mu A_\nu - \partial_\nu \partial^\mu A_\mu = 0\,,
\end{equation}
a system of four equations for the four components of $A_\mu$.

The $\nu=t$ component of \eqref{eq:maxwellexpanded} contains no second time derivative of any $A_\mu$,
\begin{equation}
\label{eq:gauss}
\partial_i \left( \partial^i A^t - \partial^t A^i \right) = 0\,,
\end{equation}
which is precisely the Gauss constraint $\partial_i E^i = 0$. It constrains the initial data $(A_\mu, \partial_t A_\mu)\big|_{t=0}$ rather than evolving them, so the data cannot be chosen freely. Conversely, the remaining three equations do not fix the four components of $A_\mu$ uniquely, an avatar of the gauge freedom
\begin{equation}
\label{eq:gaugefreedom}
A_\mu \; \longrightarrow \; A_\mu + \partial_\mu \lambda\,,
\end{equation}
which leaves $F_{\mu\nu}$ invariant for any spacetime-dependent $\lambda$. Taking $\lambda$ to have compact support away from the $t=0$ surface yields infinitely many solutions sharing identical initial data but differing at later times, all describing the same $F_{\mu\nu}$.

The standard resolution is to fix a gauge. In Lorenz gauge $\partial^\mu A_\mu = 0$, equation \eqref{eq:maxwellexpanded} reduces to the manifestly hyperbolic wave equation
\begin{equation}
\label{eq:lorenzwave}
\Box A_\nu = 0\,,
\end{equation}
whose IVP is well posed. The gauge choice is preserved by the evolution because the gauge condition itself obeys a wave equation on-shell, $\Box\left(\partial^\mu A_\mu\right) = 0$. Hence, if $\partial^\mu A_\mu = 0$ and $\partial_t\left(\partial^\mu A_\mu\right) = 0$ on the $t=0$ surface, the latter being precisely the Gauss constraint \eqref{eq:gauss} in this gauge, then $\partial^\mu A_\mu = 0$ everywhere. Prescribing initial data $(A_\mu, \partial_t A_\mu)$ satisfying the Gauss constraint and the Lorenz condition, and evolving with \eqref{eq:lorenzwave}, thus produces a solution of the full Maxwell equations which remains in Lorenz gauge throughout.

\subsection{Cauchy problem}
\label{sec:cauchyproblem}

Resolving the two tensions above is considerably harder in gravity, owing to the non-linearity of the Einstein equations. It is therefore remarkable that general relativity nonetheless admits a well-posed IVP, as established in the celebrated work of Choquet-Bruhat and Geroch \cite{foures1952theoreme,Choquet-Bruhat:1969ywq}. We merely record the outcome here, see \cite{Wald:1984rg,hawking_ellis_1973,Isenberg:2013iva} for details.

The initial data consist of a three-dimensional manifold $\Sigma$ carrying a Riemannian metric $h_{ij}$ together with a symmetric rank-two tensor $K_{ij}$, subject to the constraint equations
\begin{equation}
\label{eq:constraints}
R^{(3)} + K^2 - K_{ij}K^{ij} = 0\,, \qquad D_i\left(K^{ij} - h^{ij} K\right) = 0\,,
\end{equation}
where $R^{(3)}$ and $D_i$ are the Ricci scalar and covariant derivative of $h_{ij}$, and $K \equiv h^{ij}K_{ij}$, see appendix \ref{app:hyper basic} for further details. Given any pair $(h_{ij}, K_{ij})$ solving \eqref{eq:constraints}, there exists a spacetime metric $g_{\mu\nu}$ solving \eqref{eq:einstein} in which $\Sigma$ is embedded as a hypersurface with induced metric $h_{ij}$ and extrinsic curvature $K_{ij}$. From the bulk perspective, $h_{ij}$ plays the role of the initial field profile and $K_{ij}$ that of its initial velocity. Uniqueness holds only up to diffeomorphism, as it must. This is moreover a local statement, applying to a small region covered by a single coordinate chart. The sharp global statement, the existence of a unique maximal globally hyperbolic development, is due to \cite{Choquet-Bruhat:1969ywq}.

A well-posed IVP allows one to parametrise the space of solutions by the more controlled space of initial data, thereby trading questions about solutions for questions about their data. A remarkable use of this idea is the proof of the non-linear stability of Minkowski spacetime \cite{Christodoulou:1993uv}. Any initial data sufficiently close to those of Minkowski space evolve to a geodesically complete spacetime, extending infinitely into the far future and past while remaining asymptotically flat.

\subsection{Boundary conditions}
\label{sec:boundaryconditions}

Motivated by the well-posed IVP formulation, one may ask whether general relativity on a manifold with boundary admits well-posed BVP or IBVP formulations, and which boundary conditions achieve this. In Euclidean signature, where \eqref{eq:einstein} is elliptic, the boundary conditions are imposed on the boundary of the manifold. In Lorentzian signature, where \eqref{eq:einstein} is hyperbolic, they are imposed on a timelike boundary and supplemented by standard Cauchy data on a spacelike slice. The first mathematical analysis of this question is the seminal work of Friedrich and Nagy \cite{Friedrich:1998xt} (see also \cite{Friedrich:1995vb} for asymptotically anti-de Sitter spacetime, where the timelike boundary sits at spatial infinity), which established boundary conditions leading to a well-posed IBVP, albeit only after a certain gauge choice. In what follows we consider only geometric boundary conditions, namely those built from the induced metric and extrinsic curvature of the boundary. Not covered in this thesis are totally geodesic \cite{Fournodavlos:2020wde} and umbilic \cite{Fournodavlos:2021eye} boundary conditions, which admit a well-posed IBVP \cite{Fournodavlos:2020wde,Fournodavlos:2021eye}, and more recently developed gauge-dependent boundary conditions \cite{Kreiss:2007zz,Kreiss:2009jia,Fournodavlos:2019ckr,An:2025cbs}.

Below we report on the status of two boundary conditions, Dirichlet boundary conditions (DBCs) and conformal boundary conditions (CBCs).

\paragraph{Dirichlet boundary conditions.} These hold fixed the induced metric of the boundary,
\begin{equation}
\label{eq:DBC}
h_{ij}\big|_\Gamma = \text{fixed}\,.
\end{equation}

In Euclidean signature, it was proven in \cite{Anderson_2008} that the Einstein equations with DBCs do not admit a well-posed BVP formulation, both existence and uniqueness failing, see also \cite{Witten:2018lgb}. There is, however, a class of Euclidean solutions for which the problem effectively exhibits existence and uniqueness, namely those for which the boundary quantity $K_{ij} - K h_{ij}$, left unfixed by the Dirichlet data, has either positive- or negative-definite eigenvalues.

In Lorentzian signature, the Dirichlet IBVP was likewise proven to be generically ill-posed \cite{An:2025gvr}, see \cite{An:2021fcq,Anninos:2022ujl} for the zero modes rendering the linearised problem non-unique around a Minkowski background with a flat timelike boundary. As in the Euclidean case, DBCs display better well-posedness properties for solutions whose timelike boundary has $K_{ij} - K h_{ij}$ of definite sign, an example being Minkowski space with a spatially spherically symmetric timelike boundary. To the best of our knowledge, continuous dependence on the initial data remains unexplored for DBCs, and would be worth studying especially in situations where DBCs are effectively well-posed.

\paragraph{Conformal boundary conditions.} These hold fixed the conformal structure of the induced metric together with the trace of the extrinsic curvature of the boundary,
\begin{equation}
\label{eq:CBC}
\left[ h_{ij} \right]_{\text{conf}} = \text{fixed}\,, \qquad K = \text{fixed}\,.
\end{equation}
In practice, fixing the conformal structure of the induced metric can be done by fixing its conformal representative. As such, the Weyl factor $\boldsymbol{\omega}$ of the boundary is left unfixed.\footnote{Fixing these data has also appeared as a method for isolating freely specifiable initial data solving the Hamiltonian and momentum constraints on a Cauchy slice \cite{JMPA_1944_9_23__37_0, Choquet-Bruhat:1969ywq}.}

In contrast to DBCs, Euclidean general relativity with CBCs was proven to admit a well-posed BVP formulation \cite{Anderson_2008}. In Lorentzian signature, the IBVP with CBCs was conjectured to have good existence and uniqueness properties \cite{An:2021fcq}. The conjecture stimulated interest in CBCs, leading for instance to an analogue of the Brown-York tensor \cite{Odak:2021axr} and to a one-parameter generalisation of CBCs \cite{Liu:2024ymn}.

A study of linearised gravitational dynamics subject to CBCs reveals perturbations taking the form of physical diffeomorphisms, namely those acting non-trivially at the timelike boundary \cite{Anninos:2023epi,Liu:2024ymn}.\footnote{Mathematically speaking, a diffeomorphism of $\mathcal{M}$ is a smooth bijective map $\mathcal{M} \to \mathcal{M}$. When $\partial \mathcal{M}\neq \emptyset$, such a map must preserve the boundary as a set, i.e. $\partial \mathcal{M} \to \partial \mathcal{M}$. Infinitesimally, this requires the diffeomorphism vector field to be tangent to the boundary. For instance, if the boundary is located at $x=0$, the linearised diffeomorphism $\xi^\mu$ must satisfy $\xi^x|_{x=0}=0$. The physical diffeomorphisms referred in this thesis are those with a non-zero component normal to the boundary. Hence, they move the boundary and are not regarded as diffeomorphism of $\mathcal{M}$ as described above. However, by viewing $\mathcal{M}$ as embedded in an ambient manifold $\tilde{\mathcal{M}}$, it is possible to interpret the physical diffeomorphism as a diffeomorphism of $\tilde{\mathcal{M}}$ which maps $\mathcal{M}$ to a different region.} It subsequently inspired a mathematical work \cite{An:2025rlw}, which establishes existence and uniqueness for the linearised equations around an arbitrary background, leading to a notion of well-posed \emph{initial boundary corner value problem} of general relativity. One fixes CBCs at the timelike boundary $\Gamma$, standard Cauchy data on the Cauchy surface $\Sigma$, and the corner angle at the intersection $C \equiv \Gamma \cap \Sigma$, together with a corner compatibility condition.

Continuous dependence has been considered recently in \cite{Liu:2025xij}, which established that the linearised Einstein equations with CBCs around a class of backgrounds do not depend continuously on the initial data. The key ingredient is the existence of large tangential momentum modes growing exponentially in time \cite{Anninos:2023epi,Liu:2024ymn,Liu:2025xij}. From these modes one builds a sequence of initial data at $t=0$ approaching the trivial data, while the corresponding solutions at $t>0$ do not, similar to the Hadamard example presented in section \ref{sec:waveequation}. It is worth noting that, from the effective field theory perspective, where general relativity is merely the leading term in a derivative expansion, the existence of exponentially growing modes at parametrically large angular momentum must be treated with care, see chapter \ref{sec:stretchedhorizon} further details.

General relativity with finite boundaries therefore sits at the intersection of mathematical and theoretical physics. On the mathematical side, the generic failure of existence and uniqueness for DBCs, together with the loss of continuous dependence exhibited by CBCs, leaves a well-posed IBVP of general relativity an important open problem. On the physical side, a finite boundary offers a promising route towards the quantum gravity of spacetime without asymptotia, with the boundary serving as an observatory collecting experimental data. This thesis sets out to make progress on both sides by performing a systematic analysis of the linearised dynamics and thermodynamics of general relativity subject to CBCs across all signs of the cosmological constant. We hope the results established here prove useful to both fields.

\section{Publications}
\label{sec:outline}
 
 
This thesis is based on the following published work.
\begin{itemize}
\item D. Anninos, D.A. Galante and C. Maneerat, \emph{Gravitational observatories}, JHEP \textbf{12} (2023) 024 [2310.08648].
\item D. Anninos, D.A. Galante and C. Maneerat, \emph{Cosmological observatories}, Class. Quant. Grav. \textbf{41} (2024) 165009 [2402.04305].
\item D. Anninos, R. Arias, D.A. Galante and C. Maneerat, \emph{Gravitational observatories in AdS$_4$}, JHEP \textbf{47} (2025) 234 [2412.16305].
\end{itemize}
The author has also published the following work during their PhD programme, which does not form part of this thesis.
\begin{itemize}
\item D.A. Galante, C. Maneerat and A. Svesko, \emph{Conformal boundaries near extremal black holes}, Class. Quant. Grav. \textbf{42} (2025) 195003 [2504.14003].
\item D. Anninos, D.A. Galante, S. Georgescu, C. Maneerat and A. Svesko, \emph{The Stretched Horizon Limit}, JHEP \textbf{05} (2026) 234 [2512.16738].
\item W. Abou Hamdan and C. Maneerat, \emph{Undulating Conformal Boundaries in 3D Gravity}, [2605.08058].
\end{itemize}

I acknowledge the use of AI software (\href{https://claude.ai}{Claude}) to rewrite, rephrase and/or paraphrase parts of this thesis to ensure the quality and standard of the English used, consistent with the King's College London guidance on generative AI in doctoral assessment; this thesis remains a genuine account of the research I have undertaken and its content can still be considered my own words.

\chapter{Zero cosmological constant}\label{chap: flat}

\section{Introduction}
\label{section: Introduction}

Depending on the physical setting, the asymptotic structure of spacetime can vary considerably. In certain circumstances, one is afforded a spacetime that asymptotes to spatial or null infinity where the dynamical behaviour of the metric can be significantly tamed. A prime example is an asymptotically anti-de Sitter spacetime whereby the metric field admits a well-posed Dirichlet condition \cite{fefferman1985conformal,KICHENASSAMY2004268} at the conformal boundary, accompanied by standard Cauchy data across a complete spacelike slice. This sets the gravitational stage \cite{deHaro:2000vlm,Skenderis:2002wp,Anderson:2004yi} for the AdS/CFT correspondence, and provides a rigid boundary coordinate system with respect to which one can characterise the dynamical features of the physical processes in the interior spacetime. The null boundary of an asymptotically Minkowski spacetime, though less rigid than that of anti-de Sitter space, is sufficiently structured to permit the gravitational scattering of asymptotic  states.\footnote{In four spacetime dimensions, the gravitational $S$-matrix suffers from infrared issues \cite{Weinberg:1965nx,Prabhu:2022zcr}, but is in the least sensible order by order in a perturbative expansion.}

There are, however, circumstances where such asymptotic structures are not available. For instance, as is often the case in cosmological models, the Cauchy surface might be a compact space such as an $S^3$. Alternatively, for spacetimes expanding at a sufficiently rapid rate, an observer may be surrounded by a cosmological horizon and hence out of causal contact from any asymptotic regions even if they were available in the global spacetime. This is of particular relevance to an asymptotically de Sitter space in which observers are surrounded by a cosmological event horizon. In such a circumstance, it is natural to ask whether one can construct a more quasi-local framework. As an auxiliary step in this direction one can imagine creating a quasi-artificial, finite-size, timelike boundary in spacetime. We can view this as a type of thickened wordline $\mathcal{M}$ \cite{Anninos:2011af,Coleman:2021nor,Banihashemi:2022jys,Witten:2023qsv,Blacker:2023oan,Loganayagam:2023pfb} in the midst of spacetime whose boundary data comprise a type of reference system. {Relatedly, in the stretched horizon picture one  imagines a timelike surface that approaches the horizon of a black hole \cite{Damour:1978cg,znajek1978electric,tHooft:1984kcu,Price:1986yy,Susskind:1993if,Bredberg:2011xw, Freidel:2023bnj} or a cosmological horizon \cite{Banks:2003cg,Banks:2006rx,Anninos:2011zn,Susskind:2021omt,Anninos:2021ihe,Shaghoulian:2021cef,Shaghoulian:2022fop,Banihashemi:2022htw}.} Notably, a worldtube  perspective appears as an important ingredient in recent literature incorporating methods of AdS/CFT to analyse the de Sitter static patch in two \cite{Anninos:2017hhn,Anninos:2018svg,Svesko:2022txo,Anninos:2022hqo} and three \cite{Coleman:2021nor,Shyam:2021ciy} dimensional models (see \cite{Galante:2023uyf} for a recent review). Moreover, the presence of a worldline plays an important role in the definition of a sensible von Neumann entropy for quantum fields weakly coupled to gravity in the static patch \cite{Chandrasekaran:2022cip}.

Given  a gravitational worldtube $\mathcal{M}$, a mathematical problem of interest is  one of well-posedness for the data on the timelike boundary $\Gamma$ of $\mathcal{M}$, where the spatial section of $\Gamma$ is taken to have finite size. In the general relativity literature, this is often referred to as an initial boundary value problem \cite{Sarbach:2012pr}. This is to be contrasted with the standard initial value problem of general relativity long established to be well-posed \cite{foures1952theoreme,Choquet-Bruhat:1969ywq}. To first approximation, one might imagine a standard Dirichlet-type boundary value problem whereby one specifies the induced metric $g_{m n}$ on $\Gamma$,  along with Cauchy data on a spacelike slice $\Sigma$ that intersects $\Gamma$. This problem has been explored for both Euclidean \cite{Anderson:2006lqb,Witten:2018lgb} and Lorentzian \cite{Friedrich:1998xt,Fournodavlos:2020wde,Fournodavlos:2021eye,An:2021fcq,Anninos:2022ujl} signature. Somewhat remarkably, and in contrast to the Klein-Gordon and Yang-Mills equations, the second order nature of the Einstein constraint equation significantly restricts the type of boundary data one can place on $\Gamma$. In particular, generic real valued Dirichlet data  do not satisfy the Einstein constraint equations at $\Gamma$ in either signature \cite{Anderson:2006lqb,An:2021fcq}. To make the discussion concrete, take the Lorentzian four-dimensional Einstein constraint equation projected along $\Gamma$ 
\begin{equation}
	\mathcal{R} - K^2 + K_{m n}K^{m n} \, |_\Gamma \= 0 \, ,
\end{equation}
where $\mathcal{R}$  denotes the Ricci scalar with respect to the metric $g_{m n}$ induced at $\Gamma$, and $K_{mn}$ denotes the second fundamental form or extrinsic curvature at $\Gamma$, with trace $K = g^{mn}K_{mn}$. If we imagine boundary data $g_{m n}$ that solves the above constraint, and vary slightly away by $\delta g_{m n}$, to linear order in $\delta g_{m n}$ one must satisfy
\begin{equation}\label{ec}
	\delta \mathcal{R} - 2 K \delta K + 2 K_{m n} \delta K^{m n} + 2 K^{m}{}_{r} K^{r n}\delta g_{m n}  \, |_\Gamma \= 0 \, .
\end{equation}
Upon taking into account tangential diffeomorphisms at $\Gamma$, the space of deformations $\delta g_{m n}$ at $\Gamma$ is captured by three independent functions. Therefore, the constraint (\ref{ec}) might only be satisfied for a subset of the full space of $\delta g_{m n}$. This reasoning has been proven for Euclidean signature \cite{Anderson:2006lqb}, as well as for a standard Minkowski corner with vanishing unperturbed $K_{mn}$ \cite{An:2021fcq}.  (In three spacetime dimensions, one has a single independent $\delta g_{m n}$ so the argument does not apply.) Similarly, one can argue that the Neumann boundary condition, whereby one instead fixes the second fundamental form $K_{m n}$ along $\Gamma$, does not have good existence properties. The presence of a non-vanishing cosmological constant $\Lambda$ does not affect the preceding reasoning, provided the boundary lies away from asymptotic boundaries. Further to the existence issues raised above, at least for certain choices of Dirichlet data on $\Gamma$ there is also a question of (non)-uniqueness that must be addressed. At least for flat boundaries there is an infinite class of physical diffeomorphisms that render the Dirichlet problem non-unique in Euclidean \cite{Anderson:2006lqb,Witten:2018lgb} and Lorentzian \cite{An:2021fcq,Anninos:2022ujl} signature.

It is useful to  contrast the generic absence of solutions to the Einstein constraint equation for a finite size boundary with the situation in a four-dimensional asymptotically AdS spacetime. There, the induced metric at the AdS boundary permits a finite neighbourhood of deformations near some reference value \cite{fefferman1985conformal,KICHENASSAMY2004268,Friedrich:1995vb}.  The reason one can more easily satisfy the constraint equation (\ref{ec}) is that near the AdS boundary one has the behaviour
\begin{equation} 
\frac{ds^2}{\ell^2} = d\rho^2 + \left( e^{2\rho} g^{(0)}_{m n} (x^m) + g^{(2)}_{m n}(x^m) +  e^{-\rho} g^{(3)}_{m n}(x^m) + \ldots \right) dx^m dx^n~,
\end{equation}
with $\rho \to \infty$, and $m$, $n$ range over the boundary coordinates $x^m$. As such, near the boundary $K_{m n} \approx e^{2\rho}  g^{(0)}_{m n} \ell$ is  parameterically large. To leading order at large $\rho$,  the cosmological extension of (\ref{ec}) is automatically satisfied for arbitrary $g_{m n}$. This can be viewed as a variant of the umbilic boundary conditions which are shown to be well-posed in \cite{Fournodavlos:2020wde, Fournodavlos:2021eye}. One can then continue solving the constraint order by order in a small $e^{-\rho}$ expansion which converges \cite{fefferman1985conformal,KICHENASSAMY2004268}. From the perspective of AdS/CFT, $g_{m n}$ constitutes a source for the dual CFT stress tensor, and is crucial to define boundary stress tensor correlation functions. More generally, from a CFT perspective one often considers coupling the CFT to a curved metric $g_{m n}$ (ideally of non-negative curvature) thereby turning on a small but finite source for the stress tensor. In Euclidean signature, the bulk solution with  prescribed boundary metric $g_{m n}$ is a real valued saddle of the Einstein equations with $\Lambda<0$. For instance, when $g_{m n}$ is the round $S^3$, we can fill the space with the Euclidean AdS$_4$ spacetime of curvature $-12/\ell^2$ and metric 
\begin{equation}
\frac{ds^2}{\ell^2} = d\rho^2 + \sinh^2 \rho \, d\Omega_3^2~,
\end{equation}
with $\rho \ge 0$. One can subsequently consider small positive curvature deviations away from the round $S^3$ and find real valued bulk solutions filling the interior. Similarly, in Lorentzian AdS one considers small but finite deformations of the boundary metric \cite{Friedrich:1995vb} leading to novel solutions in the interior. 

Perhaps all this is an indication that for boundaries of finite extent (particularly in the Euclidean case) one should consider complexified solutions, as often happens in a saddle point analysis. However, before delving into the complex plane, it is important to note that there is a set of better behaved `conformal' boundary conditions on a finite size $\Gamma$. In Euclidean signature  \cite{Anderson:2006lqb}, it has been shown that fixing the conformal class $\left[g_{m n}|_{\Gamma} \right]$ of the induced metric at the boundary and the trace $K$ of the second fundamental form $K_{m n}$ at $\Gamma$ leads to an elliptic problem (see also \cite{Figueras:2011va}). In Lorentzian signature,  it is conjectured \cite{An:2021fcq} that the same boundary conditions accompanied by standard Cauchy data on $\Sigma$ lead to a well-posed hyperbolic problem. The Lorentzian conformal boundary conditions have been shown to be well posed at the linearised level \cite{Anninos:2022ujl}. Moreover, for timelike surfaces parameterically near a black hole \cite{Bredberg:2011xw}  or cosmological  \cite{Anninos:2011zn} event horizon the conformal boundary conditions permit a rich solution space governed by a variant of the non-relativistic incompressible Navier-Stokes equation. 

This paper explores the conformal boundary conditions of \cite{Anderson:2006lqb,An:2021fcq} at the level of linearised general relativity in four spacetime dimensions with $\Lambda=0$. We consider Lorentzian boundaries of the type $\Gamma = \boldsymbol{\sigma} \times \mathbb{R}$ with $\boldsymbol{\sigma}$ a  spatial two-manifold and $\mathbb{R}$ denoting the time direction. As boundary data we append $\boldsymbol{\sigma}$ with an induced metric that is either flat or positively curved and consider a variety of extrinsic data $K$. The linearised problem is analysed through the  Kodama-Ishibashi formalism \cite{Kodama:2000fa,Kodama:2003jz}, adapted to manifolds with a boundary. The general setup is discussed in section \ref{sec:framework}. Our linearisation procedure is described in section \ref{lineargr}. Spatial boundaries of vanishing curvature with vanishing $K$ are analysed in section \ref{flatwall}, where it is shown that the perturbations are linearly stable. Spatial boundaries appended with  round two-sphere induced metric of size $\mathfrak{r}$ and $K = 2 \, \mathfrak{r}^{-1}$ are considered in \ref{sec:spherical}. It is shown that there exist linear perturbations that grow exponentially for generic initial Cauchy data, which is the conformal boundary condition counterpart of the instabilities discussed in \cite{Andrade:2015gja} for the Dirichlet problem. It is also shown that the non-uniqueness issues that appear in the flat case for the Dirichlet boundary conditions \cite{An:2021fcq,Anninos:2022ujl}  are alleviated for a spherical spatial boundary. In section \ref{sec: euclidean bh} we consider general relativity in Euclidean signature on a manifold with $S^2\times S^1$ boundary subject to conformal boundary conditions and generalise the considerations of York \cite{York:1986it} for the thermodynamics of a Schwarzschild black hole. We show that while entropy of the black hole retains the Bekenstein-Hawking form $\mathcal{S}_\text{BH} = A/4G_N$,  the specific heat and energy are modified as compared to the Dirichlet problem. Various technical details are provided in the appendices.

\section{General framework} \label{sec:framework}

\tikzset{every picture/.style={line width=0.75pt}} 

We consider vacuum solutions to general relativity in four spacetime dimensions and zero cosmological constant. In Lorentzian signature its action $I$ is given by
\begin{equation}\label{eqn: action}
	I \= \frac{1}{16 \pi G_N}\int_{\mathcal{M}} d^4 x \sqrt{-g}\, R + I_B \, ,
\end{equation}
where $G_N$ is the Newton's constant and $I_B$ is a boundary term that we will shortly specify and is needed for the variational principle to be well-defined. The equations of motion are the Einstein field equation
\begin{equation}
	R_{\mu \nu}-\frac{1}{2} g_{\mu \nu}R \= 0 \,, \label{ee}
\end{equation}
regardless of the choice of the boundary term. In what follows, we will use Greek indices, $\mu, \nu = 0,...,d$, for $(d+1)$-dimensional spacetime indices.

\begin{figure}[t]
	\centering

\tikzset{every picture/.style={line width=0.75pt}} 

\begin{tikzpicture}[x=0.75pt,y=0.75pt,yscale=-1,xscale=1]

\draw  [fill={rgb, 255:red, 208; green, 2; blue, 27 }  ,fill opacity=0.46 ] (271.2,260.8) .. controls (267.34,252.12) and (285.93,247.16) .. (291.75,248.25) .. controls (297.56,249.35) and (308.47,244.98) .. (335.93,258.25) .. controls (363.38,271.53) and (365.38,264.07) .. (383.38,267.53) .. controls (401.38,270.98) and (408.35,275.48) .. (406.4,285.2) .. controls (404.45,294.93) and (382.45,285.17) .. (340.95,283.67) .. controls (299.45,282.17) and (275.06,269.48) .. (271.2,260.8) -- cycle ;
\draw  [fill={rgb, 255:red, 65; green, 117; blue, 5 }  ,fill opacity=0.75 ] (271.2,260.8) .. controls (267.34,252.12) and (285.93,247.16) .. (291.75,248.25) .. controls (297.56,249.35) and (308.47,244.98) .. (335.93,258.25) .. controls (363.38,271.53) and (365.38,264.07) .. (383.38,267.53) .. controls (401.38,270.98) and (408.35,275.48) .. (406.4,285.2) .. controls (404.45,294.93) and (382.45,285.17) .. (340.95,283.67) .. controls (299.45,282.17) and (275.06,269.48) .. (271.2,260.8) -- cycle ;
\draw  [fill={rgb, 255:red, 151; green, 190; blue, 232 }  ,fill opacity=0.57 ] (268,80.82) .. controls (265.14,76.71) and (301.14,53.63) .. (333.14,54.43) .. controls (365.14,55.23) and (402.72,35.57) .. (405.11,37.69) .. controls (407.5,39.81) and (408.06,59.34) .. (380.86,71.29) .. controls (353.66,83.23) and (343.43,85.86) .. (310.57,90.43) .. controls (277.71,95) and (270.86,84.93) .. (268,80.82) -- cycle ;
\draw  [fill={rgb, 255:red, 232; green, 151; blue, 158 }  ,fill opacity=1 ] (272.4,78.8) .. controls (266.43,72.83) and (304,49.2) .. (336,50) .. controls (368,50.8) and (395.71,28.07) .. (402.4,40.4) .. controls (409.09,52.73) and (374.53,72.13) .. (331.42,80.13) .. controls (288.31,88.13) and (278.37,84.77) .. (272.4,78.8) -- cycle ;
\draw  [fill={rgb, 255:red, 215; green, 81; blue, 96 }  ,fill opacity=1 ] (402.6,40.6) .. controls (402.86,39.21) and (437,87.8) .. (416.6,128.6) .. controls (396.2,169.4) and (403.8,203.8) .. (415.8,227) .. controls (427.8,250.2) and (402.29,293.71) .. (406.6,285.4) .. controls (410.91,277.09) and (390.2,265.8) .. (370.2,267.4) .. controls (350.2,269) and (334.2,253.8) .. (313.4,250.2) .. controls (292.6,246.6) and (267.72,251.39) .. (271.4,261) .. controls (275.08,270.61) and (251,217.4) .. (275.8,171.4) .. controls (300.6,125.4) and (268.67,74.27) .. (272.6,79) .. controls (276.53,83.73) and (299.06,89.27) .. (349.4,76.2) .. controls (399.74,63.13) and (402.34,41.99) .. (402.6,40.6) -- cycle ;
\draw  [fill={rgb, 255:red, 74; green, 144; blue, 226 }  ,fill opacity=0.44 ] (405.11,37.69) .. controls (403.75,35.06) and (440.89,89.36) .. (418.67,133.8) .. controls (396.44,178.24) and (410.13,206.33) .. (420,230.24) .. controls (429.87,254.16) and (409.56,286.82) .. (407.56,290.69) .. controls (405.56,294.56) and (394,292.69) .. (388,292.47) .. controls (382,292.24) and (370,288.69) .. (362,288.24) .. controls (354,287.8) and (304.89,286.47) .. (287.56,278.47) .. controls (270.22,270.47) and (268.85,262.69) .. (269.33,264.24) .. controls (269.82,265.8) and (246.4,219.22) .. (271.2,173.22) .. controls (296,127.22) and (266.29,77.64) .. (268,80.82) .. controls (269.71,84) and (286.76,104.18) .. (349.56,81.8) .. controls (412.35,59.42) and (406.47,40.32) .. (405.11,37.69) -- cycle ;
\draw    (408.57,162.7) -- (427.65,167.08) ;
\draw [shift={(430.57,167.75)}, rotate = 192.93] [fill={rgb, 255:red, 0; green, 0; blue, 0 }  ][line width=0.08]  [draw opacity=0] (3.57,-1.72) -- (0,0) -- (3.57,1.72) -- cycle    ;

\draw (254.5,149.9) node [anchor=north west][inner sep=0.75pt]  [color={rgb, 255:red, 74; green, 144; blue, 226 }  ,opacity=1 ]  {$\Gamma $};
\draw (329.5,295.9) node [anchor=north west][inner sep=0.75pt]  [color={rgb, 255:red, 65; green, 117; blue, 5 }  ,opacity=1 ]  {$\Sigma $};
\draw (319.5,26.4) node [anchor=north west][inner sep=0.75pt]  [color={rgb, 255:red, 208; green, 2; blue, 27 }  ,opacity=1 ]  {$\mathcal{M}$};
\draw (433,160.6) node [anchor=north west][inner sep=0.75pt]    {$\hat{n}_\mu$};

\end{tikzpicture}

	\caption{Illustration of a four-manifold $\mathcal{M}$ with a timelike boundary $\Gamma$, a Cauchy surface $\Sigma$, and a unit normal vector field $\hat{n}_\mu$.}
	\label{tube}
\end{figure}
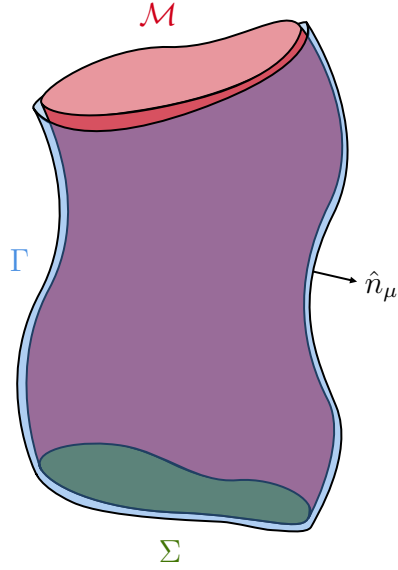

We are interested in finding solutions to (\ref{ee}), for a theory placed on a spacetime manifold $\mathcal{M}$ with a non-empty boundary $\partial \mathcal{M} \neq \emptyset$. The boundary is further composed of a spacelike boundary, that we will denote by $\Sigma$, and a timelike boundary, $\Gamma$. 
Latin indices $m,n,...$ denote  spacetime indices tangent to the timelike boundary. The general setup is depicted in figure \ref{tube}.

Following \cite{An:2021fcq}, we consider conformal boundary conditions on $\Gamma$. Concretely, we fix the conformal class of the induced metric and the trace of the extrinsic curvature on the timelike boundary
\begin{eqnarray}\label{eqn: conformal bdry data}
	\text{Conformal boundary conditions} &:&  \{ \left[g_{mn}|_{\Gamma} \right]_\text{conf}\, , \, K|_{\Gamma}  \} \quad = \quad \text{fixed} \,.
\end{eqnarray}
The above  boundary data is said to be geometric in the sense that it is constructed from quantities that are  defined on $\Gamma$ alone. The trace of the extrinsic curvature $K$ is defined as,
\begin{equation}\label{eqn: trace K}
	K \= g^{mn} K_{mn} \quad , \quad K_{mn} \= \frac{1}{2}\mathcal{L}_{\hat{n}} g_{mn} \, ,
\end{equation}
where $\hat{n} \= \hat{n}^\mu \partial_\mu$ is a unit normal vector associated to the boundary and pointing outward, and $\mathcal{L}_{\hat{n}}$ denotes a Lie derivative with respect to $\hat{n}^\mu$. We can write $K$ as a normal derivative of the boundary local volume, $K= \mathcal{L}_{\hat{n}} \log \sqrt{-\text{det}\, g_{mn} }$. The normal-normal and normal-tangential components of the Einstein field equation \eqref{ee}  projected onto $\Gamma$ are given by
\begin{equation}\label{eqn: constraints from einstein}
	\mathcal{R} - K^2 + K_{mn}K^{mn} \, |_\Gamma \= 0 \, , \qquad \qquad \mathcal{D}_m K^m{}_n - \mathcal{D}_n K \, |_\Gamma \= 0 \, ,
\end{equation}
where $\mathcal{R}$ and $\mathcal{D}_m$ denote the Ricci scalar and covariant derivative with respect to the induced metric $g_{mn}$. 
 
We emphasise that the projection of the Einstein equation on $\Gamma$, given in \eqref{eqn: constraints from einstein}, act as constraints for the boundary data of the gravitational dynamics. Note they are a non-linear combination of the induced metric $g_{mn}|_\Gamma$ and the extrinsic curvature $K_{mn}|_\Gamma$. As discussed in the introduction, specifying generic boundary data on $\Gamma$ will not satisfy these equations, causing the existence problem for the initial boundary value problem in general relativity \cite{An:2021fcq}.
 
Given the boundary conditions \eqref{eqn: conformal bdry data}, the boundary action is given by,
\begin{equation}\label{IB}
	I_B  = \frac{1}{24 \pi G_N} \int_\Gamma d^3 x \sqrt{-\text{det}\,g_{mn}}\, K \, ,
\end{equation}
which is a third of the standard Gibbons-Hawking-York term \cite{yorkbdy, Gibbons:1976ue} used for the Dirichlet problem. 
\section{Linearised gravity}\label{lineargr}

In this section, we consider perturbation about empty Minkowski spacetime, namely,
\begin{equation}
	g_{\mu \nu} \= \eta_{\mu \nu} + \epsilon \, h_{\mu \nu}  \, ,\qquad | \,\epsilon \,| \ll 1 \, .
\end{equation}
It will be convenient to write the background flat metric in Gaussian normal coordinates, that are adapted to the timelike boundary
\begin{equation}\label{eqn: choice of coordinate}
	\eta_{\mu \nu} dx^\mu dx^\nu \= dx_\perp^2 + \bar{g}_{mn} dx^m dx^n \, ,
\end{equation}
where the location of the timelike boundary $\Gamma$ is taken to be at a constant $x_\perp$, and the unit normal vector is $\hat{n} \=\partial_\perp$. As detailed in appendix \ref{app:normal}, many expressions simplify by using this metric. From now onwards, we will use this background metric to raise and lower indices. The corresponding boundary data is given by the conformal class of the three-dimensional metric $\left[g_{mn}|_{\Gamma} \right]_\text{conf}$ and the trace of the extrinsic curvature at the boundary $K|_{\Gamma}$.

\subsection{Linearised  boundary conditions}
Demanding that the conformal boundary data remains invariant under any perturbation implies that
\begin{eqnarray}
	h_{mn}|_\Gamma &=& \gamma(x) \bar{g}_{mn} |_\Gamma \,, \label{trace}\\
\epsilon \, \delta K(h_{\mu\nu})  & \equiv &	\left[ K(\eta_{\mu\nu}+\epsilon \, h_{\mu\nu})-K(\eta_{\mu\nu}) \right] |_{\Gamma}   = 0 \, ,
\end{eqnarray}
where $\gamma(x)$ is an arbitrary function, that will depend on the initial data of the linearised metric $h_{\mu \nu}$. By contracting (\ref{trace}) with $\bar{g}^{mn}$, one may rewrite the first condition as
\begin{equation}\label{eqn: bdry cond 1}
	h_{m n} - \frac{1}{3} \bar{g}_{m n } h^p{}_p  \, |_{\Gamma}  \= 0 \, .
\end{equation}
It is more convenient to use this expression instead of (\ref{trace}), as in this one, $\gamma(x)$ does not appear explicitly. Using \eqref{eqn: trace K}, it can be shown that the variation of the trace of the extrinsic curvature at a linearised level becomes
\begin{equation}
	2\delta K (h_{\mu\nu}) \= \partial_\perp h^m{}_m - 2 \mathcal{D}^m h_{\perp m} - Kh_{\perp \perp}  \, |_{\Gamma} = 0  \, ,\label{eqn: bdry cond 2}
\end{equation}
where $\mathcal{D}_n$ denotes the covariant derivative with respect to the boundary metric $\bar{g}_{mn}$. (Further details can be found in appendix \ref{app:normal}.) In our analysis we will take \eqref{eqn: bdry cond 1} and \eqref{eqn: bdry cond 2} as the conformal boundary conditions for linearised gravity.
 
We must also  declare boundary conditions for the space of allowed diffeomorphisms. At the linearised level, they act on the coordinates and metric respectively as
\begin{equation}
	x^\mu \rightarrow x^\mu -\epsilon \, \xi^\mu \, , \qquad\qquad
	h_{\mu \nu} \rightarrow h_{\mu \nu} + \nabla_\mu \xi_\nu + \nabla_\nu \xi_\mu \,,
\end{equation}
for any smooth vector field $\xi_\mu$. In the presence of a timelike boundary, one should require that diffeomorphisms leave the boundary data unchanged. This leads to boundary conditions for $\xi_\mu$, that are given by
\begin{eqnarray}\label{eqn: diffeo bdry cond 1}
	\left. \left(K_{mn} - \frac{K}{3}\bar{g}_{mn}\right)\xi_{\perp}  + \left(\mathcal{D}_{(m} \xi_{n)} - \frac{\bar{g}_{mn}}{3} \mathcal{D}^p \xi_p \right)\, \right|_{\Gamma} &=& 0 \,,  \\
	\left. \left(\partial_\perp K - \mathcal{D}^m \mathcal{D}_m \right)\xi_\perp + \xi_m \mathcal{D}^m K \, \right|_{\Gamma} &=& 0 \ \, .\label{eqn: diffeo bdry cond 2} 
\end{eqnarray}
Moreover, since we picked coordinates so that the timelike boundary $\Gamma$ is located at constant $x_\perp$, we need an additional restriction so that $\xi^\mu$ does not move $\Gamma$. Then, we will further impose that
\begin{equation}
	\xi^\perp |_\Gamma \= 0 \, . \label{eqn: diffeo bdry cond perp}
\end{equation}
The full set of boundary conditions on allowed diffeomorphisms is then given by \eqref{eqn: diffeo bdry cond 1}, \eqref{eqn: diffeo bdry cond 2}, and \eqref{eqn: diffeo bdry cond perp}. For generic background spacetimes $g_{\mu \nu}$, these reduce to setting $\xi_\mu |_\Gamma = 0$. For special sets of boundary data, the boundary conditions can allow certain non-trivial diffeomorphisms to act non-trivially on the boundary. These are precisely a set of conformal Killing vectors of the timelike boundary $\Gamma$ preserving the trace of the extrinsic curvature $K$. 
 
So far, we computed conformal boundary conditions on the timelike boundary for linearised gravity. To study different geometric configurations, it will be convenient to work in the harmonic gauge, which we discuss next. {In section \ref{sec:spherical} we will consider the gauge-invariant Kodama-Ishibashi formalism \cite{Kodama:2000fa,Kodama:2003jz}.}

\subsection{Harmonic gauge}

The advantage of working in the harmonic (or de Donder) gauge is that the linearised Einstein equation become wave equations, 
\begin{equation}\label{eqn: harmonic EOM}
-\nabla^\rho \nabla_\rho h_{\mu \nu}  \=0 \, .
\end{equation}
In the case of a non-vanishing cosmological constant, there will be an extra mass term proportional to it. At the linearised level, the harmonic gauge condition is given by imposing
\begin{equation}\label{eqn: harmonic gauge cond}
	T_\nu(h_{\mu\nu}) \,\equiv\,  \nabla^\mu h_{\mu \nu} - \frac{1}{2}\nabla_\nu h^\mu{}_\mu  \= 0 \, .
\end{equation}
These gauge conditions\footnote{For an arbitrary gravitational perturbation $\hat{h}_{\mu\nu}$, not necessarily in the harmonic gauge, the conditions \eqref{eqn: harmonic gauge cond} can be obtained by making a suitable gauge transformation,
\begin{equation}
	\hat{h}_{\mu \nu}  \rightarrow \hat{h}_{\mu \nu} + \nabla_\mu \hat{\xi}_\nu + \nabla_\nu \hat{\xi}_\mu = h_{\mu \nu} \, ,
\end{equation}
with the vector field $\hat{\xi}_\mu$ solving the inhomogeneous wave equation
\begin{equation}
	\nabla^\rho \nabla_\rho \hat{\xi}_\mu = - T_\mu (\hat{h}_{\mu\nu}) \, ,
\end{equation}
and obeying the boundary conditions \eqref{eqn: diffeo bdry cond 1}, \eqref{eqn: diffeo bdry cond 2}, and \eqref{eqn: diffeo bdry cond perp}.
}  
can be expressed in components tangent and normal to the timelike boundary as
\begin{eqnarray}\label{eqn: harmonic gauge cond 1}
	T_n(h_{\mu\nu}) & \equiv & \mathcal{D}^m h_{mn} - \frac{1}{2} \mathcal{D}_nh^m{}_m + \left(K+ \partial_\perp \right)h_{\perp n}-\frac{1}{2}\mathcal{D}_n h_{\perp \perp} = 0 \, ,\\ 
	T_\perp(h_{\mu\nu}) & \equiv & -K^{mn}h_{mn} - \frac{1}{2}\partial_{\perp} h^m{}_m + \mathcal{D}^m h_{\perp m} + K h_{\perp \perp} + \frac{1}{2}\partial_\perp h_{\perp\perp} = 0 \, . \label{eqn: harmonic gauge cond 2}
\end{eqnarray}
Since the gauge conditions must be satisfied at every point in the manifold $\mathcal{M}$ including the boundary $\partial \mathcal{M} = \Gamma \cup \Sigma$, the above equations can be regarded as providing additional boundary conditions on the metric perturbation,   $T_n(h_{\mu\nu}) |_{\Gamma} = T_\perp (h_{\mu\nu}) |_{\Gamma}=0$. So, in total, we obtain a set of four boundary conditions for the conformal boundary problem in the harmonic gauge. These are given explicitly by (\ref{eqn: bdry cond 1}) and (\ref{eqn: bdry cond 2}), as well as the two gauge conditions \eqref{eqn: harmonic gauge cond 1} and \eqref{eqn: harmonic gauge cond 2}. All in all, we have the following boundary conditions
\begin{equation}
\begin{cases}
h_{mn}-\frac{1}{3} \bar{g}_{mn}h^p{}_p \, |_{\Gamma} \= 0 \, ,\\ 
	\frac{2}{3}K h^m{}_m - \left(K+\partial_\perp\right)h_{\perp \perp} \, |_{\Gamma} \= 0 \,, \\
 	-\frac{1}{6}\mathcal{D}_n h^m{}_m + \left(K+\partial_\perp \right)h_{\perp n}-\frac{1}{2}\mathcal{D}_n h_{\perp \perp} |_{\Gamma} \=  0 \,, \\
	-\left(\frac{1}{3}K + \frac{1}{2} \partial_\perp \right) h^m{}_m + \mathcal{D}^m h_{\perp m}  + \left( K  + \frac{1}{2}\partial_\perp \right) h_{\perp \perp}  |_{\Gamma} \=0 \,.
\end{cases}
\label{eqn: harmonic bdry cond 3}
\end{equation}
We will now proceed to analyse properties of linearised gravity for a variety of boundaries subject to the above conditions.

\section{Flat boundaries}\label{flatwall}

In this section, we study the linearised gravitational problem which preserves conformal boundary data on a flat timelike boundary. More precisely, we set the timelike boundary to be locally three-dimensional Minkowski spacetime with $K=0$ and then study the behaviour of gravitational fluctuations which become conformally flat and have $\delta K(h_{\mu\nu})=0$ at the boundary. We work in the harmonic gauge (\ref{eqn: harmonic gauge cond}) in this section.
 
We consider two types of flat boundaries: a non-compact and a compact version. The former problem is equivalent to studying gravitational waves on the half-plane with conformal boundary conditions; 
in the latter case, spacetime is confined to a finite box in all spatial dimensions.\footnote{One could also consider a similar treatment for a toroidal boundary. However,  at a technical level the simplest such setup would modify the asymptotic structure of spatial infinity.}

\subsection{Non-compact}

For flat boundaries, it is convenient to use Cartesian coordinates to describe the background metric,
\begin{equation}
	ds^2 \= -dt^2 + dx^2 +dy^2 + dz^2 \, , \qquad x\geq 0 \,, \qquad t,y,z \in \mathbb{R} \,.
\end{equation}
In this setup, we consider $x_\perp = -x$, the timelike boundary $\Gamma$ to be located at $x\= 0$, and $\bar{g}_{mn}=\eta_{mn}$,. The unit normal vector associated to the timelike boundary is $\hat{n} \= -\partial_x$, namely it is pointing in an outward direction. 
 
The induced metric and the extrinsic curvature at the timelike boundary are given by
\begin{equation}\label{eqn: flat wall}
	ds^2 |_{x=0} \= -dt^2  +dy^2 + dz^2 \, , \qquad K_{mn} \= 0 \, .
\end{equation}
In the harmonic gauge, the linearised Einstein field equation is simply given by
\begin{equation}\label{eqn: eom harmonic}
	\left(-\partial_t^2 + \partial_x^2 + \partial_y^2 + \partial_z^2 \right) h_{\mu \nu}(t,\vec{x})  \= 0 \, ,
\end{equation}
where $\vec{x} \equiv (x,y,z)$. To solve this, we use the following ansatz,
\begin{equation}
	h_{\mu \nu}(t,\vec{x}) \= f_{\mu \nu}(x) e^{i k_m x^m} \, , \qquad i k_m x^m =  -i\omega t + i k_y y + i k_z z \, , \qquad k_m k^m = -k_x^2 \leq 0 \,,
\end{equation}
where we separated the $x$-dependence of $h_{\mu\nu}(t, \vec{x})$ into the function $f_{\mu \nu}(x)$. Substituting this ansatz back into (\ref{eqn: eom harmonic}), and solving the differential equation for $f_{\mu\nu}(x)$, we obtain a general metric perturbation for a given value of $\omega$, $k_y$, and $k_z$,
\begin{equation}\label{eqn: flat non-compact sol 1}
	h_{\mu \nu}(t,\vec{x}) \= 
	\begin{cases}
	 \alpha_{\mu \nu} \cos(k_x x) e^{i k_m x^m} + \beta_{\mu \nu} \sin(k_x x) e^{i k_m x^m} \, & ,  \quad  \text{if} \, \, k_nk^n \= - k_x^2 \neq 0 \,,  \\
	 \tilde{\alpha}_{\mu \nu} \, e^{i k_m x^m}+ \tilde{\beta}_{\mu \nu}\, x \,e^{i k_m x^m}\,  \, &, \quad  \text{if} \, \, k_nk^n \= - k_x^2 = 0 \,,
	 \end{cases}
\end{equation}
where $\alpha_{\mu\nu}$, $\tilde{\alpha}_{\mu\nu}$, $\beta_{\mu\nu}$, $\tilde{\beta}_{\mu\nu}$ are $k_m$-dependent functions to be determined later, and it is understood that the real part of $h_{\mu \nu}(t,\vec{x})$ must be taken. We note that the solution in the second line exhibits a behaviour linear in $x$ that might seem problematic in the $x\to \infty$ limit. However, this will not generate physical solutions with a growing mode, as shown in  section \ref{sec:kx0}. 
 
Next, we consider the conformal boundary conditions \eqref{eqn: bdry cond 1} and \eqref{eqn: bdry cond 2} in the harmonic gauge. For a flat boundary they simply become, 
\begin{eqnarray}\label{eqn: flat non-compact bdry cond 1}
	h_{m n} - \frac{1}{3} \eta_{m n} h^p{}_p \, |_{x=0} \= 0  \,, \qquad 
	\partial_x h_{x x}\, |_{x=0} \=  0\, .
\end{eqnarray}
Notice that the above equations imply that we are specifying Dirichlet boundary conditions for the non-diagonal components of $h_{mn}$ and Neumann boundary conditions on $h_{xx}$.
 
Imposing the harmonic gauge conditions further require
\begin{eqnarray}
	T_n (h_{\mu\nu}) & = & \partial^m h_{mn} - \frac{1}{2}\partial_n h^m{}_m + \partial_x h_{x n} - \frac{1}{2}\partial_n h_{x x} \= 0 \, , \label{eqn: gauge cond harmonic 1} \\
	T_x (h_{\mu\nu})  & = & -\frac{1}{2}\partial_x h^m{}_m + \partial^m h_{x m} + \frac{1}{2}\partial_x h_{x x} \= 0 \, . \label{eqn: gauge cond harmonic 2}
\end{eqnarray}
Finally, residual diffeomorphisms are described by vector fields $\xi_\mu$, which satisfy the wave equation
\begin{equation}\label{eqn: flat non-compact diff eom}
	\left(-\partial_t^2 + \partial_x^2 + \partial_y^2 + \partial_z^2 \right)\xi_\mu \= 0 \, .
\end{equation}
Analogously to the metric perturbation, solutions to this equation become,
\begin{eqnarray} \label{eqn: flat non-compact residual gauge 1} 
\xi_{\mu}(t,\vec{x}) \= 
	\begin{cases}
	 \alpha_{\mu} \cos(k_x x) e^{i k_m x^m} + \beta_{\mu} \sin(k_x x) e^{i k_m x^m} \, & ,  \quad  \text{if} \, \, k_m k^m \= - k_x^2 \neq 0 \,,  \\
	 \tilde{\alpha}_{\mu} \, e^{i k_m x^m}+ \tilde{\beta}_{\mu}\, x \,e^{i k_m x^m}\,  \, &, \quad  \text{if} \, \, k_m k^m \= - k_x^2 = 0 \,,
	 \end{cases}
\end{eqnarray}
where now the $\alpha_\mu, \tilde{\alpha}_\mu, \beta_\mu, \tilde{\beta}_\mu$ are the $k_m$-dependent functions to be determined later.
The boundary conditions for linearised diffeomorphisms \eqref{eqn: diffeo bdry cond 1} and \eqref{eqn: diffeo bdry cond 2} together with \eqref{eqn: diffeo bdry cond perp} now become
\begin{eqnarray}\label{eqn: flat non-compact diffeo bdry cond 1}
	\frac{1}{2}\left( \partial_n \xi_m + \partial_m  \xi_n \right) - \frac{1}{3}\eta_{nm} \partial^p \xi_p |_{x=0}  =  0 \,, \qquad
	\xi^x |_{x=0} =  0 \, .
\end{eqnarray}
Collecting all of these together, we can find physical  solutions to the gravitational perturbation problem with conformal boundary conditions in the presence of flat boundaries. For this, the strategy can be summarised as follows: (i) we start with the solutions \eqref{eqn: flat non-compact sol 1} (ii) we impose boundary conditions \eqref{eqn: flat non-compact bdry cond 1}, which will determine some of the coefficients of $\alpha_{\mu\nu}$, $\beta_{\mu\nu}$ (iii) we  impose the gauge constraints \eqref{eqn: gauge cond harmonic 1}, \eqref{eqn: gauge cond harmonic 2}, to further determine the remaining functions (iv) we  use residual diffeomorphisms \eqref{eqn: flat non-compact residual gauge 1} with boundary conditions \eqref{eqn: flat non-compact diffeo bdry cond 1} to fully determine the solution. What remain are  physical gravitational degrees of freedom   preserving conformal boundary data. Consistent initial data is supposed to fix the solution uniquely. We will repeat this procedure both for the $k_x\neq0$ and the $k_x \=0$ case.

 \subsubsection{$k_x\neq0$ solutions} \label{sec:kx!=0}
Recall that the metric perturbation \eqref{eqn: flat non-compact sol 1} is given by
\begin{equation}
	h_{\mu \nu}(t,\vec{x}) \= \alpha_{\mu \nu} \cos(k_x x) e^{i k_m x^m} + \beta_{\mu \nu} \sin(k_x x) e^{i k_m x^m} \, ,
\end{equation}
where it is understood that we are taking the real part. Using the above solution and the fact that $k_x \neq 0$, the boundary conditions \eqref{eqn: flat non-compact bdry cond 1} become
\begin{eqnarray}
	\alpha_{m n} - \frac{1}{3}\eta_{mn} \alpha^{p}{}_p = 0 \,, \qquad 
	\beta_{x x} = 0\, . \label{bdy_cond222}
\end{eqnarray}
Next, we need to impose the gauge constraints. Using again that $k_n k^n = -k_x^2 \neq 0$, we can define two orthogonal three-dimensional vectors that are also orthogonal to $k^n$ and have unit norm. Let us call those vectors $u^{(1)n}$ and  $u^{(2)n}$, so that
\begin{equation}
	u^{(1)m} k_m \= u^{(2)m} k_m  \= u^{(1)m} u^{(2)}_m \= 0 \, , \qquad \qquad u^{(1)m} u^{(1)}_m \= u^{(2)m} u^{(2)}_m \= 1 \,. 
\end{equation}
Now, we can first project the gauge condition $T_m(h)=0$ onto the $k^m$, $u^{(1)m}$, and $u^{(2)m}$ directions to obtain
\begin{eqnarray}\label{eqn: dummy 1}
	ik^n k^m h_{mn} + k^n \partial_x h_{x n} + i \frac{ k_x^2}{2} \left(h^m{}_m + h_{x x}\right) = 0 \,, \qquad 
	u^{(1,2)m} \left( ik^n  h_{mn} + \partial_x h_{x m} \right)= 0 \,. 
\end{eqnarray}
The other gauge condition, $T_x (h)=0$, gets simplified by acting with $\partial_x$ and imposing the equation of motion $\partial_x^2 h_{\mu \nu} \= -k_x^2 h_{\mu \nu}$, so that
\begin{eqnarray}
	k^n \partial_x h_{n x} + \frac{ik_x^2}{2} \left(- h^m{}_m +h_{x x}\right) &=& 0 \,. \label{eqn: dummy 3}
\end{eqnarray}
Inserting \eqref{eqn: flat non-compact sol 1} into \eqref{eqn: dummy 1} and \eqref{eqn: dummy 3}, and combining with the boundary conditions \eqref{bdy_cond222}, we obtain the following set of restrictions on the coefficients $\alpha_{\mu\nu}, \beta_{\mu\nu}$,
\begin{equation} \label{eqn: gauge cond kx!=0}
\begin{dcases}
               \alpha_{mn} \= 0 \,, & \left( k^m k^n + k_x^2 \eta^{mn} \right)\beta_{mn} \= 0 \,, \\
               2ik^n \beta_{x n} - k_x \alpha_{x x} = 0  \,, &  k_x  \beta^m{}_m -2ik^n \alpha_{x n} \= 0 \,, \\
               u^{(1)m} k_x \beta_{x m}  \= 0 \,,          &   i u^{(1)m} k^n \beta_{mn} - u^{(1)m} k_x \alpha_{x m}  \= 0 \,, \\
               u^{(2)m} k_x \beta_{x m}  \= 0 \,,          &   i u^{(2)m} k^n \beta_{mn} - u^{(2)m} k_x \alpha_{x m}  \= 0 \,.
        \end{dcases}
\end{equation}
Notice that the first line implies that the linearised induced metric becomes identically zero, just as if we were imposing Dirichlet boundary conditions. This is merely a coincidence for the flat boundary case. In general, it needs not to be zero, and a generic gravitational fluctuation can lead to a non-zero Weyl factor on the boundary, see, for instance, section \ref{sec:kx0}. Moreover, while this problem is well-posed, it is known that imposing Dirichlet boundary conditions on the flat boundary does violate the geometric uniqueness of the linearised solution \cite{An:2021fcq, Anninos:2022ujl}, see also section \ref{sec:dirichlet}.
 
Finally, we can use residual diffeomorphisms to completely gauge-fix the solution. Recall from \eqref{eqn: flat non-compact residual gauge 1} that we have
\begin{equation}
	\xi_\mu \= \alpha_\mu \cos(k_x x)e^{i k_n x^n} + \beta_{\mu} \sin(k_x x)e^{i k_n x^n} \, .
\end{equation}
The boundary condition for the diffeomorphisms fixes
\begin{eqnarray}
	k_n \alpha_m + k_m \alpha_n - \frac{2}{3} \eta_{n m} k^p \alpha_p &=& 0 \,, \qquad 
	\alpha_x = 0 \, .
\end{eqnarray}
For $k_n \neq 0$ (and $k_nk^n \neq 0$), the first equation implies that $\alpha_n =0$. Hence the allowed diffeomorphisms are the ones which act trivially on the timelike boundary at $x=0$. Using these residual diffeomorphisms, the remaining components of the metric perturbation transform as,
\begin{eqnarray}
\begin{cases}
	\beta_{m n} & \rightarrow \quad \beta_{mn} + i k_m \beta_n + i k_n \beta_m \,, \\
	\beta_{x n} & \rightarrow \quad \beta_{x n}  + i k_n \beta_x \,,  \\
	\alpha_{x n} & \rightarrow \quad \alpha_{x n}  + k_x  \beta_n \,,  \\
	\alpha_{x x} & \rightarrow \quad \alpha_{x x} + 2 k_x \beta_x \, .
	\end{cases}
\end{eqnarray}
We can use the last two lines to choose $\beta_\mu$ so as to fix the gauge in which $ \alpha_{x \mu}\, \= 0$. As a result, the gauge-fixed metric perturbation can be written in a compact form as 
\begin{eqnarray}\label{eqn: gauge-fixed sol kx!=0}
	h_{\mu \nu}dx^\mu dx^\nu = \left[\beta^{(+)} \left(u^{(1)}_n u^{(1)}_m - u^{(2)}_n u^{(2)}_m\right) + 2 \beta^{(\times)} u^{(1)}_n u^{(2)}_m \right]  \sin(k_x x) e^{i k_n x^n} dx^n dx^m \,,
\end{eqnarray}
where $\beta^{(+)}$ and $\beta^{(\times)}$ are built from the remaining $\beta_{mn}$ and represent the two polarisations of the gravitational fluctuation that preserve the conformal boundary data at $x=0$. We provide more details in appendix \ref{app:derivation}. Note that both gravitational modes $\beta^{(+)}$ and $\beta^{(\times)}$ are odd under $x\rightarrow -x$. Since there is no notion of a region outside the boundary with $x<0$, this property should be thought of as a mathematical feature rather than a physical symmetry of the solution.
 
\noindent \textbf{Example.} Consider the solution with $k_y\=k_z\=0$, or equivalently,
\begin{equation}
	k^n \partial_n \= \omega \partial_t \, ,
\end{equation}
where $\omega = |k_x|$. The vectors $u^{(1)n}$ and $u^{(2)n}$ can be chosen as
\begin{equation}
	u^{(1)n} \partial_n \= \partial_y \, ,\qquad \qquad u^{(2)n} \partial_n \= \partial_z \,.
\end{equation}
Then, the general metric perturbation obeying the conformal boundary conditions in this case is simply given by
\begin{equation}
	h_{\mu \nu}dx^\mu dx^\nu = \left[ \beta^{(+)}  \left( dy^2-dz^2 \right)  + 2 \beta^{(\times)} dydz\right] \sin(k_x x) e^{i k_n x^n}\, .
\end{equation}

 \subsubsection{$k_x=0$ solutions}\label{sec:kx0}
Now consider the second solution in \eqref{eqn: flat non-compact sol 1}, for which $k_x = 0$,
\begin{equation}
	h_{\mu \nu}(t,\vec{x}) \= \tilde{\alpha}_{\mu \nu} e^{i k_m x^m}+ \tilde{\beta}_{\mu \nu}\, x \,e^{i k_m x^m}\, . \label{kx0}
\end{equation}
In this case, the vector $k_n$ becomes null with respect to the boundary metric. Let $q^n$ and $u^n$ be a null and a spacelike vector, respectively, which satisfy the following properties,
\begin{equation}
	q^n k_n \= 1 \, , \qquad \qquad q^n u_n  \= k^n u_n  \= 0 \,.
\end{equation}
It is convenient to project the gauge condition $T_\mu (h_{\mu\nu})=0$ onto $k^n$, $q^n$, $u^n$ and the normal vector. Inserting the solution \eqref{kx0}, the gauge conditions become,
\begin{equation}
\begin{dcases}
             i k^m k^n \tilde{\alpha}_{mn} + k^n \tilde{\beta}_{x n} = 0\, , & k^m k^n \tilde{\beta}_{mn} \= 0 \,,  \\
	- i u^m u^n \tilde{\alpha}_{mn} + 2 q^n \tilde{\beta}_{x n} - i\alpha_{x x} = 0\,, &  2u^m u^n \tilde{\beta}_{mn}-\tilde{\beta}_{x x} \=0 \,, \\
	i k^m u^n \tilde{\alpha}_{mn} + u^n \tilde{\beta}_{x n} = 0 \,, &  k^m u^n \tilde{\beta}_{mn}\=0 \,, \\
	k^m \tilde{\alpha}_{x m} + \frac{i}{2} \left(\tilde{\beta}^m{}_m - \tilde{\beta}_{x x}\right) = 0 \,, &  k^n \tilde{\beta}_{x n}\=0 \,.
        \end{dcases}
\end{equation}
The conformal boundary conditions take a similar form to the ones in the previous subsection,
\begin{eqnarray}
	\tilde{\alpha}_{m n} - \frac{1}{3}\eta_{mn} \tilde{\alpha}^p{}_p \= 0 \, , \qquad
	\tilde{\beta}_{xx} \= 0 \, .
\end{eqnarray}
To solve the remaining gauge condition, let us look at the residual diffeomorphisms in \eqref{eqn: flat non-compact residual gauge 1},
\begin{equation}
	\xi_\mu \= \tilde{\alpha}_\mu e^{i k_n x^n} + \tilde{\beta}_{\mu} \, x \, e^{i k_n x^n} \, .
\end{equation}
Similar to the $k_x\neq 0$ case, the boundary conditions for the diffeomorphisms impose that
\begin{equation}
	k_m \tilde{\alpha}_n + k_n \tilde{\alpha}_m -\frac{2}{3}\eta_{nm} k^p \tilde{\alpha}_p \= 0 \, , \qquad \tilde{\alpha}_x \= 0 \,,
\end{equation}
which lead to $\tilde{\alpha}_\mu \= 0$. The transformation of the metric perturbation under the allowed diffeomorphisms is given by
\begin{eqnarray}
\begin{cases}
	\tilde{\alpha}_{x n} &\rightarrow \quad \tilde{\alpha}_{x n}+ \tilde{\beta}_n  \,, \nonumber\\
	\tilde{\alpha}_{x x} &\rightarrow \quad \tilde{\alpha}_{x x}+2 \tilde{\beta}_x \,, \nonumber\\
	\tilde{\beta}_{m n} &\rightarrow \quad \tilde{\beta}_{m n}+i k_m \tilde{\beta}_n + i k_n \tilde{\beta}_m  \,, \nonumber\\
	\tilde{\beta}_{x n} &\rightarrow \quad \tilde{\beta}_{x n}+  i k_n \tilde{\beta}_x \, .
	\end{cases}
\end{eqnarray}
We can use the residual diffeomorphisms to choose $\tilde{\beta}_\mu$ such that $q^m \tilde{\beta}_{m \mu} \= 0$. As a result, the general gauge-fixed metric perturbation becomes
\begin{equation}
	h_{\mu \nu}dx^\mu dx^\nu \= \tilde{\alpha}^{(+)} e^{ik_nx^n} \left(\eta_{nm}dx^n dx^m - dx^2\right) + 2 \left(\tilde{\alpha}^{(\times)} u_n + i\tilde{\alpha}^{(\perp)} k_n\right) e^{ik_n x^n}  dx^n dx \, ,
\end{equation}
where $\tilde{\alpha}^{(+)}$, $\tilde{\alpha}^{(\times)}$, and $\tilde{\alpha}^{(\perp)}$ are remaining gravitational degrees of freedom preserving the conformal boundary data at $x\=0$. Interestingly, the modes $\tilde{\alpha}^{(\times)}$ and $\tilde{\alpha}^{(\perp)}$ are odd under parity $x\rightarrow -x$ while $\tilde{\alpha}^{(+)}$ is even. In particular, the even mode becomes non-zero when evaluated at the boundary,
\begin{equation}
	h_{\mu \nu} dx^\mu dx^\nu |_{x=0} \= \tilde{\alpha}^{(+)}  \left(-dt^2 + dy^2 + dz^2\right) e^{ik_nx^n} \, ,
\end{equation}
serving as an example of a fluctuating Weyl factor that cannot be eliminated by a coordinate transformation.
 
The modes $\tilde{\alpha}^{(+)}$ and $\tilde{\alpha}^{(\times)}$ can be interpreted as two polarisations of the gravitational perturbation as they give rise to a non-vanishing Riemann tensor. On the other hand, the extra mode $\tilde{\alpha}^{(\perp)}$ leads to a vanishing Riemann tensor. In particular, it can be written locally as
\begin{equation}
	h_{\mu\nu} \= \partial_\mu \xi_\nu + \partial_\nu \xi_\mu \, , \qquad \qquad \xi_\mu \= \eta_{\mu x} \tilde{\alpha}^{(\perp)} e^{ik_n x^n} \, .
\end{equation}
Note, however, that $\xi_\mu$ is a diffeomorphism that preserves the conformal boundary data \eqref{eqn: diffeo bdry cond 1} and \eqref{eqn: diffeo bdry cond 2} but not the location of the timelike boundary, \eqref{eqn: diffeo bdry cond perp}. Consequently, it cannot be gauged away and should be treated as physical. 
 
The existence of the physical diffeomorphism mode $\tilde{\alpha}^{(\perp)}$ features a striking difference to the gravitational dynamics without boundary. In fact, by relaxing the harmonic gauge choice, one can make an arbitrary gauge transformation away from the timelike boundary, $x=0$, so that the physical diffeomorphism mode $\tilde{\alpha}^{(\perp)}$ is localised arbitrarily close to the boundary. Hence, one may call it a boundary mode.
 
\noindent \textbf{Example.} Consider the particular solution with $k_z=0$ and $k_y=\omega$. The vectors $k^n$, $q^n$, and $u^n$ are then given by
\begin{eqnarray}
	k^n \partial_n = \omega\left(\partial_t + \partial_y \right) \,, \qquad
	q^n \partial_n = \frac{1}{2 \omega}\left(-\partial_t + \partial_y \right) \,, \qquad
	u^n \partial_n = \partial_z \, .
\end{eqnarray}
The corresponding metric perturbation now reads
\begin{equation}
	h_{\mu \nu}dx^\mu dx^\nu \= \left[\tilde{\alpha}^{(+)}  \left(-dt^2+dy^2+dz^2 - dx^2\right) + 2\tilde{\alpha}^{(\times)}  dx dz + 2i\tilde{\alpha}^{(\perp)} \omega \left(-dt dx+dy dx\right)\right] e^{-i\omega(t-y)}\, .
\end{equation}

\subsection{Modifying $\delta K$}

In this section, we consider slightly changing the boundary conditions to include a non-vanishing (but small) variation of the trace of the extrinsic curvature at the timelike boundary. Consequently, the boundary conditions \eqref{eqn: flat non-compact bdry cond 1} now become,
\begin{eqnarray}\label{eqn: flat non-compact bdry cond deltaK}
	h_{m n} - \frac{1}{3} \eta_{m n} h^p{}_p \, |_{x=0} \= 0  \,, \qquad 
	\partial_x h_{x x}\, |_{x=0} \=  2 \, \delta K(t,y,z)\, ,
\end{eqnarray}
where $\delta K(t,y,z)$ is an external function that will act as the new conformal boundary data. We will now solve this problem using the solutions we found for the vanishing $\delta K$ case.
 
Consider a metric perturbation that consists of the sum of two terms
\begin{equation}\label{eqn: h_munu with delta K}
	h_{\mu \nu} \= \bar{h}_{\mu \nu} + \partial_\mu \xi_\nu + \partial_\nu \xi_\mu \,.
\end{equation}
The condition that $h_{\mu\nu}$ is in the harmonic gauge implies that $\bar{h}_{\mu\nu}$ satisfies \eqref{eqn: eom harmonic} and $T_\mu (\bar{h}_{\mu\nu})=0$, and $\xi_\mu$ satisfies \eqref{eqn: flat non-compact diff eom}. 
 
Next, we impose that, at the boundary, $\bar{h}_{\mu\nu}$ obeys the boundary conditions with $\delta K=0$, namely boundary conditions \eqref{eqn: flat non-compact bdry cond 1}. Instead, $\xi_\mu$ will obey the following boundary conditions,
\begin{equation}\label{eqn: diffeo remove delta K}
	\xi_m |_{x=0} \= 0 \, , \qquad \qquad \xi_x (t,y,z) |_{x=0} \= v(t,y,z)\, ,
\end{equation}
where the scalar function $v(t,y,z)$ satisfies the inhomogeneous wave equation
\begin{equation}\label{eqn: flat wall inhom}
	\partial_m \partial^m v(t,y,z) \= \delta K (t,y,z) \, .
\end{equation}
From here, it becomes evident that $\bar{h}_{\mu\nu}$ is the metric perturbation with vanishing trace of the extrinsic curvature previously studied and the second term, $\partial_\mu \xi_\nu + \partial_\nu \xi_\mu$, is responsible for the effect of adding $\delta K$.
 
The fact that equation \eqref{eqn: flat wall inhom} is a wave equation means that a unique $v$ can be obtained once initial data $\{v|_{t=0}, \partial_t v|_{t=0}\}$ is specified. Given such $v$, the problem of finding $\xi_\mu$ reduces to the problem of solving the wave equation of a four-dimensional vector field \eqref{eqn: flat non-compact diff eom} with Dirichlet boundary conditions \eqref{eqn: diffeo remove delta K} which is known to be well-posed.
 
Note also that this $\xi_\mu$ does not obey the boundary conditions for allowed diffeomorphisms \eqref{eqn: diffeo bdry cond perp}, so it cannot be removed by residual diffeomorphisms and consequently, it is physical.
 
We can also describe the effect of adding $\delta K$ in terms of a change in the location of the timelike boundary. Given \eqref{eqn: h_munu with delta K}, consider the following infinitesimal coordinate transformation,
\begin{equation}
	x^\mu \rightarrow x'^\mu \= x^\mu + \epsilon \, \xi^\mu(x^\mu) \, ,
\end{equation}
so that the term $\partial_\mu \xi_\nu + \partial_\nu \xi_\mu$ is removed from the metric perturbation \eqref{eqn: h_munu with delta K}. In the new coordinates $x'^\mu$, the location of the timelike boundary $\Gamma$ is mapped to $x'^\mu \= \xi^\mu|_{x=0}$ or, equivalently, 
\begin{equation}
	x' \= \epsilon \, \xi^x|_{x=0} \= \epsilon \, v(t,y,z) \, .
\end{equation}
Therefore, the general metric perturbation preserving boundary data $\delta K$ and conformally flat can be described by
\begin{equation}
	h_{\mu \nu} \= \bar{h}_{\mu\nu} \, ,
\end{equation}
on a Minkowski spacetime restricted to $ x \geq \epsilon \, v(t,y,z)$. 
 
In summary, adding a small variation $\delta K$ to the flat boundary problem amounts to imposing the dynamics \eqref{eqn: flat wall inhom} to the two-dimensional boundary of the spacelike boundary $\partial\Sigma$, which we now call a corner. To fully obtain the location of the corner, one must specify the initial location $v|_{t=0}$ and the initial velocity $\partial_tv|_{t=0}$.
 
\noindent \textbf{Example.} Finally, let us provide one simple example. Consider perturbing the trace of the extrinsic curvature by some constant $\delta K$. One solution to \eqref{eqn: flat wall inhom} is given by
\begin{equation}
	v \= -\delta K \frac{t^2}{2} \, ,
\end{equation}
which, to the linearised order, implies that the timelike boundary $\Gamma$ is moving with a constant acceleration set by $\epsilon \, \delta K$. Working in the harmonic gauge, one can extend $\xi_\mu$ into the bulk spacetime using \eqref{eqn: flat non-compact diff eom} as
\begin{equation}
	\xi_\mu dx^\mu \= -\delta K \left(\frac{t^2-x^2}{2}\right) dx \, .
\end{equation}
 
Another solution to \eqref{eqn: flat wall inhom}, with different initial data, is given by
\begin{equation}
	v \= \frac{\delta K}{2}\frac{y^2+z^2}{2} \, .
\end{equation}
This solution can be interpreted as describing the spacetime region near a timelike boundary which, at a constant time, is a two-dimensional sphere of radius $\mathfrak{r} = \frac{2}{\epsilon \, \delta K}$. Likewise, $\xi_\mu$ can be extended into the bulk as
\begin{equation}
	\xi_\mu dx^\mu \= \frac{\delta K}{2} \left(\frac{y^2+z^2}{2}-x^2\right)dx \, .
\end{equation}
 
We will study linearised solutions with spherical  boundaries with non-vanishing trace of the extrinsic curvature in more detail in section \ref{sec:spherical}.

\subsection{Dirichlet flat boundary} \label{sec:dirichlet}
 
Before continuing, it is instructive to briefly summarise the equivalent problem but with Dirichlet boundary conditions, as studied in \cite{Anninos:2022ujl}. 
  
If instead of imposing conformal boundary conditions, we impose Dirichlet boundary conditions, we would need to satisfy
\begin{equation}
	h_{mn} |_{x=0} \= 0 \,.
\end{equation}
In the $k_x\neq0$, this is equivalent to setting $\alpha_{mn} \= 0$ in \eqref{eqn: flat non-compact sol 1}. As a consequence, the harmonic gauge conditions become 
\begin{eqnarray}
\begin{dcases}
	-2ik^n \beta_{x n} +k_x \alpha_{x x}  = 0 \,, & 2ik^n \alpha_{x n} + k_x \beta_{x x} - k_x \eta^{mn} \beta_{mn} = 0 \, , \\
	u^{(1)n} \beta_{x n}  = 0 \,, & u^{(1)n} k_x \alpha_{x n} -  i u^{(1)m} k^n \beta_{m n} = 0 \, ,\\
	u^{(2)n} \beta_{x n}  = 0 \,, & u^{(2)n} k_x \alpha_{x n} -  i u^{(2)m} k^n \beta_{m n} = 0 \, ,\\
	\left(k^m k^n + k_x^2 \eta^{m n }\right) \beta_{mn} = 0 \,,  & 
	\end{dcases}
\end{eqnarray}
where $u^{(1)m}$ and $u^{(2)m}$ are defined in section \ref{sec:kx!=0}. The allowed diffeomorphisms in this case are the same as in \eqref{eqn: flat non-compact residual gauge 1},
\begin{equation}
	\xi_\mu \= \alpha_\mu \cos(k_x x)e^{i k_n x^n} + \beta_{\mu} \sin(k_x x)e^{i k_n x^n} \, .
\end{equation}
Requiring that $\xi_\mu$ preserves the induced metric and the location of the timelike boundary leads to the following boundary conditions,
\begin{equation}
	k_m \alpha_n + k_n \alpha_m \= 0 \, , \qquad \qquad \alpha_x \= 0 \, ,
\end{equation}
which imply that $\alpha_m=0$ for $k_x \neq 0$. Using these residual diffeomorphism, the metric perturbation transforms as
\begin{eqnarray}
\begin{cases}
	\beta_{m n} & \rightarrow \quad \beta_{mn} + i k_m \beta_n + i k_n \beta_m \,, \\
	\beta_{x n} & \rightarrow \quad \beta_{x n}  + i k_n \beta_x \,,  \\
	\alpha_{x n} & \rightarrow \quad \alpha_{x n}  + k_x  \beta_n \,,  \\
	\alpha_{x x} & \rightarrow \quad \alpha_{x x} + 2 k_x \beta_x \, .
	\end{cases}
\end{eqnarray}
Using the residual diffemorphism $\beta_\mu$, we can choose the gauge fixing conditions
\begin{equation}\label{eqn: gauge choice Dirichlet}
	k^m\beta_{mn} \= 0 \, , \qquad \qquad \beta_{xx} \= i \alpha_{xx} \, . 
\end{equation}
As a result, the gauge-fixed metric perturbation is given by
\begin{eqnarray}\label{eqn: sol Dirichlet}
	h_{\mu \nu}dx^\mu dx^\nu &=& \left[\beta^{(+)} \left(u^{(1)}_m u^{(1)}_n - u^{(2)}_m u^{(2)}_n\right) + 2 \beta^{(\times)} u^{(1)}_m u^{(2)}_n \right]  \sin(k_x x) e^{i k_n x^n} dx^m dx^n  \nonumber \\
	& & \quad + 2 i \alpha^{(\perp)} e^{i k_x x + i k_n x^n} \left(k_x dx^2 + k_n dx^n dx\right)\, ,
\end{eqnarray}
where $\beta^{(+)}$, $\beta^{(\times)}$, and $\alpha^{(\perp)}$ are induced metric-preserving gravitational degrees of freedom after removing gauge ambiguity. The two polarisations of the gravitational perturbation are represented by $\beta^{(+)}$ and $\beta^{(\times)}$, while $\alpha^{(\perp)}$ corresponds to a physical diffeomorphism in the same spirit as in section \ref{sec:kx0}. The main difference here is that the physical diffeomorphism mode $\alpha^{(\perp)}$ for Dirichlet problem exists even when $k_x \neq 0$. 

\noindent \textbf{Geometric uniqueness.} The existence of multiple solutions satisfying the same Dirichlet boundary conditions can be seen as follows. We can choose initial data to set $\beta^{(+)} = \beta^{(\times)} = 0$, so that the metric perturbation is governed by the $\alpha^{(\perp)}$ mode. This mode takes the form of a right-moving wave in the $x$-direction for $k_x \geq 0$. Following the argument in \cite{Anninos:2022ujl}, it is possible to construct waves which are initially localised in the $(y,z)$-direction and have compact support in the negative-$x$ region. The existence of such solutions therefore spoils the geometric uniqueness of the linearised Einstein field equation with Dirichlet boundary conditions.
 
Note also that since the non-uniqueness is directly connected to the mode $\alpha^{(\perp)}$ which is a physical diffeomorphism, any results obtained from naive gauge-invariant quantities need to be handled carefully as it sees such mode as physically trivial.
 
Finally, we emphasise that although the existence of right-moving solutions might seem as an artifact of choosing the gauge \eqref{eqn: gauge choice Dirichlet}, there exist gauge-invariant arguments of non-uniqueness of this Dirichlet problem (in Euclidean signature) \cite{Witten:2018lgb}.

\subsection{Compact boundary}

In this section, we consider a similar conformal boundary problem but in a finite spatial region. Namely, we consider spacetime inside a timelike box, whose boundary is composed of six flat timelike boundaries. 
 
Starting with the standard Minkowski metric,
\begin{equation}
	ds^2 \= -dt^2 + dx^2 + dy^2 + dz^2 \, ,
\end{equation}
the manifold $\mathcal{M}$ of interest is a region restricted to $x \in [0,L_x]$, $y \in [0,L_y]$, and $z \in [0,L_z]$ while $t \in \mathbb{R}$. Its timelike boundary is given by $\Gamma = \partial \mathcal{M} = \bigcup_{i=1}^6 \Gamma_i $, where $\Gamma_i$ is a previously studied flat timelike boundary located at $x=0$, $x=L_x$, $y=0$, $y=L_y$, $z=0$, or $z=L_z$. It follows that each side of the box $\Gamma_i$ has the same boundary data, {i.e.} conformally flat and vanishing $K$. For simplicity, we will set the length of all sides to be equal, $L_x=L_y=L_z=L$. 
 
Working in the harmonic gauge, gravitational dynamics in the interior of the box are similarly governed by the wave equation \eqref{eqn: eom harmonic} and the gauge constraints \eqref{eqn: gauge cond harmonic 1} and \eqref{eqn: gauge cond harmonic 2}, as in the non-compact version. The difference here is that now, the boundary conditions \eqref{eqn: flat non-compact bdry cond 1} have to be imposed on all the six sides of the box.  
 
 Let us first consider the case when the momentum of the gravitational perturbation is non-zero in all components. Let $n_{(i)}$ be the normal vector of the side $\Gamma_i$. We consider gravitational fluctuation with the momentum $k^\mu n_{(i)\mu} \neq 0$ for $i=1,2,...,6$. 
  
 Near each timelike boundaries $\Gamma_i$, the metric perturbation behaves according to the result found in section \ref{sec:kx!=0}. This suggests that we should look for a solution which behaves like either $\sin$ or $\cos$ functions in all $x,y,z$ coordinates.
  
 For example, the boundary conditions \eqref{eqn: flat non-compact bdry cond 1} combining with the gauge constraints, \eqref{eqn: gauge cond harmonic 1} and \eqref{eqn: gauge cond harmonic 2}, lead to
 \begin{equation}
 	h_{\mu \nu}\left(\delta^\mu_\rho- n^\mu_{(i)} n_{(i)\rho}\right)\left(\delta^\nu_\sigma- n^\nu_{(i)} n_{(i)\sigma}\right)\Big|_{\Gamma_i} \= 0 \, , \qquad \qquad n^\rho_{(i)} \partial_\rho \left(h_{\mu \nu}n^{\mu}_{(i)} n^{\nu}_{(i)} \right)\Big|_{\Gamma_i} \= 0 \, ,
 \end{equation}
 for $i=1,2,...,6$. The first condition means that the metric perturbation projected onto the timelike boundary $\Gamma_i$ obeys the Dirichlet boundary condition. Similarly, the second condition imposes the Neumann boundary condition on the normal-normal component of the metric perturbation. Boundary conditions on the remaining components then follow from the gauge constraints $T_\mu(h_{\mu\nu})\=0$.
  
As an example, let us look at a particular component $h_{xx}$. The boundary conditions imposed on $h_{xx}$ are given by
 \begin{equation}
 	h_{xx}|_{y=0 \cup y=L \cup z=0 \cup z=L} \= 0 \, , \qquad \qquad \partial_x h_{xx}|_{x=0 \cup x=L} \= 0 \, .
 \end{equation}
 A solution to the equation \eqref{eqn: eom harmonic} subject to these boundary conditions is 
 \begin{equation}
 	h_{xx} \= \alpha_{xx} \cos\left(k_x x\right) \sin\left(k_y y\right) \sin\left(k_z z\right) e^{-i \omega t}\, , \qquad \left(k_x,k_y,k_z\right) \= \frac{ \pi}{L} \left(n_x,n_y,n_z\right) \, ,
 \end{equation}
 where $n_x,n_y,n_z = 1,2,3,...$ are non-zero integers, $\omega = \sqrt{  k_x^2 + k_y^2 +k_z^2 }$, and $\alpha_{xx}$ is a $k_\mu$-dependent function. Note that, as expected for the dynamics of waves in a confined region, the gravitational fluctuation in the conformal box is characterised by a discrete set of frequencies $\omega$. 
  
 To obtain the gauge-fixed metric perturbation, we consider residual diffeomorphisms $\xi_\mu$. The harmonic gauge condition require $\xi_\mu$ to satisfy the wave equation \eqref{eqn: flat non-compact diff eom}. In section \ref{sec:kx!=0}, we showed that the boundary conditions \eqref{eqn: flat non-compact diffeo bdry cond 1} in the harmonic gauge simplify to
 \begin{equation}
 	\xi_\mu |_{\Gamma_i} \= 0 \, ,
 \end{equation}
 for $i=1,2,...,6$. Therefore, the residual diffeomorphism is given by
 \begin{equation}
 	\xi_\mu \= \beta_{\mu} \sin(k_x)\sin(k_y)\sin(k_z)e^{-i\omega t} \, , \qquad \left(k_x,k_y,k_z\right) \= \frac{ \pi}{L} \left(n_x,n_y,n_z\right) \, ,
 \end{equation}
 where $\beta_{\mu}$ are $k_\mu$-dependent functions. We then use $\beta_\mu$ to further fix some components of $h_{\mu\nu}$ to zero.
  
 For the sake of completeness, we present here the gauge-fixed gravitational perturbation subject to the conformal boundary conditions of the timelike box:
 \begin{eqnarray}
 	h_{\mu \nu} &=& h_{\mu\nu}^{(+)} + h_{\mu\nu}^{(\times)} \, , \\
	h_{\mu\nu}^{(+)}dx^\mu dx^\nu &=& \beta^{(+)}e^{-i \omega t}\Bigg( \frac{k_yk_z}{\omega^2} \cos(k_x x)\sin(k_y y)\sin(k_z z) dt dx - \frac{k_xk_y}{\omega^2} \sin(k_x x)\sin(k_y y)\cos(k_z z) dt dz  \nonumber \\
	&& + \frac{ik_z}{\omega} \cos(k_x x)\cos(k_y y)\sin(k_z z) dx dy - \frac{ik_x}{\omega} \sin(k_x x)\cos(k_y y)\cos(k_z z) dy dz \Bigg)  \, , \\
	h_{\mu\nu}^{(\times)} dx^\mu dx^\nu &=& \beta^{(\times)}e^{-i \omega t}\Bigg( \frac{k_yk_z}{\omega^2} \cos(k_x x)\sin(k_y y)\sin(k_z z) dt dx - \frac{k_xk_z}{\omega^2} \sin(k_x x)\cos(k_y y)\sin(k_z z) dt dy  \nonumber \\
	&& + \frac{ik_y}{\omega} \cos(k_x x)\sin(k_y y)\cos(k_z z) dx dz - \frac{ik_x}{\omega} \sin(k_x x)\cos(k_y y)\cos(k_z z) dy dz \Bigg)  \, ,
 \end{eqnarray}
 where 
 \begin{equation}
 	(k_x,k_y,k_z) \= \frac{\pi}{L}(n_x,n_y,n_z) \, , \qquad \omega \= \sqrt{k_x^2+k_y^2+k_z^2}\, .
 \end{equation}
 $\beta^{(+)}$ and $\beta^{(\times)}$ are $k_\mu$-dependent functions representing gravitational polarisations. No gravitational mode gives rise to a non-vanishing Weyl factor when $k^\mu n_{(i)\mu} \neq 0$ for all $i=1,2,...,6$.
  
 Now we consider the case when the momentum $k^\mu$ is pointing parallel to, at least, one side of the box, namely $k^\mu n_{(i)\mu}=0$ for some $i$. For concreteness, we set $k_x\=0$ while $k_y,k_z \neq 0$. It follows that near the boundaries $x\=0,L$, the metric perturbation now behaves as in section \ref{sec:kx0}, that is the ansatz is taken to be a function linear in $x$ and $\sin$ and $\cos$ in $y,z$.
  
 In the harmonic gauge, the residual diffeomorphism $\xi_\mu$ is a solution to \eqref{eqn: flat non-compact diff eom} subject to boundary conditions
 \begin{equation}
 	\xi_\mu |_{\Gamma_i} \= 0 \, ,
 \end{equation}
 on all sides $i=1,2,...,6$. Since the $x$-dependence of $\xi_\mu$ is given by a linear function, there is no non-trivial way to satisfy these boundary conditions at both $x=0$ and $x=L$. Hence, the only residual diffeomorphism is the trivial one $\xi_\mu \= 0$, and no further gauge fixing is needed.
  
Thus, for $k_x\=0$, a gauge-fixed metric perturbation preserving conformal boundary data of the timelike box is given by $h_{\mu \nu} = h_{\mu\nu}^{(\times)} + \partial_{\mu}\xi_{\nu}+ \partial_{\nu}\xi_{\mu}$, with
   \begin{eqnarray}
 \begin{cases}
	h_{\mu\nu}^{(\times)} dx^\mu dx^\nu = \tilde{\beta}^{(\times)}e^{-i \omega t}\Bigg( \frac{ik_y}{\omega} \sin(k_y y)\cos(k_z z) dx dz - \frac{ik_z}{\omega} \cos(k_y y)\sin(k_z z) dx dy \Bigg)  \, , \\
	\xi_\mu dx^\mu = \left(\tilde{\alpha}^{(\perp)} +\tilde{\beta}^{(\perp)} x\right) e^{-i\omega t}\sin(k_y y) \sin(k_z z) dx \, ,
	\end{cases}
 \end{eqnarray}
 where $\tilde{\beta}^{(\times)}$, $\tilde{\alpha}^{(\perp)}$, and $\tilde{\beta}^{(\perp)}$ are $k_\mu$-dependent functions. The first mode $\tilde{\beta}^{(\times)}$ represents a gravitational polarisation confined to a box, while $\tilde{\alpha}^{(\perp)}$, and $\tilde{\beta}^{(\perp)}$ are physical diffeomorphisms. 
 
\noindent \textbf{Geometric uniqueness.} Similar to the non-compact boundary setup, the same argument can be applied here to show that there exist multiple solutions exhibiting the same initial and Dirichlet boundary data. As an example, let us consider the following solution.
\begin{equation}
	h_{\mu\nu} \= \partial_\mu \xi_\nu + \partial_\nu \xi_\mu \, , \qquad \xi_\mu dx^\mu \= \alpha^{(\perp)}e^{-i\omega t + ik_x x} \sin(k_y y)\sin(k_z z) dx \, ,
\end{equation}
where $\omega \= \sqrt{k_x^2 + k_y^2 + k_z^2}$ and $\left(k_y , k_z\right) \= \frac{\pi}{L}\left(n_y,n_z\right)$. The momentum $k_x$ is not restricted to a multiple of integer. This mode takes the form of a wave moving in an $x$-increasing direction. As a result, one can construct waves which initially have support outside the box. Then, at a later time, these waves will move into the interior of the box and therefore spoil the geometric uniqueness of the linearised Einstein field equation with Dirichlet boundary conditions.
 
\section{Spherical boundaries} \label{sec:spherical}

In this section, we study the linearised gravitational problem which preserves conformal boundary data on a timelike boundary with spherical symmetry. In particular, we set the timelike boundary to be a three-dimensional timelike tube of constant radius endowed with a metric of a three-dimensional cylinder and $K \= $ constant. We then study gravitational fluctuations which leave the cylinder's conformal structure unchanged and have $\delta K \=0$.
 
More precisely, by working in standard spherical coordinates,
\begin{equation}\label{eqn: flat spherical coord}
	ds^2 \= -dt^2 + dr^2 + r^2 d\Omega^2 \,, \qquad d\Omega^2 \,\equiv\, d\theta^2 + \sin^2\theta d\phi^2 \, ,
\end{equation}
we consider $x_\perp \= r$, the timelike tube $\Gamma$ to be located at constant $r=\mathfrak{r}$, and 
\begin{equation}
	\bar{g}_{mn}dx^m dx^n \= -dt^2 + r^2 d\Omega^2 \, .
\end{equation}
We are interested in the spacetime region inside the timelike tube, namely we consider $r \in \left(0, \mathfrak{r}\right)$, $t \in \mathbb{R}$, and $\theta \in (0,\pi)$, $\phi \sim \phi+2\pi$ to be angular coordinates parameterising the unit two-sphere. 
 
The induced metric and extrinsic curvature associated to the tube are given by
\begin{equation}\label{eqn: spherical wall}
	ds^2 |_{r=\mathfrak{r}} \= -dt^2  +\mathfrak{r}^2 d\Omega^2 \, , \qquad K_{mn}dx^mdx^n \= \mathfrak{r} \, d\Omega^2 \, ,
\end{equation}
implying that $K={2}/{\mathfrak{r}}$ is a positive constant. The size of the tube $\mathfrak{r}$ is thus set by the trace of the extrinsic curvature $K$.
 
We note that there is another natural choice of spacetime corresponding to the region outside of the tube, $r \in \left(\mathfrak{r},\infty\right)$. For this, the normal vector is pointing in an opposite direction which results in a negative sign of the extrinsic curvature $K_{mn}$.\footnote{The convention we use here is that the normal vector is pointing outward with respect to the region of interest.} Although we will not consider this case here in detail, but briefly discuss it in subsection \ref{l2modes}.
 
Given the spherical symmetry, it is convenient to decompose the metric perturbation $h_{\mu\nu}$ into irreducible representations of the two-sphere. This can be done by using the spherical harmonics decomposition of a rank-2 symmetric tensor. Consequently, the general metric perturbation $h_{\mu \nu}$ is organised by angular momentum $l \in \mathbb{N}_0$.
 
For lower angular momentum, $l\=0,1$, we work in the harmonic gauge.
 
For $l \geq 2$, we use the Kodama-Ishibashi formalism, which we review in Appendix \ref{sec: Kodama-Ishibashi}. This formalism has the benefit of dealing with gauge invariant quantities.
However, as we already encountered in the case of flat boundaries (see section \ref{sec:kx0}), there can be physical gravitational fluctuations that can be described as pure gauge solutions, so we must consider them separately.
 
Regardless of the choice of gauge and angular momentum $l$, the metric perturbations are required to obey the conformal boundary conditions \eqref{eqn: bdry cond 1} and \eqref{eqn: bdry cond 2}. For gravitational perturbations inside the worldtube, we also require the perturbations to be regular at the origin. The only exception is a time-independent, spherically symmetric perturbation with $l=0$ which, as we will see, can be treated as describing a small black hole.

\subsection{$l=0$ mode}\label{sec: l=0}

In this section, we study spherically symmetric metric perturbations. The general metric perturbation preserving spherical symmetry can be written as
\begin{equation}
	h_{\mu \nu}dx^\mu dx^\nu \= h_{tt} dt^2 + h_{r r} dr^2 + 2 h_{t r} drdt + \gamma \, r^2 d\Omega^2  \, ,
\end{equation}
where $h_{tt}$, $h_{tr}$, $h_{rr}$, and $\gamma$ are functions of $t$ and $r$ only. In the harmonic gauge, the linearised Einstein field equation is given in 
\eqref{eqn: harmonic EOM}, that in this coordinate system becomes
\begin{eqnarray}
\begin{cases}
	\left(-\partial_t^2 + \partial_r^2 + \frac{2}{r} \partial_r  \right)h_{tt} &= 0 \, , \\
	\left(-\partial_t^2 + \partial_r^2 + \frac{2}{r} \partial_r - \frac{2}{r^2}\right) h_{ tr} &= 0 \, , \\
	\left(-\partial_t^2 + \partial_r^2 + \frac{2}{r} \partial_r \right) \left(h_{rr}+2\gamma\right) &= 0 \, , \\
	\left(-\partial_t^2 + \partial_r^2 + \frac{2}{r} \partial_r - \frac{6}{r^2} \right)\left(h_{rr}-\gamma \right) &= 0  \, .
	\end{cases}
\end{eqnarray}
Note that they can all be recast in the form
\begin{equation}\label{eqn: spherical Bessel diff eqn}
	\left(-\partial_t^2 + \partial_r^2 + \frac{2}{r}\partial_r - \frac{q(q+1)}{r^2}\right) F_q (t,r) \= 0 \, ,
\end{equation}
for some function $F_q (t,r)$ with $q \in \mathbb{N}_0$. To solve this equation, it is convenient to use a time Fourier transform ansatz,
\begin{equation}
	F_q(t,r) \= f_q(r) e^{-i \omega t},
\end{equation}
where the $r$-dependence is solely encoded in the unknown functions $f_q(r)$. Plugging this ansatz back into \eqref{eqn: spherical Bessel diff eqn} and solving the differential equation for $f_n(r)$, we obtain a general solution for a given $\omega$,
\begin{eqnarray} \label{eqn: spherical bessel diff sol}
F_q(\alpha, \beta; t,r) \= 
	\begin{cases}
	 \alpha \, J_q (r \omega) e^{- i \omega t} + \beta \, Y_q (r \omega) e^{-i\omega t} \, & ,  \quad  \text{if} \, \, \omega \, \neq \, 0 \,,  \\
	 \tilde{\alpha} \, r^q + \tilde{\beta} \, r^{-1-q}  \,  \, &, \quad  \text{if} \, \, \omega \=  0 \,,
	 \end{cases}
\end{eqnarray}
where $\alpha$, $\tilde{\alpha}$, $\beta$ and $\tilde{\beta}$ are $\omega$-dependent functions to be determined later. The functions $J_q (r\omega)$ and $Y_q (r\omega)$ are the spherical Bessel functions of first and second kind, respectively. 
The different metric perturbations are therefore given by 
\begin{equation}\label{eqn: spherical wall l=0 metric sol}
	h_{tt} \= F_0(\alpha_{tt},\beta_{tt}) \, , \quad h_{tr} \= F_1(\alpha_{tr},\beta_{tr}) \, , \quad h_{rr}+2\gamma \= F_0(\alpha_{+},\beta_{+}) \, , \quad h_{rr}-\gamma \= F_2(\alpha_{-},\beta_{-}) \, ,
\end{equation}
where $\alpha_{tt}$, $\alpha_{tr}$, $\alpha_{+}$, $\alpha_{-}$, $\beta_{tt}$, $\beta_{tr}$, $\beta_{+}$, $\beta_{-}$ are $\omega$-dependent functions. We omitted the $(t,r)$ dependence here for brevity. When $\omega=0$, the solutions are the same but with tilde coefficients.
 
Note that near the origin $r\rightarrow 0$, $Y_q (r \omega)$ becomes singular while $J_q (r \omega)$ remains regular for $q >0$. So,  regularity near the origin fixes all the $\beta$'s to zero.

Next, we consider the conformal boundary conditions \eqref{eqn: bdry cond 1} and \eqref{eqn: bdry cond 2}. In the harmonic gauge, they are given by 
\begin{equation}\label{eqn: spherical wall bdry cond}
	h_{tt} + \gamma \, |_{r=\mathfrak{r}} \= 0 \, , \qquad \qquad \frac{4}{\mathfrak{r}}\gamma - \left(\frac{2}{\mathfrak{r}} +\partial_r \right)h_{r r}   \, |_{r=\mathfrak{r}} \= 0\, ,
\end{equation}
where we have used the data from \eqref{eqn: spherical wall}.
 
Additionally, we need to impose the harmonic gauge conditions $T_\mu (h_{\mu\nu})\=0$. In this case, the only non-trivial components become 
\begin{eqnarray}
\begin{cases}
	T_t (h_{\mu \nu}) =& - \frac{1}{2}\partial_t \left(h_{tt}+2 \gamma \right)  + \left(\frac{2}{r}+\partial_r\right) h_{t r} - \frac{1}{2}\partial_t h_{r r} \= 0  \, , \\
	T_r (h_{\mu \nu}) =& - \frac{2}{r}\gamma + \frac{1}{2}\partial_r \left(h_{tt}-2\gamma\right) - \partial_t h_{tr} + \left(\frac{2}{r}+\frac{1}{2}\partial_r\right)h_{rr} \= 0 \, . \label{eqn: spherical wall gauge constraint}
	\end{cases}
\end{eqnarray}
 
Lastly, we consider a residual diffeomorphism $\xi_\mu$. In the spherically symmetric sector, the general diffeomorphism is given by
\begin{equation}
	\xi_\mu dx^\mu \= \xi_t dt + \xi_r dr \, ,
\end{equation}
where $\xi_t$ and $\xi_r$ are functions of $t$ and $r$ only. The harmonic gauge constraints impose that they satisfy a vector Laplacian in spherical coordinates,
\begin{eqnarray}
\begin{cases}
	\left(-\partial_t^2 + \partial_r^2 + \frac{2}{r}\partial_r \right)\xi_t &= 0 \, , \\
	\left(-\partial_t^2 + \partial_r^2 + \frac{2}{r}\partial_r - \frac{2}{r^2}\right) \xi_r &=0 \, .
\end{cases}\label{eqn: l=0 eom of diffeo}
\end{eqnarray}
These equations are of the type \eqref{eqn: spherical Bessel diff eqn}, and so the residual diffeomorphism is given by
\begin{equation}\label{eqn: sphere l=0 diffeo sol}
	\xi_t \= F_0(\alpha_t,\beta_t) \, , \qquad \xi_r \= F_1 (\alpha_r,\beta_r) \, ,
\end{equation} 
where $\alpha_t$, $\alpha_r$, $\beta_t$, and $\beta_r$ are $\omega$-dependent functions. The boundary conditions for the linearised diffeomorphism read
\begin{equation} \label{eqn: spherical diffeo bdry cond}
	\partial_t \xi_t |_{r=\mathfrak{r}} \= 0 \, , \qquad \qquad \xi^r |_{r=\mathfrak{r}} \= 0 \,,
\end{equation}
which can be derived from \eqref{eqn: diffeo bdry cond 1}, \eqref{eqn: diffeo bdry cond 2}, and \eqref{eqn: diffeo bdry cond perp}.

Under this diffeomorphism, the metric perturbation transforms as
\begin{eqnarray}
\begin{cases}
	h_{tt} & \rightarrow \quad h_{tt} + 2 \partial_t \xi_t \, , \\
	h_{tr} & \rightarrow \quad h_{tr} + \partial_t \xi_r + \partial_r \xi_t \, , \\
	h_{rr} & \rightarrow \quad h_{rr} + 2 \partial_r \xi_r \, , \\
	\gamma & \rightarrow \quad \gamma + \frac{2}{r}\xi_r \, .
	\end{cases}
\end{eqnarray}

\subsubsection{Time-dependent solutions}\label{sec: l=0 time dependent}

We first consider the time-dependent solution, namely $\omega \,\neq\, 0$. Requiring that solution is smooth everywhere in the interior restricts the solution \eqref{eqn: spherical wall l=0 metric sol} to be
\begin{eqnarray}
\begin{cases}
	h_{tt} &=\quad \alpha_{tt} J_0 (r\omega) e^{-i \omega t} \, , \\
	h_{tr} &=\quad \alpha_{tr} J_1 (r\omega) e^{-i \omega t}  \, , \\
	h_{rr} &=\quad \left(\frac{\alpha_+}{3}J_0(r\omega) + \frac{2\alpha_-}{3}J_2(r\omega)\right)e^{-i\omega t} \, , \\
	\gamma &=\quad \left(\frac{\alpha_+}{3}J_0(r\omega) - \frac{\alpha_-}{3}J_2(r\omega)\right)e^{-i \omega t} \, .
	\end{cases}
\end{eqnarray}
 
Next, we impose the harmonic gauge constraints. A direct computation shows that the harmonic constraints \eqref{eqn: spherical wall gauge constraint} yield
\begin{equation}
	\alpha_- \= - \alpha_+ \, , \qquad \qquad i\alpha_{tr} \= \frac{\alpha_+ + \alpha_{tt}}{2} \, .
\end{equation}
As a result, the general spherically symmetric time-dependent metric perturbation can be written locally as a pure gauge solution,
\begin{equation}\label{eqn: l=0 omega !=0 sol}
	h_{\mu \nu} \= \nabla_\mu \xi_\nu + \nabla_\nu \xi_\mu \, ,  \qquad \qquad \xi_\mu dx^\mu \= \left(\frac{i\alpha_{tt}}{2 \omega} J_0(r\omega) dt + \frac{\alpha_+}{2 \omega} J_1(r\omega)dr \right) e^{-i \omega t}\, .
\end{equation}
 
This result is consistent with the fact that, in the spherically symmetric sector, every smooth metric perturbation can be written locally as a pure gauge solution, a perturbative version of Birkhoff's theorem \cite{Kodama:2003jz}.  However, in the presence of the timelike boundary, this does not neccesarily mean that the gravitational dynamics is trivial, as shown in section \ref{sec:kx0} when considering a flat timelike boundary. 
 
We still need to impose the conformal boundary conditions. Inserting the solution \eqref{eqn: l=0 omega !=0 sol} into \eqref{eqn: spherical wall bdry cond}, the boundary conditions become
\begin{equation}
	\alpha_{tt} J_0(\mathfrak{r} \, \omega)+\frac{\alpha_{+}}{\mathfrak{r} \, \omega}J_1(\mathfrak{r} \, \omega) \= 0 \, , \qquad \qquad \left(2+\mathfrak{r}^2 \omega^2\right) \alpha_{+}J_1(\mathfrak{r} \, \omega) \= 0 \, .
\end{equation}
To satisfy these equations, we need a relation between $\alpha_{tt}$ and $\alpha_+$ and a condition restricting admissible values of the frequencies $\omega$. We find three inequivalent ways of satisfying those equations. Namely,
\begin{eqnarray}
\begin{cases}
	\alpha_{+} \= 0 &\text{and}\quad J_0(\mathfrak{r} \, \omega) \= 0 \, , \\
	\alpha_{tt} \= 0 &\text{and}\quad J_1(\mathfrak{r} \, \omega) \= 0 \, , \\
	\alpha_{tt} \= - \frac{ J_1(\mathfrak{r} \, \omega)}{\mathfrak{r} \, \omega J_0(\mathfrak{r} \, \omega)}\alpha_+ &\text{and}\quad \mathfrak{r}^2 \omega^2 \= - 2 \, .
	\end{cases}
\end{eqnarray}
The first and second cases fix the frequencies to be zeros of $J_0(\mathfrak{r} \, \omega)$ and $J_1(\mathfrak{r} \, \omega)$, respectively. 
In terms of $\xi_\mu$, it can be seen that they also obey the boundary conditions for the allowed diffeomorphism \eqref{eqn: spherical diffeo bdry cond}. Hence, these solutions can be gauged away using an appropriate gauge transformation. The last case restricts $\omega$ to be purely imaginary implying that the associated metric perturbation has exponential behaviour. As opposed to the first two cases, this solution does not obey the boundary conditions \eqref{eqn: spherical diffeo bdry cond} and so, it should be treated as physical. 
 
Hence, the most general metric perturbation is given by $h_{\mu \nu} = \nabla_\mu \xi_\nu + \nabla_\nu \xi_\mu$, with
\begin{eqnarray}
	\xi_\mu dx^\mu &=& \alpha^{(+)}\left(-\frac{J_1(\sqrt{2}i)}{\sqrt{2} J_0(\sqrt{2}i)} J_0(i \sqrt{2} r/\mathfrak{r})dt + J_1(i\sqrt{2}r/\mathfrak{r})dr\right)e^{\sqrt{2} t/\mathfrak{r}}  \nonumber \\
	&&+  \alpha^{(-)}\left(\frac{J_1(-\sqrt{2}i)}{\sqrt{2} J_0(-\sqrt{2}i)} J_0(-i \sqrt{2} r/\mathfrak{r})dt + J_1(-i\sqrt{2}r/\mathfrak{r})dr\right)e^{-\sqrt{2} t/\mathfrak{r}} \, , \label{eq_diff}
\end{eqnarray}
where $\alpha^{(+)}$ and $\alpha^{(-)}$ represent gravitational modes which preserve conformal structure and the trace of the extrinsic curvature at the boundary. By evaluating \eqref{eq_diff} at the boundary, we find that this solution also gives rise to a non-vanishing Weyl factor,
\begin{equation}
	h_{\mu \nu}dx^\mu dx^\nu |_{r=\mathfrak{r}} \= \left(\frac{2 \alpha^{(+)}}{\mathfrak{r}}J_1(i\sqrt{2})e^{\sqrt{2}t/\mathfrak{r}} - \frac{2 \alpha^{(-)}}{\mathfrak{r}}J_1(-i\sqrt{2})e^{-\sqrt{2}t/\mathfrak{r}} \right)\left(-dt^2 + \mathfrak{r}^2 d\Omega^2\right) \, .
\end{equation}
 
It is interesting to note that although the $\alpha^{(+)}$ mode grows exponentially, the Riemann tensor remains zero at all times since the solution can be written as a pure diffeomorphism.

\subsubsection{Time-independent solutions}

Now we consider the solution \eqref{eqn: spherical wall l=0 metric sol} with $\omega \= 0$,
\begin{eqnarray}
\begin{cases}
	h_{tt} &=\quad \tilde{\alpha}_{tt}  + \frac{\tilde{\beta}_{tt}}{r} \, , \\
	h_{tr} &=\quad \tilde{\alpha}_{tr} r + \frac{\tilde{\beta}_{tr}}{r^2}  \, , \\
	h_{rr} &=\quad \frac{\tilde{\alpha}_+}{3} + \frac{2\tilde{\alpha}_-}{3}r^2 + \frac{\tilde{\beta}_+}{3r} + \frac{2\tilde{\beta}_-}{3r^3} \, , \\
	\gamma &=\quad \frac{\tilde{\alpha}_+}{3} - \frac{\tilde{\alpha}_-}{3}r^2 + \frac{\tilde{\beta}_+}{3r} - \frac{\tilde{\beta}_-}{3r^3}  \, .
	\end{cases}
\end{eqnarray}
In this case, the harmonic gauge constraints \eqref{eqn: spherical wall gauge constraint} become algebraic equations for the eight constants of integration, $\tilde{\alpha}_{tt}$, $\tilde{\alpha}_{tr}$, $\tilde{\alpha}_{+}$, $\tilde{\alpha}_{-}$, $\tilde{\beta}_{tt}$, $\tilde{\beta}_{tr}$, $\tilde{\beta}_{+}$, and $\tilde{\beta}_{-}$. They can be solved by setting
\begin{equation}\label{eqn: sphere l=0 gauge constraints}
	\tilde{\alpha}_- \= 0 \, , \qquad \tilde{\alpha}_{tr} \= 0 \, , \qquad \tilde{\beta}_{tt} \= \frac{\tilde{\beta}_+}{3} \, .
\end{equation}
 
Next, we consider the boundary conditions \eqref{eqn: spherical wall bdry cond}, that become
\begin{equation}
	\tilde{\alpha}_{tt} + \frac{\tilde{\alpha}_+}{3} - \frac{\tilde{\alpha}_-}{3}\mathfrak{r}^2 + \frac{\tilde{\beta}_{tt}}{\mathfrak{r}} + \frac{\tilde{\beta}_+}{3\mathfrak{r}}-\frac{\tilde{\beta}_-}{3\mathfrak{r}^3} \= 0 \, , \qquad 
	\frac{2\tilde{\alpha}_+}{3\mathfrak{r}}-4\tilde{\alpha}_-\mathfrak{r} + \frac{\tilde{\beta}_+}{\mathfrak{r}^2} - \frac{2 \tilde{\beta}_-}{3\mathfrak{r}^4} \= 0 \,.
\end{equation}
Combined with the gauge constraints \eqref{eqn: sphere l=0 gauge constraints}, we find that
\begin{equation}\label{eqn: sphere l=0 bdry cond}
	\tilde{\alpha}_+ \= -\frac{3\tilde{\beta}_+}{2\mathfrak{r}}+\frac{\tilde{\beta}_-}{\mathfrak{r}^3} \, , \qquad \qquad \tilde{\alpha}_{tt} \= - \frac{\tilde{\beta}_+}{6\mathfrak{r}} \, .
\end{equation}
Thus, imposing gauge constatins \eqref{eqn: sphere l=0 gauge constraints} and boundary conditions \eqref{eqn: sphere l=0 bdry cond} reduce the number of degrees of freedom down to three.
 
Now we consider a time-independent residual diffeomorphism \eqref{eqn: sphere l=0 diffeo sol}, 
\begin{eqnarray}
	\xi_t = \tilde{\alpha}_{t}  + \frac{\tilde{\beta}_{t}}{r} \, , \qquad
	\xi_r = \tilde{\alpha}_{r} r + \frac{\tilde{\beta}_{r}}{r^2} \, .
\end{eqnarray}
The boundary conditions \eqref{eqn: diffeo bdry cond 1}, \eqref{eqn: diffeo bdry cond 2}, and \eqref{eqn: diffeo bdry cond perp} impose that 
\begin{equation}
	\tilde{\beta}_{r} \= - \tilde{\alpha}_r \mathfrak{r}^3 \, .
\end{equation}
Under this residual diffeomorphism, the metric perturbation transforms as 
\begin{eqnarray}
\begin{cases}
	\tilde{\beta}_{tr} &\rightarrow \quad \tilde{\beta}_{tr} - \tilde{\beta}_t  \, , \\
	\tilde{\alpha}_+ &\rightarrow \quad \tilde{\alpha}_+ + 6 \tilde{\alpha}_r  \, , \\
	\tilde{\beta}_{-} &\rightarrow \quad \tilde{\beta}_{-} + 6 \tilde{\alpha}_r \mathfrak{r}^3  \, .
	\end{cases}
\end{eqnarray}
Then, we can use $\tilde{\beta}_t$ and $\tilde{\alpha}_r$ to further fix the gauge $\tilde{\beta}_{tr}\=\tilde{\beta}_-\=0$. So, we are left with only one degree of freedom.
 
Note that there is a constant diffeomorphism $\tilde{\alpha}_t$ which does not change the metric perturbation, and therefore, can be omitted. In terms of the change of coordinates, it corresponds precisely to a global time translation by an infinitesimal constant $t \rightarrow t + \epsilon \, \tilde{\alpha}_t$.
 
Putting everything together, the gauge-fixed time-independent metric perturbation preserving the conformal boundary data of the timelike tube is given by
\begin{equation}\label{eqn: small bh}
	h_{\mu \nu}dx^\mu dx^\nu \= \frac{\tilde{\beta}_+}{3}\left[\left(\frac{1}{r}-\frac{1}{2\mathfrak{r}}\right)dt^2+\left(\frac{1}{r}-\frac{3}{2\mathfrak{r}}\right)\left(dr^2+r^2 d\Omega^2\right)\right]  \, .
\end{equation}
The ${1}/{r}$ behaviour suggests that the solution describes a small black hole with a singularity at $r\=0$. Its event horizon is located at a zero of $-1+ \epsilon \, h_{tt}$ which is, to first order in $\epsilon$,
\begin{equation}
	r_\text{horizon} \= \epsilon \, \frac{\tilde{\beta}_+}{3} \, .
\end{equation}
 
The appearance of the constant terms in \eqref{eqn: small bh} is crucial in maintaining the right conformal boundary data at $r=\mathfrak{r}$. In the absence of the timelike tube, one would be able to remove such terms by an appropriate rescaling of $t$ and $r$. 
 
By evaluating at $r=\mathfrak{r}$, we find that
\begin{equation}
	h_{\mu \nu}dx^\mu dx^\nu \= -\frac{\tilde{\beta}_+}{6}\left(-dt^2 + \mathfrak{r}^2 d \Omega^2\right) \, ,
\end{equation}
which implies that the size of the tube is reduced as we add a black hole in its interior. We will further explore this feature in section \ref{sec: euclidean bh} when we consider full non-linear (Euclidean) black hole solutions inside the timelike tube.

\subsection{$l=1$ mode}\label{sec: spherical l=1 mode}

Now we consider gravitational perturbations with angular momentum $l=1$. In the absence of a finite timelike boundary, any $l=1$ mode is known to be pure gauge.\footnote{To show this, start from a general $l=1$ solution to the wave equation $
	\Box h_{\mu \nu} = 0$. Then, imposing the harmonic gauge condition $T_{\mu}(h_{\mu\nu})=0$ fixes the metric perturbation to be pure gauge.}
	\, This means that the general metric perturbation in this sector can be written as $h_{\mu \nu} \= \nabla_\mu \xi_\nu + \nabla_\nu \xi_\mu \,$, with
\begin{equation}\label{eqn: l=1 ansatz}
 \xi_\mu dx^\mu \= \mathcal{T}_t \, \mathbb{S} \, dt + \mathcal{T}_r \, \mathbb{S} \, dr + \left(\mathcal{L}^S \mathbb{S}_i + \mathcal{L}^V \mathbb{V}_i \right) r \, d\Omega^i \, ,
\end{equation}
where $d \Omega^i \= (d\theta,d\phi)$. The harmonic tensors $\mathbb{S}$, $\mathbb{S}_i$, and $\mathbb{V}_i$ are  specific angular-dependent tensors that ensure that $h_{\mu\nu}$ transforms in the $l=1$ representation, see appendix \ref{sec: two-sphere}. The metric coefficients $\mathcal{T}_t$, $\mathcal{T}_r$, $\mathcal{L}^S$, and $\mathcal{L}^V$ are functions of $t$ and $r$ only.
 
The harmonic gauge constraint $T_\mu(h_{\mu\nu})\=0$ can be satisfied by requiring that $\xi_\mu$ obeys a vector Laplacian equation in spherical coordinates, leading to
\begin{eqnarray}
\begin{cases}
	\left(-\partial_t^2 + \partial_r^2 + \frac{2}{r} \partial_r - \frac{2}{r^2}  \right) \mathcal{T}_t &= 0 \, , \\
	\left(-\partial_t^2 + \partial_r^2 + \frac{2}{r} \partial_r - \frac{2}{r^2}  \right) \mathcal{L}^V &= 0 \, , \\
	\left(-\partial_t^2 + \partial_r^2 + \frac{2}{r} \partial_r \right) \left(\sqrt{2} \mathcal{T}_r - 2 \mathcal{L}^S \right) &= 0 \, , \\
	\left(-\partial_t^2 + \partial_r^2 + \frac{2}{r} \partial_r - \frac{6}{r^2} \right)\left( \sqrt{2} \mathcal{T}_r + \mathcal{L}^S \right) &= 0  \, .
	\end{cases}
\end{eqnarray}
These equations are again of the type \eqref{eqn: spherical Bessel diff eqn} and can be satisfied by the functions $F_q$ defined in \eqref{eqn: spherical bessel diff sol}. We also require that the metric perturbation with angular momentum $l=1$ is smooth everywhere in the bulk spacetime. As a result, we find that
\begin{equation}\label{eqn: spherical l=1 sol}
	\mathcal{T}_t \= F_1(\alpha_t,0) \, , \quad \mathcal{L}^V \= F_1(\alpha_v,0) \, , \quad \sqrt{2}\mathcal{T}_r - 2 \mathcal{L}^S \= F_0(\alpha_-,0) \, , \quad \sqrt{2}\mathcal{T}_r + \mathcal{L}^S \= F_2(\alpha_+,0) \, ,
\end{equation}
where $\alpha_t$, $\alpha_v$, $\alpha_-$, and $\alpha_+$ are $\omega$-dependent functions.
 
The conformal boundary conditions \eqref{eqn: bdry cond 1} and \eqref{eqn: bdry cond 2} further require that
\begin{equation}\label{eqn: spherical wall l=1 bdry cond}
	\left. \partial_t \mathcal{L}^V \right|_{r=\mathfrak{r}} \= 0 \, , \,
	\left. \partial_t^2 \mathcal{T}_r \right|_{r=\mathfrak{r}} \= 0 \, , \,
	\left. \left( r \, \partial_t \mathcal{L}^S - \sqrt{2} \mathcal{T}_t \right) \right|_{r=\mathfrak{r}} \= 0 \, , \,
	\left. \left( \sqrt{2}\mathcal{T}_r + \mathcal{L}^S + \sqrt{2} \, r\, \partial_t \mathcal{T}_t \right) \right|_{r=\mathfrak{r}} \= 0 \, .
\end{equation}
Inserting solutions \eqref{eqn: spherical l=1 sol}, these equations give relations among the different $\alpha$ functions. Solving them, we then obtain a general metric perturbation in $l=1$ sector preserving the conformal boundary data on the timelike tube at $r=\mathfrak{r}$.
 
Recall that any solution that also satisfies \eqref{eqn: diffeo bdry cond perp} can be gauged away by an allowed residual diffeomorphism. In this case, this condition becomes just $\mathcal{T}^r |_{r=\mathfrak{r}} \= 0$, 
so to obtain physical modes, it is sufficient to look only at solutions with 
\begin{equation}
	\mathcal{T}^r |_{r=\mathfrak{r}} \neq 0.
\end{equation} 

The second boundary condition in \eqref{eqn: spherical wall l=1 bdry cond} then implies that we must set $\omega \= 0$, {i.e.}, only time-independent solutions are physical.\footnote{The only exception is when $\mathcal{T}_r|_{r=\mathfrak{r}}$, $\mathcal{L}^S|_{r=\mathfrak{r}}$ are linear in $t$ and $\mathcal{T}_t|_{r=\mathfrak{r}}$ is a constant. By imposing the harmonic gauge constraint and the conformal boundary conditions, this diffeomorphism simply becomes a generator of a Lorentz boost, which leads to a vanishing metric perturbation.} This turns all the $\alpha$'s into $\tilde{\alpha}$'s. Then, the most general smooth time-independent solution \eqref{eqn: spherical l=1 sol} is given by
\begin{equation}\label{eqn: spherical l=1 time-independent sol}
	\mathcal{T}_t \= \tilde{\alpha}_t r \, , \qquad \mathcal{L}^V \= \tilde{\alpha}_v r \, , \qquad \mathcal{T}_r \= \sqrt{2} \tilde{\alpha}_+ r^2 - \frac{\tilde{\alpha}_-}{\sqrt{2}} \, , \qquad \mathcal{L}^S \= \tilde{\alpha}_+ r^2 + \tilde{\alpha}_-  \, .
\end{equation}
The boundary conditions \eqref{eqn: spherical wall l=1 bdry cond} in the time-independent case are simplified to
\begin{equation}\label{eqn: spherical wall l=1 time-independent bdry cond}
	\left. \mathcal{T}_t \right|_{r=\mathfrak{r}} \= 0 \, , \qquad \qquad \left. \sqrt{2}\mathcal{T}_r +\mathcal{L}^S \right|_{r=\mathfrak{r}} \= 0 \, .
\end{equation}
From here we obtain that $\tilde{\alpha}_t \= \tilde{\alpha}_+ \= 0$. Also, $\tilde{\alpha}_v$ can be set to zero by a residual diffeomorphism. 
 
As a result, the physical metric perturbation in $l=1$ sector obeying the conformal boundary condition is described just by the $\tilde{\alpha}_-$ mode,
\begin{equation}
	\xi_\mu dx^\mu \= -\frac{\tilde{\alpha}_-}{3} \left(\frac{\mathbb{S}}{\sqrt{2}} \, dr - \mathbb{S}_i \, r \,d\Omega^i \right) \, .
\end{equation}
However, for this solution, we find that 
$	h_{\mu \nu} \= \nabla_\mu \xi_\nu + \nabla_\nu \xi_\mu \= 0 $, so the metric perturbation identically vanishes.

\noindent \textbf{Example.} Consider the case in which $\mathbb{S}$ and $\mathbb{S}_i$ are $\phi$-independent. Their explicit expressions are given by
\begin{equation}
	\mathbb{S} \= \sqrt{\frac{3}{4 \pi}} \cos\theta \, , \qquad \mathbb{S}_\theta \= \sqrt{\frac{3}{8 \pi}} \sin\theta \, , \qquad \mathbb{S}_\phi \= 0 \, .
\end{equation}
The $\tilde{\alpha}_-$ mode now reads
\begin{equation}
	\xi_\mu dx^\mu \= -\tilde{\alpha}_- \sqrt{\frac{3}{8 \pi}}\left( \cos\theta dr - r\sin\theta d\theta \right) \= - \tilde{\alpha}_- \sqrt{\frac{3}{8 \pi}} d \left(r \cos\theta\right)\,,
\end{equation}
which corresponds to an infinitesimally global translation. Then, we can simply ignore this mode.
 \newline\newline
In summary, there is no gravitational fluctuation around flat spacetime in the $l=1$ sector which preserves the conformal boundary data on the timelike tube. The question of whether the $\tilde{\alpha}_-$ mode can become physical when a black hole or a cosmological constant is included remains an interesting open question.\footnote{We note that, in the case of Dirichlet boundary conditions investigated in \cite{Andrade:2015gja}, similar physical diffeomorphisms occur in $l=1$. In the case where there is a black hole in the interior and Dirichlet boundary conditions, the associated metric perturbation is not vanishing. The resulting linearised solutions then describe the motion of the black hole centre of mass relative to the timelike boundary.}

\subsection{$l\geq2$ modes}\label{l2modes}

Now we consider gravitational fluctuations with angular momentum $l \geq 2$. We first study gravitational perturbations in the Kodama-Ishibashi formalism. From a local point of view, these are gauge-invariant perturbations. In the presence of a boundary, one could also have physical diffeomorphisms. 

We show in a gauge-invariant way that there are no physical diffeomorphisms with angular momentum $l\geq2$  preserving the conformal boundary data of the timelike tube.
 
 \textbf{Kodama-Ishibashi formalism.}  
For a given $l \geq 2 $, the two polarisations of the gravitational fluctuations can be divided into vector and scalar 
perturbations which are encoded in master fields $\Phi^{V}(t,r)$ and $\Phi^{S}(t,r)$, respectively. As detailed in appendix \ref{sec: Kodama-Ishibashi}, let us decompose our metric perturbation into a scalar and vector perturbation as follows
\begin{equation}
    h_{\mu \nu} \= h_{\mu \nu}^{(V)} + h_{\mu \nu}^{(S)} \, ,
\end{equation}
where
\begin{eqnarray}
\begin{cases}
    h_{a b}^{(V)} &=\, 0 \, , \\
    h_{a i}^{(V)} &=\, rf^V_a \,\mathbb{V}_i \, , \\
    h_{ij}^{(V)} &=\, 2r^2H^V \,\mathbb{V}_{ij} \, ,  
    \end{cases}
    \qquad \qquad 
\begin{cases}
    h_{a b}^{(S)} &=\, f_{ab}\,\mathbb{S} \, , \\ 
    h_{a i}^{(S)} &=\, rf^S_a \,\mathbb{S}_i  \, , \\
    h_{ij}^{(S)} &=\, 2r^2 H^S \,\mathbb{S}_{ij} + 2r^2\gamma \, \sigma_{ij} \mathbb{S}\, .
 \end{cases}
\end{eqnarray}
Here $a,b = \{t,r \}$ and $i,j = \{ \theta,\phi \}$, whilst  $\mathbb{S}$, $\mathbb{S}_i$, $\mathbb{S}_{ij}$, $\mathbb{V}_i$, and $\mathbb{V}_{ij}$ are angle-dependent tensors, defined in appendix \ref{sec: Kodama-Ishibashi}, ensuring that $h_{\mu\nu}$ transforms in the $l$ representation of $SO(3)$. The coefficients of the perturbation, $f_{ab}$, $f_{a}^{V/S}$, $H^{V/S}$, and $\gamma$, are functions of the  coordinates $(t,r)$. Finally, $\sigma_{ij}$ are the metric components of the round two-sphere.

To explicitly express the metric perturbation $h_{\mu\nu}$ in terms of the master fields, it is useful to fix the following gauge:
\begin{equation}\label{eqn: sphere l>2 gauge fixing}
	H^V \= H^S \= f_t^S \=  f_{tt} + 2 \gamma \= 0  \, .
\end{equation}
Using \eqref{eqn: sphere l>2 gauge fixing}, \eqref{eqn: KI gauge invariant 1}, and \eqref{eqn: KI gauge invariant 2}, the metric perturbation is given in terms of the master fields by
\begin{eqnarray}
\begin{cases}
	h_{mn} &= - \bar{g}_{mn} \frac{1}{r}\left[\frac{l(l+1)}{2}+r^2\partial_t^2+r\partial_r\right]\Phi^S \mathbb{S} + \left(\delta_{m}^i \delta_{n}^t+\delta_{n}^i \delta_{m}^t\right) \partial_r \left(r \, \Phi^V\right)\mathbb{V}_i \, , \\
    h_{rr} &= -\frac{1}{r}\left[\frac{3l(l+1)}{2}+ 3r^2\partial_t^2 + \left(l(l+1) + 1 + r^2\partial_t^2\right) r\partial_r\right]\Phi^S \mathbb{S} \, , \\
    h_{tr} &=  -\frac{1}{2}\partial_t \left[l(l+1) - 2 + r^2\partial_t^2\right]\Phi^S \mathbb{S} \, ,  \\
    h_{ri} &=  \frac{\sqrt{l(l+1)}}{2}\left[l(l+1)+r^2\partial_t^2+2r\partial_r\right]\Phi^S \mathbb{S}_i - r \partial_t \Phi^V \mathbb{V}_i \, ,	
    \end{cases}\label{eqn: spherical l>2 ansatz}
\end{eqnarray}
where $m,n=\{t,\theta,\phi\}$.
Using the gauge-fixed metric perturbation \eqref{eqn: spherical l>2 ansatz}, the linearised Einstein field equation becomes
\begin{equation}\label{eqn: sphere master field}
	\left(- \partial_t^2 + \partial_r^2 - \frac{l(l+1)}{r^2}\right)\Phi^{V/S} \= 0 \, ,
\end{equation}
for both $\Phi^V$ and $\Phi^S$. To solve this, we use the time-Fourier transform ansatz,
\begin{equation}
	\Phi^{V/S}(t,r) \= r\omega \phi^{V/S}(r) e^{-i \omega t},
\end{equation}
where $\phi^{V/S}(r)$ are functions of $r$, and $\omega$ is an arbitrary constant. Plugging back to equation \eqref{eqn: sphere master field} and solving a differential equation of $\phi^{V/S}$, we obtain a general solution of master fields,
\begin{equation}\label{eqn: spherical l>2 sol}
	\Phi^{V/S} \= r \omega \Big( \alpha^{V/S} J_l(r\omega) + \beta^{V/S} Y_l (r\omega) \Big) e^{-i \omega t} \, ,
\end{equation}
where $\alpha^{V/S}$, $\beta^{V/S}$ depend on $\omega$ and $l$. The $J_l (r\omega)$ and $Y_l (r\omega)$ are spherical Bessel functions of the first and second kind of order $l$, respectively. Requiring that the metric perturbation is smooth everywhere in the interior sets $\beta^V \= \beta^S \= 0$. 
 
\begin{figure}[h!]
        \centering
         \subfigure[Vector, $l = 2$]{
                \includegraphics[scale=0.55]{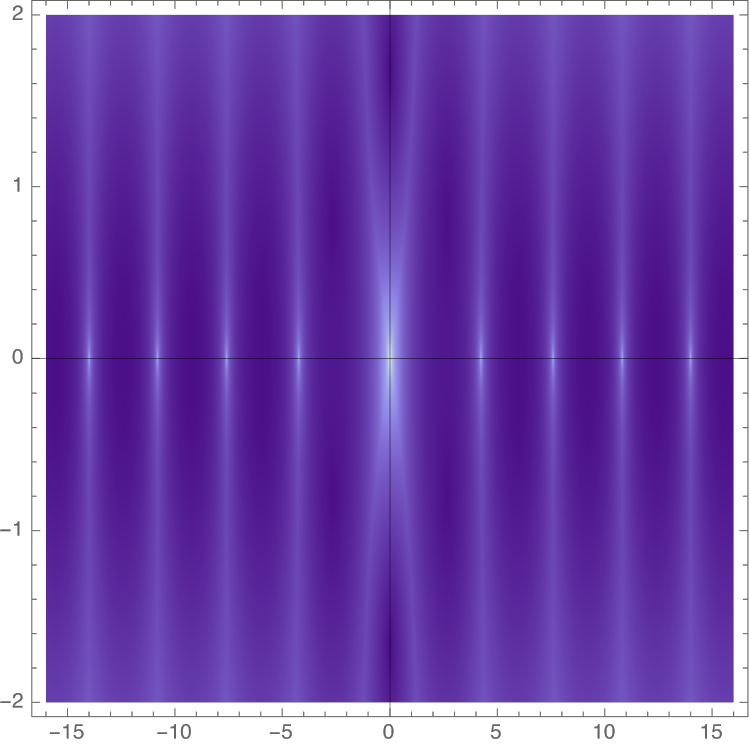}\label{q_2_4}}  \quad\quad
                 \subfigure[Vector, $l = 4$]{
                \includegraphics[scale=0.55]{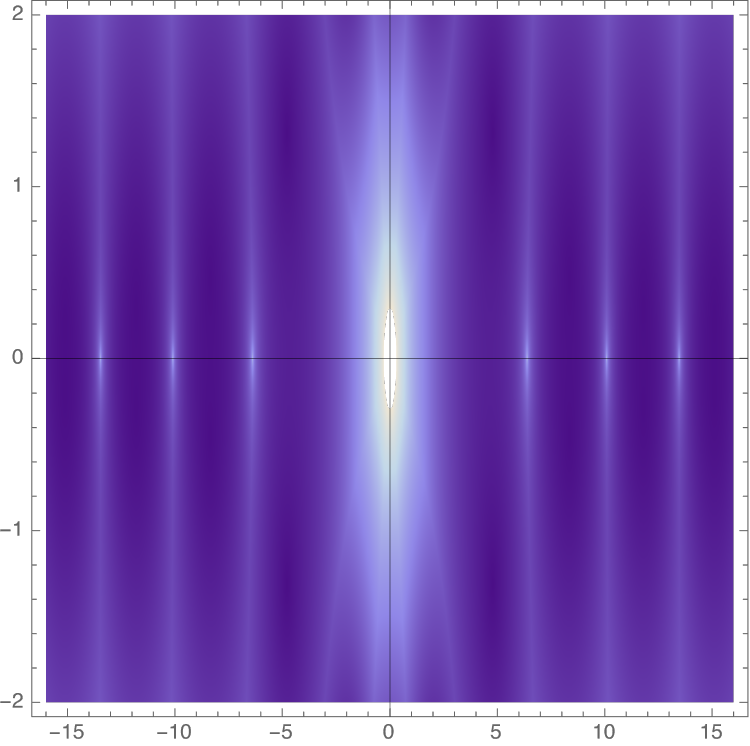} \label{q_2_4i}}  \quad\quad
        \subfigure[Scalar, $l = 2$]{
                \includegraphics[scale=0.55]{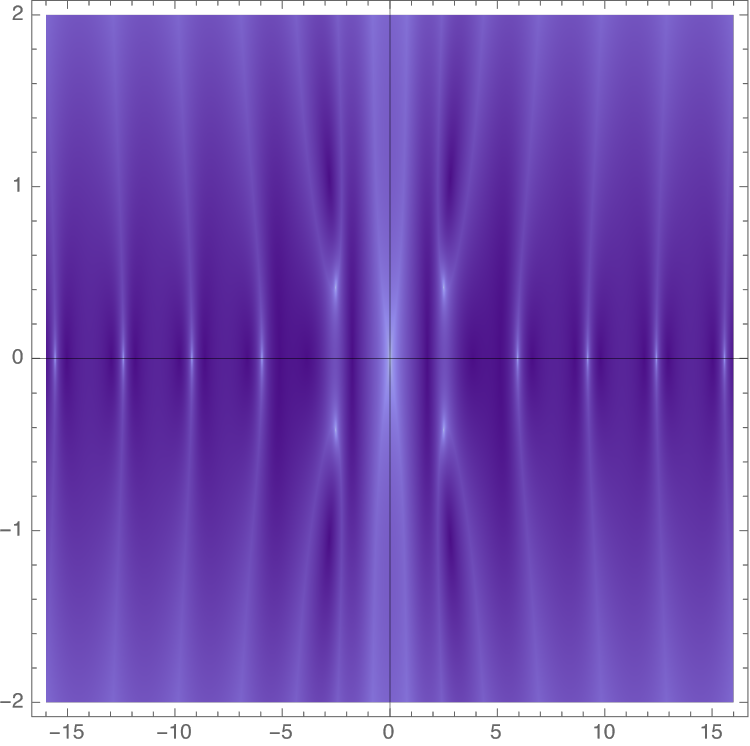} \label{q_4_8}}  \quad\quad
             \subfigure[Scalar, $l = 4$]{
                \includegraphics[scale=0.55]{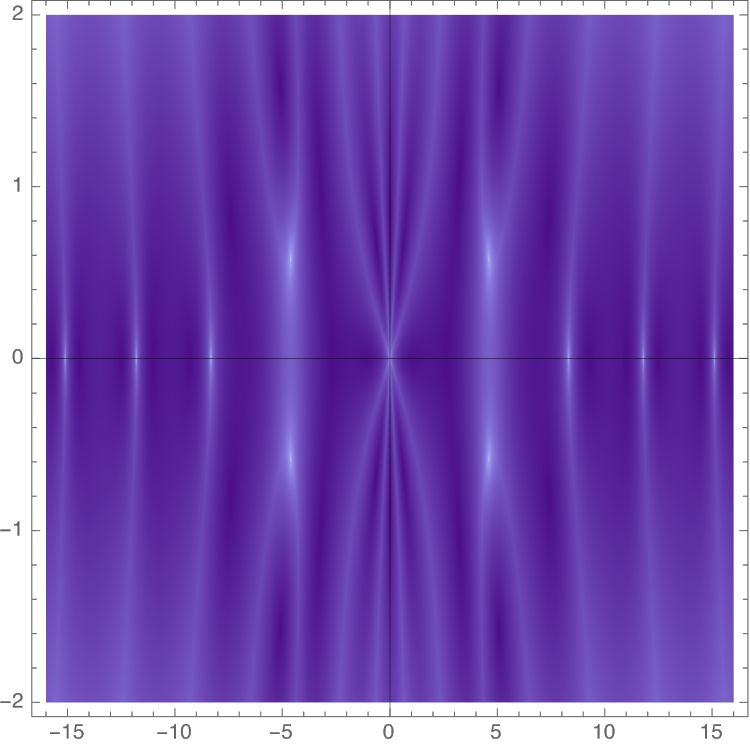} \label{q_4_8i}}                         
                \caption{Density plot of absolute value of $\log \mathcal{F}^{V/S}_l(\mathfrak{r} \, \omega)^2$ in the complex $\mathfrak{r} \, \omega$ plane for $l=2$ and $l=4$. In the scalar case, we further multiply $\mathcal{F}^{S}_l$ by  $|\mathfrak{r} \, \omega|^{-3}$ to highlight the position of the zeros. Note that the scalar function (for both $l$'s) has four zeros that are not real.} \label{fig: omega density plot}
\end{figure}

Next, we impose the conformal boundary conditions. As a consequence of the gauge \eqref{eqn: sphere l>2 gauge fixing}, the conformal boundary conditions \eqref{eqn: bdry cond 1} and \eqref{eqn: bdry cond 2} become
\begin{equation}\label{eqn: sphere vector bdry cond}
	\frac{\Phi^V}{r} + \partial_r \Phi^V |_{r=\mathfrak{r}} \= 0 \, ,
\end{equation}
\begin{equation}\label{eqn: sphere scalar bdry cond}
	\Bigg(\frac{l(l+1) \left(2l(l+1)-3\right)}{r^4}+\frac{4l(l+1)-4}{r^2} \partial_t^2 + 2 \partial_t^4\Bigg) \Phi^S + \Bigg(\frac{3l(l+1)-4}{r^2}+2\partial_t^2\Bigg)\frac{\partial_r\Phi^S}{r} \Big|_{r=\mathfrak{r}} \= 0 ,
\end{equation}
where we have imposed the equation of motion \eqref{eqn: sphere master field} in order to eliminate terms containing $\partial_r^2 \Phi^S$. Since the two master fields $\Phi^V$ and $\Phi^S$ are decoupled both in  the equation of motion \eqref{eqn: sphere master field} as well as the boundary conditions \eqref{eqn: sphere vector bdry cond} and \eqref{eqn: sphere scalar bdry cond}, they can be analysed independently. 
\newline\newline
\textbf{Vector perturbation.} For the vector perturbation, by using in the general solution \eqref{eqn: spherical l>2 sol}, the boundary condition \eqref{eqn: sphere vector bdry cond} gives rise to the following transcendental equation for $\mathfrak{r} \, \omega$:
\begin{equation}\label{eqn: F^V = 0}
	\mathcal{F}^V_l(\mathfrak{r} \, \omega) \,\equiv\, (l+2)J_l(\mathfrak{r} \, \omega) - \mathfrak{r} \, \omega J_{l+1}(\mathfrak{r} \, \omega) \= 0 \, .
\end{equation}
This equation restricts the allowed values of the frequencies $\omega$ of the vector perturbation to a discrete set, that we call $\Omega_{V,l}$. We show those cases for $l=2,4$ in figures \ref{q_2_4} and \ref{q_2_4i}, respectively.
 
Given the set $\Omega_{V,l}$, one can write down the general solution of master field $\Phi^V$ for a fixed $l \geq 2$ as
\begin{equation}\label{eqn: sol vector perturbation}
	\Phi^V(t,r) \= \sum_{\omega \in \Omega_{V,l}} \alpha_l^V (\mathfrak{r} \, \omega) r\omega J_l (r\omega) e^{-i \omega t} + \text{c.c.} 
\end{equation}
Plugging back into \eqref{eqn: spherical l>2 ansatz} and setting $\Phi^S\=0$, we obtain a general gauge-fixed vector perturbation for a given angular momentum $l\geq 2$ and preserving the conformal boundary data of the timelike tube at $r\=\mathfrak{r}$. We observe that the solution does not affect the background Weyl factor on the timelike tube.
\newline\newline
\textbf{Scalar perturbation.} Similarly, inserting the solution \eqref{eqn: spherical l>2 sol} into the boundary condition for the scalar perturbation \eqref{eqn: sphere scalar bdry cond}, we obtain the following condition for $\mathfrak{r} \, \omega$,
\begin{multline} \label{eqn: F^S = 0}
	\mathcal{F}^S_l(\mathfrak{r} \, \omega) \equiv \left(-4-4l+5l^2+7l^3+2l^4-2(l(3+2l)-1)\mathfrak{r}^2 \omega^2 + 2 \mathfrak{r}^4 \omega^4\right) J_l(\mathfrak{r} \, \omega)  \\ + \mathfrak{r} \, \omega \left(4-3l(l+1)+2\mathfrak{r}^2 \omega^2\right) J_{l+1}(\mathfrak{r} \, \omega)   = 0 \, . 
\end{multline}
Let $\Omega_{S,l}$ be the set of zeros of $F^S_l(\mathfrak{r} \, \omega)$, then the general solution for the master field $\Phi^S$ for a fixed $l \geq 2$ is given by
\begin{equation}\label{eqn: sol scalar perturbation}
	\Phi^S(t,r) \= \sum_{\omega \in \Omega_{S,l}} \alpha_l^S (\mathfrak{r} \, \omega) r\omega J_l (r\omega) e^{-i \omega t} + \text{c.c.} 
\end{equation}
Plugging back into \eqref{eqn: spherical l>2 ansatz} and setting $\Phi^V\=0$, we obtain a general gauge-fixed scalar perturbation for a given angular momentum $l\geq 2$ and preserving the conformal boundary data of the timelike tube at $r\=\mathfrak{r}$. A scalar perturbation of fixed $l$ gives rise to a non-vanishing Weyl factor on the timelike tube,
\begin{eqnarray}
	 && h_{\mu \nu}dx^\mu dx^\nu |_{r=\mathfrak{r}} \=  \\
	 &&  \text{Re} \left[ \sum_{\omega \in \Omega_{S,l}} \alpha_l^S (\mathfrak{r} \, \omega) \, \omega \Bigg(2 \, \mathfrak{r} \, \omega J_{l+1}(\mathfrak{r} \, \omega) - \left(2+l(l+3)-2 \, \mathfrak{r}^2\omega^2\right)J_l(\mathfrak{r} \, \omega)\Bigg) \, e^{-i \omega t} \, \mathbb{S}    \right] \left(-dt^2 + \mathfrak{r}^2 d \Omega^2\right) \, . \nonumber
\end{eqnarray}
 
\begin{figure}[h!]
        \centering
         \subfigure[$l = 2$]{
                \includegraphics[scale=0.55]{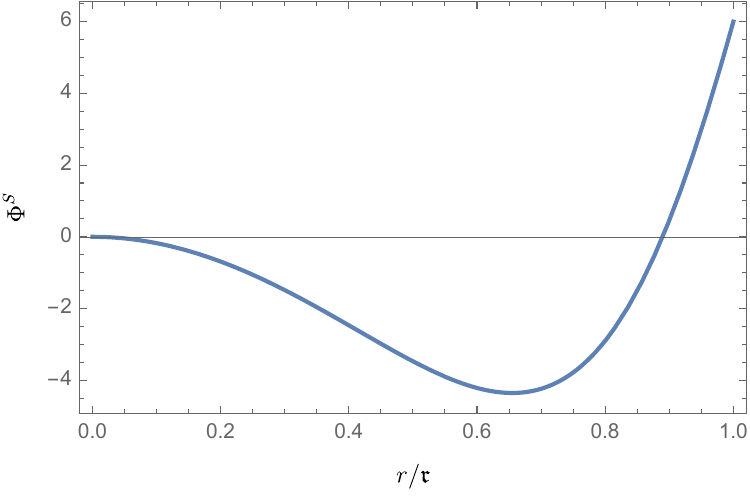}\label{fig_a}}  \quad\quad
                 \subfigure[$l = 4$]{
                \includegraphics[scale=0.55]{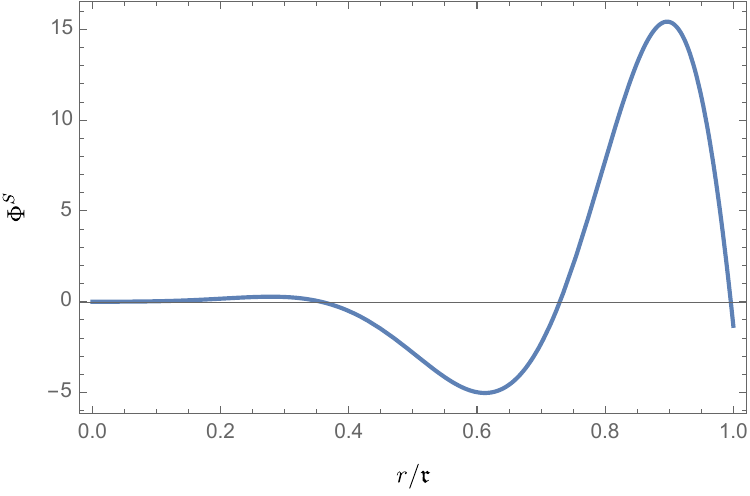} \label{fig_b}}  \quad\quad        
                \caption{Radial profile of $\Phi^S$ as a function of $r/\mathfrak{r}$, at fixed time $t/\mathfrak{r} = 1$ and angular momentum $l$. Only the two frequencies that have positive imaginary part are shown in the plots. Profiles are smooth for any time $t/\mathfrak{r}$.} \label{fig: radial}
\end{figure}

Unlike the set of frequencies for the vector perturbation $\Omega_{V,l}$, the scalar perturbation modes contain two pairs of complex conjugate frequencies. These frequencies can be computed numerically. For the lower angular momenta, for instance, we obtain
\begin{equation}
	\mathfrak{r} \, \omega \=
	\begin{cases}
		\pm 2.50459 \pm 0.413246 i \, ,&  l=2 \, , \\
		\pm 4.60402 \pm 0.579649 i \, ,&  l=4 \, , \\
		\pm 6.64021 \pm 0.664807 i \, ,&  l=6 \, .
	\end{cases}
\end{equation}
See figures \ref{q_4_8} and \ref{q_4_8i} for density plots for $l=2,4$, respectively. In the large-$l$ limit, we numerically find that the four complex frequencies behave roughly as $\mathfrak{r}\omega \approx \pm l \pm i \, l^{1/3}$. For modes with positive imaginary part, the corresponding gravitational perturbation grows exponentially in time. As such,  the perturbative method  breaks down at late times, and a non-linear analysis is required. At any given time though, the radial profile is smooth, see figure \ref{fig: radial}. The situation here is in  contrast to the problem with Dirichlet boundary conditions where it was shown in \cite{Andrade:2015gja} that the interior modes in flat space are linearly stable. On the other hand, it might be interesting to compare the effect to the AdS instability discussed in \cite{Bizon:2011gg}.

We note that asymptotic solutions to \eqref{eqn: F^V = 0} and \eqref{eqn: F^S = 0} at large real values of $\mathfrak{r} \, \omega$ can be obtained analytically. For fixed $l\geq2$, they are given by
\begin{equation}
	\mathfrak{r} \, \omega \= \begin{cases}
	\pm \frac{\pi}{2} \left(2n - l-1 \right)   &  \text{,}  \quad \text{vector perturbation}   \\
	\pm \frac{\pi}{2} \left( 2n-l \right)    & \text{,} \quad \text{scalar perturbation} 
	\end{cases} \,, \qquad
	n \in \mathbb{N} \,.
\end{equation}

\textbf{Physical diffeomorphism.} We now consider a gravitational perturbation which can be written as a pure diffeomorphism, 
\begin{equation}
	h_{\mu \nu} \= \nabla_\mu \xi_\nu + \nabla_\nu \xi_\mu \, , \qquad \qquad \xi_\mu dx^\mu \= \mathcal{T}_t \mathbb{S} dt + \mathcal{T}_r \mathbb{S} dr + \left(\mathcal{L}^S \mathbb{S}_i + \mathcal{L}^V \mathbb{V}_i \right) r d\Omega^i \, ,
\end{equation}
for $l\geq 2$. The conformal boundary conditions \eqref{eqn: diffeo bdry cond 1} and \eqref{eqn: diffeo bdry cond 2} impose that
\begin{eqnarray}
\begin{cases}
	\left. 2r\partial_t \mathcal{T}_t  + \sqrt{l(l+1)}\mathcal{L}^S+2\mathcal{T}_r  \right|_{r=\mathfrak{r}} & \=0  \,, \\
	\left. r\partial_t \mathcal{L}^S - \sqrt{l(l+1)}\mathcal{T}_t \right|_{r=\mathfrak{r}} & \= 0 \,, \\
	\left. \mathcal{L}^S \right|_{r=\mathfrak{r}} & \= 0  \,, \\
	\left. \left(-r^2\partial_t^2 + 2-l(l+1) \right)\mathcal{T}_r  \right|_{r=\mathfrak{r}} & \= 0  \,, \\
	\left. \mathcal{L}^V \right|_{r=\mathfrak{r}} & \= 0  \,,
	\end{cases}
\end{eqnarray}
for $l\geq 2$. It follows that $\mathcal{L}^S|_{r=\mathfrak{r}} \= \mathcal{L}^V|_{r=\mathfrak{r}}\=0$. Then, the remaining conditions become
\begin{equation}
	\left. r \, \partial_t \mathcal{T}_t + \mathcal{T}_r \right|_{r=\mathfrak{r}}\= 0 \, , \qquad \left.\mathcal{T}_t\right|_{r=\mathfrak{r}} \= 0 \,, \qquad \left. \left(-r^2 \partial_t^2 + \left( 2-l(l+1)\right) \right)\mathcal{T}_r  \right|_{r=\mathfrak{r}}  \= 0 \, .
\end{equation}
These conditions give rise to $\mathcal{T}_t |_{r=\mathfrak{r}} \= \mathcal{T}_r |_{r=\mathfrak{r}} \= 0$. Clearly, the last condition is equivalent to the condition that the given diffeomorphism does not move the location of the timelike boundary \eqref{eqn: diffeo bdry cond perp}. Therefore, any linearised diffeomorphism $\xi_\mu$ with $l \geq 2$ and preserving the conformal boundary data of the timelike tube is a trivial diffeomorphism which can be gauged away.
 
We note that since for this argument we have not imposed any gauge condition, the result applies to any gauge choice.
\newline\newline
\textbf{Brief remark on outgoing modes.} Though we are not focusing on the modes exterior to the wordtube that propagate toward null infinity, we would like to make a brief comment. The structure of the linearised solutions is as for the interior modes. The difference is that we no longer need to impose that the modes are non-singular at the origin since that region is excised from the spacetime.\footnote{This leads to the possibility of including a portion of the negative mass Schwarzschild solution, which would have a naked timelike singularity at the origin.} Near null infinity we will generically have a linear combination of incoming and outgoing spherical waves. We further imagine imposing that the waves are purely outgoing for each angular momentum $l$. The essential difference between the exterior and interior modes is that once an exterior mode is outgoing, it will eventually reach null infinity rather than returning back.

As an example, we consider the fixed-$l$ scalar solution
\begin{equation}
\Phi^S(t,r) =  \sum_\omega \alpha_l^S (\mathfrak{r} \, \omega) \, r  \omega  \left( J_l (r\omega) + i \, Y_l(r\omega) \right) e^{-i \omega t} + \text{c.c.}  \,,
\end{equation}
that is purely outgoing in the radial direction. To preserve the conformal boundary conditions, we can restrict to initial data on $\Sigma$, so that the purely outgoing modes have support away from the intersection of $\Sigma$ with $\Gamma$. 

More general configurations will also have ingoing modes and in such a situation one must impose the boundary condition (\ref{eqn: sphere scalar bdry cond}) at $r=\mathfrak{r}$. The condition is a mild generalisation of (\ref{eqn: F^S = 0}), since we are also allowed to have modes built from the $Y_l(r\omega)$. Nevertheless, at least for the modes built exclusively from the $J_l(r\omega)$, the boundary condition is equivalent to (\ref{eqn: F^S = 0}), and consequently, a subset of configurations will also exhibit exponential growth. As another example, let us consider the purely outgoing condition equipped with conformal boundary conditions at $r=\mathfrak{r}$, and take $l=2$. In this case, we find two growing modes with values $\mathfrak{r}\omega= \pm 1.78355 + 0.372158 i$. 

\subsection{Modifying $\delta K$}

We now consider conformal boundary conditions for which the trace $K$ of the extrinsic curvature is varied slightly away from $2/\mathfrak{r}$ while the conformal class of the boundary metric remains intact. This is achieved through the following boundary conditions on the metric perturbation:
\begin{equation}\label{eqn: spherical modified K}
	h_{mn}-\frac{1}{3}\bar{g}_{mn} h^p{}_p |_{r=\mathfrak{r}} \= 0 \, , \qquad \partial_r h^m{}_m - 2 \mathcal{D}^m h_{rm}- \frac{2}{\mathfrak{r}}h_{rr} |_{r=\mathfrak{r}} \= 2\delta K(t,\theta,\phi) \, ,
\end{equation}
where $\delta K (t,\theta,\phi)$ is an arbitrary function of the boundary coordinates so that the trace of the extrinsic curvature of the timelike boundary becomes $K = {2}/{\mathfrak{r}}+ \epsilon \, \delta K (t,\theta,\phi)$.

In order to find metric perturbation obeying modified boundary conditions, we decompose $\delta K$ using the spherical harmonic functions $\mathbb{S}$,
\begin{equation}
	\delta K (t,\theta,\phi) \= \sum_{l=0}^\infty \sum_{m=-l}^l \delta K_{l,m}(t) \mathbb{S}(\theta,\phi) \, .
\end{equation}
We then consider metric perturbations for each $l$ separately. Since the $m$-dependence is not important for the following calculation, we will neglect it for the rest of the section.
\subsubsection{$l=0$ mode}
We first consider the modified boundary conditions in the spherically symmetric sector. Specifically, we consider the following gravitational perturbation:
\begin{equation}
	h_{\mu \nu} \= \nabla_\mu \xi_\nu + \nabla_\nu \xi_\mu \, ,  \qquad \qquad \xi_\mu dx^\mu \= \xi_t(t,r) dt + \xi_r(t,r) dr \, ,
\end{equation}
where $\xi_t(t,r)$ and $\xi_r (t,r)$ are some arbitrary functions of $t$ and $r$. This perturbation automatically satisfies the linearised Einstein field equation as it is a pure diffeomorphism. Working in harmonic gauge imposes that it satisfies \eqref{eqn: l=0 eom of diffeo}.

Imposing the boundary conditions \eqref{eqn: spherical modified K}, we find
\begin{equation}\label{eqn: spherical modified K bdry l=0}
	\partial_t \xi_t + \frac{\xi_r}{r} |_{r=\mathfrak{r}} \= 0 \, , \qquad \left(\partial_t^2 - \frac{2}{r^2}\right) \xi_r |_{r=\mathfrak{r}}\= \delta K_{l=0}(t) \, .
\end{equation}
Viewing the second boundary condition as a differential equation on the boundary value $ \xi_r |_{r=\mathfrak{r}}$, one can write down the general solution as
\begin{equation}\label{eqn: modified spherical bdry xi_r}
	\xi_r |_{r=\mathfrak{r}} \= \left(\alpha^{(+)} + \mathfrak{r} \int_0^t dt' e^{-\sqrt{2} t'/\mathfrak{r}} \frac{\delta K_{l=0}(t')}{2\sqrt{2}}\right)e^{\sqrt{2}t/\mathfrak{r}} + \left(\alpha^{(-)} - \mathfrak{r}\int_0^t dt' e^{\sqrt{2} t'/\mathfrak{r}} \frac{\delta K_{l=0}(t')}{2\sqrt{2}}\right)e^{-\sqrt{2}t/\mathfrak{r}} \, , 
\end{equation}
where $\alpha^{(+)}$ and $\alpha^{(-)}$ are reminiscent of the exponentially growing and decaying modes found in section \ref{sec: l=0 time dependent}. We then use the first boundary condition from \eqref{eqn: spherical modified K bdry l=0} to obtain the boundary value of the time component $ \xi_t |_{r=\mathfrak{r}}$. Lastly, we use the harmonic gauge constraints \eqref{eqn: l=0 eom of diffeo} to radially extend the solution into the interior and calculate the metric perturbation as $h_{\mu\nu}= \nabla_\mu \xi_\nu + \nabla_\nu \xi_\mu$.

From \eqref{eqn: modified spherical bdry xi_r}, it can be seen that turning on $\delta K_{l=0} (t)$ for some finite time interval $\left[ t_i , t_f \right]$, can result in changing the exponentially growing/decaying behaviour of the solution. Thus, there is a sense in which the exponentially growing mode $\alpha^{(+)}$ can be tamed by turning on an appropriate $\delta K_{l=0}(t)$ function.

\textbf{Example.} Consider a Dirac delta function located at some time $t_0>0$ with strength $\kappa$, $\delta K_{l=0}(t) \= \kappa \, \delta (t-t_0)$. Expression \eqref{eqn: modified spherical bdry xi_r} now becomes
\begin{equation}
	\xi_r |_{r=\mathfrak{r}} \= \left(\alpha^{(+)} + \kappa \, \mathfrak{r} \frac{e^{-\sqrt{2} t_0/\mathfrak{r}}}{2\sqrt{2}}\Theta(t-t_0)\right)e^{\sqrt{2}t/\mathfrak{r}} + \left(\alpha^{(-)} - \kappa \, \mathfrak{r} \frac{e^{\sqrt{2} t_0/\mathfrak{r}}}{2\sqrt{2}}\Theta(t-t_0)\right)e^{-\sqrt{2}t/\mathfrak{r}} \, ,
\end{equation}
where $\Theta(t)$ is the Heaviside step function. By tuning $\kappa \, \mathfrak{r} \= -2\sqrt{2} e^{\sqrt{2}t_0} \alpha^{(+)}$, we find that
\begin{equation}\label{eqn: modified spherical bdry xi_r 2}
	\xi_r |_{r=\mathfrak{r}} \= \alpha^{(+)}\left(1-\Theta(t-t_0)\right)e^{\sqrt{2} t/\mathfrak{r}} + \left(\alpha^{(-)} + \alpha^{(+)} e^{2\sqrt{2}t_0/\mathfrak{r}}\Theta(t-t_0)\right) e^{-\sqrt{2}t/\mathfrak{r}} \, .
\end{equation}
This solution does not exhibit exponential growth at late times anymore. In fact, it is exponentially decaying after time $t_0$. Given that there is no exponentially growing mode for $l=1$, we now directly analyse the case $l\geq 2$.
\subsubsection{$l\geq2$ modes} Recall that upon fixing the gauge \eqref{eqn: sphere l>2 gauge fixing}, the general metric perturbation for fixed $l$ can be written in terms of the master fields $\Phi^V$ and $\Phi^S$, see \eqref{eqn: spherical l>2 ansatz}. The master fields satisfy \eqref{eqn: sphere master field}, and hence the general solution is given by \eqref{eqn: spherical l>2 sol}. Regularity of the solution in the interior still requires $\beta^{V/S} = 0$.

Imposing the modified boundary conditions \eqref{eqn: spherical modified K} leads to
\begin{equation}
	\frac{\Phi^V}{r} + \partial_r \Phi^V |_{r=\mathfrak{r}} \= 0 \, ,
\end{equation}
\begin{equation}\label{eqn: spherical modified K l >= 2}
	\Bigg(\frac{l(l+1) \left(2l(l+1)-3\right)}{r^4}+\frac{4l(l+1)-4}{r^2} \partial_t^2 + 2 \partial_t^4\Bigg) \Phi^S + \Bigg(\frac{3l(l+1)-4}{r^2}+2\partial_t^2\Bigg)\frac{\partial_r\Phi^S}{r} \Big|_{r=\mathfrak{r}} \= -\frac{4\delta K_{l}(t)}{\mathfrak{r}^2}\, .
\end{equation}
Since the vector master field $\Phi^V$ obeys the same boundary condition, its solution is the same as in the previous subsection. 

To solve the boundary condition \eqref{eqn: spherical modified K l >= 2}, we apply the time-Fourier transform 
\begin{equation}
\delta K_{l}(t) \= \int_\mathbb{R} \frac{d \omega}{2 \pi} e^{-i \omega t}\delta K_{l}(\omega) \, ,
\end{equation}
after which the boundary condition \eqref{eqn: spherical modified K l >= 2} becomes
\begin{equation}
	\mathcal{F}^S_l(\mathfrak{r} \, \omega)\alpha^S (\mathfrak{r}\, \omega) \= - \frac{4 \delta K_l(\omega)}{\mathfrak{r}^2} \, .
\end{equation}
This  fixes $\alpha^S(\mathfrak{r} \, \omega)$ in terms of $\delta K_l(\omega)$. Inserting this back into \eqref{eqn: spherical l>2 sol} and integrating over $\omega$, one finds that the inhomogeneous part of the solution is given by
\begin{equation}
	\Phi^S(t,r) \= - \int_\mathcal{C} \frac{d \omega}{2\pi} \frac{4 \, \delta K_l (\omega)}{\mathfrak{r}^2 \mathcal{F}^S_l (\mathfrak{r} \, \omega)}  r \omega J_l(r \omega) e^{-i \omega t} + \text{c.c.}\, , 
\end{equation}
where $\mathcal{C}$ is a contour in the complex $\omega$-plane chosen such that it lies above all the zeros of $\mathcal{F}^S_l(\mathfrak{r} \, \omega)$. To write down a complete solution, one must include a homogeneous solution which is $\Phi^S$ obeying the boundary condition with $\delta K_l \= 0$. So, the general $\Phi^S$ preserving the conformal class of the spherical boundary and having $K \= \frac{2}{\mathfrak{r}}+\delta K_l(t)\,\mathbb{S}(\theta,\phi)$ for fixed $l\geq 2$ is given by
\begin{equation}\label{eqn: Phi^S integrated over contour}
	\Phi^S(t,r) \= - \int_\mathcal{C} \frac{d \omega}{2\pi} \frac{4 \, \delta K_l (\omega)}{\mathfrak{r}^2 \mathcal{F}^S_l (\mathfrak{r} \, \omega)}  r \omega J_l(r \omega) e^{-i \omega t} + \sum_{\omega' \in \Omega_{S,l}} \alpha_l^S (\mathfrak{r} \, \omega') r\omega' J_l (r\omega') e^{-i \omega' t} + \text{c.c.}\, , 
\end{equation}
where $\alpha^S_l (\mathfrak{r}\, \omega')$ is an arbitrary function.

\textbf{Example.} Consider $\delta K_{l}(t) \= \kappa_l \, \delta(t-t_0)$, with $\kappa_l$ constant. Its Fourier transform is given by $\delta K_l (\omega) \= \kappa_l e^{i \omega t_0}$. Plugging this into \eqref{eqn: Phi^S integrated over contour}, we find
\begin{equation}
	\Phi^S (t,r) \= - \int_\mathcal{C} \frac{d \omega}{2\pi} \frac{4 \, \kappa_l \, e^{-i \omega(t-t_0)}}{\mathfrak{r}^2\mathcal{F}_l^S(\mathfrak{r} \, \omega)} r \omega J_{l} (r\omega) + \sum_{\omega' \in \Omega_{S,l}} \alpha_l^S (\mathfrak{r} \, \omega') r\omega' J_l (r\omega') e^{-i \omega' t} + \text{c.c.}\, , 
\end{equation}
When $t<t_0$, the contour can be closed in the upper half plane. Since there is no pole lying above the contour, the integral simply leads to zero. As a result, information from turning on $\delta K_l(t)$ at time $t_0$ does not propagate back in time. For time $t>t_0$, the contour can be closed in the lower half plane. Using the residue theorem, one finds that the integral reduces to a sum over the zeros of $\mathcal{F}^S_l(\mathfrak{r}\, \omega)$. Combining with the homogeneous part of the solution, one arrives at
\begin{equation}
	\Phi^S (t,r) \= \sum_{\omega \in \Omega_{S,l}} \left(\frac{4 i \kappa_l e^{i \omega t_0}}{\mathfrak{r}^2 \left(\mathcal{F}^S_l (\mathfrak{r} \, \omega) \right)' } \Theta(t-t_0) + \alpha_l^S(\mathfrak{r} \, \omega) \right)r \omega J_l (r\omega)e^{-i\omega t} + \text{c.c.} \, ,
\end{equation}
where $\left( \mathcal{F}^S_l (x) \right)' \,\equiv\, \left.\frac{d\mathcal{F}^S_l}{dx}\right|_x$. The resulting solution implies that turning on $\delta K_l(t)$ shifts the amplitude of the gravitational modes $\alpha^S_l(\mathfrak{r} \, \omega)$ in the bulk by a factor that depends on $\kappa_l$. It follows that a gravitational mode associated to a frequency with positive imaginary part can be eliminated by choosing appropriate parameters $\kappa_l$ and $t_0$. If there are more than one of these frequencies, then a slightly more involved $\delta K_l (t)$ would be needed. 

As a consequence, and similarly to the spherically symmetric sector case, we can judiciously select a $\delta K(t,\theta,\phi)$ for some finite time interval such that the metric perturbation does not exhibit exponential growth at late times.

\subsection{Dirichlet spherical boundary}

In this section, we consider Dirichlet boundary conditions instead and show that, in this case, the geometric uniqueness of the linearised Einstein equation is preserved.

Linearised gravitational modes  subject to Dirichlet boundary conditions which are not local diffeomorphisms were studied in \cite{Andrade:2015gja}, using the Kodama-Ishibashi formalism. 
Here, we study metric perturbations that are locally diffeomorphisms. These take the form $h_{\mu \nu} \= \nabla_\mu \xi_\nu + \nabla_\nu \xi_\mu $, where
\begin{equation}\label{eqn: spherical Dirichlet ansatz}
	\xi_\mu dx^\mu \= 
	\begin{cases}
	\mathcal{T}_t \, \mathbb{S} \, dt + \mathcal{T}_r \, \mathbb{S} \, dr  &, \qquad l\=0 \, ,  \\
	\mathcal{T}_t \, \mathbb{S} \, dt + \mathcal{T}_r \, \mathbb{S} \, dr + \left(\mathcal{L}^S \, \mathbb{S}_i + \mathcal{L}^V \mathbb{V}_i \right) r d\Omega^i \, &, \qquad l \geq 1 \, .
	\end{cases}
\end{equation}
Working in harmonic gauge requires that
\begin{equation}
	\nabla^\nu \nabla_\nu \xi_\mu \= 0\, . 
\end{equation}
 
The induced metric on the timelike boundary is given by 
\begin{equation}
	\left. ds^2 \right|_{r=\mathfrak{r}} \= -dt^2  +\mathfrak{r}^2 d\Omega^2 \, . 
	\end{equation} 
 
Instead of imposing conformal boundary conditions on this timelike boundary, we impose Dirichlet boundary conditions,
\begin{equation}\label{eqn: spherical Dirichlet bdry cond}
	\left. h_{m n} \right|_{r=\mathfrak{r}} \= 0 \, .
\end{equation}
Inserting the ansatz \eqref{eqn: spherical Dirichlet ansatz} into the boundary conditions \eqref{eqn: spherical Dirichlet bdry cond}, one obtains a set of boundary conditions in terms of $\mathcal{T}_t$, $\mathcal{T}_r$, $\mathcal{L}^S$, and $\mathcal{L}^V$, that depend on $l$. We are interested in physical diffeomorphisms, so we require that $\xi_\mu$ does not obey the condition \eqref{eqn: diffeo bdry cond perp}, and hence
\begin{equation}\label{eqn: Dirichlet sphere physical diffeo cond}
	\left. \mathcal{T}^r \right|_{r=\mathfrak{r}} \neq 0 \, .
\end{equation}
 
Using the metric perturbation constructed from \eqref{eqn: spherical Dirichlet ansatz} that satisfies both \eqref{eqn: spherical Dirichlet bdry cond} and \eqref{eqn: Dirichlet sphere physical diffeo cond}, we now show that it is not possible to build waves from such solutions which have compact support outside the timelike tube at some initial time while subsequently moving into the interior of the tube.
 
Let us first consider the $l=0$ case, the boundary conditions \eqref{eqn: spherical Dirichlet bdry cond} become 
\begin{eqnarray}
\begin{cases}
	\left. \partial_t \mathcal{T}_t \right|_{r=\mathfrak{r}} & \=0  \,, \\
	\left. \mathcal{T}_r \right|_{r=\mathfrak{r}} & \= 0 \, .
	\end{cases}
\end{eqnarray}
It is straightforward to see that the second condition contradicts \eqref{eqn: Dirichlet sphere physical diffeo cond}. Hence, there is no physical diffeomorphism with $l=0$.
 
For $l=1$, the boundary conditions \eqref{eqn: spherical Dirichlet bdry cond} become
\begin{eqnarray}
\begin{cases}
	\left. \partial_t \mathcal{T}_t \right|_{r=\mathfrak{r}} & \=0  \,, \\
	\left. r\partial_t \mathcal{L}^S - \sqrt{2}\mathcal{T}_t \right|_{r=\mathfrak{r}} & \= 0 \,, \\
	\left. \mathcal{L}^S +\sqrt{2}\mathcal{T}_r  \right|_{r=\mathfrak{r}} & \= 0  \,.
	\end{cases}
\end{eqnarray}
Imposing the condition \eqref{eqn: Dirichlet sphere physical diffeo cond}, one finds that the solution must either be constant\footnote{This solution was studied in \cite{Andrade:2015gja}, where it was shown using symplectic structure in the phase space that it is unphysical when no black hole is present.} or linear in time. By radially extending this to the bulk, a physical diffeomorphism with $l\=1$ which preserves the Dirichlet boundary data is obtained. However, as the resulting diffeomorphism linearly grows in time, it is fixed once we impose the initial condition.
 
Lastly, we consider solutions with $l\geq2$. The Dirichlet boundary conditions \eqref{eqn: spherical Dirichlet bdry cond} now impose that 
\begin{eqnarray}
\begin{cases}
	\left. \partial_t \mathcal{T}_t \right|_{r=\mathfrak{r}} & \=0  \,, \\
	\left. r\partial_t \mathcal{L}^S - \sqrt{l(l+1)}\mathcal{T}_t \right|_{r=\mathfrak{r}} & \= 0 \,, \\
	\left. \mathcal{L}^S \right|_{r=\mathfrak{r}} & \= 0  \,, \\
	\left. \sqrt{l(l+1)}\mathcal{L}^S + 2\mathcal{T}_r  \right|_{r=\mathfrak{r}} & \= 0  \,, \\
	\left. \mathcal{L}^V \right|_{r=\mathfrak{r}} & \= 0  \,.
	\end{cases}
\end{eqnarray}
Note that combining the third and fourth boundary conditions above, contradicts  \eqref{eqn: Dirichlet sphere physical diffeo cond}. Hence, there is no physical diffeomorphism with $l\geq2$ \cite{Andrade:2015fna}.
 
In summary, when $l=0$ and $l\geq2$, there is no metric built from a pure diffeomorphism satisfying both Dirichlet boundary conditions \eqref{eqn: spherical Dirichlet bdry cond} and the condition for being physical \eqref{eqn: Dirichlet sphere physical diffeo cond}, at the same time. For $l=1$, there exists such solutions, however the boundary conditions force them to grow linearly in time which implies that they are uniquely fixed once the initial conditions are specified. 

As a result, and in contrast to the standard corner analysis of \cite{An:2021fcq,Anninos:2022ujl}, geometric uniqueness appears to be preserved for spherical boundaries with Dirichlet boundary conditions. It may be interesting to relate this observation to the improved Euclidean Dirichlet problem in those cases where the extrinsic curvature obeys certain positivity conditions \cite{Anderson:2006lqb,Witten:2018lgb}. In any case, it is tempting to conjecture that for the case of a spherical  spatial boundary, the gravitational Dirichlet problem is non-linearly well-posed, but further analysis is required.

\section{Black hole conformal thermodynamics}\label{sec: euclidean bh}

The problem of a Euclidean black hole in a finite size box with fixed induced metric at the boundary $\Gamma$ was considered in \cite{York:1986it} by York. In this section,  we consider an extension of the York setup for a finite size space subject to the conformal boundary conditions of \cite{An:2021fcq}. Aspects of this problem were also considered in \cite{Odak:2021axr}. The purpose of the section is to assess how the thermodynamic quantities of the black hole are affected by the alternative choice of boundary conditions.

Concretely, we consider a Euclidean black hole solution preserving the following boundary data,
\begin{equation}
	ds^2 |_\Gamma \= e^{2 \boldsymbol{\omega}} \left(d\tau^2 + \mathfrak{r}^2 d\Omega^2\right) \, , \qquad \qquad K \= \text{constant} \, ,
\end{equation}
for some unspecified function $\boldsymbol{\omega}$. We take the topology of $\Gamma$ to be $S^1 \times S^2$.  The Euclidean time coordinate $\tau \sim \tau +\beta$ parameterises the $S^1$ factor. The parameter $\mathfrak{r}$ characterises the size of the $S^2$. Given that only the conformal structure of the boundary $\Gamma$ is specified, only the dimensionless parameter $\beta / \mathfrak{r}$ is geometrically meaningful. The trace of the extrinsic curvature $K$ has dimensions of inverse length, which provides a length scale for the problem.
 
\subsection{Black hole solution} 
 
Assuming stationarity, the Euclidean black hole solution obeying the given boundary data is 
\begin{equation}\label{bhs}
ds^2 \= \frac{e^{2 \boldsymbol{\omega}}}{1-\frac{2M}{r}}dr^2 + e^{2 \boldsymbol{\omega}}\left(\frac{1-\frac{2M}{r}}{1-\frac{2M}{\mathfrak{r}}}d\tau^2 + r^2 d\Omega^2\right) \, ,
\end{equation}
where the Weyl factor is a constant given by
\begin{equation}
	e^{\boldsymbol{\omega}} \= \frac{1-\frac{3M}{2\mathfrak{r}}}{\sqrt{1-\frac{2M}{\mathfrak{r}}}}\frac{2}{K\mathfrak{r}} \, .
\end{equation}
The boundary $\Gamma$ is located at $r=\mathfrak{r}$. To map the solution to more standard Schwarzschild coordinates, one can perform the following coordinate transformation:
\begin{equation}
	r \rightarrow r_\text{Sch} \= e^{\boldsymbol{\omega}} r \, \quad , \quad \,
	\tau \rightarrow \tau_\text{Sch} \=  \frac{e^{\boldsymbol{\omega}}\tau}{\sqrt{1-\frac{2M}{\mathfrak{r}}}} \, ,
\end{equation}
where $r_\text{Sch}$ and $\tau_\text{Sch}$ are the Euclidean Schwarzschild coordinates. One thus identify the mass of the black hole and the physical radius of the tube as 
\begin{eqnarray}
	M_\text{bh}  \equiv e^{\boldsymbol{\omega}} M \, \quad , \quad \, 
	\mathfrak{r}_\text{tube} \equiv& e^{\boldsymbol{\omega}} \mathfrak{r} \, .
\end{eqnarray}
The above solution has a conical singularity at the Euclidean horizon $r_\text{Sch} = 2M_\text{bh}$ or $r = 2M$ unless one identifies
\begin{equation}\label{eqn: roots of bh mass}
	\frac{\beta}{\mathfrak{r}} \= \frac{ 8 \pi M}{\mathfrak{r}}\sqrt{1-\frac{2M}{\mathfrak{r}}} \=   \frac{8 \pi M_\text{bh}}{\mathfrak{r}_\text{tube}}\sqrt{1-\frac{2M_\text{bh}}{\mathfrak{r}_\text{tube}}} \, .
\end{equation}
Since the Weyl factor $e^{\boldsymbol{\omega}}$ is a constant, the parameter $\beta/\mathfrak{r}$ can be interpreted as a ratio between the physical radii of $S^1$ and $S^2$ of the boundary. This equation implies that the dimensionless parameter $\beta/\mathfrak{r}$ controls the size of the black hole relative to the tube. Following \cite{York:1986it,Brown:1992br}, there are two different black hole solutions for a given $\beta/\mathfrak{r}$ if 
\begin{equation}\label{eqn: upper bound of beta/R}
	\frac{\beta}{\mathfrak{r}} \leq \frac{8 \pi}{\sqrt{27}} \, ,
\end{equation}
and only one black hole solution when the equality holds. When $\beta/\mathfrak{r}$ violates the inequality, no black hole of real mass exists. We remark that the upper bound of $\beta/\mathfrak{r}$ in which the black hole exists is independent of the value of $K$. We will denote $M_+$ and $M_-$ the larger and smaller positive roots of  \eqref{eqn: roots of bh mass}, respectively.

\subsection{Thermodynamic quantities}

According to the Gibbons-Hawking prescription for black hole thermodynamics, one can evaluate the on-shell action on the Euclidean black hole solution to compute the leading contribution to the thermal partition function in a canonical ensemble \cite{Gibbons:1976ue}. Here, we would like to consider the problem subject to conformal rather than Dirichlet boundary conditions. To leading order, the modified partition function is related to the Euclidean version of the on-shell action (\ref{eqn: action}) with conformal boundary term  (\ref{IB}) as
\begin{equation}
	\mathcal{Z}\left(\frac{\beta}{\mathfrak{r}},K\right) \approx e^{-I(g^*_{\mu \nu})}~.
\end{equation}

The metric $g_{\mu \nu}^*$ is a classical solution to the Einstein field equation subject to the boundary data $\beta/\mathfrak{r}$ and $K$ on the boundary $\Gamma$ of topology $S^1 \times S^2$. If there are many classical solutions obeying the same boundary data, then we sum all of them when computing $Z$. Since the inverse temperature $\beta$ alone is not fixed but its ratio with the size of the two-sphere $\beta/\mathfrak{r}$ is, we will refer to this as the conformal canonical ensemble. For notational simplicity in what follows we define $\tilde{\beta} \equiv \beta/\mathfrak{r}$.
 
From the previous sub-section, certain regimes in the space  of boundary data permit several classical solutions. Specifically, upon setting $G_N=1$, we find
\begin{eqnarray}
\begin{cases}
	I^{\text{(flat)}} &= -\frac{4 \tilde{\beta}}{3 K^2} \, , \\
	I^{\text{(large bh)}} &= -\frac{4\pi}{3K^2} \frac{M_+}{\mathfrak{r}}\frac{\left(2-\frac{3M_+}{\mathfrak{r}}\right)^3}{1-\frac{2M_+}{\mathfrak{r}}} \, , \\
	I^{\text{(small bh)}} &= -\frac{4\pi}{3K^2} \frac{M_-}{\mathfrak{r}}\frac{\left(2-\frac{3M_-}{\mathfrak{r}}\right)^3}{1-\frac{2M_-}{\mathfrak{r}}} \, . 
	\end{cases}
\end{eqnarray}
We emphasise that $M_\pm/\mathfrak{r}$ should be thought of as a function of $\tilde{\beta}$. The first solution corresponds to the flat Euclidean solution which exists for all values of $\tilde{\beta}$. The latter two correspond to the large and small Euclidean black holes (\ref{bhs}). Both exist only when the inequality \eqref{eqn: upper bound of beta/R} holds, which gives an upper bound for $\tilde{\beta}$. All three Eulidean solutions are defined for any positive value of $K$.
 
Accordingly, one can define the conformal free energy 
\begin{equation}
	\mathcal{F}_\text{conf} \, \equiv \, {I}/{\tilde{\beta}} \, .
\end{equation} 
The Euclidean flat space then has the conformal free energy $\mathcal{F}_\text{conf}^\text{(flat)}=-4/3K^2$ independent of $\tilde{\beta}$. In the low temperature regime, the flat Euclidean solution has the lowest conformal free energy and hence dominates the partition function. At the critical $\tilde{\beta}_c = 32 \pi/27$, the large black hole contribution competes with the flat solution. Then, the large black hole becomes the dominant contribution for high temperatures, $0<\tilde{\beta}<\tilde{\beta}_c$. The small black hole solution is sub-dominant for all $\tilde{\beta}$.
 
One can also define the conformal energy $E_\text{conf}$ and conformal entropy $\mathcal{S}_\text{conf}$ as
\begin{eqnarray}
	E_\text{conf} \, \equiv \, \left. {\partial_{\tilde{\beta}} I} \right|_K \, \quad , \quad \, 
	\mathcal{S}_\text{conf} \, \equiv \, \left. \left(\tilde{\beta} \partial_{\tilde{\beta}} - 1 \right) I \right|_K \, .
\end{eqnarray}
Given these definitions, one can show that all three thermal solutions satisfy a first law relation
\begin{equation}
	\tilde{\beta} \, dE_\text{conf} \=  d \mathcal{S}_\text{conf} \, , \qquad \qquad K \text{ fixed} \, .
\end{equation}
The flat Euclidean solution gives vanishing conformal entropy due to its linear dependence on $\tilde{\beta}$. For both of the black holes, the conformal energy and conformal entropy are given by
\begin{eqnarray}
\begin{cases}
	E_\text{conf} &\= - \frac{4 \left(1-\frac{3M}{2\mathfrak{r}}\right)^2\left(1-\frac{3M}{\mathfrak{r}}\right)}{3K^2 \left(1-\frac{2M}{\mathfrak{r}}\right)^{3/2}}  \= \frac{\left(\frac{3M_\text{bh}}{\mathfrak{r}_\text{tube}}-1\right)\mathfrak{r}_\text{tube}^2}{3\sqrt{1-\frac{2M_\text{bh}}{\mathfrak{r}_\text{tube}}}}  \, , \\
	\mathcal{S}_\text{conf} &\= \frac{16 \pi \left(\frac{M}{\mathfrak{r}}\right)^2 \left(1-\frac{3M}{2\mathfrak{r}}\right)^2}{K^2 \left(1-\frac{2M}{\mathfrak{r}}\right)} \= 4 \pi M_\text{bh}^2 \, .
	\end{cases}
\end{eqnarray}
The final expression of $\mathcal{S}_\text{conf}$ agrees precisely with the Bekenstein-Hawking entropy $A_\text{horizon}/4 G_N$. We note that $\sqrt{1-2M/\mathfrak{r}} \, E_\text{conf}/\mathfrak{r}_\text{tube}$ is equal to the conserved charge in the canonical formalism investigated in \cite{Odak:2021axr}, prior to the addition of a regularisation term.
 
In the high temperature limit or $\tilde{\beta} \rightarrow 0$, the tube approaches the black hole horizon $M/\mathfrak{r} \rightarrow 1/2$, and the entropy scales as
\begin{equation}
	\mathcal{S}_\text{conf} \=  \pi  \left(\frac{3 E_\text{conf}}{K}\right)^{2/3}  \qquad \text{with}  \qquad E_\text{conf} \rightarrow \infty \, .
\end{equation}
 
Finally, one can study the sensitivity of the system as we change $\tilde{\beta}$ while keeping $K$ fixed. This is the heat capacity at constant $K$, which we define as
\begin{equation}
	C_K \, \equiv \, \left. -\tilde{\beta}^2 {\partial_{\tilde{\beta}}^2}  \, I  \right|_K  \, .
\end{equation}
When the large black hole dominates, the heat capacity is given by 
\begin{eqnarray}
	C_K \= - \frac{16 \pi \left(\frac{M_+}{\mathfrak{r}}\right)^2 \left(1-\frac{3M_+}{2\mathfrak{r}}\right)\left(2-\frac{8M_+}{\mathfrak{r}}+\frac{9M_+^2}{\mathfrak{r}^2}\right)}{K^2 \left(1-\frac{2M_+}{\mathfrak{r}}\right)\left(1-\frac{3M_+}{\mathfrak{r}}\right)} \, > \, 0 \, \, \, \, \, \text{for } \, \, \, 0< \tilde{\beta}<\tilde{\beta}_c \, .
\end{eqnarray}
Therefore, when the large black hole dominates the system becomes thermally stable. We also note that in the high temperature limit, the heat capacity diverges quadratically in the temperature, namely
\begin{equation}\label{highT}
	\lim_{ \tilde{\beta} \rightarrow 0} C_K \=  {\frac{8 \pi^3}{G_N K^2 \tilde{\beta}^2}}\,  .
\end{equation}

The thermal phase structure with an unstable small black hole, and a large black hole that dominates over the thermal vacuum at high enough temperatures is reminiscent of the situation for  black holes in anti-de Sitter space \cite{Hawking:1982dh}. Moreover, the quadratic temperature dependence in (\ref{highT}) is similar to that of a large black hole in AdS$_4$ -- or equivalently  a three-dimensional conformal field theory. From this we deduce that the number of putative degrees of freedom scales as $N_{\text{d.o.f.}} \approx \tfrac{1}{G_N K^2}$.\footnote{Note that near the boundary of AdS$_4$ we have $K = 3/\ell$ giving the familiar AdS$_4$/CFT$_3$ expression for $N_{\text{d.o.f.}}$.}  It is straightforward to generalise the high-temperature behaviour to $D$-dimensions, for which the high temperature behaviour goes as $\tilde{\beta}^{2-D}$. The temperature dependence is  different from the high temperature behaviour of a thermally stable black hole in a Dirichlet box \cite{York:1986it} (see also \cite{Alessio:2020lpk,Asante:2021blx}) which instead goes as ${\beta}^2$ for all $D\geq 4$.
\begin{center}
\pgfornament[height=5pt, color=black]{83}
\end{center}
\vspace{5pt}

It is of interest to explore more refined observables subject to conformal boundary conditions in either Lorentzian or Euclidean signature. Of particular interest are  gravitational correlation functions with points anchored on $\Gamma$. The structure of such observables around black hole backgrounds or the vacuum, with vanishing or non-vanishing $\Lambda$, for Lorentzian or Euclidean signature, with the possibility of incorporating matter fields may provide a more quasi-local framework to assess properties of spacetime.

This appears to be a particularly valuable framework for the dynamical characterisation of spacetimes which do not possess an asymptotic region where observables can be naturally anchored. 

For the case of $\Lambda>0$, a more complete characterisation of the dynamical features of the de Sitter horizon  is of particular interest \cite{Anninos:2011zn,Anninos:2018svg,Anninos:2022ujl,Susskind:2021esx,Chapman:2022mqd,Aalsma:2022eru}. According to a theorem of Gao and Wald  \cite{Gao:2000ga}, the Penrose diagram of de Sitter space generally stretches vertically in response to the presence of null-energy preserving excitations. This is in sharp contrast to the behaviour of a black hole, whose Penrose diagram stretches horizontally. Such dynamical effects are tied to the chaotic nature of horizons \cite{Shenker:2013pqa}, now perceived as holographic liquids. Examining how dynamical features of the de Sitter horizon (such as those appearing in the theorem by Gao and Wald) are encoded from a worldtube perspective seems like a  well-motivated exercise.

For the case of $\Lambda<0$, our considerations naturally connect to the AdS/CFT correspondence in the presence of a finite cutoff \cite{McGough:2016lol, Taylor:2018xcy, Hartman:2018tkw, Coleman:2020jte}. 

As a final remark, our worldtube $\Gamma$ might be viewed as an auxiliary part of a more complete picture. Take for instance the Euclidean path-integral for general relativity with $\Lambda>0$,
\begin{equation}
\mathcal{Z}\left[ g_{m n}, K \right] = \int \left[ \mathcal{D} g_{\mu\nu}  \right] \, e^{-I[g_{\mu\nu}]}~,
\end{equation}
on a manifold $\mathcal{M}$ with $S^2\times S^1$ boundary subject to the well-posed conformal data $\{ [g_{m n}]_{\text{conf}},K\}$. We may then ask whether there exists some operation on $\mathcal{Z}$ that retrieves the ordinary Gibbons-Hawking Euclidean $S^4$ path integral \cite{Gibbons:1976ue,Anninos:2020hfj} for general relativity with $\Lambda>0$.  Such an operation may involve path-integrating over  the conformal boundary metric \cite{Alishahiha:2004md,Anninos:2021ydw,Blacker:2023oan}. We have our work cut out for us.

\chapter{Positive cosmological constant}\label{chap: dS}

\section{Introduction}
\label{subsection: Introduction}

In the absence of an asymptotic spatial or null boundary, the construction of gauge invariant observables subject to the constraints of diffeomorphism redundancies in a theory of gravity becomes a challenging task. One is often led to relational notions \cite{geheniau1956invariants,Komar:1958ymq,Bergmann:1961wa} ({for a review on recent work see \cite{Tambornino:2011vg}}) whereby a given physical phenomenon is measured in relation to some other semiclassical feature. For instance, in inflationary models of the early Universe, we can measure the time-dependence of physical phenomena with respect to the slow classical roll of the background inflaton field. From a more quasilocal perspective, one might imagine decorating spacetime with a worldline \cite{fermi,Pirani:1956tn,Manasse:1963zz,Unruh:1976db}, perhaps slightly thickened into a worldtube, and use this as a reference frame for ambient phenomena. This perspective appears to be of particular value for an asymptotically de Sitter spacetime \cite{Anninos:2011af,Anninos:2017hhn,Anninos:2018svg,Coleman:2021nor,Witten:2023qsv,Blacker:2023oan,Loganayagam:2023pfb,Kudler-Flam:2023qfl}, where not only are the Cauchy spatial slices potentially compact, but quasilocal entities are moreover surrounded by a cosmological event horizon rendering most of the expanding portion of spacetime physically obscure. A drawback of the relational approach is that it often necessitates the presence of a semiclassical feature in spacetime, making the general picture away from the semiclassical or perturbative regime difficult to control. 

An alternative, complementary, route may be to study the gravitational theory on a manifold endowed with a quasi-auxiliary timelike boundary $\Gamma$, much like we do when considering gravitational physics in anti-de Sitter space, and try to make sense of general relativity in such a setting. This setup has been the focus of recent work in mathematical relativity \cite{Friedrich:1998xt,Anderson_2008,Sarbach:2012pr,Fournodavlos:2020wde,An:2021fcq,Fournodavlos:2021eye}, accompanied by \cite{Witten:2018lgb,Anninos:2022ujl,Anninos:2023epi} (see also \cite{Figueras:2011va, Adam:2011dn} for related work). Of particular interest to our work is the proposal  of \cite{Anderson_2008,An:2021fcq} that  in four spacetime dimensions certain conformal data along $\Gamma$ lead to a well-posed initial boundary value problem. In \cite{Anderson_2008,An:2021fcq} it is further established that generically, the Dirichlet problem in general relativity---whereby one fixes the induced metric along $\Gamma$---suffers from potential existence and non-uniqueness issues for both Euclidean and Lorentzian signature. Concretely, the conformal boundary conditions of interest fix the conformal class of the induced metric, $[g_{mn}|_\Gamma]_{\text{(conf)}}$, and the trace of the extrinsic curvature, $K$, along $\Gamma$ whilst also specifying standard Cauchy data along a spalike surface $\Sigma$ intersecting $\Gamma$ at its boundary.

In this paper, we explore the conformal boundary conditions of \cite{Anderson_2008,An:2021fcq} for general relativity with a positive cosmological constant $\Lambda$ \cite{Spradlin:2001pw,Anninos:2012qw,Galante:2023uyf}. We consider the problem in both Euclidean and Lorentzian signatures. In Euclidean signature, our main goal is to compute the semiclassical approximation of the gravitational path integral and consider its interpretation from the point of view of Euclidean gravitational thermodynamics \cite{Gibbons:1976ue,Gibbons:1977mu}. In the absence of a boundary, the natural Euclidean geometry for general relativity with $\Lambda>0$ is the sphere \cite{Gibbons:1976ue}, void of any external data such as the size of a thermal circle, and one is led to path integrate fields on top of it. In adding a boundary to our Euclidean manifold, as pointed out in \cite{Hayward:1990zm,Coleman:2021nor,Banihashemi:2022jys} among other places, one has the possibility of providing a novel perspective to the sphere path integral and its rich though elusive physical content \cite{Anninos:2020hfj}.\footnote{More speculatively, as suggested in \cite{Anninos:2021eit}, the presence of boundaries in the $\Lambda>0$ Euclidean path integral might be necessitated in parallel to the non-perturbative necessity of boundaries on the two-dimensional worldsheet of closed strings \cite{Polchinski:1994fq} upon trying to make sense of the sum over topologies. From a Lorentzian perspective, we may imagine the creation of a long-lived heavy particle from the vacuum. These events are Boltzmann suppressed but can occur. In the semiclassical limit, such timelike features become parametrically long-lived and may warrant a treatment involving timeilke boundaries.} In Lorentzian signature, one is led to the question of dynamical features of de Sitter space, known to be dynamically stable at the classical level \cite{Friedrich:1986qfi}, in the presence of a timelike boundary. There are two timelike surfaces of particular interest. One of these is the worldline limit, whereby the spatial size of $\Gamma$ becomes small in units of $\Lambda$. The other is the cosmological horizon limit, whereby $\Gamma$ approaches the cosmological de Sitter horizon. The former limit is of interest in describing the theory of a quasilocal entity, whilst the latter is of interest if one wishes to describe physics from the perspective of a stretched horizon \cite{Susskind:1993if,banks2012holographic,Shaghoulian:2021cef,Shaghoulian:2022fop,Narovlansky:2023lfz}.

\subsubsection*{Organisation and summary of main results}

In subsection \ref{sec: general}, we present the general framework, and provide an explicit definition of the conformal boundary conditions. As noted, our gravitational theory is endowed with a positive cosmological constant $\Lambda = +(D-1)(D-2)/2\ell^2$ in $D$ spacetime dimensions.

In subsections \ref{sec: 3d conformal thermo} and  \ref{sec: 4d}, we consider the problem in Euclidean signature for $D=3$ and $D=4$ spacetime dimensions respectively. The boundary of our manifold is taken to have an $S^1\times S^{D-2}$ topology. In the standard treatment of semiclassical black hole thermodynamics \cite{Gibbons:1976ue} with Dirichlet boundary conditions, one defines the canonical ensemble by fixing the size of the boundary $S^1$ to be the inverse temperature $\beta$ and the radius of the spatial sphere to be fixed to some size   $\frakr$. In our treatment, we will instead fix the conformal class of the boundary metric. As such, we fix a conformal version of the inverse temperature, $\tilde{\beta} \equiv \beta/\frakr$. The other boundary data we fix is the trace of the extrinsic curvature, $K$. We refer to this ensemble as the conformal canonical ensemble. Gravitational solutions in the conformal canonical ensemble include patches with no horizons, referred to as {pole patches}, patches with cosmological horizons, referred to as {cosmic patches}, and patches with black hole horizons, referred to as {black hole patches}. Upon tuning $\tilde{\beta}$ and $K$, one finds that the space is filled with a pure de Sitter spacetime, and we refer to this as a {pure de Sitter patch}.

The complete thermal phase space of static and spherically symmetric solutions at a given  $\tilde{\beta} > 0$ and $K \ell \in \mathbb{R}$ is provided in both $D=3$ and $D=4$ spacetime dimensions. Below we summarise the main results, with an emphasis on the conformal thermodynamics of the pure de Sitter patch.

\textbf{Three-spacetime dimensions.} Our analysis in $D=3$ spacetime dimensions naturally builds on recent developments on dS$_3$ \cite{Coleman:2021nor,Coleman:2020jte,Shyam:2021ciy,Anninos:2021ihe}. We find that upon imposing conformal boundary conditions:

\begin{itemize}
\item Both a pole and a cosmic patch exist at any value of $\tilde{\beta}$ and $K\ell$.

\item The entropy of the cosmic patch is given by the Gibbons-Hawking entropy of the cosmological horizon, and the specific heat is positive for all $\tilde{\beta}$ and $K\ell$.

\item The thermodynamic quantities take the form of a two-dimensional conformal field theory. Viewed as such, we identify a $c$-function
\begin{equation}
   \frak{c}_{\text{conf}} = \frac{3\ell}{4G_N}\left(\sqrt{K^2\ell^2+4}-K\ell\right)\,,
\end{equation}
which decreases monotonically as one goes from the worldline limit, where $K\ell\to-\infty$, to the stretched horizon, where $K\ell \to +\infty$.

\item There is a phase transition at $\tilde{\beta}_c = 2\pi$ (for all values of $K \ell$). At $\tilde{\beta}>\tilde{\beta}_c$, the cosmic patch is the thermally preferred solution. In the stretched horizon limit, the cosmic patch is thermally stable, while in the worldline limit it is metastable.

\end{itemize}

The thermodynamic picture in $D=3$ contrasts that of the canonical ensemble, obtained by imposing Dirichlet boundary conditions. In the canonical ensemble, the specific heat of the cosmic patch is always negative.

\textbf{Four-spacetime dimensions.} We now summarise the situation in $D=4$ spacetime dimensions, which naturally builds on previous work in \cite{Hayward:1990zm,Wang:2001gt,Draper:2022ofa, Banihashemi:2022jys,Banihashemi:2022htw}. We find that upon imposing conformal boundary conditions:
\begin{itemize}
\item A pole patch solution exists for any value of $\tilde{\beta}$ and $K\ell$. On the other hand, cosmic/black hole horizon patch solutions only exist for certain values of $\tilde{\beta}$ and $K\ell$. When they do exist, we identify three distinct solutions for a given $\tilde{\beta}$ and $K \ell$---one pole patch and two horizon patches,  which can be of the cosmic or black hole type.

\item The entropy of the solutions with horizons is always given by their respective horizon area formulas. The solution with the larger horizon area always has positive specific heat, while the one with a smaller horizon area always has negative specific heat.

\item When the cosmic patches have positive specific heat, one can take a large temperature limit. In this limit, the entropy goes as $ N_{\text{d.o.f.}}/\tilde{\beta}^{2}$, and thus resembles that of a conformal field theory in three dimensions. We identify
\begin{equation}
    N_{\text{d.o.f.}} = \frac{32 \pi^3 \ell^2}{81 G_N} \left(\sqrt{K^2\ell^2+9}-K\ell\right)^2 \,.
\end{equation}
Further taking the large $K\ell$ limit of the above expression yields $N_{\text{d.o.f.}} \approx \tfrac{8 \pi^3}{G_N K^2}$, a behaviour identified in \cite{Anninos:2023epi} for black holes in Minkowski space subject to conformal boundary conditions.

\item We identify pure dS$_4$ patches with positive specific heat. These solutions exist when the tube is positioned sufficiently near the cosmological horizon, starting from $\frakr_{\text{tube}} \approx 0.259 \ell$. Depending on $K\ell$, these solutions are either metastable or globally stable. The pure dS$_4$ patch is thermally stable in the stretched horizon limit, at least among the spherically symmetric sector. Near the worldline regime, pure dS$_4$ patches  have negative specific heat. The full phase diagram is presented in figure \ref{fig: phase4d}. 
\end{itemize}
We can contrast the thermodynamic behaviour above to that of the canonical thermal ensemble stemming from Dirichlet boundary conditions \cite{Hayward:1990zm,Wang:2001gt,Draper:2022ofa, Banihashemi:2022jys,Banihashemi:2022htw}. In the latter, the specific heat of the pure dS$_4$ patch is always negative.

In subsection \ref{sec: dynamics}, we consider the four-dimensional Lorentzian picture. The theory is placed on a manifold with timelike boundary of $\mathbb{R} \times S^{2}$ topology. In the Lorentzian case, one must further supply standard Cauchy data along the initial time spatial slice $\Sigma$. Employing the Kodama-Ishibashi method \cite{Kodama:2000fa}, we present the linearised gravitational dynamics about the pure de Sitter solution. The linearised solutions split into vector and scalar modes concerning their transformation properties under $SO(3)$. We are mainly interested in two limits: one in which the boundary is close to the worldline observer, which we call the worldline limit; and a second one, in which the boundary becomes close to the cosmological horizon, which we call the stretched horizon limit. In each case, the main results of our analysis are:
\begin{itemize}
\item In the worldline limit of the cosmic patch, we retrieve a set of modes that approximate the quasinormal modes of the static patch \cite{Lopez-Ortega:2006aal}, whilst also uncovering a family of modes in the scalar sector with a negative imaginary part. The latter modes have a Minkowskian analogue uncovered in \cite{Anninos:2023epi}.
\item In the stretched horizon limit, our modes degenerate into a variety of modes. The low-lying vector modes match a set of modes identified in \cite{Anninos:2011zn} as a type of linearised shear mode for an incompressible non-relativistic Navier-Stokes equation. The scalar modes take either the form of a sound mode with diverging speed of sound as we approach the horizon limit, or a pair of modes with $\omega \ell = \pm i$. We provide an understanding of these two modes from a purely Rindler perspective and note that in a local inertial frame the exponential behavior becomes polynomial. 
\end{itemize}

Additional technical details are provided in the various appendices.

\section{General framework} \label{sec: general}

We consider vacuum solutions to general relativity with positive cosmological constant $\Lambda= +(D-1)(D-2)/2\ell^2$ in $D=3$, and $D=4$ spacetime dimensions. In Euclidean signature, the action $I_E$ is given by
\begin{equation}
	I_E \= -\frac{1}{16 \pi G_N} \int_{\mathcal{M}} d^D x \sqrt{\text{det} \, g_{\mu\nu}} \,\left(R-2\Lambda\right) - \frac{\alpha_{\text{b.c.}}}{(D-1) 8 \pi G_N} \int_\Gamma d^{D-1}x \sqrt{\text{det}\,g_{mn}} \, K  \, , \label{euclidean_action}
\end{equation}
where $G_N$ is the Newton's constant, $\Gamma = \partial \mathcal{M}$, $g_{mn}$ denotes the induced metric at $\Gamma$, and the trace of the extrinsic curvature $K$ is given by
\begin{equation}\label{eqn: def of K}
	K \= g^{mn} K_{mn} \, , \qquad K_{mn} \= \frac{1}{2}\mathcal{L}_{\hat{n}}g_{mn} \, .
\end{equation}
Here, $\hat{n} \= \hat{n}^\mu \partial_\mu$ is an outward pointing unit normal vector associated with the boundary, and $\mathcal{L}_{\hat{n}}$ denotes a Lie derivative with respect to $\hat{n}^\mu$. We adopt the notation in which Greek indices $\mu  \= 0,...,D-1$ are used for spacetime indices and $m \=0,...,D-2$ are used for spacetime indices tangent to the boundary.

The constant $\alpha_{\text{b.c.}}$ in (\ref{euclidean_action}) depends on the choice of boundary conditions. In most of the paper we will consider conformal boundary conditions, in which we fix the conformal class of the induced metric and the trace of the extrinsic curvature at the boundary, 
\begin{equation}
	\text{Conformal boundary conditions} \quad : \quad \{\left[g_{mn}|_\Gamma\right]_\text{conf} \, , K|_\Gamma \} \quad=\quad \text{fixed} \, .
\end{equation}
With this set of boundary conditions, the initial boundary value problem in general relativity is proven to be well-posed in Euclidean signature \cite{Anderson_2008,Witten:2018lgb} and conjectured to be well-posed in Lorentzian signature \cite{An:2021fcq, Anninos:2022ujl}. This is in contrast to Dirichlet or Neumann boundary conditions, where general relativity does not permit a well-posed initial boundary value problem for generic boundary data. On occasion, it will be useful to contrast results between different boundary conditions. One has
\begin{equation}
\alpha_{\text{b.c.}} = \begin{cases}
1 & \, \, \text{for conformal boundary conditions} \,, \\
(D-1) & \, \, \text{for Dirichlet boundary conditions} \,, \\
\frac{(4-D) (D-1)}{2} & \, \, \text{for Neumann boundary conditions} \,,
\end{cases}
\end{equation}
to ensure the variational principle is well-defined. Note that for Dirichlet boundary conditions, this gives the standard Gibbons-Hawking-York term \cite{Gibbons:1976ue, yorkbdy} and that conformal and Neumann boundary conditions have the same action in $D=3$ \cite{Odak:2021axr}.

Regardless of the choice of the boundary term, the equations of motion satisfied in the interior manifold are the Einstein field equations
\begin{equation}\label{eqn: Einstein field}
	R_{\mu \nu} - \frac{1}{2}g_{\mu\nu}R + \Lambda  g_{\mu \nu}  \= 0 \, .
\end{equation}

\subsection{Conformal thermodynamics}

Following \cite{Anninos:2023epi}, we would like to study the thermodynamic behaviour of solutions subject to conformal boundary conditions, but now in the presence of $\Lambda>0$.

For this, we take the topology of the boundary to be $S^1 \times S^{D-2}$, and consider the following boundary data,
\begin{equation}\label{eqn: euclidean conf bdry cond}
	\left.ds^2\right|_{\Gamma} \= e^{2 \boldsymbol{\omega}}\left(d\tau^2 + \frakr^2 d\Omega^2_{D-2}\right)\, , \qquad\qquad K \= \text{constant} \, ,
\end{equation}
where $\boldsymbol{\omega}$ is an unspecified function that in principle could depend on boundary coordinates,\footnote{In most of the paper, we will consider solutions that have constant $\boldsymbol{\omega}=\omega$, but it is also possible to find non-static solutions where $\boldsymbol{\omega}$ depends on the boundary coordinates. For instance, consider the case where the metric in the bulk is purely de Sitter in $D=3$, with radius $\ell$. Fixing the trace of the extrinsic curvature at $r=\frakr$ to be a constant, $K$, imposes the following differential equation on $\boldsymbol{\omega}$ (assuming it is only a function of boundary time $\tau$),
\begin{equation}\label{key}
    \frakr^2\partial_\tau^2\boldsymbol{\omega}  = 1-(\frakr\partial_\tau \boldsymbol{\omega})^2-\frac{2\frakr^2e^{2\boldsymbol{\omega}}}{\ell^2}-K\frakr e^{\boldsymbol{\omega}} \sqrt{1-(\frakr\partial_\tau \boldsymbol{\omega})^2 - \frac{\frakr^2 e^{2\boldsymbol{\omega}}}{\ell^2}} \,.
\end{equation}
This equation may have solutions apart from $\boldsymbol{\omega} = \omega$ constant (a preliminary numerical analysis indeed suggests solutions periodic in $\tau$). A similar phenomenon occurs in Lorentzian signature, for higher dimensional cases, and also for $\Lambda = 0$. We leave a full analysis of these solutions for future work. Just as a simple concrete example, one can consider Euclidean de Sitter solutions in $D=3$. One solution is simply given by choosing $\boldsymbol{\omega}$ constant, which gives $K \ell$ as in (\ref{poleK}). One could also consider the (Euclidean) de Sitter slicing. In this case, the bulk metric can be conveniently written as 
\begin{equation}
    \frac{ds^2}{\ell^2} = d\rho^2 + \frac{\sin^2\rho}{\frakr^2\cosh^2\tau/\frakr} \left( d\tau^2 + \frakr^2 d\phi^2 \right) \,,
\end{equation}
which at a constant $\rho=\rho_0$, has the same boundary conditions as in (\ref{eqn: euclidean conf bdry cond}), but now $\boldsymbol{\omega}$ depends on $\tau$. The trace of the extrinsic curvature at the boundary is given by $K \ell = 2 \cot \rho_0$, so one can choose $\rho_0$ so that both solutions have the same boundary data. Nonetheless, the time-symmetric spatial slice with $\tau = 0$, which has vanishing extrinsic curvature, has a different proper area than the constant $\boldsymbol{\omega}$ solution. Moreover, the above metric is not periodic in $\tau$. A Lorentzian version of these configurations is obtained by taking $\tau \to i t$. A subset of solutions to (\ref{key}) will appear at the linearised level, and we analyse them in appendix \ref{sec: l0/l1 modes}. These additional solutions need not spoil the uniqueness properties of the Lorentzian conformal boundary conditions, as they have distinguishable Cauchy data.} 
and $d\Omega^2_{D-2}$ is the round metric of the unit $(D-2)$-sphere. The Euclidean time coordinate $\tau \sim \tau + \beta$ parameterises the $S^1$ factor. The parameter $\frakr$ characterises the size of the $S^{D-2}$. Given that only the conformal class of the metric is specified, only the dimensionless parameter $\tilde{\beta}\equiv\beta/\frakr$ is geometrically meaningful. 

To define the conformal canonical ensemble, we consider a partition function $\mathcal{Z}(\tilde{\beta},K)$ as
\begin{equation}\label{eqn: defining partition function}
	\mathcal{Z}(\tilde{\beta}, K) \,\equiv\, \sum_{g_{\mu\nu}^*} e^{-I_E [g_{\mu\nu}^*]} \, ,
\end{equation}
where $g_{\mu\nu}^*$ are Euclidean metrics satisfying the Einstein field equation \eqref{eqn: Einstein field} and obeying the boundary conditions \eqref{eqn: euclidean conf bdry cond}. Note that if there is more than one solution with the same boundary data, we sum all of them. 

According to the Gibbons-Hawking prescription \cite{Gibbons:1976ue}, we interpret $\mathcal{Z}(\tilde{\beta},K)$ as a leading contribution to the thermodynamics partition function in the $G_N \rightarrow 0$ limit. Since we do not fix the Euclidean time periodicity but rather the dimensionless ratio $\tilde{\beta}$, we interpret this as a thermal system in a conformal canonical ensemble at a fixed conformal temperature $\tilde{\beta}^{-1}$.

Given the partition function $\mathcal{Z}(\tilde{\beta},K)$, one can compute different thermodynamic quantities. For instance, the conformal energy, conformal entropy, and specific heat at fixed $K$ are given by
\begin{equation}\label{eqn: defining thermo quant}
	E_\text{conf} \,\equiv\, - \left.\partial_{\tilde{\beta}}\right|_K \log{\mathcal{Z}} \, , \qquad \mathcal{S}_\text{conf} \,\equiv\, \left.\left(1-\tilde{\beta}\partial_{\tilde{\beta}}\right)\right|_K \log{\mathcal{Z}} \, , \qquad C_K \,\equiv\, \left.\tilde{\beta}^2 \partial^2_{\tilde{\beta}}\right|_K \log{\mathcal{Z}} \, .
\end{equation}

\noindent \textbf{Regular Euclidean solutions.} For certain ranges of $\tilde{\beta}$ and $K$, the Einstein field equation may give rise to a solution $g_{\mu\nu}^*$ which contains a Euclidean horizon. In analogy to the Dirichlet case, requiring the solution to be regular at the horizon fixes its size in terms of $\tilde{\beta}$ and $K$. We will consider $g^*_{\mu\nu}$ that are both static and spherically symmetric, taking the explicit form
\begin{equation}\label{mini}
	ds^2 \= e^{2\omega} \left(\frac{f(r)}{f(\frakr)}d\tau^2 + \frac{dr^2}{f(r)} + r^2 d\Omega^2_{D-2}\right)\, ,
\end{equation}
where $\omega$ is a constant and $f(r)$, a function of $r$ only. (Although it would be interesting to explore the existence of saddles subject to conformal boundary conditions with less restrictive symmetry properties than (\ref{mini}), we will postpone such an analysis to future work.) Note that at the boundary $r=\frakr$ with an outward\footnote{It is also possible to consider solutions with the inward normal vector. In such cases, the trace of the extrinsic curvature at the boundary will have an additional minus sign. For our cases of interest, such choice will give rise to the pole and black hole patches in the next two subsections.} normal vector $\hat{n} \= \sqrt{f(r)}\partial_r$, 
\begin{equation}\label{eqn: conf bdry cond example}
	\left.ds^2\right|_{\Gamma} \= e^{2 \omega}\left(d\tau^2 + \frakr^2 d\Omega^2_{D-2}\right)\, , \qquad\qquad \left.K\right|_{\Gamma}  \= \frac{f'(\frakr)}{2 e^{\omega}\sqrt{f(\frakr)}} + \frac{D-2}{e^{\omega}\frakr}\sqrt{f(\frakr)} \, ,
\end{equation}
so this metric satisfies conformal boundary conditions \eqref{eqn: euclidean conf bdry cond}. We further assume that $f(r)$ has a simple root at $r=r_+$, so that
\begin{equation}
	f(r) \= (r-r_+)f'(r_+) + \mathcal{O}\left(r-r_+\right)^2\, .
\end{equation}
Then, close to $r_+$, its near horizon geometry (to leading order) is given by,
\begin{equation}
	ds^2 \= \rho^2 \left(\frac{f'(r_+)}{2\sqrt{f(\frakr)}}\right)^2 d\tau^2 + d\rho^2 + e^{2\omega}r_+^2 d\Omega^2_{D-2} \, ,
\end{equation} 
where $\rho \,\equiv\, 2 e^\omega \sqrt{\frac{r-r_+}{f'(r_+)}}$. This geometry has a conical singularity near $r\=r_+$ unless one identifies $\tau \,\sim\, \tau + \beta$ with
\begin{equation}\label{eqn: ref beta}
	\beta \= \frac{4\pi\sqrt{f(\frakr)}}{\left|f'(r_+)\right|} \, .
\end{equation} 
Thus, regularity near the horizon fixes the horizon radius $r_+$ in terms of boundary data $\tilde{\beta}$ and $K$.

\section{dS$_3$ conformal thermodynamics}\label{sec: 3d conformal thermo}

We first study conformal thermodynamics of three-dimensional gravity with $\Lambda = + 1/\ell^2 > 0$. A family of static Euclidean solutions to \eqref{eqn: Einstein field} is given by
\begin{equation}\label{eqn: euclidean sol 3D}
	ds^2 \= e^{2\omega} \left(\frac{f(r)}{f(\frakr)}d\tau^2 + \frac{dr^2}{f(r)} + r^2 d\phi^2 \right)\, , \qquad\qquad f(r) \= \frac{\frakr_{\text{c}}^2 - e^{2\omega} r^2}{\ell^2} \, ,
\end{equation} 
where $\tau \,\sim\, \tau \, + \, \beta$ and $\phi \,\sim\, \phi \, + \, 2\pi$. The choice for this particular parameterisation of the solution will soon become evident. The parameter $\omega$ is an unspecified constant and directly controls the physical size of the boundary, namely $\frakr_\text{tube} \= e^{\omega}\frakr$. The cosmological horizon is located at $r\=e^{-\omega}\frakr_{\text{c}}$ and has a physical radius $\frakr_{\text{c}}>0$. There is no black hole horizon in the present setup.\footnote{Note, however, that related theories of three-dimensional gravity with a gravitational Chern-Simons term admit black hole solutions \cite{Anninos:2009jt}. Moreover, quantum black holes have recently been constructed in dS$_3$ \cite{Emparan:2022ijy,Panella:2023lsi}.} However, for $\frakr_{\text{c}} \,\neq\, \ell$, there is a conical defect located at the origin $r\=0$.

Note that one can recover the standard dS static patch coordinates $\left(\tau_\text{static}, r_\text{static}, \phi_\text{static}\right)$ 
via the identification
\begin{equation}
	\tau_\text{static} \= \frac{e^\omega \tau}{\sqrt{f(\frakr)}} \, , \qquad\qquad r_\text{static} \= e^\omega r \, , \qquad\qquad  \phi_\text{static} \= \phi \, .
\end{equation}

It is straightforward to verify that due to the choice of parameterisation \eqref{eqn: euclidean sol 3D}, the metric automatically satisfies the first boundary condition in \eqref{eqn: euclidean conf bdry cond} at $r\=\frakr$. Requiring that the trace of the extrinsic curvature at the boundary is constant, further fixes the parameter  $\omega$ in terms of the boundary data,
\begin{equation} \label{omega}
	e^{2\omega} \= \frac{\frakr_{\text{c}}^2}{2\frakr^2}\frac{\sqrt{K^2\ell^2+4}\pm K\ell}{\sqrt{K^2\ell^2+4}} \,,
\end{equation}
where the $\pm$ corresponds to two different spacetime regions of interest, which we discuss below. We also note that, assuming that $\omega$ does not depend on $\tau$, we find that \eqref{eqn: euclidean sol 3D} is the most general solution to the Einstein field equation in three dimensions.

\begin{figure}[h!]
        \centering
        \includegraphics[scale=0.25]{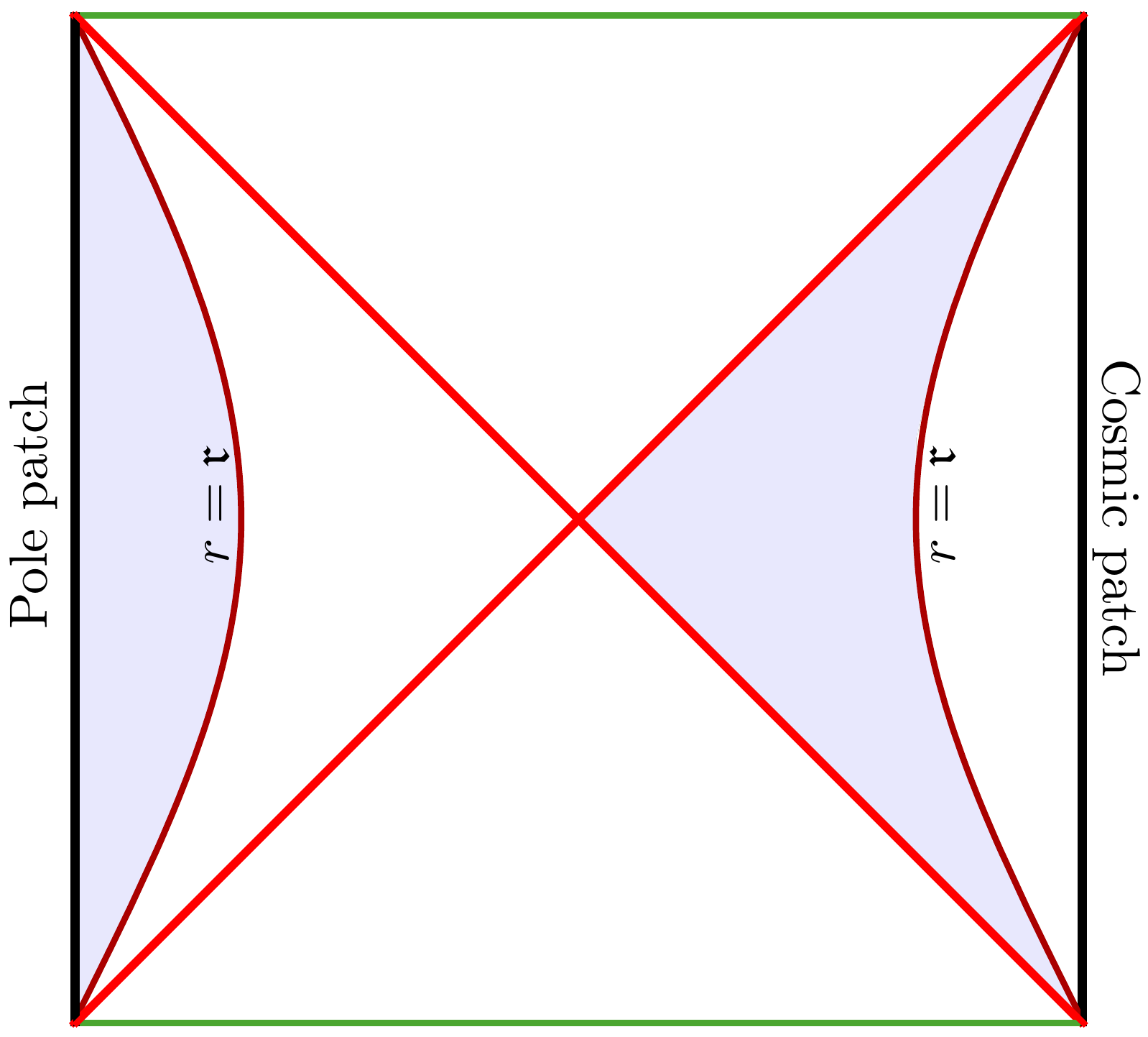}      
                \caption{Penrose diagram of dS space, with a timelike boundary at $r=\frakr$. On the left static patch, the shaded region corresponds to the pole patch, while on the right, it corresponds to the cosmic patch.} \label{fig: cosmic3d}
\end{figure}

We consider two classes of solutions, which we call the pole and the cosmic patch \cite{Coleman:2021nor}. The first one is a patch of spacetime which does not contain the cosmological horizon, while the second one does. In Lorentzian signature they would correspond to the regions shown in figure \ref{fig: cosmic3d}. Below we study the two solutions and their corresponding thermodynamic quantities, separately.

\subsection{Pole patch} 
In the first class of solutions, the cosmological horizon is fixed to be at $\frakr_{\text{c}} \= \ell$, leading to the absence of the conical defect. The spacetime region of interest is $r \,\in\, \left[0,\frakr\right]$, with the boundary at $r \= \frakr \,\leq\, \ell$. We call this spacetime the pole patch of dS. This solution can be obtained by choosing the minus sign in (\ref{omega}).

Imposing that the boundary has a constant trace of the extrinsic curvature $K$ leads to
\begin{equation} \label{poleK}
	K\ell \= \frac{\ell^2-2\frakr_\text{tube}^2}{\frakr_\text{tube}\sqrt{\ell^2-\frakr^2_\text{tube}}} \,,
\end{equation}
which  can be inverted  to obtain
\begin{equation}
	\frakr_\text{tube}^2 \= \frac{\ell^2}{2}\frac{\sqrt{K^2\ell^2+4}-K\ell}{\sqrt{K^2\ell^2+4}} \, .
\end{equation}
The dimensionless parameter $K\ell \,\in\, \mathbb{R}$ controls the size of the boundary. For $K\ell \,\to\, +\infty$ the boundary locates near the origin, whilst for $K\ell \,\to\, -\infty$ it is located near the cosmological horizon. When $K \ell \=0$, the boundary is located exactly at $\frakr_\text{tube} \= \ell/\sqrt{2}$.

Since the cosmological horizon is not part of the pole patch, the parameter $\tilde{\beta}$ is free, and there is a pole patch solution for all values of $\tilde{\beta}$ and $K$.

\textbf{Pole patch thermodynamics.} By evaluating \eqref{euclidean_action} with $\alpha_{\text{b.c.}}=1$ and $D=3$ on the pole patch solution, the on-shell Euclidean action in terms of the boundary data becomes,
\begin{equation}\label{eqn: thermal dS action 3d}
	I_E^{(\text{pole})} \= -\frac{\tilde{\beta}\ell}{16G_N}\left(\sqrt{K^2\ell^2+4}-K\ell\right) \, .
\end{equation}
Since the action depends linearly on $\tilde{\beta}$, one immediately finds that $\mathcal{S}_\text{conf}\=C_K\=0$ and that
\begin{equation}\label{eqn: thermal dS energy 3d}
	E_\text{conf} \= \frac{I_E^{(\text{pole})}}{\tilde{\beta}} =  -\frac{\ell}{16G_N}\left(\sqrt{K^2\ell^2+4}-K\ell\right) \, ,
\end{equation}
which is independent of $\tilde{\beta}$. Note that small fluctuations of the energy can be written as,
\begin{equation}
	\delta E_\text{conf} \= \frac{\frakr_\text{tube}^2}{8G_N} \,\delta K \,.
\end{equation}
Following \cite{York:1986it}, we treat the pole patch of dS as a reference configuration and the on-shell action \eqref{eqn: thermal dS action 3d} as a subtraction term. Therefore, the conformal energy \eqref{eqn: thermal dS energy 3d} plays the role of a vacuum energy. From now onwards, we will compute subtracted quantities such that the energy of the pole patch solution with trace of extrinsic curvature $K$ vanishes.

\subsection{Cosmic patch}
We now consider the second class of static geometries \eqref{eqn: euclidean sol 3D}, which contain the cosmological horizon and hence are dubbed as cosmic patches of dS. 
This is achieved by choosing the plus sign on \eqref{omega} and considering the region $r \,\in\, \left[\frakr,e^{-\omega}\frakr_{\text{c}}\right]$. As a consequence, the conical defect at $r=0$ is not part of the cosmic patch. 

Regularity of the geometry near the cosmological horizon imposes that the inverse conformal temperature of the cosmic patch is given by
\begin{equation}\label{eqn: cosmo ds3 beta}
	\tilde{\beta} \= \frac{2\pi\ell\sqrt{\frakr_{\text{c}}^2 - \frakr_{\text{tube}}^2}}{\frakr_{\text{c}} \frakr_\text{tube}} \,, 
\end{equation}
which is always greater than zero. The conformal temperature $\tilde{\beta}^{-1}$ becomes zero as the boundary approaches the origin. On the other hand, the conformal temperature diverges to infinity as the boundary approaches the cosmological horizon.

Requiring that the boundary has a constant trace of the extrinsic curvature $K$, fixes
\begin{equation}\label{eqn: cosmo ds3 K}
	K \ell \= -\frac{\frakr_{\text{c}}^2-2\frakr_\text{tube}^2}{\frakr_\text{tube}\sqrt{\frakr_{\text{c}}^2-\frakr^2_\text{tube}}} \,,
\end{equation}
which can take any real value. Contrary to the pole patch, the limit of $K \ell$ going to positive and negative infinity now corresponds to the limit of the boundary approaching the cosmological horizon and the origin, respectively. This is expected as the normal vector now points in the opposite direction. 

Using \eqref{eqn: cosmo ds3 beta} and \eqref{eqn: cosmo ds3 K}, we can express $\frakr_{\text{c}}$ and $\frakr_\text{tube}$ in terms of the boundary data $\tilde{\beta}$ and $K\ell$,
\begin{equation}
	\frakr_{\text{c}} \= \frac{\pi\ell}{\tilde{\beta}}\left(\sqrt{K^2\ell^2+4}-K\ell\right) \, , \qquad\qquad \frakr_\text{tube} \= \frac{\sqrt{2}\pi\ell}{\tilde{\beta}}\sqrt{\frac{\sqrt{K^2\ell^2+4}-K\ell}{\sqrt{K^2\ell^2+4}}} \, .
\end{equation}
Interestingly, both $\frakr_{\text{c}}$ and $\frakr_\text{tube}$ depend linearly on the conformal temperature $\tilde{\beta}^{-1}$. This fact implies that the cosmological horizon $\frakr_{\text{c}}$ is a monotonically increasing function of the conformal temperature, 
which contrasts with the Dirichlet problem, where one finds an opposite behaviour, see appendix \ref{sec:3d_Dirichlet}. 

\begin{figure}[h!]
        \centering
        \includegraphics[width=9 cm]{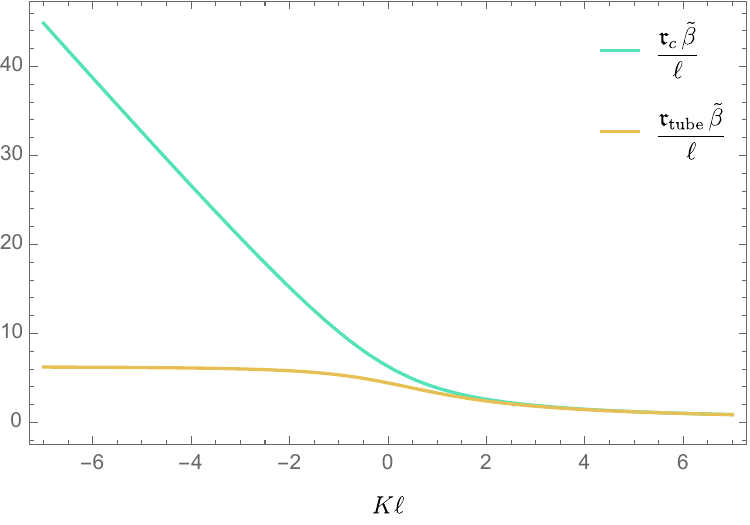}      
                \caption{Plot of $\frac{\frakr_{\text{c}}\tilde{\beta}}{\ell}$ (green) and $\frac{\frakr_\text{tube}\tilde{\beta}}{\ell}$ (yellow) as functions of $K\ell$. For $K \ell \to \infty$, the tube becomes very close to the cosmological horizon; for $K \ell \to - \infty$, the tube approaches $\frac{\frakr_\text{tube}\tilde{\beta}}{\ell} = 2\pi$.} \label{fig: rc rt versus K 3d}
\end{figure}

Additionally, for any positive $\tilde{\beta}$ and real $K$, one always finds that $0\,<\,\frakr_\text{tube}\,<\, \frakr_{\text{c}}$. We show $\frac{\frakr_{\text{c}} \tilde{\beta}}{\ell}$ and $\frac{\frakr_\text{tube} \tilde{\beta}}{\ell}$ as functions of $K\ell$ in figure \ref{fig: rc rt versus K 3d}. 

\begin{figure}[h!]
        \centering
        \includegraphics[width=9 cm]{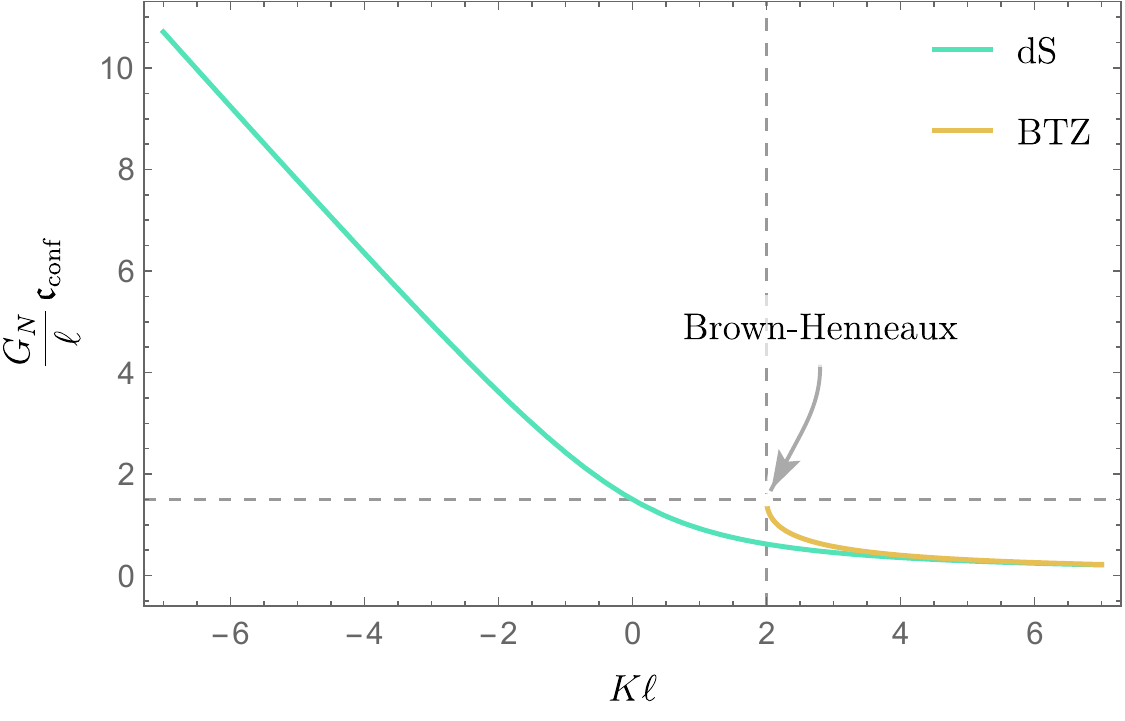}      
                \caption{The quantity $\frak{c}_{\text{conf}}$ as a function of $K \ell$. This central charge decreases monotonically as a function of $K \ell$. As comparison, we included the plot of the analogous $\frak{c}_{\text{conf}}^{\text{BTZ}}$, for the BTZ black hole with conformal boundary conditions. The dS and AdS radii are chose to be equal. In the limit where $K \ell \to \infty$ both central charges coincide, as the tube in both cases gets very close to the horizon and they both exhibit Rindler behaviour. In dashed lines, we show the position of the conformal boundary in the AdS case, where we recover the Brown-Henneaux central charge.
              } \label{fig: central charge 3d}
\end{figure}

 \textbf{Cosmic patch thermodynamics.} To compute thermodynamic quantities, we evaluate the cosmic patch solution on-shell in the Euclidean action \eqref{euclidean_action}. It is useful to define a regulated quantity by subtracting the pole patch action \eqref{eqn: thermal dS action 3d} with the same boundary data. Then,  the associated regulated action corresponding to the cosmic patch solution is given by
\begin{equation}\label{eqn: I_E^cosmo 3d}
	I_{E\text{, reg}}^{(\text{cosmic})} (K) \,\equiv\, I_E^{(\text{cosmic})} - I_E^{(\text{pole})} \= - \frac{\pi \ell}{8G_N}\left(\frac{2\pi}{\tilde{\beta}}-\frac{\tilde{\beta}}{2\pi}\right) \left(\sqrt{K^2\ell^2+4}-K\ell\right)\, .
\end{equation}
We emphasise that the pole action that we subtract has the same trace of extrinsic curvature $K$. We further note that $I_{E\text{, reg}}^{(\text{cosmic})} (K)$ is invariant under $\frac{\tilde{\beta}}{2\pi} \rightarrow - \frac{2\pi}{\tilde{\beta}}$.
The conformal energy and conformal entropy are given by
\begin{equation}\label{eqn: conformal energy entropy cosmic 3d}
	\begin{cases}
		E_\text{conf} &=\, \frac{\pi \ell}{8 G_N \tilde{\beta}} \left(\frac{2\pi}{\tilde{\beta}} + \frac{\tilde{\beta}}{2\pi}\right)\left(\sqrt{K^2\ell^2+4}-K\ell\right) \, , \\
		\mathcal{S}_\text{conf} &=\, \frac{\pi^2 \ell}{2G_N \tilde{\beta}} \left(\sqrt{K^2\ell^2+4}-K\ell\right) \, .
	\end{cases}
\end{equation}
The entropy $\mathcal{S}_\text{conf}$ agrees with the Gibbons-Hawking entropy $A_\text{horizon}/4G_N$ of the cosmological horizon. 
We can also compute the specific heat at constant $K$, which is given by
\begin{equation}\label{eqn: specific heat cosmic 3d}
	C_K \= \frac{\pi^2\ell}{2G_N\tilde{\beta}} \left(\sqrt{K^2\ell^2+4}-K\ell\right) \, .
\end{equation}
The specific hear is positive for all allowed values of $\tilde{\beta}$ and $K$, which means that this configuration is thermally stable under small thermal fluctuations. This is in contrast to the specific heat of the cosmic patch with Dirichlet boundary conditions, which is always negative, as reviewed in appendix \ref{sec:3d_Dirichlet}.
Moreover, note that $C_K$ grows linearly with the conformal temperature. This behaviour resembles that of a two-dimensional conformal field theory at finite temperature. This observation can be sharpened 
upon expressing $\tilde{\beta}$ in terms of $E_\text{conf}$, whereby the conformal entropy and specific heat can be written as
\begin{equation}\label{eqn: entropy cardy}
	\mathcal{S}_\text{conf} \= 2 \pi \sqrt{\frac{\frak{c}_{\text{conf}}}{3}\left(E_\text{conf}-\frac{\frak{c}_{\text{conf}}}{12}\right)} \, \ , \qquad\qquad \, C_K  \= \frac{2\pi^2\frak{c}_{\text{conf}}}{3\tilde{\beta}}  \, ,
\end{equation}
where
\begin{equation}\label{eqn: central charge 3d}
	\frak{c}_{\text{conf}} \equiv \frac{3\ell}{4G_N}\left(\sqrt{K^2\ell^2+4}-K\ell\right) \, ,
\end{equation}
which is a monotonically-decreasing function of $K \ell$ displayed in figure \ref{fig: central charge 3d}.\footnote{A similar computation can be done for the BTZ black hole in AdS$_3$ with conformal boundary conditions \cite{Shyam:2021ciy,shaghoulian}. The resulting value for this function is now given by $\frak{c}_{\text{conf}}^{\text{BTZ}} = \frac{3 \ell_{\text{AdS}}}{4 G_N} (K\ell_{\text{AdS}}-\sqrt{K^2\ell_{\text{AdS}}^2-4})$. This is also a monotonically decreasing function of $K \ell_{\text{AdS}}$. It approaches its maximal real value as $K \ell_{\text{AdS}} \to 2$, which corresponds to the AdS conformal boundary, and for which we recover the standard Brown-Henneaux central charge \cite{Brown:1986nw}.} Note that the entropy in \eqref{eqn: entropy cardy} resembles the Cardy formula, describing the growth of states of energy $E_\text{conf}$ in a two-dimensional conformal field theory of central charge $\frak{c}_{\text{conf}}$ \cite{Cardy:1986ie}. By considering large positive/negative $K\ell$ limits, 
\begin{equation}
	\frak{c}_\text{conf} \,\rightarrow\, 
	\begin{cases}
		\frac{3}{2G_N K} \, , & \text{as} \quad K \ell \rightarrow \infty \, , \\
		\frac{3|K|\ell^2}{2G_N} \, , & \text{as} \quad K\ell \rightarrow - \infty \, .
	\end{cases}
\end{equation}

Finally, using \eqref{eqn: entropy cardy}, we find a first-law type of relation in which
\begin{equation} 
	\delta \left(E_\text{conf}-\frac{\frak{c}_{\text{conf}}}{12}\right) \= \tilde{\beta}^{-1}\, \delta \mathcal{S}_\text{conf} - \mu_K \,\delta K\, , 
\end{equation}
where $\mu_K$ can be understood as the chemical potential associated to $K$, 
\begin{equation}
	\mu_K \,\equiv\,  -\frac{\frakr_\text{tube}^2}{8G_N} \, .
\end{equation}

\subsection{Pure dS$_3$ patch}

We now discuss a particular solution, denoted as the pure dS$_3$ solution. The solution arises from tuning the conformal temperature of the system such that $\frakr_{\text{c}} \= \ell$, leading to a pure dS$_3$ with a boundary. For the cosmic patch, this happens at the conformal temperature $\tilde{\beta} = \tilde{\beta}_{\text{dS}}$, which is a function of $K \ell$ given by
\begin{equation}\label{eqn: dS temp 3d}
	\tilde{\beta}_{\text{dS}} (K \ell)  \,\equiv\, \pi \left(\sqrt{K^2\ell^2+4}-K \ell\right) \= 2\pi\left(\frac{2 G_N \, \frak{c}_{\text{conf}}}{3 \ell}\right) \, .
\end{equation}

By using \eqref{eqn: cosmo ds3 K} with $\frakr_{\text{c}} \= \ell$, we can re-express this in terms of the physical size of the boundary $\frakr_\text{tube}$ as
\begin{equation}
	\tilde{\beta}_{\text{dS}}(\frakr_\text{tube}) \= 2\pi \sqrt{\frac{\ell^2}{\frakr_\text{tube}^2}-1} \, .
\end{equation}
We can recover the standard dS temperature when the tube becomes small, 
\begin{equation}
	\tilde{\beta}_{\text{dS}} \frakr_\text{tube} \,\rightarrow\, 2\pi\ell \qquad\qquad \text{as} \qquad \frakr_\text{tube}\,\rightarrow\, 0\,.
\end{equation}
This is the worldline limit of the solution. It also corresponds to a small conformal temperature. On the other hand, we can consider the stretched horizon limit, which is defined by taking $\frakr_\text{tube} \rightarrow \ell$. This is equivalent to a high conformal temperature limit,
\begin{equation}
	\tilde{\beta}_{\text{dS}} \,\rightarrow\, 0 \qquad\qquad \text{as} \qquad \frakr_\text{tube}\,\rightarrow\, \ell\,.
\end{equation}

Now we can use \eqref{eqn: conformal energy entropy cosmic 3d} and \eqref{eqn: specific heat cosmic 3d} to calculate the thermodynamic properties of pure dS$_3$. The conformal entropy and specific heat at constant $K$ are equal and independent of $\frakr_\text{tube}$,
\begin{equation}
	\mathcal{S}_\text{conf} \= C_K \= \frac{\pi \ell}{2G_N} \, .
\end{equation}
We stress again that the specific heat $C_K$ is positive for all values of $\frakr_{\text{tube}}$.
The conformal energy reads
\begin{equation}
	E_\text{conf}  \= \frac{\frak{c}_{\text{conf}}}{12} \left(1-\frac{\frakr_\text{tube}^2}{\ell^2}\right)^{-1} \, .
\end{equation}
Note that in the worldline limit, the energy reduces to the energy $\frak{c}_{\text{conf}}/12$.
One may wonder why the energy does not go to zero as we take worldline limit. The reason is that we are considering subtracted energies. 

To reproduce the full static patch thermodynamics from a cosmic patch solution, we must include the pole patch to complete a full static patch. Note that this pole patch is not the same that we are using to define the regulated action, as this one has the same $\tilde{\beta}$ but the opposite trace of the extrinsic curvature, $-K$. Now it is straightforward to check that, for $\tilde{\beta}\=\tilde{\beta}_{\text{dS}}$, a combination of a pure de Sitter patch with $K$ and a pole patch with $-K$ (without the subtraction terms) indeed reproduces the Gibbons-Hawking result,
\begin{equation}
 \mathcal{Z}^{\text{(cosmic)}} (\tilde{\beta}_{\text{dS}},K) \mathcal{Z}^{\text{(pole)}} (\tilde{\beta}_{\text{dS}},-K) = \exp \frac{\pi \ell}{2 G_N}\, ,
\end{equation}
as the right-hand side is the exponential of the Gibbons-Hawking entropy for the de Sitter horizon. For the energy, we can define a regulated action for this pole patch with opposite $K$,
\begin{equation}
	I_{E\text{, reg}}^{(\text{pole})}(-K) \, \equiv \, I_{E}^{(\text{pole})}(-K) - I_{E}^{(\text{pole})}(K) \= - \frac{\tilde{\beta} K\ell^2}{8G_N} \, .
\end{equation}

In the pure dS$_3$ solution, the conformal energy of the pole patch with $-K$ is negative and it is exactly opposite to the conformal energy of the cosmic patch with $K$ such that 
\begin{equation}
	\left.\left(E_\text{conf}^{(\text{pole})} (-K) - \frac{\frak{c}_\text{conf}}{12}\right) + \left(E_\text{conf}^{(\text{cosmic})} (K) - \frac{\frak{c}_\text{conf}}{12}\right)\right|_{\tilde{\beta} \= \tilde{\beta}_{dS}} \= 0,
\end{equation}
as we expect for the full static patch of de Sitter space.

\subsection{Phase diagram}

\begin{figure}[h!]
        \centering
        \includegraphics[width=9 cm]{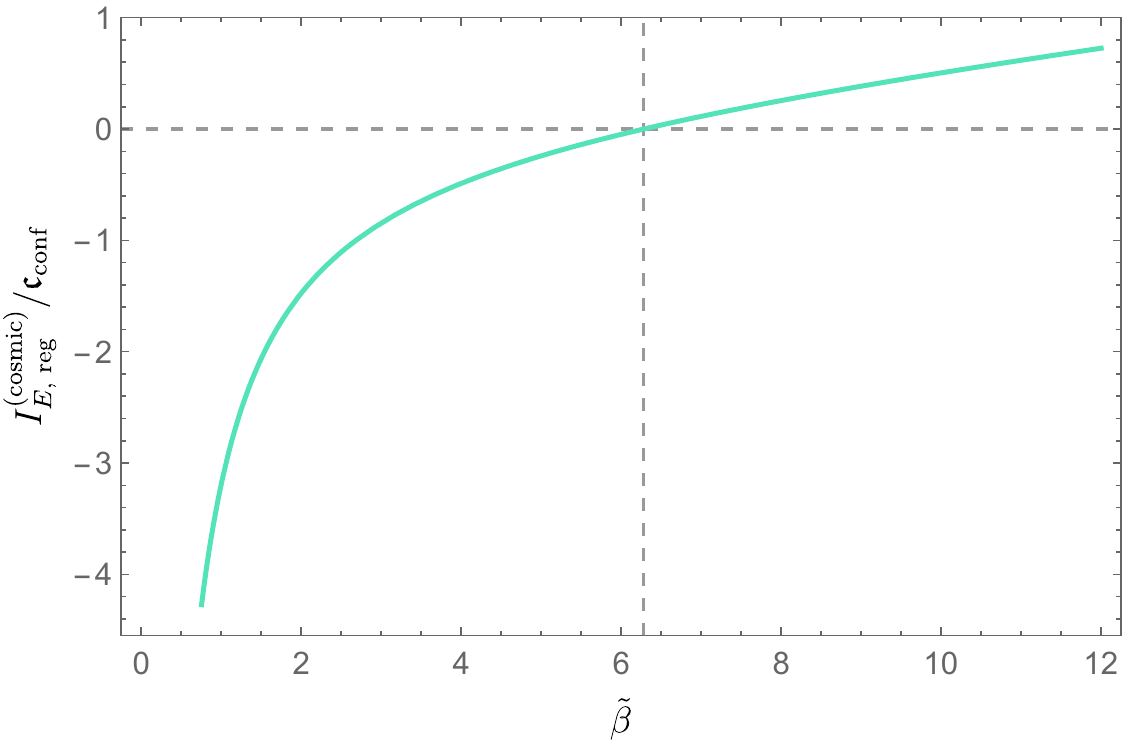}      
                \caption{The regulated on-shell action for the cosmic patch solution, as a function of boundary data $\tilde{\beta}$. The dashed vertical line indicates the critical inverse temperature $\tilde{\beta}_c=2\pi$.} \label{fig: free energy 3d}
\end{figure}

Now we can combine the results from pole patch and cosmic patch thermodynamics. Recall that in $D=3$, both solutions exist for all values of $\tilde{\beta}$ and $K$. This means that one may write the partition function of the total system in the semiclassical limit as
\begin{equation}
	\mathcal{Z} (\tilde{\beta},K) \= \text{max}\left(e^{-I_{E\text{, reg}}^{(\text{cosmic})}(\tilde{\beta},K)}, 1\right) e^{-I_E^{(\text{pole})}(\tilde{\beta},K)} \, ,
\end{equation}
where $I_E^{(\text{pole})}(\tilde{\beta},K)$ and $I_{E\text{, reg}}^{(\text{cosmic})}(\tilde{\beta},K)$ are given by \eqref{eqn: thermal dS action 3d} and \eqref{eqn: I_E^cosmo 3d}, respectively. The sign of $I_{E\text{, reg}}^{(\text{cosmic})}(\tilde{\beta},K)$ therefore determines which configuration is stable/meta-stable. It can be shown that, independently of $K\ell$, there is a critical inverse temperature $\tilde{\beta}\= 2\pi \equiv \tilde{\beta}_c $, for which $I_E^{(\text{cosmic})}(\tilde{\beta},K)$ changes sign, see figure \ref{fig: free energy 3d}. There is a first-order phase transition at $\tilde{\beta} = \tilde{\beta}_c$.

\begin{itemize}
    \item In the low-temperature regime with $\tilde{\beta} \,>\, \tilde{\beta}_c$, $I_{E\text{, reg}}^{(\text{cosmic})}$ is positive for all values of $K \ell$ which implies that the pole patch is thermodynamically favoured. Given it has positive specific heat, the cosmic patch is then metastable.
    \item In the high-temperature regime with $\tilde{\beta} \,<\, \tilde{\beta}_c $, $I_{E\text{, reg}}^{(\text{cosmic})}$ becomes negative, so it becomes thermodynamically favored and a stable configuration.
\end{itemize}

It is interesting to analyse the phase structure of conformal thermodynamics at fixed $K \ell$. Consider the conformal energy along the path of lowest free energy at fixed $K \ell$. Using \eqref{eqn: thermal dS energy 3d} and \eqref{eqn: conformal energy entropy cosmic 3d} and expressing them in terms of the central charge $\mathfrak{c}_{\text{conf}}$, we find that
\begin{equation}
	E_\text{conf} \= 
	\begin{cases}
		\frac{\mathfrak{c}_{\text{conf}}}{12}\left(1+\frac{\tilde{\beta}_c^2}{\tilde{\beta}^2}\right) \, ,& \qquad \tilde{\beta}\,<\, \tilde{\beta}_c\, , \\
		0\, , & \qquad \tilde{\beta} \,\geq\,  \tilde{\beta}_c \, .
	\end{cases}
\end{equation}
The discontinuity of $E_\text{conf}$ at $\tilde{\beta} \= \tilde{\beta}_c$ reflects a first-order phase transition. 

For the conformal entropy, we find a behaviour similar to the Hawking-Page transition of the AdS black hole \cite{Hawking:1982dh, Witten:1998zw}. For temperatures lower than $\tilde{\beta}^{-1}_c$, the conformal entropy is zero. There is a discontinuity in the entropy at the critical temperature $\tilde{\beta}^{-1}_c$, after which, in the high temperature regime, the entropy is precisely given by the Gibbons-Hawking entropy. 

\textbf{Pure dS$_3$ phase structure.} For the pure dS$_3$ solution, we  must  constrain the inverse temperature to the dS inverse temperature \eqref{eqn: dS temp 3d}. In this case, $K \ell\=0$ corresponds to $\frakr_{\text{tube}} \= \ell/\sqrt{2}$.

\begin{figure}[h!]
        \centering
        \includegraphics[scale=0.35]{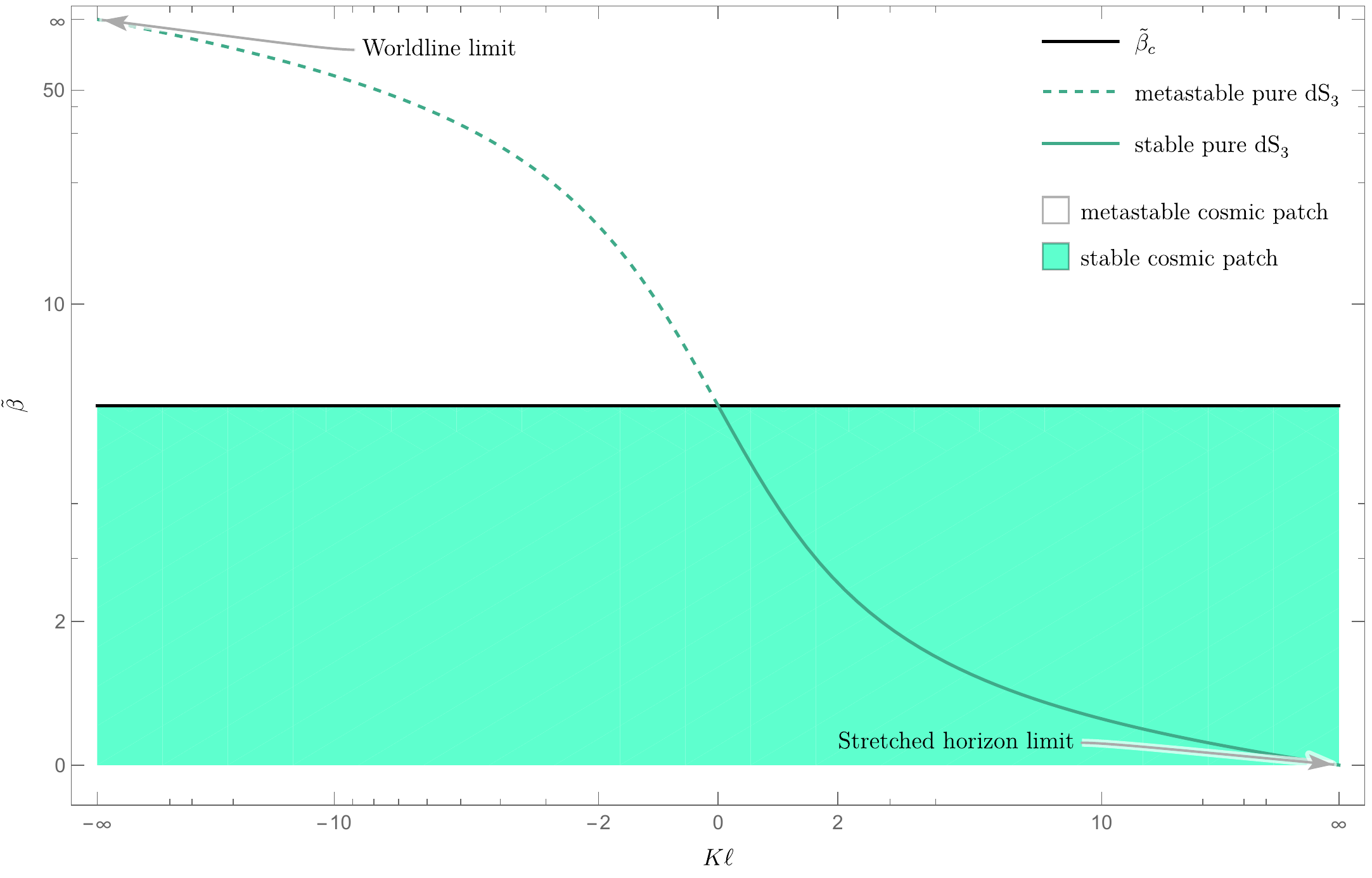}      
                \caption{Phase diagram of conformal dS$_3$ thermodynamics. In each point of the diagram, there co-exist a pole and a cosmic patch solution. There is a critical conformal inverse temperature at $\tilde{\beta}_c=2\pi$, marked in black. For $\tilde{\beta}< \tilde{\beta}_c$ (shaded in green), the cosmic patch is the most favourable configuration. For $\tilde{\beta}> \tilde{\beta}_c$ (in white), the cosmic patch is metastable. The darker green curve shows pure stable (solid) and metastable (dashed) dS$_3$ solutions, that follow relation (\ref{eqn: dS temp 3d}). Worldline and stretched horizon limits of pure dS$_3$ are further indicated. }\label{fig: phase3d}
\end{figure}

\begin{itemize}
    \item $K \ell \,>\,0$ implies that $\frak{c}_{\text{conf}} \,<\,\frac{3\ell}{2G_N}$. In this regime, the dS temperature is higher than the critical temperature, $\tilde{\beta}_{\text{dS}}\,<\,\tilde{\beta}_c$. 
As a consequence, in this regime, the pure dS$_3$ has free energy lower than the pole patch and hence is thermodynamically favoured. 
    \item For $K \ell <0$ or $\frak{c}_{\text{conf}}>\frac{3\ell}{2G_N}$, we find that $\tilde{\beta}_{\text{dS}}\,>\,\tilde{\beta}_c$, 
so the pure dS solution is only metastable. Lastly, at $\frak{c}_{\text{conf}} \= \frac{3\ell}{2G_N}$, the phase transition and dS$_3$ temperature coincide. 
\end{itemize}
The full phase diagram, including the curve of pure dS$_3$ solutions, is depicted in figure \ref{fig: phase3d}.

\subsection{A two-sphere perspective}

As a final remark before moving on to the four-dimensional case, we consider a two-sphere rather than toroidal boundary topology. As our coordinate system, we take
\begin{equation}
    ds^2 \= \frac{\ell^2 d\rho^2}{\ell^2-\rho^2} + \frac{\ell^2-\rho^2}{\ell^2}\left(d\theta^2 + \sin^2\theta d\phi^2\,\right) \, ,
\end{equation}
where for the full three-sphere we have $\rho\,\in\, \left(-\ell,\ell\right)$, $\theta \in (0, \pi)$, and $\phi \,\sim\, \phi+2\pi$. We take the conformal boundary to be located at constant $\rho \= \rho_0$. The induced metric has the conformal structure of the unit $S^2$ metric. Requiring further that the boundary has a constant $K$ fixes
\begin{equation}
    K\ell \= -\frac{2\rho_0}{\sqrt{\ell^2-\rho_0^2}} \, ,
\end{equation}
where we have used a unit normal vector $\hat{n} \= \frac{\ell}{\sqrt{\ell^2-\rho^2}}\partial_\rho$. By computing the on-shell action \eqref{euclidean_action} for  this solution, we can approximate the path integral as
\begin{equation}
    \mathcal{Z}[{S^2}] \approx  \left( \frac{ K \ell - 2i}{\sqrt{K^2 \ell^2 + 4}}  \right)^{\frac{i}{3} \times \frac{3  \ell}{2 G_N}}~. 
\end{equation}
The above expression is real valued. Unlike the thermodynamic expression (\ref{eqn: conformal energy entropy cosmic 3d}) and (\ref{eqn: entropy cardy}), the above expression does not immediately take the form of the two-sphere path integral of a two-dimensional conformal field theory of central charge $\frak{c}_{\text{conf}}$. A similar observation will hold for Euclidean AdS$_3$ with a two-sphere boundary subject to conformal boundary conditions.

\section{dS$_4$ conformal thermodynamics} \label{sec: 4d}

In this subsection, we study conformal thermodynamics of four-dimensional gravity with $\Lambda \= + 3/\ell^2$ for the following family of static and spherically symmetric Euclidean solutions
\begin{equation}\label{eqn: euclidean sol 4D}
	ds^2 \= e^{2\omega} \left(\frac{f(r)}{f(\frakr)}d\tau^2 + \frac{dr^2}{f(r)} + r^2 \left( d\theta^2 + \sin^2{\theta} d\phi^2 \right) \right)\, , \qquad f(r) \= 1- \frac{2\mu}{r} - e^{2\omega} \frac{r^2}{\ell^2}\, ,
\end{equation} 
where $\tau \,\sim\, \tau + \beta$, $\theta \,\in\, (0, \pi)$, and $\phi \,\sim\, \phi + 2\pi$. Similarly to the three-dimensional case, the parameter $\omega$ controls the size of the boundary, namely $\frakr_{\text{tube}} \= e^\omega \frakr$. 

In $D=4$, one further has the Euclidean Schwarzschild-de Sitter solution corresponding to a black hole placed inside the cosmological horizon. The parameter $\mu$ is related to the size of the cosmological horizon $\frakr_{\text{c}}$ through
\begin{equation}
	\mu \= \frac{e^{-\omega}\frakr_{\text{c}}}{2}\left(1-\frac{\frakr_{\text{c}}^2}{\ell^2}\right) \, .
\end{equation}
For $\mu \= 0$, we obtain an empty de Sitter solution with  cosmological horizon at $\frakr_{\text{c}} \= \ell$. For $\mu \,>\, 0$, there is a black hole horizon located at $r\=e^{-\omega}\frakr_{\text{bh}}$ with horizon radius $\frakr_{\text{bh}}$. The cosmological horizon in this case is smaller than the one without the black hole, $\frakr_{\text{c}}\, <\, \ell$. Note that the size of the two horizons are related by
\begin{equation}\label{eqn: relation between rbh and rc}
	\frakr_{\text{bh}} \= \frac{1}{2}\left(\sqrt{4\ell^2-3\frakr_{\text{c}}^2}-\frakr_{\text{c}}\right)  \, .
\end{equation}
Both horizon sizes coincide when $\frakr_{\text{c}} \= \frakr_{\text{bh}} \= \frac{\ell}{\sqrt{3}}$. This is known as the Nariai radius, which also serves as a lower bound of $\frakr_{\text{c}}$.  For $\mu \,<\,0$, there is a naked singularity located at $r \= 0$, and the cosmological horizon is greater than the de Sitter length, $\frakr_{\text{c}} \,>\,\ell$.

As in dS$_3$, we note that the de Sitter static patch coordinates $\left(\tau_\text{static}, r_\text{static}, \theta_\text{static},\phi_\text{static}\right)$ can be recovered 
via the rescaling
\begin{equation}
	\tau_\text{static} \= \frac{e^\omega \tau}{\sqrt{f(\frakr)}} \, , \qquad\qquad r_\text{static} \= e^\omega r \, , \qquad\qquad  \theta_\text{static} \= \theta\, , \qquad\qquad  \phi_\text{static} \= \phi \, .
\end{equation}

We now study these solutions on a four-manifold with an $S^1 \times S^2$ boundary subject to the conformal boundary data \eqref{eqn: euclidean conf bdry cond}. Solving the Einstein equation in terms of boundary data, we obtain multiple expressions for $e^{2\omega}$. The exact expressions in the different ranges of parameters are provided in appendix \ref{sec: useful formulae for 4d thermo}.

\begin{figure}[h!]
        \centering
        \includegraphics[scale=0.35]{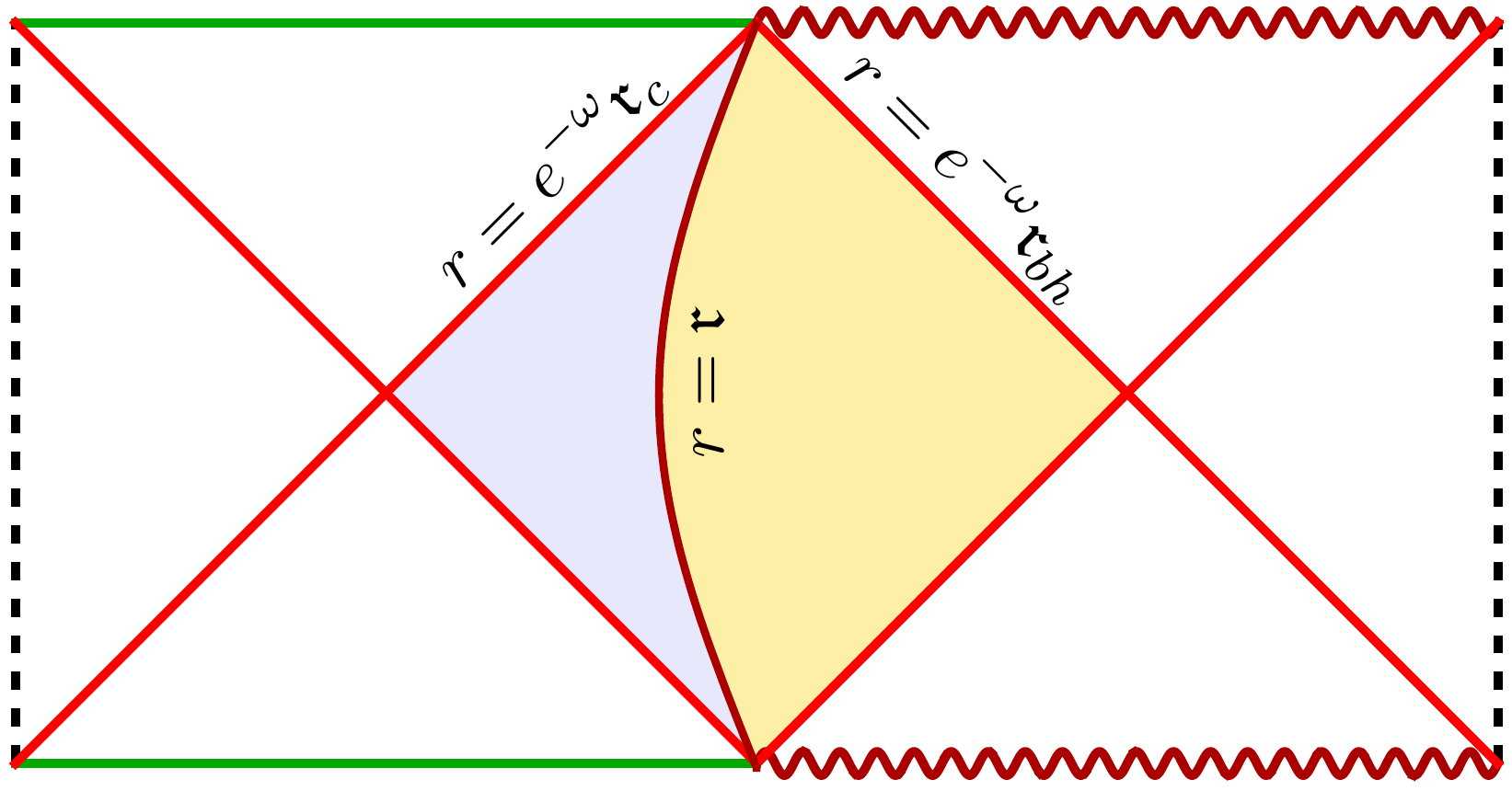}      
                \caption{Penrose diagram of the Schwarzschild de Sitter spacetime. The boundary is given by $r=\frakr$. The shaded blue area corresponds to a cosmic patch, while the yellow one, to a black hole patch. The pole and the pure dS$_4$ patches can be obtained when $\frakr_{\text{c}} = \ell$.} \label{fig: sds}
\end{figure}

There are three different classes of solutions denoted by the pole patch, the black hole patch, and the cosmic patch. Examples of them are displayed in figure \ref{fig: sds}.

\subsection{Pole patch} We begin with the solutions with $\frakr_{\text{c}} \= \ell$ and take the spacetime region of interest to be $r \,\in\, \left[0,\frakr \right]$. As a consequence, the pole patches contain the worldline at $r\=0$ without any horizon.

Imposing that the boundary has a constant trace of the extrinsic curvature $K$ leads to
\begin{equation}
	K\ell \= \frac{2\ell^2- 3\frakr_\text{tube}^2}{\frakr_\text{tube}\sqrt{\ell^2-\frakr_\text{tube}^2}} \, .
\end{equation}
By inverting this equation, we find that the physical size of the boundary can be written as a function of $K \ell$ as
\begin{equation}
	\frakr_\text{tube}^2 \= \ell^2\left(\frac{K^2\ell^2+12-K\ell \sqrt{K^2\ell^2+8}}{2 K^2\ell^2+18}\right) \, .
\end{equation}
The parameter $K\ell$ can take any real value. The limit of large positive and large negative $K\ell$ corresponds to pushing the boundary to the origin and the cosmological horizon, respectively.

Since the pole patches do not contain any horizon, the parameter $\tilde{\beta}$ is free and can take any positive value. Hence, the pole patch exists for all values of $\tilde{\beta}$ and $K$.

\textbf{Pole patch thermodynamics.}
We now evaluate \eqref{euclidean_action} with $\alpha_{\text{b.c.}}=1$ and $D=4$ on the pole patch solution. The on-shell Euclidean action in terms of the boundary data reads,
\begin{equation}\label{eqn: thermal dS action 4d}
	I_E^{(\text{pole})} \= -\frac{\tilde{\beta}\ell^2}{6G_N} \sqrt{\frac{\left(K^2\ell^2+4-K\ell\sqrt{K^2\ell^2+8}\right)\left(K^2\ell^2+12-	K\ell\sqrt{K^2\ell^2+8}\right)}{K^2\ell^2+9}} \, .
\end{equation}
By taking the limit $K\ell \rightarrow \infty$, we find that $I_E^{(\text{pole})} \rightarrow -\frac{4 \tilde{\beta}}{3 G_NK^2}$, retrieving the result in flat space obtained in \cite{Anninos:2023epi}. This is expected as this limit corresponds to a boundary size that is parameterically small compared to the cosmological horizon.

As the action depends linearly on $\tilde{\beta}$, one immediately finds that $\mathcal{S}_\text{conf}\=C_K\=0$ and that
\begin{equation}\label{eqn: thermal dS energy 4d}
	E_\text{conf} \= \frac{I_E^{(\text{pole})}}{\tilde{\beta}} \=  -\frac{\ell^2}{6G_N} \sqrt{\frac{\left(K^2\ell^2+4-K\ell\sqrt{K^2\ell^2+8}\right)\left(K^2\ell^2+12-	K\ell\sqrt{K^2\ell^2+8}\right)}{K^2\ell^2+9}}\, ,
\end{equation}
which is independent of $\tilde{\beta}$. Note that small fluctuations of the energy can be written as
\begin{equation}
	\delta E_\text{conf} \= \frac{\frakr_\text{tube}^3}{3G_N} \,\delta K \,. \label{chemical_pole}
\end{equation}
Curiously, the coefficient in front of $\delta K$ is independent of $\ell$. As in $D=3$, we treat the pole patch of de Sitter as a reference configuration, and the on-shell action \eqref{eqn: thermal dS action 4d} as a subtraction term.

\subsection{Cosmic patch} 

In this subsection, we consider a class of geometries \eqref{eqn: euclidean sol 4D} which contain the cosmological horizon. We call these cosmic patches of dS. The spacetime region of interest is taken to be $r \,\in\, \left[\frakr,e^{-\omega}\frakr_{\text{c}}\right]$. For $\ell/\sqrt{3} \, < \, \frakr_{\text{c}} \,<\, \ell$, the full solution has a black hole horizon that lies outside the boundary, so it is not present in the cosmic patch. Similarly,  
for $\frakr_{\text{c}} \, > \, \ell$, there is a naked timelike singularity at the origin in Lorentzian signature, which would be associated with the presence of negative energy. Again, since this region is not part of the cosmic patch, we also allow for $\frakr_{\text{c}} \,>\, \ell$.

Regularity of the geometry near the cosmological horizon fixes the conformal temperature of the cosmic patch to be
\begin{equation}\label{eqn: cosmic ds4 beta}
	\tilde{\beta} \= \frac{4\pi \frakr_{\text{c}}\ell\sqrt{\frakr_\text{tube}\ell^2 - \frakr_\text{tube}^3 - \frakr_{\text{c}}\ell^2 + \frakr_{\text{c}}^3}}{\frakr_\text{tube}^{3/2} \left(3 \frakr_{\text{c}}^2-\ell^2\right)} \, ,
\end{equation}
which is always greater than zero. In this case, the conformal temperature $\tilde{\beta}^{-1}$ does not have a lower bound. Specifically, for $\ell/\sqrt{3} \, < \,\frakr_{\text{c}} \, \leq \, \ell$, the zero temperature limit can be reached by setting $\frakr_{\text{c}} \= \ell$ and taking $\frakr_\text{tube}/\ell \rightarrow 0$. For $\frakr_{\text{c}} > \ell$, there are also cosmic patches with zero conformal temperature. They have the boundary located closed to the naked singularity, that is to say $\frakr_\text{tube}/\ell \rightarrow 0$.

The high conformal temperature limit $\tilde{\beta}\rightarrow 0$ can be achieved in different ways, for instance, by taking the near horizon limits, $\frakr_\text{tube} \rightarrow \frakr_{\text{bh}}$ or $\frakr_\text{tube} \rightarrow \frakr_{\text{c}}$.

Setting the trace of the extrinsic curvature at the boundary to be constant leads to
\begin{equation}\label{eqn: cosmo ds4 K}
	K \ell \= -\frac{4\frakr_\text{tube}\ell^2 - 6 \frakr_\text{tube}^3-3\frakr_{\text{c}}\ell^2 + 3 \frakr_{\text{c}}^3}{2\frakr_\text{tube}^{3/2}\sqrt{\frakr_\text{tube}\ell^2-\frakr_\text{tube}^3-\frakr_{\text{c}}\ell^2+\frakr_{\text{c}}^3}} \,.
\end{equation}
There is no upper or lower bound on $K\ell$. The limit of large positive and large negative $K\ell$ correspond to taking the boundary to be near the cosmological horizon and the black hole horizon, respectively. In the case of $\frakr_{\text{c}} \,\geq\, \ell$, there is no black hole and so the large negative limit of $K \ell$ corresponds to taking the boundary to be near the origin.

Unlike the $D=3$ case, we could not find analytic expressions for $\frakr_\text{tube}$ and $\frakr_{\text{c}}$ in terms of the boundary data $\tilde{\beta}$ and $K$, but we relegate some useful analytical expressions to appendix  \ref{sec: useful formulae for 4d thermo}. We later present examples of $\frakr_\text{tube}$ and $\frakr_{\text{c}}$ as functions of $\tilde{\beta}$ at fixed $K\ell$ in figure \ref{fig: radius versus beta}, together with the black hole patch solutions.

\textbf{Cosmic patch thermodynamics.} We start by computing the on-shell action \eqref{euclidean_action} of the cosmic patch. We define a regulated action as the on-shell action of the cosmic patch subtracted by the pole patch action with the same $\tilde{\beta}$ and $K$. By expressing it in terms of $\frakr_{\text{bh}}$ and $\frakr_\text{tube}$, we find that
\begin{equation}\label{eqn: onshell cosmic 4d}
	I_{E\text{, reg}}^{(\text{cosmic})} \= - \frac{\pi \frakr_{\text{c}} \left(4\frakr_\text{tube}\ell^2 - 3 \frakr_{\text{c}}\ell^2 - 3 \frakr_{\text{c}}^3\right)}{3G_N\left(\ell^2-3\frakr_{\text{c}}^2\right)} - I_E^{(\text{pole})} \, ,
\end{equation}
where $I_E^{\text{(pole)}}$ is given by \eqref{eqn: thermal dS action 4d}.

The conformal energy and the conformal entropy of the cosmic patch are
\begin{equation}\label{eqn: conformal energy entropy cosmic 4d}
		E_\text{conf} \= \frac{\frakr_\text{tube}^{3/2}\left(2\frakr_\text{tube}\ell^2-3\frakr_{\text{c}} \ell^2+ 3 \frakr_{\text{c}}^3\right)}{6 G_N\ell \sqrt{\frakr_\text{tube}\ell^2-\frakr_\text{tube}^3-\frakr_{\text{c}}\ell^2+\frakr_{\text{c}}^3}} - \frac{I_E^{(\text{pole})}}{\tilde{\beta}} \, , \qquad\qquad
		\mathcal{S}_\text{conf} \= \frac{\pi \frakr_{\text{c}}^2}{G_N} \, .
\end{equation}
The conformal entropy $\mathcal{S}_\text{conf}$ agrees with the Gibbons-Hawking entropy of the cosmological horizon, $A_\text{horizon}/4G_N$. 
The specific heat at constant $K$ of the cosmic patch is given by
\begin{equation}\label{eqn: spec heat cosmic ds4}
	C_K = \frac{2\pi \frakr_{\text{c}}^2 \left(-\ell^2+3 \frakr_{\text{c}}^2\right)\left(9\frakr_{\text{c}}^2\left(\frakr_{\text{c}}^2-\ell^2\right)^2+16 \frakr_{\text{c}} \left(\frakr_{\text{c}}^2-\ell^2\right)\frakr_\text{tube}\ell^2+8\frakr_\text{tube}^2\ell^4 - 4 \frakr_\text{tube}^4\ell^2\right)}{G_N\left(\ell^2+3\frakr_{\text{c}}^2\right)\left(9 \frakr_{\text{c}}^2\left(\frakr_{\text{c}}^2-\ell^2\right)^2+2\frakr_{\text{c}}\frac{\left(-9\ell^4-10\frakr_{\text{c}}^2\ell^2+15\frakr_{\text{c}}^4\right)}{\left(\ell^2+3\frakr_{\text{c}}^2\right)}\frakr_\text{tube}\ell^2+8\frakr_\text{tube}^2\ell^4-4\frakr_\text{tube}^4\ell^2\right)} \, .
\end{equation}

It is interesting to remark that in the limit where the boundary approaches the cosmological horizon, the specific heat becomes
\begin{equation}\label{eqn: spec heat cosmic 4d}
C_K \,\rightarrow\, \frac{2 \pi \frakr_{\text{c}}^2}{G_N} \,, \qquad \text{as} \quad \frac{\frakr_{\text{tube}}}{\frakr_{\text{c}}} \to 1 \,.
\end{equation}

This positive specific heat is to be contrasted with the negative specific heat that is obtained when Dirichlet boundary conditions are imposed on the cosmic patch \cite{Draper:2022ofa, Banihashemi:2022jys}. We will further discuss this fact when we consider the pure dS$_4$ solutions.

\begin{figure}[h!]
        \centering
        \includegraphics[width=9 cm]{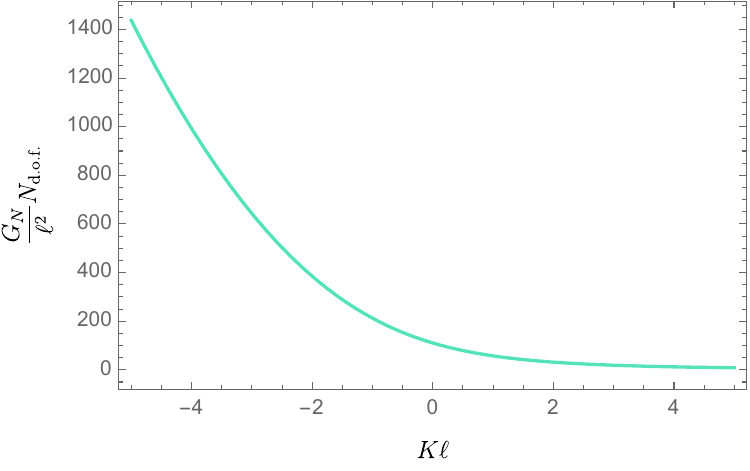}      
                \caption{The number of degrees of freedom $N_{\text{d.o.f.}}$ as a function of $K \ell$. The number of degrees of freedom decreases monotonically as a function of $K \ell$.
              } \label{fig: central charge 4d}
\end{figure}

Interestingly, the high conformal temperature limit of the specific heat \eqref{eqn: spec heat cosmic 4d} at finite $K\ell$ is given by
\begin{equation}
    C_K \,\rightarrow\, \frac{32\pi^3\ell^2}{81\tilde{\beta}^2G_N}\left(\sqrt{K^2\ell^2+9}-K\ell\right)^2\,, \qquad \text{as} \quad\tilde{\beta} \,\rightarrow\, 0 \, .
\end{equation} 
This takes the form of the specific heat of a three-dimensional conformal field theory. Under this interpretation,    the putative number of degrees of freedom goes as
\begin{equation}
    N_\text{d.o.f.} \= \frac{32 \pi^3 \ell^2}{81G_N}\left(\sqrt{K^2\ell^2+9}-K\ell\right)^2 \,,
\end{equation}
which is a monotonically decreasing function of $K\ell$, displayed in figure \ref{fig: central charge 4d}. Further taking $K \ell \to + \infty$ yields $N_\text{d.o.f.} \,\to 8\pi^3 / G_N K^2$ matching the $\Lambda=0$ result in \cite{Anninos:2023epi}.

Finally, we find that the thermodynamic quantities satisfy a first-law type of relation
\begin{equation}
    \delta E_\text{conf} \= \tilde{\beta}^{-1} \delta \,\mathcal{S}_\text{conf} - \mu_K \,\delta K \,,
\end{equation}
where $\mu_K$, similarly to the three-dimensional case, is interpreted as the chemical potential associated to $K$, 
\begin{equation}
    \mu_K \,\equiv\, - \frac{\frakr_\text{tube}^3}{3G_N} - \frac{1}{\tilde{\beta}} \left. \frac{\partial I_E^{(\text{pole})}}{\partial K}\right|_{\tilde{\beta}} \, .
\end{equation}

Note that the first term in $\mu_K$ looks identical to the one that appears for the pole patch in $D=4$, see (\ref{chemical_pole}).

\subsection{Black hole patch} \label{sec: bh patch}
We now consider a class of geometries \eqref{eqn: euclidean sol 4D} which contain the black hole horizon. We refer to these  as black hole patches of dS$_4$. These solutions exist as long as the cosmological horizon radius takes values between the dS length and the Nariai radius, $\frac{\ell}{\sqrt{3}}\,<\,\frakr_{\text{c}}\,<\,\ell$. The spacetime region of interest is then given by $r \,\in\, \left[e^{-\omega}\frakr_{\text{bh}} , \frakr\right]$ where $\frakr_{\text{bh}}$ is related to $\frakr_{\text{c}}$ through \eqref{eqn: relation between rbh and rc}. For convenience, in this subsection, we will always express $\frakr_{\text{c}}$ in terms of $\frakr_{\text{bh}}$. 

Given that these regions are complementary to the cosmic patches (see figure \ref{fig: sds}) many results for these solutions are closely related to those in the cosmic patch, upon replacing $\frakr_{\text{c}} \to \frakr_{\text{bh}}$. Here we point out the main differences between the two patches and relegate the explicit formulae to appendix \ref{black hole patch app}.

It is easy to obtain the boundary data from the results of the cosmic patch. The inverse conformal temperature is the same as in \eqref{eqn: cosmic ds4 beta}, but with $\frakr_{\text{c}} \to \frakr_{\text{bh}}$ and an extra minus sign. The trace of the extrinsic curvature is minus the expression that appears in \eqref{eqn: cosmo ds4 K}, again with $\frakr_{\text{c}} \to \frakr_{\text{bh}}$. As opposed to the cosmic patch, in this case, the conformal temperature $\tilde{\beta}^{-1}$ has a lower bound $\tilde{\beta}^{-1}_{\text{min}} \= 2\pi$, which occurs in the Nariai limit, by setting $\frakr_\text{tube} \= \ell/\sqrt{3}$ and taking $\frakr_{\text{bh}} \rightarrow \ell/\sqrt{3}$ from below.

Below this conformal temperature, the black hole patch solution does not exist. For larger conformal temperatures, $\tilde{\beta}^{-1} \, > \, \tilde{\beta}^{-1}_{\text{min}}$, there is a one-parameter family of black hole patches.

Regarding the trace of the extrinsic curvature, the limit of $K \ell$ approaching negative infinity corresponds to pushing the boundary to be near the cosmological horizon. For $K\ell$ going to positive infinity, the boundary is pushed near the black hole horizon. The behaviour of $K\ell$ as a function of $\frakr_{\text{c}}$ and $\frakr_\text{tube}$ is exactly opposite to the cosmic patch since the normal vector points in the opposite direction.

\textbf{Black hole patch thermodynamics.} Similarly, thermodynamic quantities can also be obtained from those of the cosmic patch. In particular, the regulated action $I_{E\text{, reg}}^{(\text{bh})}$ is the same as in the cosmic patch, but with $\frakr_{\text{c}} \to \frakr_{\text{bh}}$. The conformal energy is also the same as in \eqref{eqn: conformal energy entropy cosmic 4d}, but with a minus sign in front of the first term and the replacement of $\frakr_{\text{c}} \to \frakr_{\text{bh}}$. The entropy is now given by
\begin{equation}
		\mathcal{S}_\text{conf} \= \frac{\pi \frakr_{\text{bh}}^2}{G_N} \, ,
\end{equation}
which agrees with the Bekenstein-Hawking entropy $A_\text{horizon}/4G_N$ where the horizon in this formula now corresponds to the black hole horizon. The specific heat can also be obtained from \eqref{eqn: spec heat cosmic ds4}, with the replacement $\frakr_{\text{c}} \to \frakr_{\text{bh}}$. In particular, it is also positive as the tube approaches the black hole horizon. Explicit expressions and further interesting limits are shown in appendix \ref{black hole patch app}.

\subsection{Pure dS$_4$ patch}\label{sec: 4d pure dS}

As in the dS$_3$ case,  we can recover the pure dS$_4$ solution for a particular family of conformal temperatures. Using the boundary data of the cosmic patch \eqref{eqn: cosmic ds4 beta} and \eqref{eqn: cosmo ds4 K}, the conditions for having pure dS$_4$ are
\begin{equation}
	\tilde{\beta} \= \tilde{\beta}_{\text{dS}} \,\equiv\, \frac{2 \pi \sqrt{\ell^2-\frakr_\text{tube}^2}}{\frakr_\text{tube}} \, , \qquad\qquad K\ell \= - \frac{2\ell^2-3\frakr_\text{tube}^2}{\frakr_\text{tube}\sqrt{\ell^2-\frakr_\text{tube}^2}} \, .
\end{equation}
From these, it follows that $\frakr_{\text{c}} = \ell$.  One can further solve for $\tilde{\beta}_{\text{dS}}$  in terms of $K \ell$,
\begin{equation}\label{eqn: dS temp 4d}
	\tilde{\beta}_{\text{dS}} \= \frac{\pi}{2} \left(\sqrt{K^2\ell^2+8}-K\ell\right) \, .
\end{equation}

Consider now the worldline limit $\frakr_\text{tube}\,\rightarrow\,0$. The standard dS temperature is recovered when
\begin{equation}
	\tilde{\beta}_{\text{dS}} \frakr_\text{tube} \,\rightarrow\, 2\pi\ell \qquad\qquad \text{as} \qquad \frakr_\text{tube}\,\rightarrow\, 0\,.
\end{equation}
The stretched horizon limit,  corresponds to the high conformal temperature limit in which
\begin{equation}
	\tilde{\beta}_{\text{dS}} \,\rightarrow\, 0 \qquad\qquad \text{as} \qquad \frakr_\text{tube}\,\rightarrow\, \ell\,.
\end{equation}

Now we can use \eqref{eqn: conformal energy entropy cosmic 4d} and \eqref{eqn: spec heat cosmic ds4} to calculate the thermodynamic properties of pure dS$_4$. The conformal entropy and the specific heat at constant $K$ of the pure dS$_4$ are given by
\begin{equation}
	\mathcal{S}_\text{conf} \= \frac{\pi \ell^2}{G_N} \, , \qquad\qquad C_K \= - \frac{2\pi\frakr_\text{tube}\ell^2\left(2\ell^2-\frakr_\text{tube}^2\right)}{G_N\left(\ell^3 - 4 \frakr_\text{tube}\ell^2 + 2 \frakr_\text{tube}^3\right)}\, . \label{ck ds4}
\end{equation}
The conformal entropy is constant regardless of $\frakr_\text{tube}$ and is given by the Gibbons-Hawking entropy of the cosmological horizon, as expected. 

\begin{figure}[h!]
        \centering
        \includegraphics[width=9 cm]{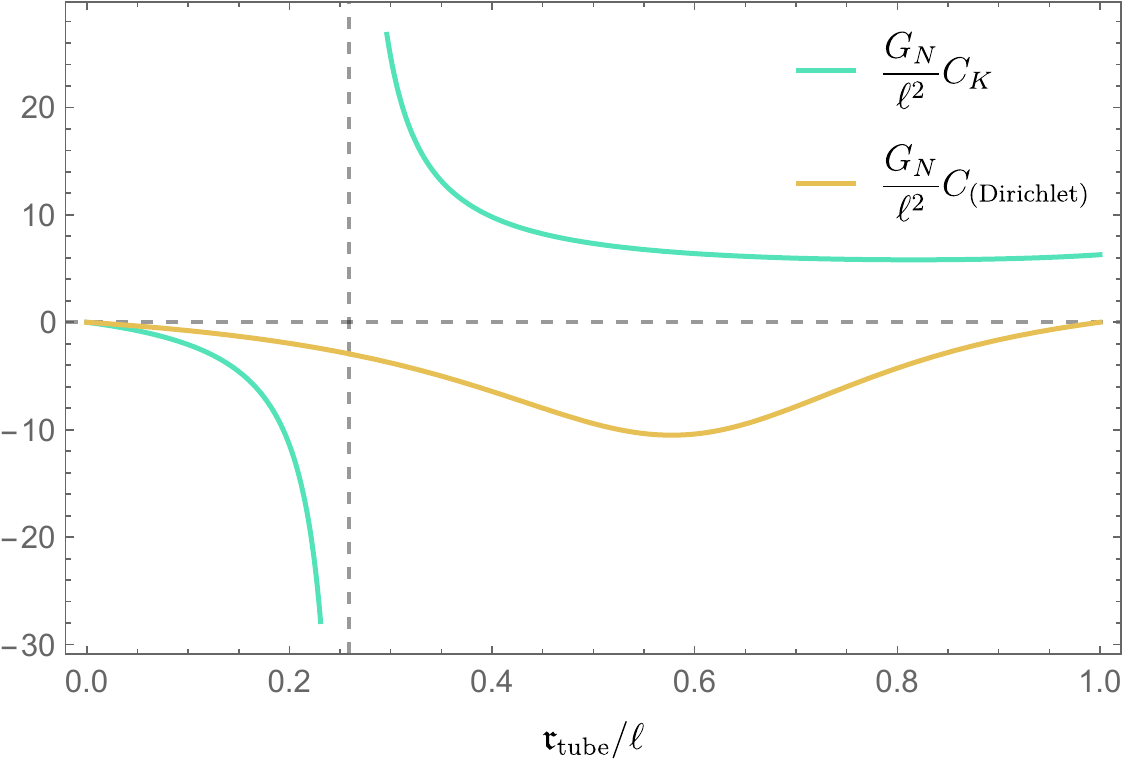}      
                \caption{A plot of the specific heat of the pure de Sitter patch for conformal (green) and Dirichlet (yellow) boundary conditions. For the Dirichlet case, the specific heat is never positive.} \label{fig: specific heat pure dS4}
\end{figure}

It is interesting to note the  behaviour of the specific heat at constant $K$. Close to the worldline, $C_K$ in (\ref{ck ds4}) is negative. In fact, in the worldline limit, the specific heat converges to zero from below, $C_K \,\rightarrow\, - 4\pi \frakr_\text{tube}\ell/G_N$. This is similar to what happens for the Dirichlet case \cite{Draper:2022ofa, Banihashemi:2022jys} where
\begin{equation}
C_{\text{(Dirichlet)}} = - \frac{2\pi \frakr_\text{tube}\ell^2\left(\ell^2-\frakr_\text{tube}^2\right)}{G_N\left(\ell^3 - 2 \frakr_\text{tube}\ell^2 + 2 \frakr_\text{tube}^3\right)}~.
\end{equation}
We plot both specific heats in figure \ref{fig: specific heat pure dS4}. A notable feature is that for the case of conformal boundary conditions, the specific heat diverges as $\frakr_\text{tube} \to \frakr_0$, with $\frakr_0$ given by\footnote{This is the only root to the cubic equation $\ell^3-4\frakr_0\ell^2+2\frakr_0^3 = 0$, satisfying $0<\frakr_0<\ell$, 
\begin{equation}
    \frac{\frakr_0}{\ell} = 2\sqrt{\frac{2}{3}}\cos\left(\frac{1}{3}\cos^{-1}\left(-\frac{3\sqrt{3}}{8\sqrt{2}}\right)-\frac{2\pi}{3}\right) \,.
\end{equation}}
\begin{equation}\label{eqn: K pure dS 4d}
	\frakr_0 \,\approx\, 0.259 \ell \, \qquad \Leftrightarrow \qquad  K \ell \approx -7.202 \,. 
\end{equation}
For $\frakr_\text{tube} \,>\, \frakr_0$, we find that the specific heat is positive and approaches a constant $C_K \,\rightarrow\, 2 \pi \ell^2/G_N$ as we take the stretched horizon limit. 
The pure dS$_4$ patch with a conformal boundary sufficiently close to the de Sitter horizon is thus thermally stable. 


The conformal energy of the pure dS$_4$ is given by
\begin{equation}\label{eqn: conf energy pure ds4}
	E_\text{conf} \= \frac{4\ell^2\left(\ell^2-\frakr_\text{tube}^2\right)}{3\frakr_\text{tube}G_N\sqrt{4\ell^2-3\frakr_\text{tube}^2}} + \frac{\frakr_\text{tube}^2\ell}{3G_N\sqrt{\ell^2-\frakr_\text{tube}^2}}\, ,
\end{equation}
which is positive for all $0\,<\,\frakr_\text{tube}\,<\, \ell$ with a lowest value given by
\begin{equation}\label{eqn: minimum energy pure ds4}
	\left.E_\text{conf}\right|_{\frakr_\text{tube}\=\sqrt{\frac{2}{3}}\ell} \= \frac{4\ell^2}{3\sqrt{3}G_N}.
\end{equation}
In terms of the boundary data, the minimum energy \eqref{eqn: minimum energy pure ds4} is obtained precisely when $K\ell\=0$. Both in the worldline and in the stretched horizon limit, the energy is divergent. In particular, $E_\text{conf} \,\rightarrow\, \frac{2\ell^3}{3\frakr_\text{tube}G_N}$, in the worldline limit and $E_\text{conf} \,\rightarrow\, \frac{\ell^2}{3\sqrt{2}G_N \sqrt{1- \frakr_\text{tube}/\ell}}$, in the stretched horizon limit. 

As in the pure dS$_3$ discussion, one could consider a pole patch which, together with the cosmic patch of pure dS$_4$, completes the full Euclidean static patch geometry, which is the four-sphere. This is achieved by considering a pole patch which has the same $\tilde{\beta}$ but an opposite trace of the extrinsic curvature $-K$. As in $D=3$, it is straightforward to check that at the saddle point level
\begin{equation}
\mathcal{Z}^{\text{(cosmic)}} (\tilde{\beta}_{dS},K) \mathcal{Z}^{\text{(pole)}} (\tilde{\beta}_{dS},-K) = \exp  \, \frac{\pi \ell^2}{G_N} ~. 
\end{equation}
In the above, the $\mathcal{Z}$ are computed with the bare action, without subtracting the pole on-shell action. The expression parallels a similar observation for two-dimensional near Nariai geometries \cite{Anninos:2022hqo}, and suggests that the thermodynamic content of the empty static patch is purely entropic. It would be interesting to understand the above expression at the one-loop level or higher.

For the black hole patch of pure dS$_4$, we find that, by setting $\frakr_{\text{bh}} \= 0$ in \eqref{eqn: conformal energy entropy bh 4d} and \eqref{eqn: spec heat bh ds4}, the thermodynamic quantities are trivial, $E_\text{conf} \= \mathcal{S}_\text{conf} \= C_K \= 0$, as can be expected.

\subsection{Nariai patch}

In addition to the pure dS$_4$ solution, there is another interesting geometry that can be reached by tuning the conformal temperature $\tilde{\beta}^{-1}$ to a particular value set by the trace of the extrinsic curvature $K\ell$. We call these Nariai patches, which exist as the size of the cosmological and black horizon become close to each other.

To find the Nariai temperature $\tilde{\beta}_N^{-1}$, we consider the following limit. Let $\rho \,\equiv\, \frac{1}{\epsilon}\left(\frakr_\text{tube}-\frac{\ell}{\sqrt{3}}\right)$ be a dimensionless parameter describing deviation of $\frakr_\text{tube}$ from the Nariai radius. The Nariai geometry is obtained by setting $\frakr_{\text{c}} \= \frac{\ell}{\sqrt{3}} + \epsilon$ and taking $\epsilon/\ell \rightarrow 0$ while keeping $\rho$ fixed. Let us first consider the Nariai solution from the cosmic patch perspective. The cosmological horizon and black hole horizons are, respectively, located at $\rho \= 1$ and $-1$. Using \eqref{eqn: cosmic ds4 beta} and \eqref{eqn: cosmo ds4 K}, the Nariai limit fixes
\begin{equation}
	\tilde{\beta}_N \,\equiv\, 2 \pi \sqrt{1-\rho^2} \, , \qquad\qquad K \ell \= \frac{\sqrt{3}\rho}{\sqrt{1-\rho^2}}\, .
\end{equation}
These equations can be inverted analytically to obtain $\tilde{\beta}_N$ in terms of $K\ell$,
\begin{equation}\label{eqn: Nariai temperature}
	\tilde{\beta}_N \= \frac{2\sqrt{3}\pi}{\sqrt{K^2\ell^2+3}} \, .
\end{equation}
We note that $\tilde{\beta}_N$ exists for all values of $K\ell\,\in\, \mathbb{R}$. This is yet another difference with the Dirichlet problem, where to obtain the Nariai solution one needs to fix the value of $\frakr_{\text{tube}}$ to be very close to the Nariai radius.

\textbf{Nariai patch thermodynamics.} Using \eqref{eqn: conformal energy entropy cosmic 4d}, the conformal energy of the Nariai solution is given by
\begin{equation}\label{eqn: rho parameterisation}
	E_\text{conf} \= \frac{\rho\ell^2}{9G_N\sqrt{1-\rho^2}} + \frac{2\ell^2}{3G_N}\sqrt{\frac{\rho^2-4+\sqrt{3}\rho\sqrt{8-5\rho^2}}{9\rho^2-12-\sqrt{3}\rho\sqrt{8-5\rho^2}}} \, .
\end{equation}
The conformal entropy and specific heat, evaluated at constant $K$ and $\tilde{\beta} = \tilde{\beta}_N$, are respectively given by
\begin{equation}
	\mathcal{S}_\text{conf} \= \frac{\pi \ell^2}{3G_N} \, , \qquad\qquad C_K \= - \frac{2\pi\ell^2}{G_N\left(2-9\rho+4\rho^3\right)} \, .
\end{equation}
The specific heat $C_K$ becomes positive (negative) as one pushes the boundary to the cosmological (black hole) horizon. In particular, $C_K$ changes sign at $\rho\=\rho_0$ where\footnote{This is given by the only solution to the cubic equation $4\rho_0^3-9\rho_0+2=0$ with $1>\rho_0>-1$, 
\begin{equation}
    \rho_0 = \sqrt{3}\cos\left(\frac{1}{3}\cos^{-1}\left(-\frac{2}{3\sqrt{3}}\right)-\frac{2\pi}{3}\right)\,.
\end{equation}}
\begin{equation}\label{eqn: K diverge Nariai}
	\rho_0 \,\approx\, 0.227 \qquad \Leftrightarrow \qquad  K \ell \approx 0.405 \,. 
\end{equation}

We can also have black hole solutions in the Nariai patch. These are simply obtained by changing $\rho \to -\rho$ in the expressions above.

The Dirichlet thermodynamics of configurations near the Nariai geometry, from the perspective of a dimensionally reduced theory, was studied in \cite{Svesko:2022txo,Anninos:2022hqo} where it was shown that the cosmic Nariai patch always has negative specific heat.

\subsection{Phase diagram}

\begin{figure}[h!]
        \centering
         \subfigure[]{
                \includegraphics[height=7.5cm]{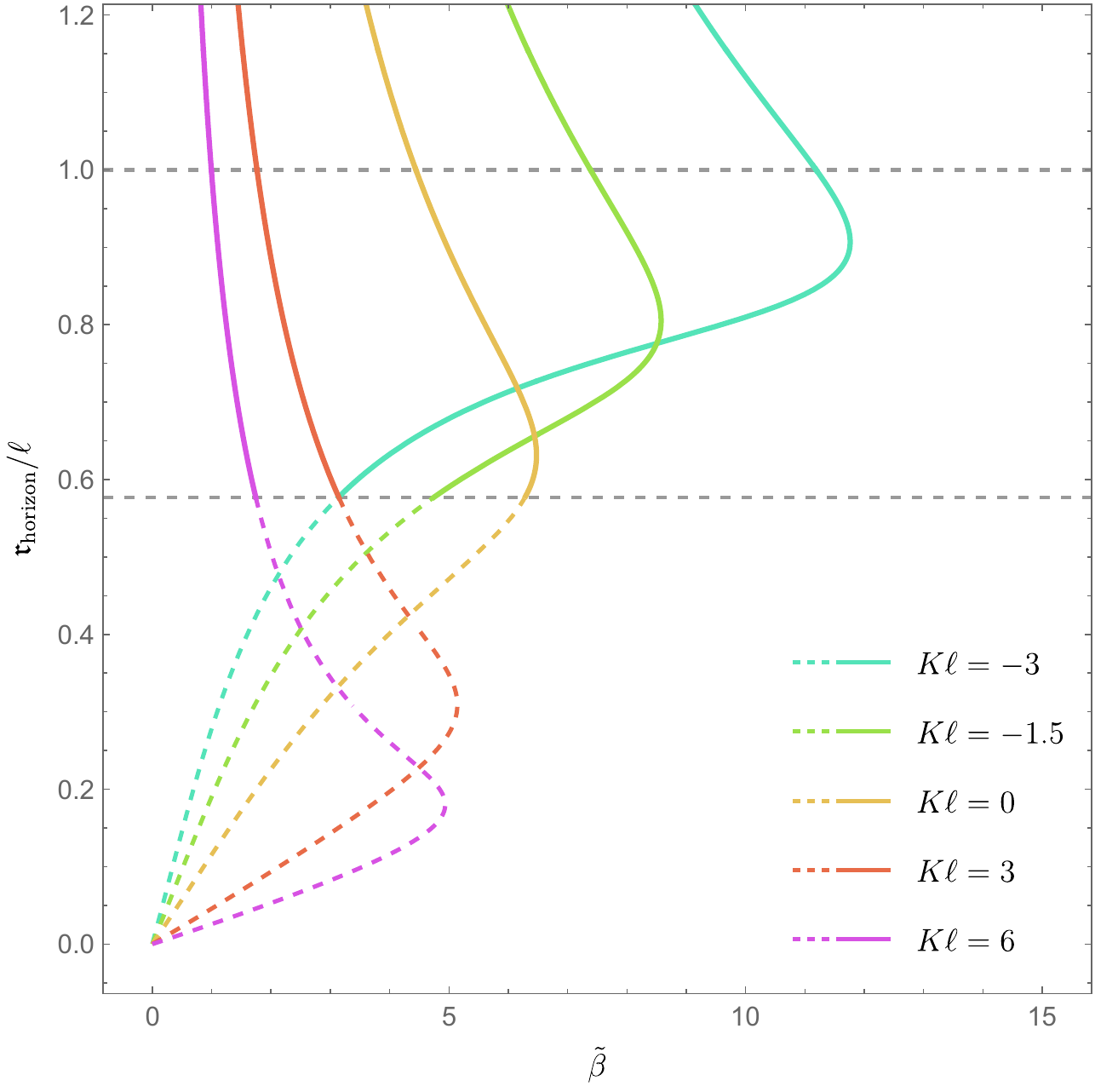}\label{fig: rHorizon 4d}}  \quad\quad
                 \subfigure[]{
                \includegraphics[height=7.5cm]{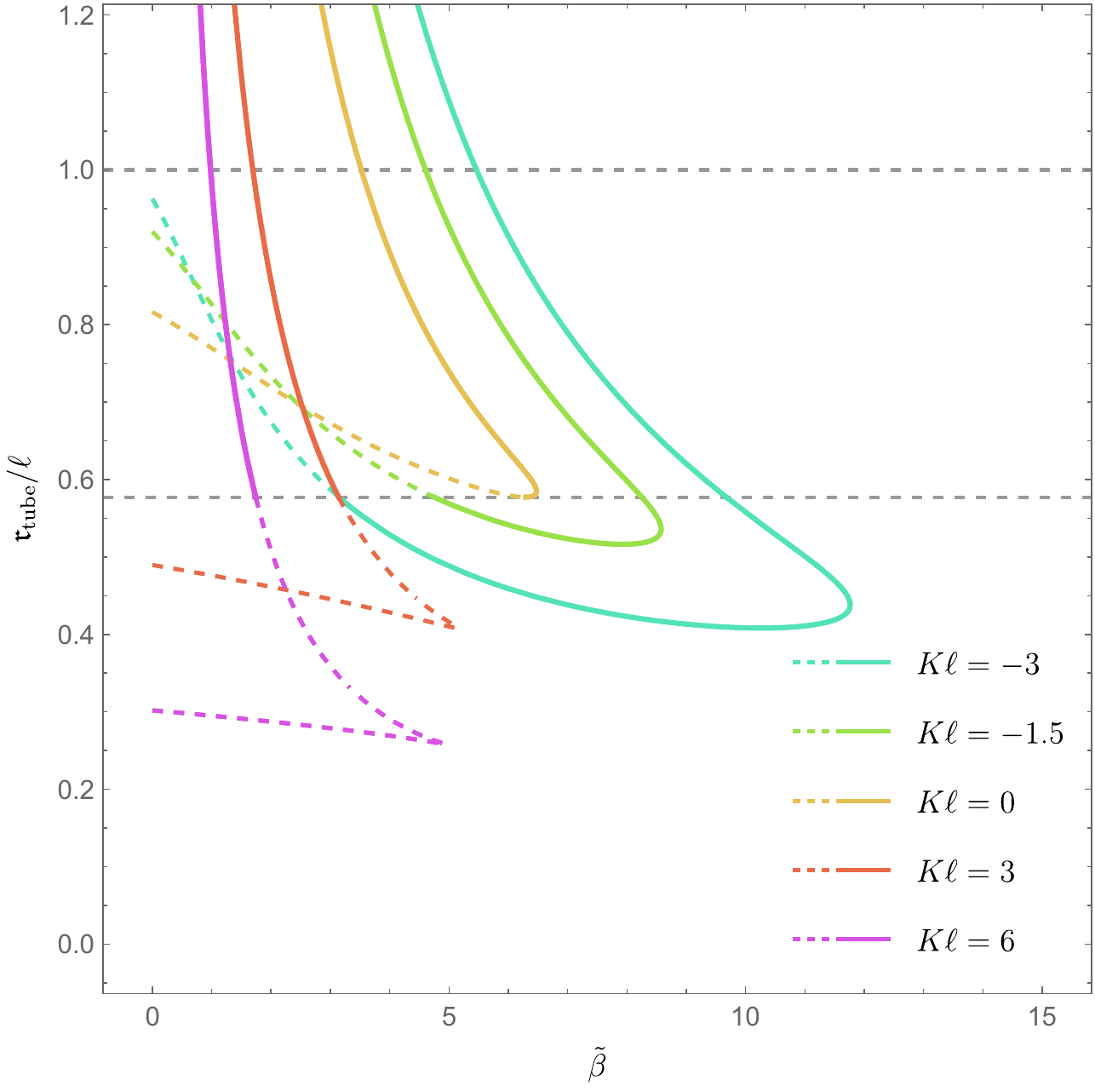} \label{fig: rTube 4d}}  \quad\quad      
                \caption{Plots of $\frakr_\text{horizon}/\ell$ and $\frakr_\text{tube}/\ell$ as a function of $\tilde{\beta}$ at fixed $K\ell$. The solid curves correspond to cosmic patches, while the dashed ones, to black hole patches. $\frakr_\text{horizon}$ denotes the radius of the black hole (cosmological) horizon when considering the black hole (cosmic) patch. The two horizontal black dashed lines correspond to the dS$_4$ patch (upper line) and Nariai patch (lower line).} \label{fig: radius versus beta}
\end{figure}

In this subsection, we discuss the thermodynamics of four-dimensional spacetime with $\Lambda \,>\,0$ by combining the results from the pole patch, cosmic patch, and black hole patch solutions. In the $G_N \rightarrow 0$ limit, the partition function of the total system is generally given by a sum of all possible patches which have the same boundary data $\tilde{\beta}$ and $K$, 
\begin{equation}\label{eqn: total partition function 4d}
	\mathcal{Z} (\tilde{\beta},K) \= e^{-I^{(\text{pole})}_{E}}\left(1 + \ldots \right) \; .
\end{equation}
A pole patch solution exists for all $\tilde{\beta} \, \in\, \mathbb{R}^+$ and $K\ell \,\in\, \mathbb{R}$. The omitted terms are additional contributions stemming from the co-existing cosmic/black hole patch solutions, i.e. $e^{-I_{E\text{, reg}}^{(\text{cosmic})}}$ or $e^{-I_{E\text{, reg}}^{(\text{black hole})}}$. The number of these terms and the details of the solutions depend on the value of the boundary data, as we will discuss below. For patches with positive specific heat, the one with lowest regulated action is thermodynamically stable; otherwise, they are thermodynamically metastable. Patches with negative specific heat are thermodynamically unstable. 

\begin{figure}[h!]
        \centering
        \includegraphics[scale=0.3]{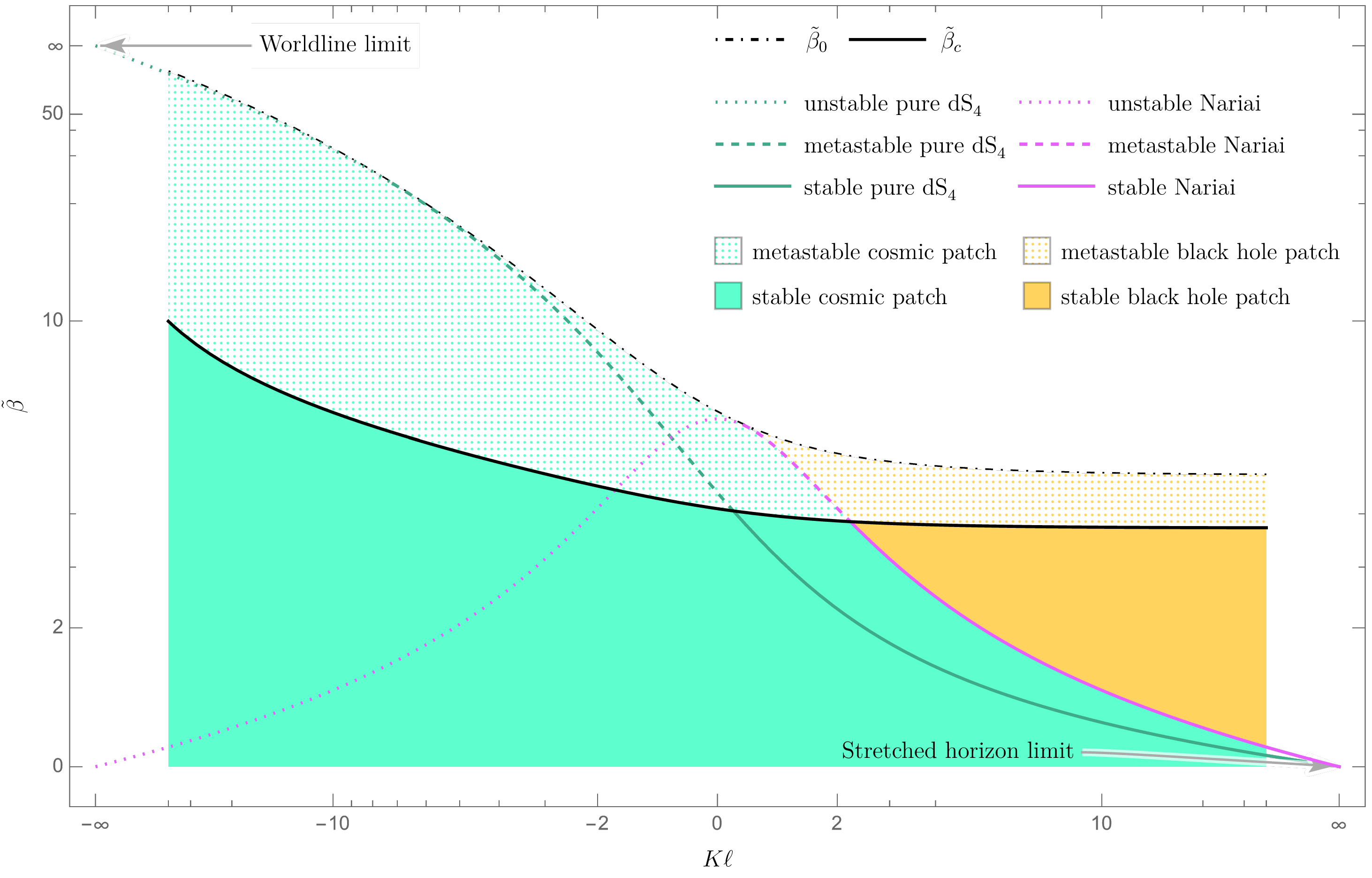}      
                \caption{Phase diagram of conformal dS$_4$ thermodynamics for static and spherically symmetric configurations. The number of different solutions co-existing at a given point in the phase diagram depends on whether the point lies above or below the $\tilde{\beta}_0$ curve (dot-dashed black curve). Above that curve, only one pole patch solution exists. Below the $\tilde{\beta}_0$ curve, apart from a pole patch solution, there co-exist two additional cosmic/black hole patches, one with negative and one with positive $C_K$. The curve of critical inverse conformal temperature $\tilde{\beta}_c$ is shown in thick black, above which the pole patch is thermodynamically preferred. In the region bounded by $\tilde{\beta}_0$ and $\tilde{\beta}_c$ curves, shaded in green (yellow) halftone, the cosmic (black hole) patch is metastable. For $\tilde{\beta}_c > \tilde{\beta}$, the cosmic (black hole) patch is stable with the associated region shaded in solid green (yellow). The dark green and purple curves represent pure dS$_4$ and Nariai patches. Both curves are divided into three segments: stable, metastable, and unstable, which are shown as thick, dashed and dotted curves, respectively. The (meta)stable Nariai curve marks the separation of the (meta)stable cosmic patch and black hole patch regions.} 
                \label{fig: phase4d}
\end{figure}

As opposed to $D=3$, solutions with horizons do not exist for all values of $\tilde{\beta}$ and $K \ell$. At low temperatures, only one pole patch solution exists. To separate the phase space regions with no horizon patches, we define the inverse conformal temperature $\tilde{\beta}_0 (K \ell)$. Note that it depends on the value of $K \ell$, so that at a given $K \ell$, horizon patches only exist for conformal temperatures such that $\tilde{\beta}\leq\tilde{\beta}_0$. The curve $\tilde{\beta}_0 (K \ell)$ can be found numerically and is shown in figure \ref{fig: phase4d}. The rest of the phase diagram can be decribed as follows:

\begin{itemize}
    \item Exactly at $\tilde{\beta}_0 (K \ell)$, there are two solutions: one pole patch and one horizon patch. The latter is a cosmic patch if $K\ell \lesssim 0.405$. Otherwise, it is a black hole patch. Note the transition happens at the $K \ell$ in which the Nariai patch specific heat changes sign. In both cases, the horizon patch has positive regulated action and, therefore, it is always sub-dominant.
    \item For lower inverse temperatures, $\tilde{\beta} < \tilde{\beta}_0 (K \ell)$, we always find three solutions for any given $\tilde{\beta}$ and $K \ell$. There is always one pole patch and two horizon patches. The horizon patches can be either cosmic or black hole patches, but always the horizon patch with larger horizon size has positive specific heat, while the one with smaller size, has negative specific heat. This can be confirmed by observing that the large (small) horizon patch has a horizon radius which is an increasing (decreasing) function of the conformal temperature, as we show in figure \ref{fig: radius versus beta}. Moreover, if the horizon radius is larger (smaller) than the Nariai radius $r_N \= \ell/\sqrt{3}$, the corresponding horizon patch is a cosmic (black hole) patch. There exists a smooth transition between black hole patch and cosmic patch as one varies the conformal temperature.
    \item At a given critical conformal temperature that depends on the value of $K \ell$, there is a first-order phase transition, similar to the Hawking-Page transition. We call this temperature $\tilde{\beta}_c (K \ell)$ and show it numerically in figure \ref{fig: phase4d}. For $\tilde{\beta} < \tilde{\beta}_c$, the large horizon patch solution dominates over the pole patch, while the opposite happens for $\tilde{\beta} > \tilde{\beta}_c$.
\end{itemize}

Consequently, for $\tilde{\beta}_0 > \tilde{\beta} > \tilde{\beta}_c$, the large horizon patch is metastable, while for $\tilde{\beta} < \tilde{\beta}_c$, the large horizon patch becomes stable and the dominant configuration. If a small horizon patch exists, then it is always subdominant.
We display various plots of the regulated action, conformal energy, and specific heat as a function of the inverse conformal temperature at fixed $K\ell$ in appendix \ref{sec: additional figures}, see figure \ref{fig: appendix figures}.

\textbf{Pure dS$_4$ phase structure.} For the pure dS$_4$ solution, we constrain the inverse conformal temperature to be given by the dS inverse temperature \eqref{eqn: dS temp 4d}. We note that $K \ell \,\approx \, 0.256$ is equivalent to $\frakr_\text{tube}/\ell \,\approx\, 0.840$.

\begin{itemize}
    \item For $K\ell\,<\,-7.202$, the dS temperature lies in the intermediate temperature regime implying that the corresponding pure dS$_4$ is sub-dominant. We find that these pure dS$_4$ have negative specific heat and are thus unstable. The worldline limit is included in this regime. At $K\ell \,\approx\, -7.202$, the dS temperature coincides with $\tilde{\beta}_0^{-1}$.
    \item For $-7.202 \, < \, K\ell \, <\, 0.256$, the dS temperature remains in the intermediate regime, but now the associated specific heat becomes positive. Therefore, these pure dS$_4$ patches are metastable. At $K\ell \,\approx\, 0.256$, the dS temperature coincides with the critical temperature, $\tilde{\beta}_c^{-1}$.
    \item For $0.256\,<\,K\ell $, the dS temperature is higher than the critical temperature. As a consequence, the pure dS$_4$ has regulated action lower than the pole patch. It has also positive specific heat. We therefore find that the pure dS$_4$, in this regime, is thermodynamically stable. We note that the strechted horizon is included in this case. 
\end{itemize}

\textbf{Nariai phase structure.} To obtain the Nariai solution, we constrain the inverse conformal temperature to the Nariai inverse temperature \eqref{eqn: Nariai temperature}.
\begin{itemize}
    \item For $K\ell \,<\, 0.405$, the Nariai temperature is higher than $\tilde{\beta}_0^{-1}$ temperature. In this regime, the Nariai solution has negative specific heat, so it is thermally unstable.
    \item For $0.405\,<\,K\ell\,<\,2.239$, the Nariai temperature lies in the intermediate temperature regime. The corresponding Nariai solution has positive specific heat and positive regulated action. This means that the Nariai soution, in this regime, is metastable.
    \item For $2.239\,<\,K\ell$, the Nariai temperature is higher than the critical temperature. We find that the Nariai solution now becomes thermodynamically stable.
\end{itemize}


\section{Linearised dynamics} \label{sec: dynamics}

So far our treatment has been largely based on a quasi-equilibrium Euclidean picture. The aim of our final subsection is to complement the Euclidean analysis with  a Lorentzian analysis. Concretely, we will consider solutions to the four-dimensional linearised Einstein equations equipped with a positive cosmological constant $\Lambda\,>\,0$. As boundary conditions, we will once again consider conformal boundary conditions for the induced metric $g_{mn}$ and mean curvature $K$ on a topologically $\mathbb{R} \times S^2$ timelike boundary $\Gamma$. Our treatment parallels that for Minkowski space \cite{Anninos:2023epi}, here extended to the case of $\Lambda\,>\,0$. A portion of our linearised analysis was already treated in \cite{Anninos:2011zn}, in the context of the fluid-gravity correspondence applied to de Sitter horizons.\footnote{The conformal boundary conditions were necessitated in \cite{Anninos:2011zn}, as well as \cite{Bredberg:2011xw}, due to an obstruction in solving the non-linear Einstein equations with Dirichlet data on a timelike surface in the near-horizon expansion. The Lorentzian problem with Dirichlet conditions was considered in \cite{Andrade:2015gja}, where exponentially growing modes were found for the cosmic patch, as well as the pole patch at sufficiently large worldtube size.}

\subsection{Basic setup}


We will consider the linearised Einstein equations about the static patch metric,
\begin{equation}\label{eqn: lorentzian background metric}
	ds^2 \= -f(r)dt^2 + \frac{dr^2}{f(r)} + r^2 d\Omega^2 \, , \qquad f(r) \= 1-\frac{r^2}{\ell^2} \, , \qquad d\Omega^2 \= d\theta^2 + \sin^2\theta d\phi^2 \, .
\end{equation}
The timelike boundary $\Gamma$ is located at $r=\frakr \in \left(0,\ell\right)$. As in \cite{Anninos:2011zn}, we are primarily interested in dynamical features of the cosmological horizon, but we also report on the dynamical features of the pole patch below. As such, the spacetime region of interest is taken to be the Lorentzian cosmological patch $r \in \left(\frakr,\ell\right)$, $t \in \mathbb{R}$, and $\theta \in \left(0,\pi\right)$, $\phi \sim \phi + 2\pi$. The induced metric on $\Gamma$ is given by
\begin{equation}\label{eqn: lorentzian induced metric}
	\left.ds^2 \right|_{r=\frakr} \= -f(\frakr) dt^2 + \frakr^2 d\Omega^2  \, .
\end{equation}
Using an inward-pointing normal vector $\hat{n}\=-\sqrt{f(r)}\partial_r$, the extrinsic curvature and its trace are given by,
\begin{equation}\label{eqn: lorentzian extrinsic}
	\left.K_{mn} dx^m dx^n\right|_{r=\frakr} \= \frac{\frakr\sqrt{f(\frakr)}}{\ell^2}\left(dt^2 + \ell^2 d\Omega^2\right)~, \quad\quad \left.K\ell\right|_{r=\frakr} \= \frac{3\frakr^2-2\ell^2}{\frakr\sqrt{\ell^2-\frakr^2}} \, .
\end{equation}
We denote linearised perturbations about the background (\ref{eqn: lorentzian background metric}) as
\begin{equation}
	g_{\mu\nu} \= \bar{g}_{\mu\nu} + \varepsilon \, h_{\mu\nu} \, , \qquad\qquad \left|\,\varepsilon\,\right| \ll 1 \, ,
\end{equation}
where the background metric $\bar{g}_{\mu\nu}$ is given in \eqref{eqn: lorentzian background metric}. The equation of motion for $h_{\mu\nu}$ is obtained by expanding \eqref{eqn: Einstein field} to first order in $\varepsilon$. Further demanding that the conformal boundary data remains invariant under arbitrary perturbation $h_{\mu\nu}$ implies that
\begin{eqnarray}\label{eqn: def bdry cond 1}
\begin{cases}
	\left.h_{mn}\right|_{r=\frakr} &= \left.\gamma(x)\bar{g}_{mn}\right|_{r=\frakr} \, , \\
	\left.\varepsilon \delta K(h_{\mu\nu})\right|_{r=\frakr} & \equiv \left.K(\bar{g}_{\mu\nu}+\varepsilon \, h_{\mu\nu}) - K(\bar{g}_{\mu \nu})\right|_{r=\frakr} \= 0 \, ,
	\end{cases}
\end{eqnarray}
where $\gamma(x)$ is an arbitrary function, which will depend on the initial data of the linearised metric $h_{\mu\nu}$, and $\bar{g}_{mn}|_{r=\frakr}$ is the induced metric \eqref{eqn: lorentzian induced metric}. By contracting the first expression in \eqref{eqn: def bdry cond 1} with $\bar{g}^{mn}$, one may write the first boundary condition in a form that does not contain $\gamma(x)$ as
\begin{equation}\label{eqn: bdry cond 1}
	\left.h_{mn} - \frac{1}{3}\bar{g}_{mn} h^p{}_p \right|_{r=\frakr} \= 0 \, .
\end{equation}
Using \eqref{eqn: def of K}, the variation of the trace of the extrinsic curvature to first order in $\epsilon$ is given by
\begin{equation}\label{eqn: bdry cond 2}
	\left.\frac{\delta K (h_{\mu\nu})}{\sqrt{f(\frakr)}}\right|_{r=\frakr} \= \left.\frac{1}{2} \partial_r h^m{}_m - \mathcal{D}^m h_{rm} - \frac{\sqrt{f(\frakr)}}{2}K h_{rr} \right|_{r=\frakr} = 0 \, ,
\end{equation}
where $\mathcal{D}_n$ denotes the covariant derivative with respect to the boundary metric $\bar{g}_{mn}$. In the following analysis, we will take \eqref{eqn: bdry cond 1} and \eqref{eqn: bdry cond 2} as the conformal boundary conditions for linearised gravity with $\Lambda>0$. We must also impose conformal boundary conditions on the space of allowed diffeomorphisms $\xi_\mu$. Finally, we  require that $\xi^r |_{r=\frak{r}} = 0$, such that the allowed diffeomorphisms do not move the location of the boundary.

{\textbf{Kodama-Ishibashi method.}} Following the treatment of Kodama and Ishibashi \cite{Kodama:2000fa,Kodama:2003jz}, due to the spherical symmetry and time-translation invariance of the background, we can split our linearised solutions into vector and scalar perturbations, denoted by $h^{(V)}_{\mu\nu}$ and $h^{(S)}_{\mu\nu}$ respectively. Our details and conventions follow directly those in appendix C of \cite{Anninos:2023epi}. As such, our treatment will be brief in what follows and mostly focused on presenting the main results for $\Lambda>0$.

The $SO(3)$ content of $h^{(V)}_{\mu\nu}$ is captured by the vectorial spherical harmonics, $\mathbb{V}_i$, which are transverse eigenfunctions of the unit two-sphere Laplacian acting on vectors, with eigenvalues $k_V = l(l+1)-1$ for $l = 1,2,\ldots$ The $SO(3)$ content of $h^{(S)}_{\mu\nu}$ is captured by the scalar spherical harmonics, $\mathbb{S}$, which are transverse eigenfunctions of the unit two-sphere Laplacian with eigenvalues $k_S = l(l+1)$ for $l = 0,1,2,\ldots$  We note that the $l=0$ and $l=1$ modes require a separate treatment and we discuss them in appendix \ref{sec: l0/l1 modes}. Together, $h^{(V)}_{\mu\nu}$ and $h^{(S)}_{\mu\nu}$ encode the two propagating degrees of freedom of the four-dimensional metric at the linearised level. 

In the absence of timelike boundaries, the Kodama-Ishibashi formalism is gauge invariant and reduces the linearised Einstein equations to a set of `master equations' governing the vectorial and scalar master fields $\Phi^{(V)}$ and $\Phi^{(S)}$ which are directly linked to $h^{(V)}_{\mu\nu}$ and $h^{(S)}_{\mu\nu}$. It proves convenient for our analysis, as it did in \cite{Anninos:2023epi}, to select a gauge where the linearised boundary conditions (\ref{eqn: bdry cond 1}) and (\ref{eqn: bdry cond 2}) act only on $h^{(V)}_{\mu\nu}$ and $h^{(S)}_{\mu\nu}$ respectively. This gauge choice is indeed possible, and in this gauge the components of our metric perturbation read
\begin{eqnarray}
\begin{cases}
	h_{mn} &= - \bar{g}_{mn} \frac{1}{2r}\left[ l\left(l+1\right)\left(1- \frac{2r^2}{\ell^2}\right)+2 r^2 \partial_t^2 + 2  \left(1-\frac{r^2}{\ell^2}\right)^2  r \partial_r\right]\Phi^{(S)} \mathbb{S}  \\
 & \quad + \left(\delta^i_m\delta^t_n + \delta^i_n \delta^t_m\right) \left(1-\frac{r^2}{\ell^2}\right)\partial_r \left(r \Phi^{(V)}\right) \mathbb{V}_i \, , \\
    h_{rr} &= -\frac{1}{r\left(1-\frac{r^2}{\ell^2}\right)^2}\Bigg[\frac{l(l+1)}{2}\left(3-\frac{7r^2}{\ell^2}+\frac{4r^4}{\ell^4}\right)+ \left(3-\frac{2 r^2}{\ell^2}\right)r^2\partial_t^2 \\
    		&\quad + \left(1-\frac{r^2}{\ell^2}\right)\left(\left(1-\frac{r^2}{\ell^2}\right)\left(l(l+1) + 1-2\frac{r^2}{\ell^2} \right)+ r^2\partial_t^2\right) r\partial_r\Bigg]\Phi^{(S)} \mathbb{S} \, , \\
    h_{tr} &=  -\frac{1}{2\left(1-\frac{r^2}{\ell^2}\right)}\partial_t \left[l(l+1)\left(1-\frac{r^2}{\ell^2}\right) - 2 + r^2\partial_t^2 - \left(1-\frac{r^2}{\ell^2}\right)\frac{r^2}{\ell^2}r\partial_r\right]\Phi^{(S)} \mathbb{S} \, ,  \\
    h_{ri} &=  \frac{\sqrt{l(l+1)}}{2\left(1-\frac{r^2}{\ell^2}\right)}\left[l(l+1)\left(1-\frac{r^2}{\ell^2}\right)+r^2\partial_t^2+\left(2-\left(3-\frac{r^2}{\ell^2}\right)\frac{r^2}{\ell^2}\right)r\partial_r\right]\Phi^{(S)} \mathbb{S}_i + \frac{r}{1-\frac{r^2}{\ell^2}} \partial_t \Phi^{(V)} \mathbb{V}_i \, .	
    \end{cases}\label{eqn: spherical l>2 ansatz}
\end{eqnarray}
In the above the indices $m$ and $n$ denote indices with respect to $(t,r)$, while the index $i$ denotes indices on the two-sphere.

\subsection{Vector perturbation}

The master equation for $\Phi^{(V)}$, for given angular momentum $l \in \mathbb{Z}^+$, is given by
\begin{equation}
\left(-\boldsymbol{\nabla}^2  + \frac{l(l+1)}{r^2} \right) \Phi^{(V)}(t,r) = 0~,
\end{equation}
where $\boldsymbol{\nabla}^2$ denotes the Laplacian on a two-dimensional de Sitter space with curvature $+2/\ell^2$. The solutions can be expressed as hypergeometric functions  (see for instance \cite{Bredberg:2011xw}). For a given frequency, they take the form
\begin{equation}
    \Phi^{(V)} \= \Re\, e^{-i\omega t}\left(1-\frac{r^2}{\ell^2}\right)^{-i\omega\ell/2} \left(\frac{r^2}{\ell^2}\right)^{i\omega\ell/2} {}_2F_1\left(-l-i\omega \ell,1+l-i\omega\ell;1-i\omega\ell;\frac{1}{2}-\frac{\ell}{2r}\right) \, .
\end{equation}
The boundary condition \eqref{eqn: bdry cond 2} is automatically satisfied while the boundary condition \eqref{eqn: bdry cond 1} imposes
\begin{equation}
    \left.\frac{\Phi^{(V)}}{r} + \partial_r \Phi^{(V)}\right|_{r=\frakr}\=0 \, .
\end{equation}
Upon scanning numerically for solutions in the complex frequency plane, we find that all vector modes satisfying the conformal boundary conditions have a negative imaginary part, and are therefore dissipative, decaying at late times. 

\textbf{Worldline limit.} In the worldline limit, where $K\ell \to -\infty$, we find two sets of modes. One set is found to be a small deformation of the quasinormal modes of the de Sitter static patch, whose analytic form (in the worldline limit) is given by \cite{Lopez-Ortega:2006aal} $\omega^{\text{qnm}} \ell = - i (l+n+1)$ where $n\in\mathbb{N}$. As for the exact quasinormal modes, the modes we find are also purely negative imaginary and their size is of the order of the de Sitter length $\ell$. The other set of modes have a real part also and are the counterpart of the vectorial Minkowski modes uncovered in \cite{Anninos:2023epi}. For each $l \ge 2$, the second set is a discrete tower of modes with increasing negative imaginary parts, and their size scales with the size of the worldtube $\frak{r}$, rather than the de Sitter length. 

\textbf{Cosmological horizon limit.} In the cosmological horizon limit, where $K\ell \to +\infty$, the structure of the modes is altered. The purely imaginary modes degenerate into a set of modes that approach 
\begin{equation}\label{shear}
    \omega^{\text{shear}} \ell  \= - i \left(l(l+1)-2\right) \frac{1}{2 K^2\ell^2} + \mathcal{O}(K^{-3}\ell^{-3}) \, .
\end{equation}
These modes were identified in \cite{Anninos:2011zn} where they were interpreted as shear modes of a linearised incompressible non-relativistic fluid dynamical behavior near the horizon, paralleling other considerations of the fluid/gravity relation \cite{Damour:1978cg,znajek1978electric,Price:1986yy,Bhattacharyya:2008kq,Bredberg:2011xw}. For each $l \ge 2$, the second set is a discrete tower of modes with increasing negative imaginary parts, and their size scales with the de Sitter length.

\subsection{Scalar perturbation}\label{sec: scalar pert}

The master equation for $\Phi^{(S)}$ is given by
\begin{equation}
\left(-\boldsymbol{\nabla}^2  + \frac{l(l+1)}{r^2} \right) \Phi^{(S)} = 0~.
\end{equation}
The solutions can be expressed as hypergeometric functions, and take the form
\begin{equation}\label{PhiS}
    \Phi^{(S)} \= \Re\, e^{-i\omega t}\left(1-\frac{r^2}{\ell^2}\right)^{-i\omega\ell/2} \left(\frac{r^2}{\ell^2}\right)^{i\omega\ell/2} {}_2F_1\left(-l-i\omega \ell,1+l-i\omega\ell;1-i\omega\ell;\frac{1}{2}-\frac{\ell}{2r}\right) \, .
\end{equation}
The boundary condition \eqref{eqn: bdry cond 1} is automatically satisfied while the boundary condition \eqref{eqn: bdry cond 2} imposes
\begin{eqnarray}\label{eqn: bdry cond scalar perturbation}
	\mathcal{F}_l(K\ell\,,  \omega\ell) \,\equiv\,\left.\left(\frac{a_1}{r^4}+\frac{a_2}{r^2}\partial_t^2 - 2 \partial_t^4\right)\Phi^S  + \left(\frac{a_3}{r^2} -2 \partial_t^2\right)\left(1-\frac{r^2}{\ell^2}\right)^2\frac{\partial_r \Phi^{(S)}}{r} \right|_{r=\frakr} \= 0 \, ,
 \end{eqnarray}
 where
 \begin{eqnarray}
     \begin{cases}
	& a_1  \= l(l+1)\left(1-\frac{r^2}{\ell^2}\right)\left(3-\frac{2r^2}{\ell^2}-2l(l+1)\left(1-\frac{r^2}{\ell^2}\right)\right) \, , \\
	& a_2  \= 4-\frac{2r^2}{\ell^2} - 4 l(l+1)\left(1-\frac{r^2}{\ell^2}\right) \, , \\
	& a_3  \= 4-\frac{2r^2}{\ell^2}-l(l+1)\left(3-\frac{2r^2}{\ell^2}\right) \, .
    \end{cases}
\end{eqnarray}
Upon scanning numerically for solutions in the complex frequency plane, we find that some scalar modes satisfying the conformal boundary conditions have a positive imaginary part. 

\textbf{Worldline limit.} In the worldline limit, where $K\ell \to -\infty$, we find two sets of allowed frequencies. One set corresponds to a small deformation of the scalar quasinormal modes in the pure static patch, which are of the order of the de Sitter length scale $\ell$. The deformed quasinormal mode frequencies have a negative imaginary part, and are hence dissipative. The other set of allowed frequencies is borrowed from the analogous modes in Minwkoski space uncovered in \cite{Anninos:2023epi} and are of the order of the worldtube size $\mathfrak{r} \ll \ell$. For this set of modes, we find a pair of allowed frequencies with positive imaginary part for each $l$. 

\textbf{Cosmological horizon limit.}  In the cosmological horizon limit, where $K\ell \to +\infty$, the structure of the modes is altered. There is still a collection of fluid-type modes, but they take a relativistic dispersion relation. Upon expanding $r$ near $\ell$ in (\ref{PhiS}), implementing the conformal boundary conditions, and scaling $\omega$ with $1/K$, we find the analytic expansion
\begin{equation}
    \omega^{\text{sound}} \ell \= \pm  \frac{1}{\sqrt{2}K\ell} \sqrt{l(l+1)} - i\frac{l(l+1)-2}{2}  \frac{1}{2K^2\ell^2}+ \mathcal{O}(K^{-3}\ell^{-3})\, .
\end{equation}
It is natural to interpret the above modes as the sound mode counterpart to the fluid dynamical shear modes in (\ref{shear}). It is worth noting, however, that they scale differently with $K\ell$, such that in the strict horizon limit only the shear modes survive. This is one of the reasons the sound modes, whose speed of sound becomes infinite in this limit, does not appear in the previous analyses of \cite{Bredberg:2011xw,Anninos:2011zn}. 

\begin{figure}[h!]
        \centering
         \subfigure[$l = 2$]{
                \includegraphics[width = 7.51 cm]{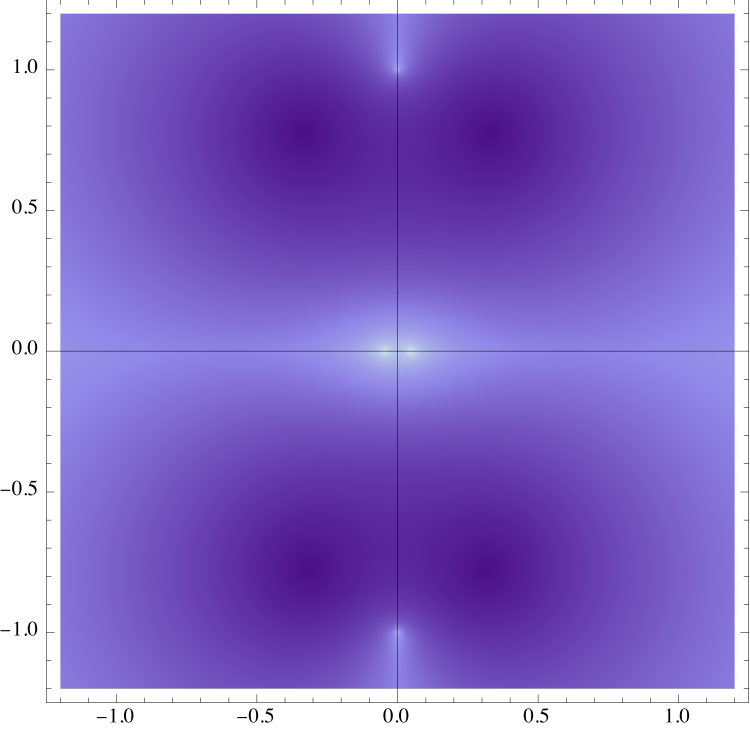}\label{fig: l2}}  \quad\quad
                 \subfigure[$l = 10$]{
                \includegraphics[width= 7.51 cm]{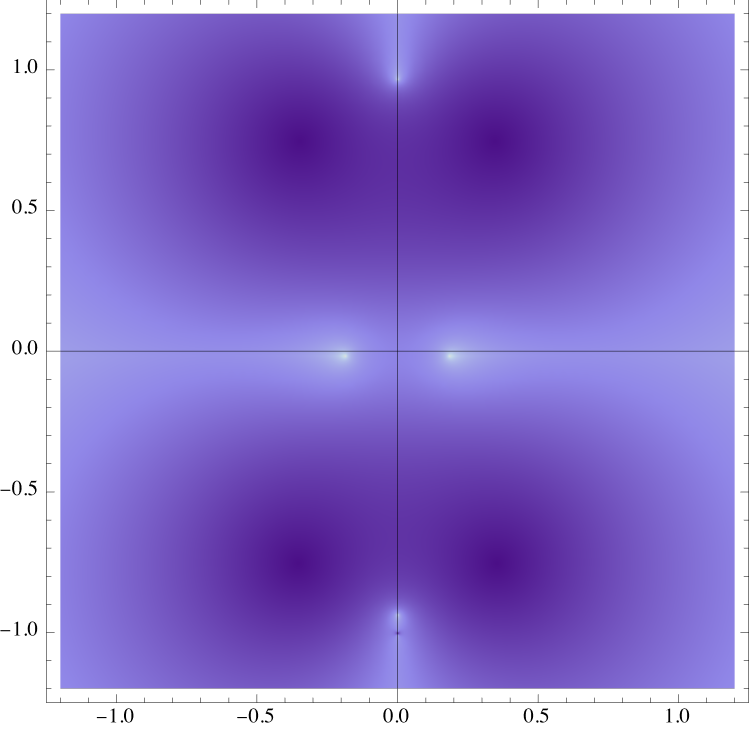} \label{fig: l10}}                           
                \caption{Density plot of absolute value of $\log \left(e^{-4 
                \omega \ell i}\mathcal{F}_l(K\ell\,,  \omega\ell)\right)$ in the complex $\omega\ell$ plane for $l=2$ and $l=10$, where $\mathcal{F}_l(K\ell\,,  \omega\ell)$ is defined in \eqref{eqn: bdry cond scalar perturbation}. In both plots, $K\ell$ is fixed to be $40$. Both the $\omega\ell \approx \pm i$ and $\omega^\text{sound}\ell$ are displayed in both plots. For $l=10$, the sound modes develop a small negative imaginary part.} \label{fig: omega density plot}
\end{figure}

In addition to the sound modes, upon taking the strict horizon limit, $K\ell \to +\infty$, of the scalar solutions (\ref{PhiS}) for each $l$,  and implementing the conformal boundary conditions, any solutions with modes with positive imaginary frequency coalesce either to $\omega \ell \= + i$ or $\omega \ell \= - i$. We show these in figure \ref{fig: omega density plot}. One can identify these modes in a Rindler analysis, subject to conformal boundary conditions. Concretely, we take the Rindler metric to be
\begin{equation}
    ds^2 \= -\frac{z^2}{z_0^2} dt^2 + dz^2 + dx^2 + dy^2~,
\end{equation}
with $(t, x,y) \,\in\, \mathbb{R}^3$ and $z \,\in\, \left(0,z_0\right]$. We then perform a straightforward analysis of the linearised Einstein equations, with vanishing $\Lambda$, subject to conformal boundary conditions at $z=z_0$. One observes that the following configuration (chosen for simplicity to have spatial momentum entirely along the $x$-direction)
\begin{equation}\label{rindler}
    \begin{cases}
        h_{tt} &=\, - \frac{z^2}{z_0^2} h_{xx} \= - \frac{z^2}{z_0^2} h_{yy} \= -\frac{z^2}{z_0^2}e^{-i\left(\omega t - k_x x\right)} J_{-i\omega z_0}(-ikz) \, , \\
        h_{zz} &=\, \frac{e^{-i\left(\omega t - k_x x\right)}}{k^2z^2} \left(\left(k^2z^2 + 2 \omega^2 z_0^2\right)  + \left(2k^2z^2-2\omega^2z_0^2\right)z\partial_z\right) J_{-i\omega z_0}(-ikz) \, , \\
        h_{zt} &=\, \frac{i\omega e^{-i\left(\omega t - k_x x\right)}}{k^2 z}\left(\left(2-k^2z^2+\omega^2z_0^2\right) - z \partial_z\right) J_{-i\omega z_0}(-ikz) \, , \\
        h_{zx} &=\, -\frac{ie^{-i\left(\omega t - k_x x\right)}}{k_x z} \left(\left(-k^2z^2 + \omega^2z_0^2\right) + z \partial_z\right)  J_{-i\omega z_0}(-ikz)\, ,
    \end{cases}
\end{equation}
solves the linearised Einstein equations with $\Lambda \= 0$, subject to conformal boundary conditions at $z=z_0$, for a selection of complex frequencies. In the limit $k z_0 \to 0$ the allowed frequencies coalesce to $\omega z_0 \= \pm i$. This mode coincides with the linearised de Sitter mode with $\omega\ell= \pm i$. Upon expressing the Rindler solution (\ref{rindler}) in terms of a local inertial time coordinate, one notes that it grows at most polynomially. As such, although growing in time, the exponential growth of (\ref{rindler}) is more tame than an ordinary unstable mode. In fact, to leading order at small $k z_0$, the Rindler mode is locally a pure diffeomorphism. 

As shown in appendix \ref{diffeoapp}, the leading contribution to the de Sitter scalar modes (\ref{PhiS}) with $\omega = \pm i\ell$, in the stretched cosmological horizon limit, are locally pure gauge. This leads to a double suppression effect whereby the physical contribution of the modes is not only small due to the linearised nature of $h_{\mu\nu}$, but also due to a suppression factor that goes as $1/K^2\ell^2$.

\subsection{Dynamics of the pole patch, briefly}

One can also consider linearised gravity subject to conformal boundary conditions in the pole patch. Here we further impose that the gravitational solutions are  smooth throughout the whole interior. The pole patch has been explored as a potential candidate for static patch holography in \cite{banks2012holographic,Coleman:2020jte,Susskind:2021esx,Shaghoulian:2021cef,Jorstad:2022mls} among other places. 

\textbf{Worldline limit.} For $K\ell \to +\infty$ the size of the timelike boundary is small in units of the de Sitter length $\ell$. Here our analysis matches the Minkowski analysis presented in \cite{Anninos:2023epi}, where it was observed that the allowed vector modes frequencies are all real-valued, whilst the scalar mode frequencies permit a subset of modes $\omega^{(S)}$ with positive real imaginary frequency for each $l$. At large $l$, these modes were numerically found to scale as $\omega^{(S)} \frak{r} \approx \pm l + i c l^{1/3}$, where $c$ is an order one number. Thus, the pole patch in the thin world tube limit mimics the Minkowskian picture. 


\textbf{Cosmological horizon limit.} For $K\ell \to -\infty$ the timelike boundary of the pole patch approaches the cosmological horizon. In this regime, the analysis differs from the thin worldline limit. Nonetheless, we observe the presence of scalar modes with complex frequencies of positive imaginary part for each $l$. These modes, similarly to the cosmic patch,  coalesce onto $\omega^{\text{pole}}\ell \= \pm i$. Moreover, we find two additional sets of modes. The first set is a pair of real valued scalar modes for each $l$. Numerically, at large $l$, they are found to be proportional to $\tfrac{l}{K\ell}$. The second set is an infinite tower of real valued modes, appearing both in the vector and scalar sector. At large $l$, they are numerically found to be evenly spaced and their magnitude is inversely proportional to $\log|\sqrt{2}K\ell|$.

\begin{center}
\pgfornament[height=5pt, color=black]{83}
\end{center}
\vspace{5pt}

We can synthesise, in short. We have provided evidence that, subjected to conformal boundary conditions, the stretched horizon limit of the pure de Sitter patch is a thermodynamically stable portion of spacetime containing a cosmological horizon. Dynamically, the majority of linearised gravitational perturbations about this portion of spacetime decay at late times, save one mode for each $l$ of total angular momentum. These modes have a purely imaginary frequency $\omega\ell = + i$, and are moreover retrieved from a Rindler analysis. We take this latter property as an indication that they are not endemic to de Sitter space, but rather a universal property of near horizon physics subjected to conformal boundary conditions. Their fate, whose behavior as measured by a local inertial clock is at most polynomial in time, is a remaining obstacle in obtaining a portion of spacetime with $\Lambda>0$ that is both thermodynamically stable, as well as dynamically stable at the linearised level. Perhaps to tame this mode we must impose an additional boundary condition, always ensuring that in doing so we do not overly restrict any interesting dynamics. Or perhaps we must relax the condition of a constant $K$. A careful examination is left to the future. 




\chapter{Negative cosmological constant}\label{chap: AdS}

\section{Introduction}
\label{subsection: Introduction}

The class of four-dimensional asymptotically anti-de Sitter solutions to the Einstein field equations with negative cosmological constant $\Lambda=-\tfrac{3}{\ell^2}$  admit the systematic expansion 
\begin{equation}\label{fg}
    \frac{ds^2}{\ell^2} = d\rho^2 + e^{2\rho}\left(g^{(0)}_{mn} + e^{-2\rho}g^{(2)}_{mn} + e^{-3\rho} g^{(3)}_{mn}  + \mathcal{O}(e^{-4\rho}) \right)dx^m dx^n \, ,
\end{equation}
first uncovered by Fefferman and Graham \cite{fefferman1985conformal}. The asymptotic AdS$_4$ boundary, $\Gamma_\infty$, is located at ${\rho} = \infty$. The asymptotic boundary, $\Gamma_\infty$, is endowed with asymptotic data  composed of an equivalence class under conformal transformations
\begin{equation}\label{Omega}
\{g^{(0)}_{mn},g^{(3)}_{mn} \} \sim \{\Omega^2(x^m) g^{(0)}_{mn},\Omega^{-1}(x^m) g^{(3)}_{mn} \}~,
\end{equation}
where $\Omega(x^m)$ is a smooth positive function of the boundary coordinates $x^m$. From the perspective of the AdS$_4$/CFT$_3$ correspondence, this is a reflection of the fact that a three-dimensional conformal field theory naturally lives on a conformal metric, which we occasionally denote as $[g^{(0)}_{mn}]$ to indicate it contains $g^{(0)}_{mn}$ as an instance. Moreover, the Einstein equations impose that $g^{(3)}_{mn}$ is transverse and traceless with respect to $g^{(0)}_{mn}$, and that $g^{(2)}_{mn}$ is fixed entirely in terms of  $g^{(0)}_{mn}$. Another general feature of the expansion (\ref{fg}), which is somewhat less often emphasised, is that the trace of the extrinsic curvature, $K(x^m)$, along $\Gamma_\infty$ takes the fixed value $K \ell = 3$ regardless of the choice $\{g^{(0)}_{mn},g^{(3)}_{mn} \}$. 

In Lorentzian signature, physically distinct configurations carrying different stress-energy are registered by changes in $g^{(3)}_{mn}$. The two independent functions in $g^{(3)}_{mn}$ encode the two locally propagating gravitational degrees of freedom. Going from one conformal frame in (\ref{Omega}) to another can be achieved by a coordinate transformation that preserves the form of (\ref{fg}). As such, $\Omega(x^m)$ does not carry physical information about the configuration --- it is pure gauge. To develop a particular initial condition, we must fix Cauchy data along a complete spatial slice $\Sigma$, which can be taken to be the set $\mathcal{C}_\Sigma = \{\tilde{g}_{ij},\tilde{K}_{ij}\}$ given by the induced metric and extrinsic curvature along the spatial slice $\Sigma$ equipped with coordinates $x^i$. The Cauchy data $\mathcal{C}_\Sigma$ must further satisfy the gravitational constraint equations along $\Sigma$, and is identified under tangential diffeomorphisms mapping $\Sigma$ to itself. The conformal structure of the boundary metric $g^{(0)}_{mn}$ is fixed data specific to a  particular theory. In Euclidean signature, one views  $g^{(0)}_{mn}$ as a source for the boundary stress-tensor \cite{Witten:1998qj}, and one seeks instead for a smooth solution of the Euclidean equations with the prescribed asymptotic data.

In this article, we are concerned with the question of how to move the asymptotic boundary $\Gamma_\infty$ into the interior of four-dimensional anti-de Sitter space (AdS$_4$). Our analysis is inherently gravitational, with the goal of acquiring sufficient theoretical ground that might guide us toward a finite-size generalisation of the AdS$_4$/CFT$_3$ dictionary. Previous works on AdS$_4$/CFT$_3$ at finite boundary include ideas about the holographic renormalisation group \cite{Akhmedov:1998vf,deBoer:1999tgo,Lee:2013dln,Heemskerk:2010hk,Skenderis:2002wp}, as well as more recent discussions on a  higher-dimensional version of the $T\bar{T}$ deformation \cite{Hartman:2018tkw,Taylor:2018xcy,Araujo-Regado:2022gvw,Silverstein:2024xnr}. 

At the face of it, and as originally envisioned by York for the Minkowskian case \cite{York:1986it}, one might be inclined to bring in the asymptotic boundary $\Gamma_\infty$ to a large but finite value of $\rho$ and fixing the entire induced metric $g_{mn}$ along a finite-size timelike boundary $\Gamma$, along with some Cauchy data along a spatial slice $\Sigma$ that intersects $\Gamma$ along a two-dimensional spatial boundary $\partial \Sigma$ \cite{Marolf:2012dr,Andrade:2015gja}. In doing so, and following \cite{An:2021fcq}, we should ensure that the Einstein constraint equation projected  along $\Gamma$ is satisfied, namely,
\begin{equation}\label{cG}
\mathcal{R} - K^2 + K_{mn} K^{mn} - 2\Lambda |_{\Gamma} = 0~,
\end{equation}
where $\mathcal{R}$ is the Ricci scalar with respect to the induced metric $g_{mn}$. 
Parameterically close to the AdS$_4$ boundary, there is no immediate obstruction to the Dirichlet problem.  An argument for this is presented in the linearised approximation about a planar AdS$_4$ background in appendix \ref{app: Dirichlet problem}.  On the other hand, it was established in \cite{Avramidi:1997sh,Anderson_2008,An:2021fcq} that when the boundary $\Gamma$ resides in an approximately Minkwoskian or Rindler corner, the constraint equation (\ref{cG}) will not be satisfied for generic $g_{mn}$. (It should be noted, that although not generic, there do exist many solutions to the Dirichlet problem as well. These include the important case of an $S^2\times \mathbb{R}$ boundary.)

Instead of fixing Dirichlet data along $\Gamma$, bearing in mind that the entire Fefferman-Graham configuration space obeys  $K\ell=3$ at $\Gamma_\infty$, one may consider \cite{York:1986lje} fixing the conformal class of the induced metric, $[g_{mn}]$, at $\Gamma$ along with the trace of the extrinsic curvature, $K(x^m)$. Such a boundary condition, which we will refer to as the conformal boundary condition, has been the subject of recent work \cite{Witten:2018lgb,Anninos:2023epi,Anninos:2024wpy,Liu:2024ymn,Banihashemi:2024yye}, whilst also having appeared previously in the fluid-gravity literature \cite{Bredberg:2011xw,Anninos:2011zn}. From the mathematical side, conformal boundary conditions were proven in  \cite{Anderson_2008} to constitute a well-posed elliptic problem in  Euclidean signature. In Lorentzian signature conformal boundary conditions, accompanied by appropriate Cauchy data along some initial time slice $\Sigma$, have been conjectured  \cite{An:2021fcq} to constitute a well-posed initial boundary value problem (see also \cite{An:2024taa,Capoferri:2024sgo} for recent developments). The corresponding gravitational phase space and Brown-York boundary stress-tensor, $T_{mn}$, have been considered in \cite{Odak:2021axr}. Interestingly, $T_{mn}$ is traceless and, in the case of constant $K$, conserved. 

In what follows we will analyse  general relativity with a negative cosmological constant $\Lambda=-\tfrac{3}{\ell^2}$ on a manifold with a finite size timelike boundary subject to conformal boundary conditions. Schematically, the location of the boundary is controlled by $K\ell$. The  shape of the boundary $\Gamma$ is instead controlled by the conformal class $[g_{mn}]$, which is fixed, along with a novel dynamical boundary degree of freedom $\bomega(x^m)$ that comes to life upon bringing in the asymptotic AdS$_4$ boundary.\footnote{The appearance of a dynamical boundary mode is somewhat reminiscent of what happens in topological quantum field theories when quantised on a manifold with a boundary \cite{Elitzur:1989nr}. In the context at hand, a linearised analysis around the Minkowski corner already indicates the existence of a boundary/corner mode that cannot be undone with a permissible diffeomorphism. This was studied in subsection 4.1.2 of \cite{Anninos:2023epi} for the de Donder gauge. Relatedly, in appendix B of \cite{Anninos:2022ujl} it was shown in the de Donder gauge and about the Minkowski corner, that upon fixing the entire initial value data set along a Cauchy surface (including the initial data of the boundary/corner mode) to vanish fixes the linearised evolution uniquely for conformal boundary conditions. These properties can be confirmed for other gauge choices also. Relatedly, for a spherical spatial boundary, it was shown in \cite{Anninos:2023epi,Anninos:2024wpy,Liu:2024ymn} that there exists a spherically symmetric physical diffeomorphism (see (5.23) of \cite{Anninos:2023epi} for the infinitesimal case) whose initial data must further be fixed to ensure a unique evolution of the initial boundary value problem. The spherically symmetric diffeomorphisms can be treated at the non-linear level also (see footnote 2 of \cite{Anninos:2024wpy} and the more detailed subsection 4.1 of \cite{Liu:2024ymn}). A complete proof of well-posedness (or lack thereof) of the conformal boundary conditions must necessarily incorporate the boundary/corner mode at the non-linear level. It remains a very interesting open problem. We would like to thank M. Anderson and L. Lehner for illuminating discussions on this point.} Interestingly, although bringing in the AdS$_4$ boundary might be envisioned as the result integrating out degrees of freedom of the dual CFT$_3$, a new degree of freedom appears in the gravitational phase space. Moreover, $\bomega(x^m)$ interacts non-trivially with the bulk gravitational degrees of freedom. From the perspective of AdS$_4$/CFT$_3$, based on our analysis we anticipate that bringing in the asymptotic boundary invokes at least two novel features. Firstly,  a new dynamical field $\bomega(x^m)$, that is not part of the original CFT$_3$, appears in the theory. Secondly, the conformal invariance of the CFT$_3$ on $S^2\times \mathbb{R}$ appears to be broken due to the presence of growing modes in the linearised spectrum about global AdS$_4$. From the gravitational side, the interaction strength between $\bomega(x^m)$ and the CFT$_3$ fields --- or, similarly, the degree of conformal symmetric breaking --- is governed by the deviation, $\delta K = K-\tfrac{3}{\ell}$, in the trace of the extrinsic curvature away from its asymptotic boundary value. An alternative somewhat more specialised boundary condition which has been shown to have good properties for $\Lambda \le 0$, is the `umbilical' boundary condition \cite{schlenker2001einstein,Fournodavlos:2021eye} whereby the induced metric at the boundary is taken to be proportional to the extrinsic curvature.

From a thermodynamic point of view, one can consider computing the horizon  entropy of a large Schwarzschild AdS$_4$ black hole subject to conformal boundary conditions, as analysed in \cite{Coleman:2020jte,Anninos:2023epi,Banihashemi:2024yye}. In a high-temperature limit, and taking $K\ell > 3$ to be constant,  the leading order contribution to the Bekenstein-Hawking relation reads \cite{Anninos:2023epi,Banihashemi:2024yye}
\begin{equation}\label{BH}
S_{\text{BH}} = \frac{16\pi^3 \ell^2}{81  G_N} \left( K \ell - \sqrt{K^2\ell^2-9} \right)^2 {\beta^{-2}}~,
\end{equation} 
where ${\beta}$ is the (conformally invariant) inverse Hawking temperature as measured by the boundary clock. As a function of $\beta$, the above behaves as the thermal entropy of a three-dimensional conformal field theory at high-temperature. The pre-factor of $\beta^{-2}$ is a monotonically decreasing function of $K\ell$, that tends to the Minkowskian value $\tfrac{4\pi^3}{G_N K^2}$ \cite{Anninos:2023epi} at large $K\ell$, and to the standard AdS$_4$/CFT$_3$ pre-factor $ \tfrac{16 \ell^2 \pi^3}{9 G_N}$ in the limit $K\ell \to 3$. The pre-factor is unaltered for the planar AdS$_4$ black brane with conformal boundary conditions also. We might view the pre-factor, then, as a generalised measure of the number of degrees of freedom, at least in some semiclassical sense, for the putative theory dual to the gravitational theory with finite size conformal boundary. Unlike a Dirichlet boundary, there is no obstruction in making the size of the horizon arbitrarily large for the case at hand.\footnote{Interestingly, a similar expression to (\ref{BH}) holds for a de Sitter horizon surrounding a timelike boundary subject to conformal boundary conditions \cite{Anninos:2024wpy}, yielding a positive prefactor irrespective of the sign of $K \ell_{\text{dS}}$. In contrast to the Dirichlet case, this indicates that the de Sitter horizon subject to conformal boundary conditions can have a positive specific heat.}

As an alternative measure of the number of degrees of freedom, we might consider the semiclassical gravitational path integral for Euclidean AdS$_4$ with an $S^3$ boundary (see for example \cite{Jafferis:2011zi}). For conformal boundary data given by the conformal class containing the round metric on $S^3$, and a $K$ which is independent of $x^m$, a straightforward computation for the saddle point approximation of the gravitational path integral yields
\begin{equation}\label{Zs3}
\log \mathcal{Z}[S^3,K] =  \frac{\pi \ell^2}{2 G_N}   \left(\frac{{K\ell}-\sqrt{K^2 \ell^2-9}}{\sqrt{K^2 \ell^2-9}}\right) \approx \frac{\pi \ell^2}{2 G_N}   \left( \sqrt{\frac{3}{2(K\ell-3)}}-1 + \ldots \right)~.
\end{equation}
We note that $\log \mathcal{Z}[S^3,K]$ is a monotonically decreasing function of $K\ell$. Once we deviate from $K\ell=3$, the divergences that are ordinarily removed by local counterterms in AdS$_4$/CFT$_3$ are subsumed into the unified expression (\ref{Zs3}). In ordinary AdS$_4$/CFT$_3$, the divergent term in the final expansion of (\ref{Zs3}) corresponds to an ultraviolet divergence that can be removed with a standard local counterterm \cite{Skenderis:2002wp}. Perhaps away from $K\ell=3$, there is a different perspective on this, that no longer allows us to discard such terms and instead they are combined into a unified picture. This might tie with the question of whether the holographic theory describing an anti-de Sitter universe with a finite-size is a non-local theory --- if so, attempting to disentangle (would be) ultraviolet divergences from other effects might be somewhat unfounded.

The outline of the paper is as follows. In subsection \ref{sec: framework}, we introduce the general framework for our analysis. We discuss the Brown-York stress energy tensor, $T_{mn}$, pertinent to the conformal boundary conditions \cite{Odak:2021axr}, and make contact with the boundary stress tensor \cite{Balasubramanian:1999re} at the asymptotic AdS$_4$ boundary. In subsection \ref{sec: lin dynamics global AdS}, we analyse the linearised gravitational equations subject to conformal boundary conditions about the global AdS$_4$ spacetime. In addition to a collection of ordinary gravitational normal modes, we uncover a boundary mode, ${\bomega(x^m)}$. Along with its time derivative, ${\bomega(x^m)}$ is part of the phase space of the gravitational theory. For sufficiently large angular momentum, we find a collection of modes with complex frequency that generalise those uncovered in the Minkowskian analysis of \cite{Anninos:2023epi}. When the boundary, $\Gamma$, is pushed closer to the asymptotic boundary, $\Gamma_\infty$, the complex modes appear at increasingly large angular momentum. In subsection \ref{sec: planar AdS}, we analyse the linearised gravitational equations subject to conformal boundary conditions about the planar AdS$_4$ spacetime. In this case, we uncover a massless boundary mode, ${\bomega(x^m)}$,  which can now be expressed (locally) as a diffeomorphism. Unlike the case of global AdS$_4$, we find that there are no exponentially growing modes for the planar case. We end the subsection by analysing the $\bomega(x^m)$ mode about an AdS$_4$ black brane background, and show it can dress the black hole in a way that modifies its energy. In subsection \ref{sec: source}, we consider the Euclidean problem. In particular, we show that the boundary datum $K({x}^m)$ acts as a source for $\bomega(x^m)$ --- extending the usual Euclidean AdS$_4$/CFT$_3$ dictionary. The output of the Euclidean gravitational path integral will now be a functional, $\mathcal{Z}[[g_{mn}(x^m)], K(x^m)]$, of the conformal structure, $[g_{mn}]$, and the mean curvature, $K(x^m)$, along $\Gamma$. Details of various computations and technical derivations, can be found in the several appendices.

\section{General framework}
\label{sec: framework}

In this subsection we give a brief overview of the general framework for the conformal boundary condition problem. Our discussion follows that of \cite{Anninos:2023epi, Anninos:2024wpy}, where a more detailed account can be found. 

\subsection{Initial boundary value problem}

Our Lorentzian spacetime action is given by
\begin{equation}\label{bulkS}
I = \frac{1}{16\pi G_N} \int_{\mathcal{M}} d^4 x \sqrt{-\det g_{\mu\nu}} \left( R  - 2\Lambda \right) + \frac{1}{24\pi G_N} \int_\Gamma d^3 x \sqrt{-\det g_{mn}} K~.
\end{equation}
The timelike boundary is denoted by $\Gamma$, and is endowed with the induced metric $g_{mn}$. The quantity $K = g^{mn} K_{mn}$, is the trace of the extrinsic curvature $K_{mn} = \tfrac{1}{2} \mathcal{L}_{n^\mu} g_{mn}$ with respect to the outward-pointing unit normal $n^\mu$. Our cosmological constant $\Lambda=-\tfrac{3}{\ell^2}$ is taken to be negative in what follows.  The boundary conditions we will impose fix the conformal class of the induced metric at the timelike boundary, $[g_{mn}]|_\Gamma$, as well as $K$ \cite{An:2021fcq}, and we denote these as conformal boundary conditions in what follows. The boundary term in (\ref{bulkS}), which differs from that of the standard Gibbons-Hawking-York term by a coefficient,  has been adjusted so as to permit for a well-defined variational problem \cite{An:2021fcq, Odak:2021axr} with respect to the conformal boundary conditions. Coordinates along the timelike boundary $\Gamma$ will often be denoted by $x^m$. 

The conformal boundary conditions along $\Gamma$ must be supplied with Cauchy data along a spatial initial time slice $\Sigma$. Part of the Cauchy data is comprised of the induced metric, $\tilde{g}_{ij}$ along $\Sigma$, as well as the extrinsic curvature, $\tilde{K}_{ij}$, with respect to the unit normal $\tilde{n}^\mu$ of $\Sigma$. The Cauchy data $\mathcal{C}_\Sigma = \{\tilde{g}_{ij},\tilde{K}_{ij}\}_{\Sigma}$ must be subjected to the gravitational constraint equations along $\Sigma$, and forms an equivalence class under tangential diffeomorphisms that preserve the conformal boundary conditions and map $\Sigma$ to itself. We must further ensure that the fields at the boundary $\partial\Sigma$ of $\Sigma$ are compatible with the boundary data along $\Gamma$ that intersects with $\partial\Sigma$.

In addition, we must specify the initial conditions for any independent boundary data residing $\partial \Sigma$. As will appear in our linearised analysis below, the boundary data is encoded by the volume element $\sqrt{\det \tilde{g}_{ij}}|_{\partial{\Sigma}}$ and the corner angle $\beta \equiv \sinh^{-1}{n^\mu \tilde{n}_\mu|_{\partial\Sigma}}$.  To properly do so, as will be detailed below, we must supplement the Cauchy data along $\Sigma$ with the initial data of a boundary mode, $\bomega(x^m)$, residing at $\partial\Sigma$ \cite{Odak:2021axr,elsewhere}.\footnote{The reason that there is independent data at $\partial\Sigma$ --- itself not part of $\mathcal{C}_\Sigma$ --- is that unlike the case of a complete Cauchy surface, we are now quotienting configurations along $\Sigma$ by a restriction of the tangential diffeomorphisms. This allows for the possibility of non-trivial contributions to the configuration space that may locally appear as a pure diffeomorphism. Explicit examples of such configurations will be presented in the following subsection.} Generically, when $\Sigma$ intersects $\Gamma$ at a non-perpendicular angle, an additional corner term (known as the Hayward term \cite{Hayward:1990zm}) must be supplemented to the action \cite{Lehner:2016vdi}. The coefficient of the Hayward term was shown to vanish for the case of conformal boundary conditions \cite{Odak:2021axr}.  A careful examination of these boundary degrees of freedom will be presented elsewhere \cite{elsewhere}, and previous work can be found in \cite{Odak:2021axr,An:2021fcq}.

\textbf{Linearised initial boundary value problem.} In much of what follows, we will often be considering the problem at the linearised level. To this end, consider a metric that is slightly perturbed away from a background $\bar{g}_{\mu\nu}$, such that
\begin{equation} \label{metric_pert_lin}
g_{\mu\nu} = \bar{g}_{\mu\nu} + \varepsilon h_{\mu\nu}~, \quad\quad |\varepsilon| \ll 1~.
\end{equation}
Unless otherwise specified, our background metric is taken to be of the form
\begin{equation}\label{gbar}
ds^2 = -f(r) dt^2 + \frac{dr^2}{f(r)} + r^2 d\Sigma_k~,
\end{equation}
where $\Sigma_k$ is a maximally symmetric two dimensional space of curvature $k= \{-1,0,1\}$. We will only focus on $k=\{0,1\}$ in what follows, leaving the case $k=-1$ for future work. In the case of $k=+1$, $d\Sigma_{+1} = d\Omega_2^2$ is the round metric on the unit two-sphere.

We take the boundary to be located along a surface of constant coordinate $r=\mathfrak{r}$. As boundary conditions, we take that the induced metric at the boundary is of the form
\begin{equation}\label{indmetric}
ds^2|_\Gamma =  e^{2\boldsymbol{\omega}(x^m)}\left(-f(\r) dt^2  + \r^2 d\Sigma_k \right)~,
\end{equation}
where $\boldsymbol{\omega}(x^m)$ is an unspecified function of the boundary coordinates, whilst the trace of the extrinsic curvature at the boundary is fixed to be $K$.

To preserve the conformal structure of the induced metric at the boundary, we must enforce that
\begin{equation}\label{eqn: bdry cond 1}
	\left.h_{mn} - \frac{1}{3}\bar{g}_{mn} h^p{}_p \right|_{\Gamma} \= 0 \, .
\end{equation}
The variation of the trace of the extrinsic curvature to first order in $\varepsilon$ is given by
\begin{equation}\label{eqn: bdry cond 2}
	\left.\frac{\delta K (h_{\mu\nu})}{\sqrt{f(r)}}\right|_{\Gamma} \equiv \left.\frac{K(\bar{g}_{\mu\nu}+\varepsilon h_{\mu\nu}) - K(\bar{g}_{\mu\nu})}{\varepsilon\sqrt{f(r)}} \right|_{\Gamma}= \left.\frac{1}{2} \partial_r h^m{}_m - \mathcal{D}^m h_{rm} - \frac{\sqrt{f(r)}}{2}K h_{rr} \right|_{\Gamma} = 0 \, ,
\end{equation}
where $\mathcal{D}_n$ denotes the covariant derivative with respect to the background metric $\bar{g}_{mn}$ induced at the timelike boundary. 

In order to fully specify the initial boundary value problem, we must also specify boundary conditions for the gauge parameter of the diffeomorphisms. At the linear level one has a vector field $\xi^\mu$ transforming
\begin{equation}
\begin{cases}
    x^\mu \to x^\mu - \varepsilon \, \xi^\mu~, \\
h_{\mu\nu} \to h_{\mu\nu} + \nabla_\mu \xi_\nu + \nabla_\nu \xi_\mu~,
\end{cases}
\end{equation}
where $\nabla_\mu$ denotes the standard covariant derivative. In order for the gauge parameter to be consistent with the conformal boundary conditions, we must impose that  
\begin{equation}\label{eqn: bdry conds diffeo}
\begin{cases}
    \left.2\left(K_{mn} - \frac{K}{3}\bar{g}_{mn}\right)\left(\sqrt{f(r)}\xi_r\right) + \left(\mathcal{D}_m \xi_n + \mathcal{D}_n \xi_m - \frac{2\bar{g}_{mn}}{3}\mathcal{D}^p \xi_p\right)\right|_{\Gamma} &=\, 0 \, , \\
    \left.\left(\sqrt{f(r)}\partial_r K - \mathcal{D}^m \mathcal{D}_m \right) \left(\sqrt{f(r)}\xi_r\right) + \xi_m \mathcal{D}^m K \right|_{\Gamma} &=\, 0 \, .
\end{cases}
\end{equation}
It is worth noting that the second term in the first equation corresponds to the conformal Killing equation with respect to the boundary metric $\bar{g}_{mn}$. Additionally, allowed diffeomorphisms should satisfy $\xi^r|_{\Gamma} = 0$, such that the location of the boundary is not disturbed.

\subsection{Conformal Brown-York boundary stress tensor} \label{sec: conf by}

One can compute the Brown-York boundary stress tensor $T_{m n}$ for our boundary value problem by varying the classical on-shell action with respect to the boundary conformal metric $[g]_{mn}|_\Gamma$, whilst keeping $K$ fixed. Here,  $[g]_{mn}$ denotes a particular representative of the conformal class $[g_{mn}]$. This exercise was undertaken in \cite{Odak:2021axr,An:2021fcq}, where it is shown that the resulting stress tensor takes the following form\footnote{The precise variation of the on-shell action with respect to the conformal metric gives the following conformal Brown-York tensor density \cite{Banihashemi:2024yye}
\begin{equation}
    T_{mn}^{\text{CBY}} \equiv -\frac{2}{\sqrt{-\text{det} [g]_{mn}}} \frac{\delta I}{\delta [g]^{mn}} = - \frac{e^{\bomega}}{8\pi G_N} \left(K_{mn} - \frac{1}{3} K g_{mn} \right) \left(-\det{[g]_{mn}}\right)^{1/6} \,,
\end{equation}
where $[g]_{mn}$ is the conformal metric at the boundary, defined via  $g_{mn} \equiv e^{2\bomega} [g]_{mn}$. Note that this definition for the stress tensor density differs from our definition by an overall factor of $\left(-\det{[g]_{mn}}\right)^{1/6}$.}
\begin{equation}\label{cby}
T_{m n} = \left. -\frac{e^{\boldsymbol{\omega}}}{8\pi G_N}\left( K_{m n} - \frac{1}{3} K g_{mn} \right) \right|_\Gamma ~.
\end{equation}
We note that $T_{mn}$ is manifestly traceless, and it transforms with conformal weight $-1$ under a Weyl transformation of the conformal representative.

The divergence of $T_{mn}$ (with respect to  (\ref{indmetric}) with $\bomega=0$) follows from the momentum constraint at $\Gamma$. One finds 
\begin{equation} \label{eq: divergence Tmn}
    \mathcal{D}^m T_{mn} = -\frac{1}{12\pi G_N}e^{3\bomega}\mathcal{D}_{n}K \, ,
\end{equation}
which renders $T_{mn}$ divergenceless when $K$ is a constant. For completeness, the Hamiltonian constraint at $\Gamma$ reads 
\begin{equation} \label{eq: Hamiltonian constraint}
    \mathcal{D}_m \mathcal{D}^m \bomega + \frac{1}{2}\ \mathcal{D}_m\bomega \mathcal{D}^m\bomega - \frac{1}{4}R - 16 \pi^2 G_N^2 T^{mn} T_{mn}e^{-4\bomega} + \frac{1}{6}\left(K^2-\frac{9}{\ell^2}\right) e^{2\bomega} = 0 \, ,
\end{equation}
where $R$ is the Ricci scalar of the conformal representative of $[g]_{mn}|_{\Gamma}$ with $\bomega=0$.\footnote{{For $(d+1)$-dimensional pure Einstein gravity with a cosmological constant $\Lambda$, the Hamiltonian constraint along a timelike boundary $\Gamma$ is given by
\begin{equation}
    \mathcal{D}_m\mathcal{D}^m \bomega + \frac{d-2}{2}\mathcal{D}_m \bomega \mathcal{D}^m \bomega - \frac{1}{2(d-1)}R - \frac{64 \pi^2G_N^2}{2(d-1)}T^{mn}T_{mn} e^{-2(d-1)\bomega} + \left(K^2 + \frac{2d}{d-1}\Lambda\right)\frac{e^{2\bomega}}{2d}=0 \, ,
\end{equation}
where the conformal Brown-York stress tensor is $T_{mn} =\left.- \frac{e^{(d-2)\bomega}}{8\pi G_N}\left(K_{mn}-\frac{1}{d}K g_{mn}\right)\right|_{\Gamma}$.}}
Both \eqref{eq: divergence Tmn} and \eqref{eq: Hamiltonian constraint} form {a hyperbolic version of Lichnerowicz-York equations} \cite{JMPA_1944_9_23__37_0, PhysRevLett.28.1082, choquet}.

Given the stress tensor \eqref{cby}, along with \eqref{eq: divergence Tmn}, one can construct a conserved quantity associated with tangential diffeomorphisms (\ref{eqn: bdry conds diffeo}) that preserve the conformal boundary conditions as follows. Let us take the case of constant $K$, and let $\xi^m$ be a conformal Killing vector of the induced metric \eqref{indmetric} with $\bomega=0$. If we further set $\xi^r$ to zero, it follows from \eqref{eqn: bdry conds diffeo} that $\xi^m$ preserves the conformal boundary conditions. The current $J_m \equiv \xi^n T_{mn}$ obeys the conservation equation $\mathcal{D}^m J_m = 0$, as implied by \eqref{eq: divergence Tmn} and the tracelessness of $T_{mn}$.\footnote{To obtain a conserved current $J_m$, one could relax the constant $K$ condition to $\xi^m \mathcal{D}_m K = 0$ for the $\xi^m$ of interest. Since the orthogonal component of $\xi^m$ is zero, this condition is equivalent to the trace of the extrinsic curvature-preserving condition \eqref{eqn: bdry conds diffeo}.} Let $u^m$ be a timelike unit-normal vector with respect to the same metric, \eqref{indmetric} with $\bomega=0$, whose explicit form is given by $u^m \partial_m = f(\r)^{-1/2}\partial_t$. It follows that 
\begin{equation}
    Q_\xi \equiv \r^2 \left. \int  d\Sigma_k  \, u^m J_m \right|_{\Gamma}\, ,
\end{equation} 
where $d\Sigma_k$ is the volume form of $\Sigma_k$, is a conserved charge with respect to $\xi^m$. Taking, for example, a time translation, $\xi^m  \partial_m = \partial_t$, the associated conserved quantity, $E_{\text{conf}}\equiv Q_{\partial_t}$, which we denote by the conformal energy, is given by
\begin{equation}
    E_\text{conf} = \left.\frac{\r^2}{\sqrt{f(\r)}} \int d\Sigma_k \, T_{tt} \right|_{\Gamma}\, .
\end{equation}
We note that $ E_\text{conf}$ is invariant under Weyl transformations of the conformal representative.

{\textbf{Fefferman-Graham limit.}} It is interesting to note here that in AdS/CFT, the Brown-York stress tensor is ordinarily computed \cite{Balasubramanian:1999re} with the standard Gibbons-Hawking-York boundary term that appears in the Dirichlet problem. Here, we have instead used an action (\ref{bulkS}) with a boundary term whose prefactor differs from the standard one by a factor of $\tfrac{2}{3}$. Let us consider a general asymptotically AdS$_4$ configuration in the following coordinates:
\begin{equation} \label{eq: fg limit}
\frac{ds^2}{\ell^2} = d\rho^2 + \frac{e^{2\rho}}{\ell^2} \left( g^{(0)}_{mn}  + \ell^2 e^{-2\rho} g^{(2)}_{mn} + \ell^3e^{-3\rho} g^{(3)}_{mn} + \ldots \right) dx^m dx^n~,
\end{equation}
where $g^{(2)}_{mn}$ is minus the Schouten tensor built from $g^{(0)}_{mn}$
\begin{equation}\label{schouten}
g^{(2)}_{mn} \equiv \frac{1}{4}R[g^{(0)}_{mn}] g^{(0)}_{mn}-R_{mn}[g^{(0)}_{mn}]~,
\end{equation}
and $g^{(3)}_{mn}$ is a transverse-traceless tensor, also with respect to $g^{(0)}_{mn}$. As a simple example, consider the case $g^{(0)}_{mn} = \eta_{mn}$. We impose conformal boundary conditions at a large $\rho = \rho_c$, such that
\begin{equation}
g_{mn}|_\Gamma =  e^{2\bomega(x^m) } \left(\eta_{mn} + \mathcal{O}\left(e^{-\rho_c}\right) \right)~, \quad\quad K \ell = 3 + \mathcal{O}\left(e^{-2\rho_c}\right)~,  
\end{equation}
where $\bomega=\rho_c$ and $[g]_{mn} = \eta_{mn}$. Given \eqref{eq: fg limit}, it is easy to compute the extrinsic curvature at the boundary (see appendix \ref{app: near infinity}). We then find that the conformal Brown-York stress tensor (\ref{cby}) evaluates to
\begin{equation}
T_{mn} = \frac{3\ell^2}{16\pi G_N } \, g^{(3)}_{mn} +... \,,
\end{equation}
as we approach the asymptotic boundary. The above expression is indeed proportional to the standard boundary stress-tensor in AdS$_4$/CFT$_3$. 

For general $g^{(0)}_{mn}$, and assuming that $R[g_{mn}^{(0)}]$ is non-vanishing, it is convenient to first bring in the boundary infinitesimally by moving $K\ell$ slightly  away from its AdS$_4$ boundary value. Details of this can be found in appendix \ref{app: near infinity}. Upon doing so, our conformal Brown-York stress tensor, to leading order in the deviation, reads
\begin{equation} \label{eq: brown-york near boundary}
    T_{mn} = -\frac{\ell^2}{8\pi G_N} \sqrt{\frac{R [g_{mn}^{(0)}]}{4(K\ell-3)}} \left( R_{mn}[g_{mn}^{(0)}] - \frac{1}{3}R [g_{mn}^{(0)}] g_{mn}^{(0)} \right) + \frac{3\ell^2}{16 \pi G_N} g^{(3)}_{mn} +...\, .
\end{equation}
The first term is a divergent quantity in the strict AdS$_4$ boundary limit, whereby $K\ell \to 3$. The divergent term is fixed in terms of the asymptotic AdS$_4$ boundary data, and is often removed via some renormalisation scheme \cite{Skenderis:2002wp}. Interestingly, the divergent term depends on the conformal structure of $g^{(0)}_{mn}$ and thus, for $K\ell \neq 3$, it contributes to the boundary stress-energy tensor (\ref{cby}) in a physically meaningful way. In particular, we will see that it can receive non-trivial contributions from the field $\bomega$, which is now part of the dynamical phase space. 

What happens in the strict asymptotically AdS$_4$ limit, is that the boundary mode effectively decouples from the remaining dynamics, and it becomes reasonable to remove it from the phase space altogether. The finite term, proportional to $g^{(3)}_{mn}$, is the standard boundary stress-tensor in AdS$_4$/CFT$_3$ \cite{Balasubramanian:1999re}. As such, the standard AdS$_4$/CFT$_3$ boundary stress-tensor is recovered from the conformal boundary conditions, even though the coefficient of the Gibbons-Hawking-York term is different from that employed in the original literature.\footnote{We would like to acknowledge H. Liu, D. Harlow, and E. Shaghoulian for valuable discussions regarding this point.} 

The difference in the approach we have discussed, as compared to the standard Dirichlet approach, is how one controls the deviation away from the asymptotic boundary. Here, instead of fixing a finite size induced metric at the boundary $\Gamma$, and varying its size, we tune the value $K\ell$. That the choice of conformal boundary conditions is argued \cite{Anderson_2008,An:2021fcq} to have, generically, better well-posedness properties than the Dirichlet one, might be viewed as adding further value to this approach. 

In any case, in order to further quantify the physical content of the conformal boundary conditions, it is important to understand their dynamical implications. We will proceed to do so,  at the linearised level, in the next subsection.

\textbf{Example: Schwarzschild AdS$_4$ mass.} To exemplify the above expressions,  we calculate the conformal energy of a Schwarszchild AdS$_4$ black hole. The  corresponding a background metric $\bar{g}_{\mu\nu}$ in (\ref{gbar}) has $k=+1$ and
\begin{equation} \label{eq: ads bh}
f(r) = 1- \frac{2 G_N M}{r} + \frac{r^2}{\ell^2}~.
\end{equation}
We take the conformal boundary conditions to be%
\begin{equation}\label{omegaM}
    \left.ds^2\right|_{r=e^\bomega\ell} = e^{2\bomega}\left(-du^2+\ell^2d\Omega_2^2\right) \, , \qquad \left.K\right|_{r=e^\bomega \ell} = \frac{3 e^{3\bomega}\ell + \left(2 e^\bomega \ell - 3 G_N M\right)}{ e^{2\bomega}\ell^2 \sqrt{f(e^\bomega \ell)}} \, ,
\end{equation}
where, for the AdS$_4$ black hole, $\bomega$ is constant. For the sake of convenience, and to make clear the comparison to the AdS/CFT literature, we have rescaled the boundary clock to the time coordinate labeled by $u\in\mathbb{R}$. For the same purpose, we also set $\r$ to $\ell$.

Given some $M>0$ and $K\ell>3$, there exists a unique real $\bomega$ that satisfies (\ref{omegaM}). Our Cauchy surface $\Sigma$ is given by the spatial slice $t=\frac{e^\bomega \, u}{\sqrt{f(e^\bomega \ell)}}=0$ that orthogonally intersects the boundary at $r=e^\bomega\ell$. In the $K\ell\to 3$ limit, the trace of the extrinsic curvature equation can be inverted, resulting in the expansion
\begin{equation}
    e^\bomega = \frac{1}{\sqrt{2 (K\ell-3)}} \left(1+ \frac{K\ell-3}{4} + \mathcal{O}(K\ell-3)^{3/2}\right)\,.
\end{equation}
The conformal Brown-York stress energy tensor (\ref{cby}) evaluates to
\begin{equation}
T_{uu} = \frac{2}{\ell^2}T_{\theta \theta} = \frac{2 }{\ell^2\sin^2{\theta}}T_{\phi \phi}  = \frac{e^\bomega \left(3 G_N M - e^\bomega \ell\right)}{12 \pi G_N \ell^2 \sqrt{f(e^\bomega \ell)}} ~,
\end{equation}
with all other components vanishing. We thus find 
\begin{equation}\label{divE}
    E_\text{conf}(M) = \ell^2  \,  \int d\Omega_2  T_{uu} = \frac{e^{\bomega}M}{\sqrt{f(e^\bomega \ell )}} - \frac{e^{2\bomega}\ell}{3 G_N \sqrt{f(e^\bomega \ell )}} \, ,
\end{equation}
where $\bomega$ is implicitly defined in terms of $K$. As $K\ell \to 3$,  we find an $\bomega$-dependent expression that diverges in the limit. This is a general feature of conserved charges for asymptotic boundaries. Instead, we can compute the difference between the black hole energy and that of empty global AdS$_4$, which yields
\begin{equation}
    E_\text{conf}(M)-E_\text{conf}(M=0) = M -\frac{5}{3}M (K\ell-3)+ \mathcal{O}(K\ell-3)^{3/2} \, .
\end{equation}
Comparing to the standard expression in AdS$_4$/CFT$_3$, as calculated for instance in \cite{Witten:1998zw,Balasubramanian:1999re}, we find agreement for the leading contribution to the energy.

\section{Linearised dynamics about global AdS$_4$}\label{sec: lin dynamics global AdS}

In this subsection, we consider linearised perturbations about a fixed global AdS$_4$ spacetime. We take the background metric to be
\begin{equation}\label{globalAdS4}
    ds^2=-\left(1+\frac{r^2}{\ell^2}\right)dt^2+\frac{dr^2}{1+\frac{r^2}{\ell^2}}+r^2d\Omega_2^2,\,\,\,\,\,\,\,\,\,\,\,\,\,\,\,\,   d\Omega_2^2\equiv d\theta^2+\sin^2\theta d\varphi^2 \,,
\end{equation}
where the coordinate ranges are $t\in \mathbb{R}$, $r \in (0,\infty)$, $\theta\in (0,\pi)$, and $\varphi\in (0,2\pi)$. 

We consider a three-dimensional timelike boundary $\Gamma$ located at $r=\r$. By appropriately selecting the sign of $K\ell$, we are able to consider gravitational dynamics for either the interior region, $r \in (0,\r)$, or for the exterior region, $r\in (\r, \infty)$. We will focus on the dynamics inside the tube. 
The induced metric at $r=\r$ is given by
\begin{equation}\label{gbar_global}
     \left. \bar{g}_{mn}dx^mdx^n \right|_\Gamma=-\left(1+\frac{\r^2}{\ell^2}\right)dt^2+\r^2 d\Omega_2^2 \,,
\end{equation}
while the extrinsic curvature is given by
\begin{equation}
  \left.  K_{mn}dx^mdx^n\right|_\Gamma=\frac{\r}{\ell^2}\sqrt{1+\frac{\r^2}{\ell^2}}\left(-dt^2+\ell^2 d\Omega_2^2\right) \,.
\end{equation}
The trace of the extrinsic curvature at $\Gamma$ is 
\begin{equation} \label{eq: K function of r}
K \ell|_{\Gamma}=\frac{2\ell^2+3\r^2}{\r\sqrt{\ell^2+\r^2}}.    
\end{equation}
Note that $K \ell$ is a positive real number satisfying $K \ell \geq 3$, which is saturated as $\r$ approaches the asymptotic boundary of AdS$_4$. In the small $\r$ limit, $K\ell$ diverges as $\tfrac{2}{\r}$, retrieving the Minkowski behavior in \cite{Anninos:2023epi}. 


We consider perturbations of the form \eqref{metric_pert_lin},
subject to conformal boundary conditions at $\Gamma$. As boundary conditions, we fix the conformal class of the metric to be (\ref{gbar_global}) and the trace of the extrinsic curvature to be constant. Furthermore, we require the leading perturbation of the trace of the extrinsic curvature to vanish at the boundary, $\delta K|_\Gamma = 0$. In addition, we must specify conformal boundary conditions for the set of allowed diffeomorphisms $\xi^\mu$. Finally, we require $\xi^r |_\Gamma=0$, so that allowed diffeomorphisms do not move the location of the boundary. 

We follow a procedure analogous to the one for $\Lambda = 0$ \cite{Anninos:2023epi} and $\Lambda>0$ \cite{Anninos:2024wpy}. Namely, we decompose the metric perturbations into irreducible representations of the $S^2$, which are labeled by the total angular momentum $l \in \mathbb{Z}^+$ and the integer valued axial angular momentum, $m \in (-l,l)$. For $l=0$, and $l=1$, only modes that are locally diffeomorphic are present. We study $l=0$ modes in subsection \ref{l0modes}, while in subsection \ref{sec: non-linear diffeos} we provide their non-linear extension. Modes with $l=1$ are studied in \ref{l1modes}. %
In subsection \ref{sec: bulk modes}, 
we analyse linearised modes with $l\geq 2$, following the Kodama-Ishibashi procedure \cite{Kodama:2000fa,Kodama:2003jz}.

\subsection{Locally diffeomorphic modes with $l=0$} \label{l0modes}

There are no local excitations of the metric field with $l=0$ at the linearised level. Nevertheless, in the presence of a boundary, we can ask whether there exist physical modes that can be expressed as diffeomorphisms in a given small neighborhood. Concretely, we are interested in linearised perturbations $h_{\mu\nu}$ taking the following form
\begin{equation}\label{eqn: physical diffeo perturbation}
    h_{\mu \nu} \= \nabla_\mu \xi_\nu + \nabla_\nu \xi_\mu \, ,
\end{equation}
for an arbitrary vector field $\xi^\mu$. The above perturbation automatically satisfies the linearised Einstein field equation.

The general coordinate transformation that preserves the spherical symmetry of the background metric is given by
\begin{equation}
\xi^\mu(t,r) \partial_\mu = \xi^r(t,r) \partial_r + \xi^t(t,r) \partial_t~.
\end{equation}
Upon imposing that the trace of the extrinsic curvature remains unchanged by (\ref{eqn: physical diffeo perturbation}) we find that the vector field components must satisfy
\begin{equation}
\left.\left( r^2 \partial_t^2 \xi^r (t,r)-\left(2+\frac{r^2}{\ell^2}\right) \xi^r(t,r) \right)\right|_{\Gamma} =0 \,.
\end{equation}
From the above we obtain
\begin{equation}
\xi^r(t,r) = e^{-i \omega^{(0)} t} f_+ (\r) + e^{i \omega^{(0)} t} f_- (\r) + f_1 (t,r) ~,
\end{equation}
where $\omega^{(0)} \r= i\sqrt{2+\frac{\r^2}{\ell^2}}$, $f_\pm(\r)$ are arbitrary constants, and $f_1 (t,r)$ is a function of both $t$ and $r$ such that at the boundary $f_1(t,r=\r)$ goes to zero. We note that at the boundary we have exponential growth/decay, as in the case of non-negative cosmological constant \cite{Anninos:2023epi,Anninos:2024wpy}, and moreover in the Minkowskian limit $\tfrac{\r}{\ell} \to 0$ we retrieve the result in \cite{Anninos:2023epi}.

Further imposing that the boundary metric is conformally equivalent to the Lorentzian cylinder, we find
\begin{equation}
\left. \left( \frac{\xi^r(t,r)}{r}  - \left(1+\frac{r^2}{\ell^2}\right) \partial_t \xi^t (t,r) \right)\right|_{\Gamma}= 0 ~,
\end{equation}
yielding
\begin{equation}\label{xit}
\xi^t(t, r) = \frac{f_+(\r) \, e^{-i \omega^{(0)} t}}{\left(1+\r^2/\ell^2\right) \sqrt{2+\r^2/\ell^2}}  -\frac{f_-(\r) \, e^{i \omega^{(0)} t}}{\left(1+\r^2/\ell^2\right) \sqrt{2+\r^2/\ell^2}} + f_2(t, r) ~,
\end{equation}
where $f_2(t, r)$ is another smooth function of $t$ and $r$, such that at the boundary it becomes independent of time, $f_2(t,r=\r) = f_2(\r)$. Now we can look at the general form of the linearised metric after imposing both boundary conditions. The correction to the Weyl factor at the boundary simply becomes,
\begin{equation}\label{omegadiff}
(\r \, \delta \bomega) |_{\Gamma} = f_+(\r) \, e^{-i \omega^{(0)} t} + f_-(\r) \, e^{i \omega^{(0)} t} \,.
\end{equation}
Note again that the correction to the Weyl factor grows exponentially with time. We can also consider the corner component of the metric at the boundary,
\begin{equation}
 \delta \beta|_{\Gamma} = h_{tr}|_{\Gamma}  =  \frac{\sqrt{2+\r^2/\ell^2} }{\r \left(1+\r^2/\ell^2\right)} \left(f_+(\r) e^{-i \omega^{(0)} t} - f_-(\r) e^{i \omega^{(0)} t} \right) -\left(1+\frac{\r^2}{\ell^2}\right) \partial_r f_2(t,\r) \,.   \label{eq: deltab} 
\end{equation} 
From (\ref{xit}) it follows that $f_2(t,r)$ can be set to zero using an ordinary diffeomorphism. The first term in \eqref{eq: deltab}, instead, is proportional to $\partial_t \delta \boldsymbol{\omega} |_{\Gamma}$. Together with $\delta \boldsymbol{\omega} |_{\Gamma}$, the pair $\{ \delta \boldsymbol{\omega}, \partial_t \delta \boldsymbol{\omega} \} |_{\Gamma}$ constitutes independent initial data along the $t=0$
 initial slice.

We observe that spherically symmetric physical diffeomorphisms can vary both the Weyl factor of the boundary metric as well as the corner component. The boundary Weyl factor and corner can be viewed as conjugate variables of the gravitational phase space at the boundary, upon imposing conformal boundary conditions \cite{Odak:2021axr}. It is worth contrasting the above with linearised perturbations about the standard Minkowski corner, as analysed in \cite{Anninos:2023epi}, where physical diffeomorphisms perturb the corner but not the boundary Weyl factor.\footnote{In \cite{Anninos:2023epi} this was shown in the harmonic gauge. It is straightforward to show this independently of the gauge choice. %
}

An analogous analysis can be performed for the $l=0$ mode about the AdS$_4$ black hole metric \eqref{eq: ads bh} in the Fefferman-Graham gauge, see appendix \ref{app: FG l=0}.



\subsection{Non-linear diffeomorphic modes with $l=0$} \label{sec: non-linear diffeos} 
One can also treat the diffeomorphism non-linearly \cite{Liu:2024ymn,Anninos:2024wpy}. Here, we seek a surface embedded in global AdS$_4$, which preserves our boundary conditions. That is, the trace of the extrinsic curvature at the boundary is fixed to $K$, where we allow $K$ to be an arbitrary function of the boundary time, and the induced metric at the boundary is conformal to \eqref{gbar_global},
\begin{equation}\label{eqn: non-linear conf rep}
    ds^2|_\Gamma = e^{2\bomega (x^m)} \left( -\left(1+\frac{\r^2}{\ell^2}\right)dt^2+\r^2 d\Omega_2^2 \right) \,.
\end{equation}
As a consequence, $\r$ is not a function of $K$ and should be treated as a part of the conformal representative of the induced metric. 

The resulting non-linear equation governing the Weyl factor upon imposing our boundary conditions is given by %
\begin{equation}\label{eqn: brane eqn}
\r^2\partial^2_t \bomega = -\left(1+\frac{\r^2}{\ell^2}\right)\left(2  + \frac{3 \r^2 e^{2\bomega}}{\ell^2}\right)- 2 \left(\r \partial_t\bomega\right)^2 + K(t) \, \r  e^\bomega\left(1+\frac{\r^2}{\ell^2}\right) \sqrt{1  + \frac{\r^2}{\ell^2} e^{2\bomega}+ \frac{\left(\r \partial_t\bomega\right)^2}{1+\frac{\r^2}{\ell^2}} }~.
\end{equation}
Upon linearising the Weyl mode about $\bomega(t)=0$ and further setting $K$ to be \eqref{eq: K function of r}, we are left with %
\begin{equation}\label{linear}
\r^2 \partial^2_t \delta \bomega(t) =  \left( \frac{\r ^2}{\ell^2}+2 \right) \delta \bomega(t)~,
\end{equation}
from which we retrieve  the exponential solutions (\ref{omegadiff}).

The conformal energy for the solutions to (\ref{omegadiff}) can be obtained using the conformal Brown-York stress energy tensor \eqref{cby}. 
Assuming that $K$ is a constant, a direct computation gives
\begin{equation}\label{Econfglob}
    E_\text{conf} = \frac{\r\sqrt{1+\frac{\r^2}{\ell^2}}}{3 G_N}\left(K\r  e^{3\bomega}- 3e^{2\bomega}\sqrt{1  + \frac{\r^2}{\ell^2} e^{2\bomega}+ \frac{\left(\r \partial_t\bomega\right)^2}{1+\frac{\r^2}{\ell^2}} }\right) \, .
\end{equation}
To obtain this expression, we use \eqref{eqn: brane eqn}. One can show that $\frac{dE_\text{conf}}{dt}=0$ by imposing \eqref{eqn: brane eqn}. {Upon using this conformal Brown-York stress tensor, we find that \eqref{eqn: brane eqn} is consistent with \eqref{eq: Hamiltonian constraint}.}

\textbf{A non-linear example.} An inherently  non-linear time-dependent solution to (\ref{eqn: brane eqn}) is given by
\begin{equation}\label{nonlinear}
e^{2\bomega(t)} = \frac{\alpha_{\text{n.l.}}}{\cos^2 \beta_{\text{n.l.}} t}~, \quad \text{where} \quad  \alpha_{\text{n.l.}} \equiv \frac{9 \ell^2}{\r^2 \left(K^2 \ell^2-9\right)} ~  , \quad \beta_{\text{n.l.}} \equiv \frac{1}{\r} \sqrt{1+\tfrac{\r^2}{\ell^2}}~.
\end{equation}

For this non-linear solution the conformal Brown-York stress-tensor vanishes identically. As such, it is distinguishable from the constant $\bomega(t)=\bomega_0$ solution to (\ref{eqn: brane eqn}) which reads
\begin{equation}
e^{2\bomega_0} = \frac{\ell^2}{\r^2} \frac{ \left(\sqrt{K^2\ell^2-8}-K\ell\right)K\ell+12}{2 \left(K^2\ell^2-9\right)}~,
\end{equation}
and has negative conformal energy. 

\textbf{Asymptotic boundary analysis.} Another interesting limit is the large Weyl factor limit. Consider $\bomega = - \frac{1}{2}\log(K\ell-3) + \delta \bomega$. Taking $K\ell \to 3$, the equation \eqref{eqn: brane eqn} becomes
\begin{equation}\label{eqn: brane dynamics near AdS bdry}
    \r^2\partial_t^2\delta\bomega = -\left(1+\frac{\r^2}{\ell^2}\right)\left(\frac{1}{2} - \frac{\r^2}{\ell^2}e^{2\delta\bomega}\right) - \frac{1}{2}\left(\r \partial_t\delta\bomega\right)^2 \, ,
\end{equation}
to leading order in the small $K\ell-3$ expansion. An exact solution to this equation is given by a constant Weyl factor $e^{2\delta\bomega_0} = \frac{\ell^2}{2\r^2}$.

We would like to compare the Brown-York stress tensor this limit with the standard one from the Fefferman-Graham expansion  \eqref{eq: brown-york near boundary}. For this, we set the Weyl factor to be $e^{2\bomega}=\tfrac{1}{K\ell-3}$ and the conformal representative to be $g^{(0)}_{mn} = e^{2\delta\bomega} \bar{g}_{mn}$ where $\bar{g}_{mn}$ is (\ref{gbar_global}) and $\delta\bomega$ is purely time-dependent.

Then, the Brown-York stress tensor \eqref{cby} in this limit becomes 
\begin{equation}
    T_{tt} = \frac{2}{\r^2}\left(1+\frac{\r^2}{\ell^2}\right)T_{\theta \theta} = \frac{2}{\r^2\sin^2\theta}\left(1+\frac{\r^2}{\ell^2}\right)T_{\phi \phi}= \frac{\left(1+\frac{\r^2}{\ell^2} \right)\left(\frac{2}{3}e^{2\delta\bomega}-\frac{\ell^2}{\r^2}\right)-(\ell\partial_t\delta\bomega)^2}{8\pi G_N\ell\sqrt{K\ell-3}}\,.
\end{equation}
Note that this term diverges in the strict $K\ell\to3$ limit. Upon implementing (\ref{eqn: brane dynamics near AdS bdry}), this  stress tensor agrees with the first term in \eqref{eq: brown-york near boundary}. Moreover, \eqref{eqn: brane dynamics near AdS bdry} is equivalent to the condition that $R[g^{(0)}_{mn}]=\tfrac{4}{\ell^2}$ for a purely time-dependent $\bomega$. We can also compute the conformal energy, which in this limit simplifies to 
\begin{equation}
E_\text{conf} = \frac{4\pi \r^2 e^{\delta\bomega}}{\sqrt{1+\frac{\r^2}{\ell^2}}} T_{tt} \,.
\end{equation}

The conformal energy takes the form of a single particle with negative kinetic term and is unbounded from below. It is interesting to note that standard positivity notions of energy may no longer hold in the presence of a timelike boundary subject to conformal boundary conditions. {It is also the case that wrong-sign kinetic terms can appear in the kinetic term of the conformal mode of the induced metric, upon performing a classical ADM analysis of gravitational dynamics. Perhaps, then, $\bomega$ is carrying some of this flavour.}

{\textbf{Euclidean uniqueness.}} As a side remark, one might suspect that the equation analogous to (\ref{eqn: brane eqn}) in  Euclidean signature could signal the violation of uniqueness for conformal boundary conditions established in the theorem of \cite{Anderson_2008}. However, we recall that the uniqueness results in Euclidean signature assume the boundary to be a compact manifold. Consider an $S^2 \times S^1$ boundary, where the Euclidean time $\tau = i t$ is a periodic variable. In this way, one cannot generically generate multiple regular Euclidean solutions given the conformal boundary data, restoring the good uniqueness properties of \cite{Anderson_2008}. For concreteness, consider the linearised expression (\ref{linear}). Taking $t\to-i\tau$ and requiring $\tau$ to have a given periodicity $\tau \sim \tau+\beta$, we see that generically there are no regular solutions to the Euclidean version of (\ref{linear}) since this equation would enforce a different periodicity in $\tau$. In Lorentzian signature, instead, care must be taken in imposing appropriate initial data at the corner to ensure a unique and well-posed evolution. Relatedly, had we considered a Euclidean manifold of with boundary of the type $S^2 \times \mathcal{I}$, where $\mathcal{I}$ is a closed interval in $\mathbb{R}$, one should be cautious about additional Euclidean zero modes \cite{Capoferri:2024sgo}.

\subsection{Locally diffeomorphic modes with $l=1$}
\label{l1modes}

For the case of diffeomorphisms built from $l=1$ spherical harmonics, we solve again for a perturbation of the form \eqref{eqn: physical diffeo perturbation}. The general diffeomorphism built from the scalar spherical harmonics with $l=1$, $\mathbb{S}_1$, can be written as%
\footnote{For $l=1$ modes, the functions $\mathbb{S}_1$ are angular-dependent functions defined by $(\tilde{\nabla}^2 +2) \mathbb{S}_1 = 0$, with $\tilde{\nabla}^2$ being the Laplacian of the unit two-sphere. From $\mathbb{S}_1$, the $\mathbb{S}_{1,i}$ functions can be simply defined by $\mathbb{S}_{1,i} = -\frac{1}{\sqrt{2}} \tilde{\nabla}_i \mathbb{S}_1$, where the $i$ represent indices on the two-sphere.}
\begin{equation}\label{eqn: l=1 sol AdS}
    \xi_\mu^{\pm} dx^\mu \, = \, \frac{\alpha_\pm \, e^{\mp i \omega^{(1)} t}}{\sqrt{1+r^2/\ell^2}} \left(\mathbb{S}_1 \, dr \pm i\, \omega^{(1)}\, r\left(1+r^2/\ell^2\right) \,\mathbb{S}_1 \, dt - \sqrt{2}\left(1+r^2/\ell^2\right) \,\mathbb{S}_{1,i} \,r \,d\Omega^i\right) + \tilde{\xi}_\mu^{\pm}dx^\mu \, ,
\end{equation}
where $\omega^{(1)}= \ell^{-1}$, $\alpha_\pm$ are constants, and $\tilde{\xi}_\mu^{\pm}$ are arbitrary diffeomorphisms that vanish at the boundary, $\tilde{\xi}_\mu^{\pm}(t,\r)=0$. As opposed to $\tilde{\xi}_\mu^{\pm}$, the $\alpha_\pm$ modes have a non-vanishing $\xi^r$ at the boundary which means that they are physical and cannot be gauged away.

To conclude, the $l=1$ diffeomorphism built from the scalar spherical harmonics yields the subset of Killing vectors on AdS$_4$ that become conformal Killing vectors (with nonvanishing conformal factor) when projected onto the boundary metric. A similar analysis holds for diffeomorphisms built from $l=1$ vectorial spherical harmonics, where up to trivial diffeomorphisms, one finds the subset of AdS$_4$ Killing vectors that project onto Killing vectors of the boundary metric itself. Thus, there are no physical perturbations of the metric that correspond to $l=1$ diffeomorphisms.

\subsection{Bulk linearised modes with $l\geq2$} \label{sec: bulk modes}

For $l\geq 2$, there are no physical diffeomorphisms (see appendix \ref{sec: phys_diffeos_l}) and the only allowed perturbations are bulk modes. We employ the Kodama-Ishibashi method to analyse them \cite{Kodama:2000fa, Kodama:2003jz}. This procedure is analogous to the one done for positive and vanishing cosmological constant in \cite{Anninos:2023epi, Anninos:2024wpy}, so we will keep the discussion brief here. We refer the reader to appendix C in \cite{Anninos:2023epi} for a detailed account of definitions and conventions of the Kodama-Ishibashi method.

Using the spherical symmetry of the background metric, we can decompose the metric perturbations into a scalar and a vector sector,
\begin{equation}
    h_{\mu \nu} = h_{\mu\nu}^{(S)} + h_{\mu\nu}^{(V)} \,.
\end{equation}

The scalar sector can be described in terms of the scalar spherical harmonics $\mathbb{S}_l$, while the vector perturbation depends on the vectorial spherical harmonics $\mathbb{V}_i$.\footnote{The spherical harmonics of total angular momentum $l$, denoted as $\mathbb{S}_l$, are given by the solutions to $(\tilde{\nabla}^2 + l (l+1)) \mathbb{S}_l = 0$. It is also useful to define $\mathbb{S}_{l,i} \equiv -(l(l+1))^{-1/2} \tilde{\nabla}_i \mathbb{S}_l$. The vectorial spherical harmonics $\mathbb{V}_i$ are defined as solutions to the equations $(\tilde\nabla^2+l(l+1)-1)\mathbb{V}_i=0$ and $\tilde\nabla_i\mathbb{V}^i=0$.} The Kodama-Ishibashi formalism reduces the linearised Einstein field equations into a set of differential equations for scalar master fields $\Phi^{(S/V)}$, from which one can directly infer a solution for the metric perturbation $h_{\mu\nu}^{(S/V)}$. In the absence of boundaries, this procedure can be performed in a gauge-invariant way. 

For our analysis, we select a gauge in which the boundary conditions only act on either $h_{\mu\nu}^{(S)}$ or $h_{\mu\nu}^{(V)}$ \cite{Anninos:2023epi, Anninos:2024wpy}, so that the general metric perturbation can be written as
\begin{eqnarray}
\begin{cases}
	h_{mn} &= - \bar{g}_{mn} \frac{1}{2r}\left[ l\left(l+1\right)\left(1+ \frac{2r^2}{\ell^2}\right)+2 r^2 \partial_t^2 + 2  \left(1+\frac{r^2}{\ell^2}\right)^2  r \partial_r\right]\Phi^{(S)} \mathbb{S}_l  \\
 & \quad + \left(\delta^i_m\delta^t_n + \delta^i_n \delta^t_m\right) \left(1+\frac{r^2}{\ell^2}\right)\partial_r \left(r \Phi^{(V)}\right) \mathbb{V}_i \, , \\
    h_{rr} &= -\frac{1}{r\left(1+\frac{r^2}{\ell^2}\right)^2}\Bigg[\frac{l(l+1)}{2}\left(3+\frac{7r^2}{\ell^2}+\frac{4r^4}{\ell^4}\right)+ \left(3+\frac{2 r^2}{\ell^2}\right)r^2\partial_t^2 \\
    		&\quad + \left(1+\frac{r^2}{\ell^2}\right)\left(\left(1+\frac{r^2}{\ell^2}\right)\left(l(l+1) + 1+2\frac{r^2}{\ell^2} \right)+ r^2\partial_t^2\right) r\partial_r\Bigg]\Phi^{(S)} \mathbb{S}_l \, , \\
    h_{tr} &=  -\frac{1}{2\left(1+\frac{r^2}{\ell^2}\right)}\partial_t \left[l(l+1)\left(1+\frac{r^2}{\ell^2}\right) - 2 + r^2\partial_t^2 + \left(1+\frac{r^2}{\ell^2}\right)\frac{r^2}{\ell^2}r\partial_r\right]\Phi^{(S)} \mathbb{S}_l \, ,  \\
    h_{ri} &=  \frac{\sqrt{l(l+1)}}{2\left(1+\frac{r^2}{\ell^2}\right)}\left[l(l+1)\left(1+\frac{r^2}{\ell^2}\right)+r^2\partial_t^2+\left(2+\left(3+\frac{r^2}{\ell^2}\right)\frac{r^2}{\ell^2}\right)r\partial_r\right]\Phi^{(S)} \mathbb{S}_{l,i} + \frac{r}{1+\frac{r^2}{\ell^2}} \partial_t \Phi^{(V)} \mathbb{V}_i \, .	
    \end{cases}\label{eqn: spherical l>2 ansatz}
\end{eqnarray}
The indices $m$ and $n$ denote indices with respect to coordinates tangential to the boundary, $(t,\theta,\phi)$, while the index $i$ denotes indices on the two-sphere. In order to satisfy the Einstein field equation, the master fields $\Phi^{(S/V)}$ should satisfy,
\begin{equation}\label{AdS2KG}
\left(-\boldsymbol{\nabla}^2  + \frac{l(l+1)}{r^2} \right) \Phi^{(S/V)}(t,r) = 0~.
\end{equation}
In the above, $\boldsymbol{\nabla}^2$ denotes the Laplacian on a two-dimensional anti-de Sitter space,
\begin{equation}
-\boldsymbol{\nabla}^2 \equiv  \left(1+\frac{r^2}{\ell^2}\right)^{-1}{\partial_t^2} - \partial_r \left(1+\frac{r^2}{\ell^2}\right) \partial_r~. 
\end{equation}
Since we are interested in the gravitational dynamics inside the worldtube, we require solutions to be regular as $r\to0$. Then, the general solution for each frequency, both for the scalar and vector master fields can be written as 
\begin{equation}\label{PhiS AdS}
    \Phi^{(S/V)} (t,r) \= -\text{Re} \left[ \, e^{-i\omega t}\, \left(\frac{r}{\ell}\right)^{l+1} \left(1+\frac{r^2}{\ell^2}\right)^{\frac{\omega \ell}{2}} \, _2F_1\left(\frac{1}{2} (l+\omega \ell
   +1),\frac{1}{2} (l+\omega \ell +2);l+\frac{3}{2};-\frac{r^2}{\ell^2}\right) \right]\, .
\end{equation}

As mentioned, the metric perturbation \eqref{eqn: spherical l>2 ansatz} is given in a gauge where the linearised conformal boundary conditions act separately on $h_{\mu\nu}^{(S)}$ and $h_{\mu\nu}^{(V)}$, so we can study each contribution independently. 

\subsubsection{Vector sector}

\begin{figure}[h!]
        \centering
         \subfigure[$K \ell = 5$]{
                \includegraphics[scale = 0.5]{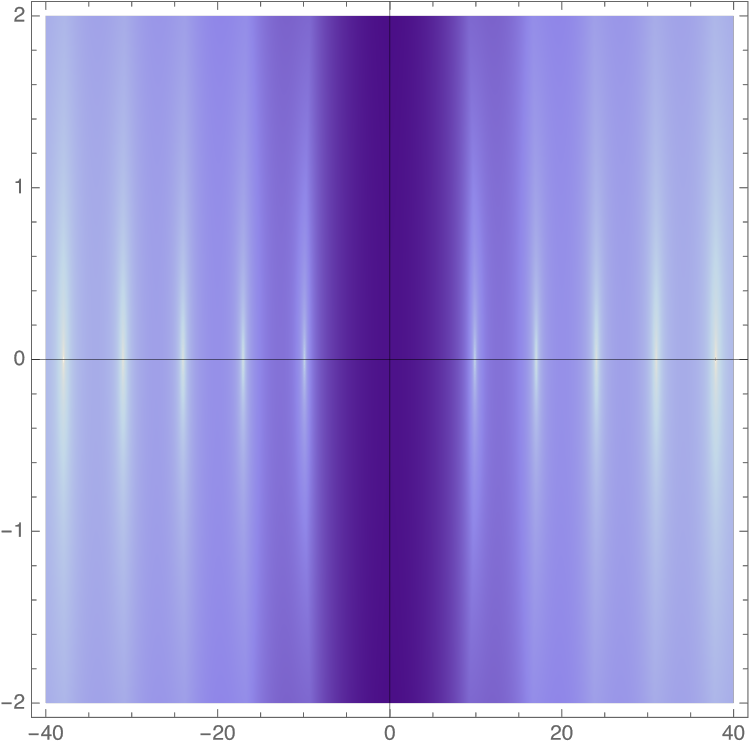}\label{fig: contour 1_V}}  \quad\quad
                 \subfigure[$K \ell = 3.01$]{
                \includegraphics[scale = 0.5]{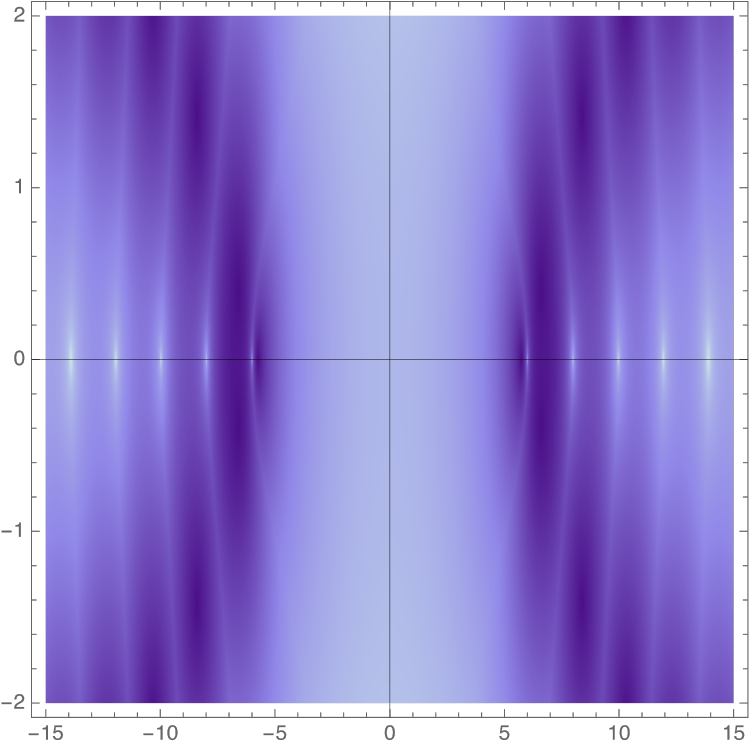} \label{fig: contour 2_V}}                           
                \caption{Density plot of absolute value of $\log \mathcal{F}^{(V)}_l(K \ell,  \omega \ell)^2$ in the complex $\omega \ell$ plane for $l=2$, at $t=0$. Note that all allowed frequencies are real.} \label{fig: contour_vector}
\end{figure}

In the case of $h_{\mu\nu}^{(V)}$, the boundary condition \eqref{eqn: bdry cond 2} is automatically satisfied while the boundary condition \eqref{eqn: bdry cond 1} imposes
\begin{equation}
 \mathcal{F}_l^{(V)} (K \ell, \omega \ell) \equiv   \left.\frac{\Phi^{(V)}}{r} + \partial_r \Phi^{(V)}\right|_{\Gamma}\=0 \,.
\end{equation}
Given the solution for the master field \eqref{PhiS AdS}, we numerically scan for solutions in the complex frequency plane for different values of $l$ and $\r$ and find that all allowed frequencies that satisfy that boundary conditions are real. Two examples for $l=2$ can be found in figure \ref{fig: contour_vector}. This behaviour mimics that of the full asymptotically AdS$_4$ case, and is essentially that of a `particle in a box'. As $K\ell \to 3$ the spectrum reproduces a subset of the normal mode frequencies,  $|\omega_n \ell| = l + 4 + 2n$ with $n \in \mathbb{Z}^+$, in global AdS$_4$. 

\subsubsection{Scalar sector}

For the scalar sector, it is straightforward to verify that \eqref{eqn: bdry cond 1} is automatically satisfied at $r=\r$. Then we only need to impose that the change in the mean curvature is vanishing at the boundary of the tube. This gives,
\begin{eqnarray}\label{eqn: deltaK scalar}
	\mathcal{F}^{(S)}_l(K\ell\,,  \omega\ell) \,\equiv\,\left.\left(\frac{a_1}{r^4}+\frac{a_2}{r^2}\partial_t^2 - 2 \partial_t^4\right)\Phi^{(S)}  + \left(\frac{a_3}{r^2} -2 \partial_t^2\right)\left(1+\frac{r^2}{\ell^2}\right)^2\frac{\partial_r \Phi^{(S)}}{r} \right|_{\Gamma} \= 0 \, ,
 \end{eqnarray}
 where we should use \eqref{eq: K function of r} to write $\r$ in terms of $K \ell$ and
 \begin{eqnarray}
     \begin{cases}
	& a_1  \= l(l+1)\left(1+\frac{r^2}{\ell^2}\right)\left(3+\frac{2r^2}{\ell^2}-2l(l+1)\left(1+\frac{r^2}{\ell^2}\right)\right) \, , \\
	& a_2  \= 4+\frac{2r^2}{\ell^2} - 4 l(l+1)\left(1+\frac{r^2}{\ell^2}\right) \, , \\
	& a_3  \= 4+\frac{2r^2}{\ell^2}-l(l+1)\left(3+\frac{2r^2}{\ell^2}\right) \, .
    \end{cases}
\end{eqnarray}

\begin{figure}[H]
        \centering
         \subfigure[$K \ell = 5$]{
                \includegraphics[scale = 0.5]{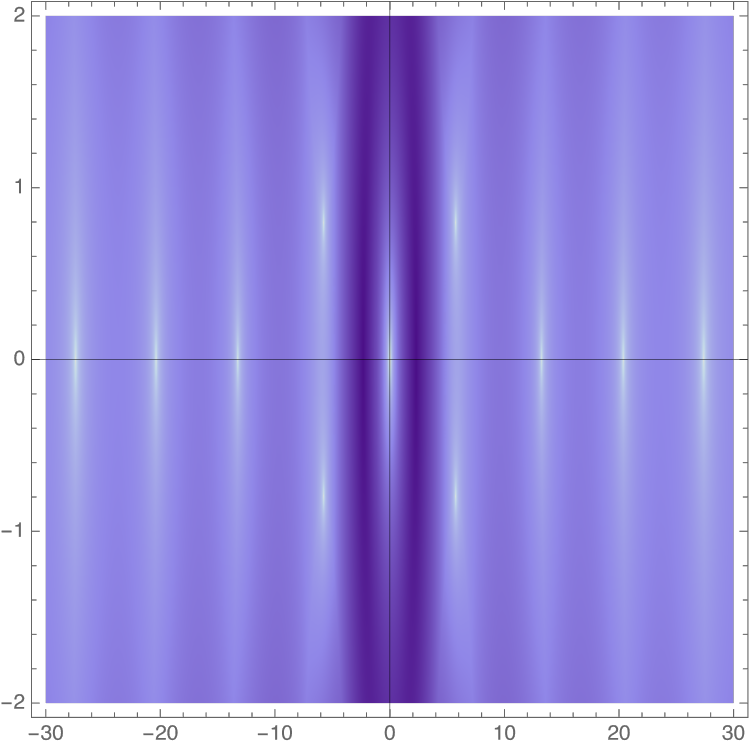}\label{fig: contour 1}}  \quad\quad
                 \subfigure[$K \ell = 3.01$]{
                \includegraphics[scale = 0.5]{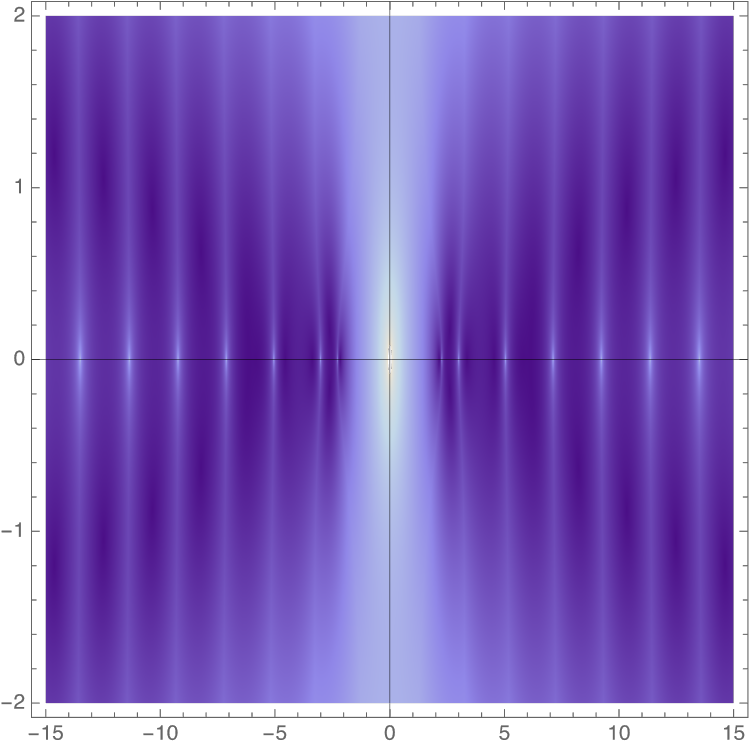} \label{fig: contour 2}}                           
                \caption{Density plot of absolute value of $\log \mathcal{F}^{(S)}_l(K \ell,  \omega \ell)^2$ in the complex $\omega \ell$ plane for $l=2$, at $t=0$. We multiply $\mathcal{F}^{(S)}_l(K \ell,  \omega \ell)$ by  $|\omega \ell|^{-2}$ to highlight the location of the zeros. Note that when $K \ell \approx 3$, all the zeros are in the real line, while for larger $K\ell$, there are, additionally, two pairs of complex conjugate frequencies.} \label{fig: contour}
\end{figure}

Equation \eqref{eqn: deltaK scalar} selects the set of allowed frequencies in the scalar sector that satisfy the conformal boundary conditions. In general these frequencies can be complex. As an example, in figure \ref{fig: contour} we provide a density plot of $\mathcal{F}^{(S)}_l (K \ell, \omega \ell)$ in the complex $\omega \ell$ plane for the $l=2$ modes for two different values of $K \ell$. In both cases, we observe a set of normal modes with real frequencies. In addition, when $K \ell$ is large enough, there are four additional complex frequencies, two of which have positive imaginary part.

The allowed modes give rise to a non-vanishing Weyl factor on the timelike boundary,
\begin{equation}
    \delta \bomega|_\Gamma = \left.-\frac{1}{4r}\left[ l\left(l+1\right)\left(1+ \frac{2r^2}{\ell^2}\right)+2 r^2 \partial_t^2 + 2  \left(1+\frac{r^2}{\ell^2}\right)^2  r \partial_r\right]\Phi^{(S)} \mathbb{S}_l\right|_{\Gamma}~.
\end{equation}
We can also compute the Brown-York stress tensor \eqref{cby} of these modes, whose $tt$-component is given by
\begin{equation}
    T_{tt}=\left.\frac{\sqrt{1+\frac{r^2}{\ell^2}}}{16 \pi G_Nr^2}\left(-\frac{4r}{3}+l(l+1)\left(\left(1+\frac{r^2}{\ell^2}\right)l(l+1)-1+r^2\partial_t^2+\left(1+\frac{r^2}{\ell^2}\right)^2r\partial_r \right)\Phi^{(S)}\mathbb{S}_l\right)\right|_{\Gamma}\,. 
\end{equation}
The conformal energy computed from this stress tensor is vanishing to leading order in the perturbative expansion. This is because, for any $l\geq 2$, the integral of any spherical harmonic $\mathbb{S}_l$ over the two-sphere  is zero. Alternatively, the energy carried by the linearised modes might follow the construction of a gravitational energy-momentum pseudo-stress tensor, as in \cite{Abbott:1981ff}, or a careful consideration of higher order terms. We leave this for future investigation. 

\begin{figure}[h!]
        \centering
         \subfigure[Re($\omega \ell$)]{
                \includegraphics[scale = 0.5]{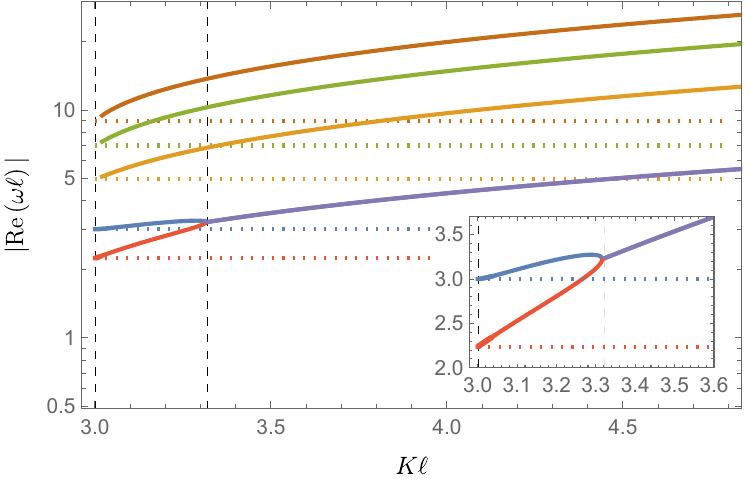}\label{fig: l2a}}  \quad\quad
                 \subfigure[Im($\omega \ell$)]{
                \includegraphics[scale= 0.5]{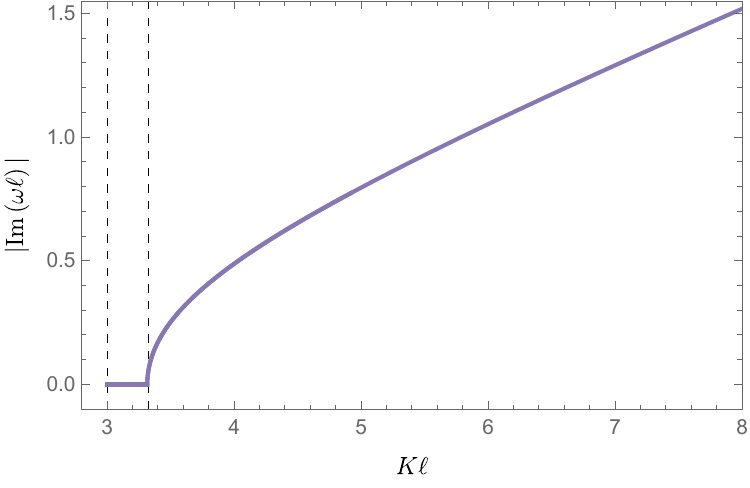} \label{fig: Re l2 plot}}                           
                \caption{Real and imaginary part of allowed frequencies in the scalar sector for $l=2$. The location of the conformal boundary of AdS$_4$ is given by $K \ell = 3$, and we observe that for $l=2$, $K_c \ell \approx 3.32$. Both are shown in vertical dashed black lines. For $K \ell < K_c \ell$, the lowest frequency asymptotes the value of $|\omega \ell| = \sqrt{5}$, followed by frequencies that asymptote to $|\omega \ell| = 3, 5, 7, 9, \cdots$. These values are marked with horizontal dotted lines in the figure. At $K\ell = K_c \ell$, the two lowest frequencies coalesce into one --- see inset in (a), which also develops an imaginary part, shown in (b). 
                After the critical value, the four lowest real frequencies, that correspond to plus and minus the blue and red curves, become four complex frequencies with $\omega \ell = \pm |\text{Re} \, (\omega \ell)| \pm i |\text{Im} \,(\omega \ell)|$, colored in purple. All the other frequencies remain real. At large enough $K \ell$, both the real and imaginary part become linear in $K \ell$, reproducing the flat space limit.
                } \label{fig: Im l2 plot}
\end{figure}

At a fixed value of $l$, the magnitude of the imaginary part of the complex frequencies decreases as $K \ell$ decreases, i.e., as the boundary of the tube moves towards the conformal boundary of AdS. In fact, for each $l$, we find a critical value of $K$, that we call $K_{c}$, such that $\text{Im} (\omega \ell) = 0$, for all $K \ell \in [3, K_{c} \ell ]$. We show this behaviour for $l=2$ in figure \ref{fig: Im l2 plot}.

Note that for large $K \ell$, both the real and imaginary parts of the allowed frequencies become linear in $K \ell$. This is consistent with the fact that in flat space, the frequencies scale with $\r$ instead of $\ell$ (for a spherical boundary in flat space, $K = \tfrac{2}{\r}$). In the limit $K \ell \to \infty$, the values of these frequencies precisely coincide with those found for the analogous problem in flat space for any $l$, including the large $l$ behavior, $\omega \r \approx \pm l \pm 0.34 i \, l^{1/3}$ \cite{Anninos:2023epi, Liu:2024ymn}. 

The imaginary part of $\omega$ is observed, numerically, to depend on $K \ell$ (see appendix \ref{app: large l}). To obtain an approximation at large $l$, we can employ a WKB approximation of (\ref{AdS2KG}). We can do this by placing (\ref{AdS2KG}) in Schr\"odinger form, by defining $x=\cos^{-1}(1+\tfrac{r^2}{\ell^2})^{-\tfrac{1}{2}}$ with $x\in(0,\tfrac{\pi}{2})$. For a given frequency, $\omega$, the Schr\"odinger equation governing the scalar mode is 
\begin{equation}
\partial^2_x \Phi^{(S)}(x) =   Q_l(x) \Phi^{(S)}(x)~, \quad\quad  Q_l(x) \equiv  \frac{l(l+1)}{\sin^{2} x}-\omega^2 \ell^2~.
\end{equation}
The WKB approximation, in the limit of parameterically large potential, is found to be 
\begin{equation}
\Phi^{(S)}(x) \approx \sqrt[4]{Q_l(x)} \exp {\int^x dx  \sqrt{Q_l(x)}}~. 
\end{equation}
The integral of $Q(x)$ can be expressed in terms of elementary functions, and we have chosen here the solution that becomes smooth near the origin. Implementing (\ref{eqn: deltaK scalar}), and expanding near the AdS$_4$ boundary by taking $K\ell \to 3$, one finds only real frequency solutions compatible with our preceding analysis. 

\begin{figure}[H]
        \centering
                \includegraphics[scale = 0.6]{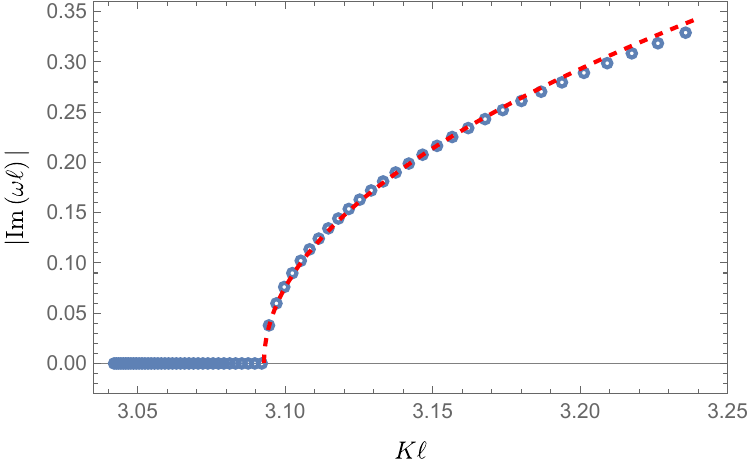}\label{fig: l4}
                \caption{Imaginary part of the allowed complex frequencies with $l=4$ in the scalar sector close to the critical value $K_c \ell \approx 3.09$. The dots are numerical, while the red dashed line shows the best fit close to the critical point, $\text{Im} (\omega \ell) = 0.89 \sqrt{K\ell - K_c \ell}$. } \label{fig: Im vs R near Rc plot}
\end{figure}

{\textbf{The transition from complex-to-real frequencies.}}
As we approach  $K \to K_{c}$, we  numerically find that for any fixed $l$,
\begin{equation}
    \text{Im} (\omega \ell) = \begin{cases}
        \alpha(l) \sqrt{K \ell - K_c \ell} + \mathcal{O}(K \ell - K_c \ell )^{3/2}~, \quad &  \quad K \ell \gtrsim K_c \ell \,, \\
        0~, \quad  &  \quad 3 \leq K \ell < K_c \ell \,.
        \end{cases}
\end{equation}
for some function $\alpha(l)$ that can be found numerically for each $l$. An example near the critical value for $l=4$ is shown in figure \ref{fig: Im vs R near Rc plot}.

\begin{figure}[H]
        \centering
                \includegraphics[scale=0.6]{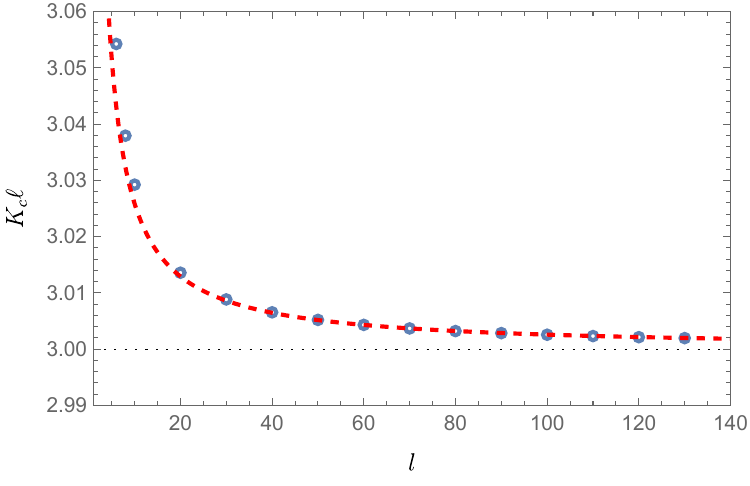}
                \caption{Critical value for the extrinsic curvature $K_c \ell$ for different values of $l$. The dots were obtained numerically for each $l$, while the dashed red curve shows the best fit at large $l$, which gives $K_c 
                \ell = 3 + 0.26 \, l^{-1}$. As a reference, we show the black dotted line of $K \ell =3$, the conformal boundary of AdS$_4$ space.} \label{fig: Rc vs l plot}
\end{figure}

We can ask how the value of $K_c \ell$ depends on the angular momentum number $l$. For this, we numerically find the value of $K_c \ell$ for different values of $l$, ranging from $l=2$ to $l = 130$. The results can be seen in figure \ref{fig: Rc vs l plot}, where we numerically find that for large $l$,
\begin{equation}
    K_c \ell = 3 + a \, l^{-1}+ \mathcal{O} (l^{-2}) \,, \quad \quad a \approx 0.26 \,.
\end{equation}
Conversely, this result can be inverted to establish that at any given $K \ell \gtrsim 3$, modes with angular momentum satisfying $l \gtrsim \tfrac{0.26}{(K\ell-3)}$ will  exhibit exponential growth in time at the linearised level.

{\textbf{Structure of the scalar normal frequencies.}} For $K < K_c$, the four complex frequencies split into four real frequencies, that we can follow as the boundary approaches the AdS$_4$ boundary. In the limit $K\ell \to 3$, the absolute value of the lowest two real frequencies is found to approach
\begin{equation}\label{eqn: bdry dispersion}
    |\omega^{(l)} \ell| = \sqrt{l(l+1) -1} \,.
\end{equation}
As an example,  in figure \ref{fig: Im l2 plot}  we plot the case of $l=2$ with $|\omega^{(2)} \ell|= \sqrt{5}$. These frequencies can be thought as near physical diffeomorphisms as the timelike boundary approaches the AdS$_4$ boundary (see appendix \ref{sec: phys_diffeos_l}).
For these modes, the variation of the boundary Weyl factor becomes,
\begin{equation}
    \delta \bomega = \left.\left(\frac{l(l+1)(l+2)(l-1)}{3l(l+1)-4}\frac{\ell}{\r}+\mathcal{O}\left(\frac{\ell}{\r}\right)^{2}\right)\Phi^{(S)}\mathbb{S}_l\right|_{\Gamma} \, .
\end{equation}

The other two real frequencies saturate to $|\omega \ell| = l+1$ as $K\ell \to 3$, and they are accompanied by a tower of normal (real) frequencies,
\begin{equation}\label{eqn: bulk dispersion}
    |\omega_n^{(l)} \ell| = (l+1) + 2n \,,\, \quad n \in \mathbb{N}_0 \,,
\end{equation}
that are the standard frequencies for metric perturbations about AdS$_4$. By evaluating the Weyl factor at the boundary for large $\r$, we find that for the frequencies (\ref{eqn: bulk dispersion}),
\begin{equation}
    \delta \bomega = \left.\left(\frac{l(l+1)(l+2)(l-1)}{8n (n+l+1)+2l+4}\frac{\ell^3}{\r^3}+\mathcal{O}\left(\frac{\ell}{\r}\right)^{4}\right)\Phi^{(S)}\mathbb{S}_l\right|_{\Gamma} \, .
\end{equation}
We note that the Weyl factor for these modes decays faster in $\r$ as compared to that of the modes in \eqref{eqn: bdry dispersion}. Given that $\sqrt{l(l+1)-1}$ is strictly less than $l+1$ for positive $l$, the lowest frequencies are always the $\pm \omega^{(l)}$ of (\ref{eqn: bdry dispersion}).

In the asymptotic limit, both sets of allowed frequencies can be found analytically by looking at the asymptotic form of $\mathcal{F}^{(S)}_l(K\ell\,,  \omega\ell)$ in \eqref{eqn: deltaK scalar}. As $K \ell \to 3$, $\tfrac{\r}{\ell} \to \infty$ and we obtain,
\begin{equation}
   \mathcal{F}^{(S)}_l(K\ell\,,  \omega\ell) \,=\,  \frac{4 \sqrt{\pi } \left(l(l+1)-1-\omega^2 \ell^2\right) \Gamma \left(l+\frac{3}{2}\right) }{\Gamma \left(\frac{1}{2} (l +1 -\omega \ell)\right) \Gamma \left(\frac{1}{2} (\omega \ell + l+1)\right)} \frac{\r}{\ell} \cos \left(\omega t\right)+\mathcal{O}\left(1\right) \,,
\end{equation}
which indeed has zeros at real frequencies with $\omega^{(l)} \ell = \pm \sqrt{l(l+1) -1}$ and $ \omega_n^{(l)} \ell =\pm ((l+1) + 2n)$, for non-negative integer $n$.

\begin{figure}[H]
        \centering
         \subfigure[]{
                \includegraphics[scale=0.55]{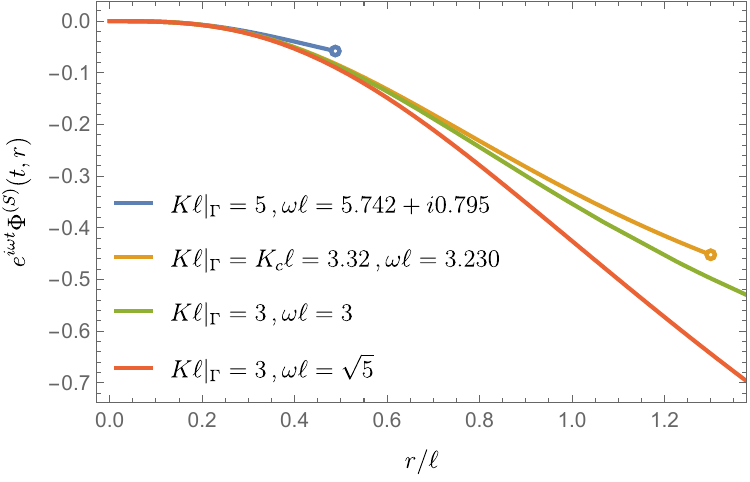}\label{fig: l2}}  \quad
                 \subfigure[]{
                \includegraphics[scale= 0.55]{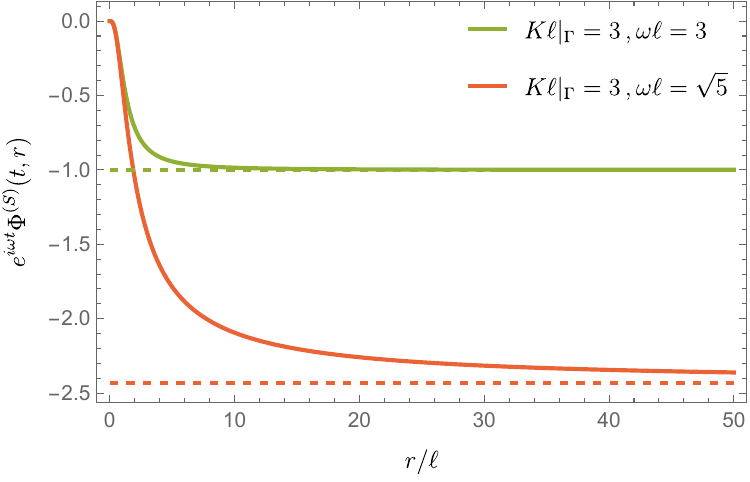} \label{fig: Re l2 scalar plot}}                           
                \caption{Radial profile of the master field with $l=2$, for different values of $K \ell$ at the boundary. In blue, we show the master field for a value of $K \ell$ that is greater than the critical; in yellow, at the critical value; and in green and red, for the asymptotic boundary of AdS$_4$. The circles indicate the position of the boundary in each case, except in the last ones where $\tfrac{\r}{\ell} \to \infty$. In all cases, the radial profiles are smooth functions of $\tfrac{r}{\ell}$. In (b), we increase the range of $\tfrac{r}{\ell}$ and only show the asymptotic form of the master field profile as the boundary approaches the conformal boundary of AdS. In dashed lines, we show the asymptotic constant values of the master field according to \eqref{asymptotic phi}.} \label{fig: Im l2 scalar plot}
\end{figure}

\textbf{Radial profiles.} It is worth noting that the radial profile of the master field at any given fixed time is smooth, even across the transition. For $l=2$, this can be seen in figure \ref{fig: Im l2 scalar plot}. When $K\ell \approx 3$, we can obtain the radial profiles analytically, finding that
\begin{eqnarray}
    e^{i \omega t} \Phi^{(S)}(t, r)|_{l=2} =
    \begin{cases}
        \frac{3 \sqrt{5} \left(4 \left(\tfrac{r}{\ell}\right)^2-3\right) \sin \left(\sqrt{5} \tan^{-1} \left(\tfrac{r}{\ell}\right) \right)+45 \left(\tfrac{r}{\ell}\right) \cos \left(\sqrt{5} \tan ^{-1} \left(\tfrac{r}{\ell}\right) \right)}{4 \left(\tfrac{r}{\ell}\right)^2} \, , \, \text{for} \,\, \omega \ell = \sqrt{5} \,, \\
        -\frac{\left(\tfrac{r}{\ell}\right)^3}{\left(\left(\tfrac{r}{\ell}\right)^2+1\right)^{3/2}} \quad , \quad \text{for} \,\, \omega \ell = 3 \,.
    \end{cases} \label{asymptotic phi 2}
\end{eqnarray}
At $r=0$ both profiles are identically zero. As $\tfrac{r}{\ell} \to \infty$, the radial profiles go to constants,
    \begin{eqnarray}
   e^{i\omega t} \Phi^{(S)}(t, r)|_{l=2} \overset{r/\ell\to \infty}{\longrightarrow}
    \begin{cases}
         3 \sqrt{5} \sin \left(\frac{\sqrt{5} \pi }{2}\right) \approx -2.43 \, , \, \text{for} \,\, \omega \ell = \sqrt{5} \,, \\
         -1 \quad , \quad \text{for} \,\, \omega \ell = 3 \,.
    \end{cases} \label{asymptotic phi}
\end{eqnarray}

\section{Linearised dynamics about planar AdS$_4$} \label{sec: planar AdS}

In this subsection, we consider linearised perturbations about a fixed planar AdS$_4$ spacetime, 
\begin{equation}
    ds^2 = \frac{\ell^2}{z^2}\left(dz^2-dt^2+dx^2+dy^2\right) \, ,
\end{equation}
where  $z>0$ and $(t,x,y)\in \mathbb{R}^{1,2}$ are coordinates on three-dimensional Minkowski space.

The timelike boundary, $\Gamma$, is taken to be a three-dimensional Minkowski spacetime located at $z=z_c > 0$. The spacetime region of interest will be $z\in\left(z_c,+\infty\right)$.

The background induced metric and the extrinsic curvature are given by
\begin{equation}\label{gbar2}
    \left.\bar{g}_{mn}dx^mdx^n\right|_\Gamma= \frac{\ell^2}{z_c^2}\eta_{mn}dx^m dx^n \,, \qquad \left.K_{mn}dx^m dx^n\right|_\Gamma = \frac{\ell}{z_c^2} \eta_{mn} dx^m dx^n \, .
\end{equation}
It follows that the trace of the extrinsic curvature is $K\ell = 3$. We note that, unlike the spherical case, the timelike boundary is independent of $z_c$ which means that, without any perturbation, every constant-$z$ surface leads to the same conformal class of the induced metric and trace of the extrinsic curvature. Indeed, there is no invariant meaning to $z_c$ as one can simply rescale coordinates to absorb it; it is nothing more than a bookkeeping device.

We consider perturbations of the form \eqref{metric_pert_lin},
subject to conformal boundary conditions at $\Gamma$. We fix the conformal class of the metric to be (\ref{gbar2}) and the trace of the extrinsic curvature to be $K = \tfrac{3}{\ell}$. As such, the perturbed Weyl factor, $\delta \bomega$, is defined as
\begin{equation}
    \left.ds^2\right|_\Gamma = (1+\varepsilon\,2 \delta \bomega) \frac{\ell^2}{z_c^2}\eta_{mn} dx^m dx^n \, .
\end{equation}
We require the leading perturbation of the trace of the extrinsic curvature to vanish at the boundary, $\delta K|_\Gamma = 0$. We also need to specify conformal boundary conditions for the set of allowed diffeomorphisms $\xi^\mu$. Furthermore, we require $\xi^z |_\Gamma=0$, so that allowed diffeomorphisms do not move the location of the boundary. 

In order to proceed one could follow an analogous procedure as in subsection \ref{sec: lin dynamics global AdS}, namely using the Kodama-Ishibashi formalism. In subsection \ref{sec: bulk planar AdS} we will work in the Fefferman-Graham coordinate choice so as to  compare our results with those in the AdS/CFT literature. 

\subsection{Locally diffeomorphic modes} \label{sec: diff planar}

We first consider linearised perturbations that can be expressed locally as a diffeomorphism,
\begin{equation}
    h_{\mu\nu}=\nabla_\mu\xi_\nu+\nabla_\mu\xi_\nu \,,
\end{equation}
for an arbitrary vector field $\xi^\mu$. The above perturbation automatically satisfies the linearised Einstein field equation. A straightforward calculation yields
\begin{eqnarray}\label{eqn: AdS planar metric diffeo}
\begin{cases}
    h_{zz} = \frac{2\ell^2}{z^2}\left(\partial_z \xi^z-\frac{\xi^z}{z}\right) \, , \\
    h_{zm} = \frac{\ell^2}{z^2}\left(\partial_m \xi^z + \eta_{mn} \partial_z \xi^n\right) \, , \\
    h_{mn} = \frac{\ell^2}{z^2}\left(\eta_{pn}\partial_m \xi^p + \eta_{pm}\partial_n \xi^p - \eta_{mn} \frac{2}{z}\xi^z\right)\, ,
    \end{cases}
\end{eqnarray}
where $\eta_{mn}$ is the three-dimensional Minkowski metric, and $\xi^z$ and $\xi^m$ are functions of both $z$ and $x^m$. Imposing that the perturbation \eqref{eqn: AdS planar metric diffeo} preserves the conformal structure of the induced metric \eqref{gbar2} at $z=z_c$ results in
\begin{equation}
    \left.\eta_{pn}\partial_m \xi^p + \eta_{pm}\partial_n \xi^p - \frac{2}{3}\eta_{mn}\partial_p \xi^p\right|_\Gamma = 0 \, .
\end{equation}
This is simply a conformal Killing equation for a three-dimensional Minkowski spacetime. Therefore this condition requires that the three-vector $\xi^m$ be given by
\begin{equation}\label{eqn: planar AdS xi^m}
    \xi^m(z,x^m) = \zeta^m(x^m) + \tilde{\xi}^m(z,x^m) \, , 
\end{equation}
where $\tilde{\xi}^m$ is an arbitrary three-vector that vanishes at $z=z_c$, namely $\tilde{\xi}^m (z_c,x^m)=0$, and $\zeta^m$ is a conformal Killing vector of three-dimensional Minkowski space,
\begin{equation}\label{eqn: Conf Killing 3D Mink}
    \zeta^m(x^m) = \mathfrak{a}^m +  x^n\mathfrak{b}_n{}^m{} + \left(2x^mx_n-x^px_p\delta^m_n\right)\mathfrak{c}^n + x^m\mathfrak{d}  \,,
\end{equation}
where $\mathfrak{a}^m$, $\mathfrak{b}_n{}^m$, $\mathfrak{c}^m$, and $\mathfrak{d}$ are arbitrary constants. Only $\mathfrak{c}^n$ and $\mathfrak{d}$ lead to a non-vanishing Weyl factor. Since $\tilde{\xi}^m$ vanishes at the timelike boundary, it is an allowed diffeomorphism.

The linearised trace of the extrinsic curvature is given by
\begin{equation}\label{eqn: planar AdS xi^z}
    \left.\delta K\right|_\Gamma = \left.\frac{z}{\ell} \partial^2 \xi^z\right|_\Gamma\,,
\end{equation}
where $\partial^2 \equiv \eta^{mn} \partial_m \partial_n$. This equation takes the form of a plane wave equation with $\delta K$ playing the role of an external source. Demanding that \eqref{eqn: AdS planar metric diffeo} preserves the trace of the extrinsic curvature at $z=z_c$ results in 
\begin{equation}\label{eqn: AdS planar sol xi^z}
    \xi^z(z,x^m) = \zeta(x^m) + \tilde{\xi}^z(z,x^m) \, ,
\end{equation}
where $\tilde{\xi}^z$ is an arbitrary function that vanishes as $z=z_c$, $\tilde{\xi}^z(z_c,x^m)=0$. Again, given that $\tilde{\xi}^z$ vanishes at the boundary, it is an allowed diffeomorphism. The $z$-independent function $\zeta$ is a solution to $\partial^2 \zeta=0$,
\begin{equation}\label{eqn: AdS planar sol zeta}
    \zeta(x^m) = \ell \, \text{Re} \int_{\mathbb{R}^2}\frac{d^2\bold{k}}{(2\pi)^2}\left(\zeta(\bold{k})e^{-i\omega t+i \bold{k} \cdot \bold{x}}\right) \, , \qquad \omega = |\bold{k}| \, ,
\end{equation}
where $\zeta(\bold{k})$ is an arbitrary function of $\bold{k}$. Combining both \eqref{eqn: Conf Killing 3D Mink} and \eqref{eqn: AdS planar sol zeta}, the linearised Weyl factor at $z=z_c$ can be written as
\begin{equation}\label{eqn: AdS planar linearised Weyl from xi^z}
    \delta \bomega(z_c ,x^m) = 2 x^m \mathfrak{c}_m + \mathfrak{d}-\frac{\ell}{z_c}\, \text{Re} \int_{\mathbb{R}^2}\frac{d^2\bold{k}}{(2\pi)^2}\left(\zeta(\bold{k})e^{-i\omega t+i \bold{k} \bold{x}}\right) \, , \qquad \omega = |\bold{k}|\, .
\end{equation}
We note that, as expected, allowed diffeomorphisms $\tilde{\xi}^\mu$ do not contribute to \eqref{eqn: AdS planar linearised Weyl from xi^z}. 

Thus, to uniquely specify a solution for $\delta \bomega|_{z=z_c}$ as a function of time, we need to specify initial data for $\delta \bomega$ at the intersubsection $\partial \Sigma$ of $\Gamma$ and the initial Cauchy surface $\Sigma$. More specifically, if we take the initial Cauchy slice to be at $t=0$, the initial dataset consists of $\{\delta \bomega,\partial_t \delta \bomega\}$ at $t=0$ and $z=z_c$. 

By setting the conformal Killing vectors \eqref{eqn: Conf Killing 3D Mink} to zero, we observe that the initial data can be expressed as
\begin{equation}\label{eqn: planar AdS Cauchy data physical diff}
    \left.\{\delta \bomega,\partial_t \delta \bomega\}\right|_{t=0,z=z_c}=\left. \left\{\frac{z^2}{4\ell^2}\left( h_{xx} + h_{yy} \right),-\frac{z}{\ell^2}h_{tz} - \frac{z}{2\ell}\left(\pi_{xx}+\pi_{yy}\right)\right\}\right|_{t=0,z=z_c} \, ,
\end{equation}
where $\pi_{xx}$ and $\pi_{yy}$ denote the linearised extrinsic curvature at $t=0$. We recall that $h_{xx}$, $h_{yy}$, $\pi_{xx}$, and $\pi_{yy}$ contribute to the standard Cauchy data of gravity. In contrast, the corner angle $\delta \beta=\left.h_{tz}\right|_{t=0,z=z_c}$  does not belong to the Cauchy data set. {One might be concerned, then, that the physical phase space is not even-dimensional or that there is a redundancy between the bulk and boundary phase space. For this reason, a better way to organise the physical Cauchy data is as follows. One can consider the standard initial dataset, namely $\mathcal{C}_\Sigma = \{h_{mn},K_{mn}\}$ on $\Sigma$, which obey the physical constraints and cannot be expressed, even locally, as diffeomorphisms of any type. Further to this, one has the boundary pair $\mathcal{C}_{\partial\Sigma} =\{\bomega, \partial_t \bomega \}$ residing at the spatial corner $\partial\Sigma$ that stems specifically from non-trivial diffeomorphisms. As such, $\mathcal{C}_{\partial\Sigma} \not\subset \mathcal{C}_\Sigma$. In asymptotically Minkowski or anti-de Sitter space we are often able to impose sufficiently stringent boundary conditions that remove $\mathcal{C}_{\partial\Sigma}$ from the physical phase space without disrupting the consistency of dynamical behaviour. For the case of a finite timelike boudary subject to conformal boundary conditions, this is no longer the case. }

Lastly, we note that by using \eqref{cby}, the linearised conformal Brown-York stress-energy tensor for \eqref{eqn: AdS planar linearised Weyl from xi^z} is given by
\begin{equation}
    T_{mn} = \left.\frac{\ell^2}{8 \pi G_N z_c}\partial_m\partial_n \delta \bomega\right|_\Gamma = -\left.\frac{\ell^2}{8 \pi G_N z_c^2}\partial_m\partial_n \zeta(x^m) \right|_\Gamma  \, .
\end{equation}
It is clear that $\partial^m T_{mn}=T^m{}_m=0$ as $\delta \bomega$ satisfies $\partial^2 \delta \bomega=0$. We note that since $K\ell=3$, we cannot directly compare to (\ref{eq: brown-york near boundary}). The contribution, as can be seen from (\ref{TmnFG}) in appendix \ref{app: near infinity}, corresponds to the divergent part of the boundary stress tensor in the standard AdS$_4$/CFT$_3$ limit. By integrating $T_{tt}$ over the spatial two-dimensional plane, we obtain the total conformal energy
\begin{equation}
    E_\text{conf} = \frac{\ell^2
    }{8 \pi G_N z_c}\int_{\mathbb{R}^2} d^2\bold{x} \, \partial_t^2 \delta \bomega(t,\bold{x}) \,.
\end{equation}
Given that $\partial^2 \delta \bomega=0$,  provided $\delta \bomega$ vanishes sufficiently fast at the boundary of $\mathbb{R}^2$ the conformal energy vanishes in the linearised approximation we are considering.

\subsection{Bulk linearised modes}\label{sec: bulk planar AdS}

Next we consider bulk modes. For this, we work in the Fefferman-Graham gauge, $h_{z\mu}=0$. It is always possible to use allowed diffeomorphisms $\tilde{\xi}^\mu$ to fix this gauge (see appendix \ref{app: FG}). Then, it is useful to parameterise the remaining components of the perturbed metric as
\begin{equation}\label{eqn: planar AdS param}
    h_{mn}(z,x^m) = \frac{\ell^2}{z^2}\gamma_{mn}(z,x^m)\, ,
\end{equation}
where $\gamma_{mn}$ is a symmetric tensor. Requiring that the perturbation \eqref{eqn: planar AdS param} preserves the conformal structure of the background induced metric at $z=z_c$ gives
\begin{equation}\label{eqn: AdS planar bdry cond 1}
    \left.\gamma_{mn}-\frac{1}{3}\eta_{mn} \gamma\right|_\Gamma = 0 \, ,
\end{equation}
where $\gamma \equiv \eta^{mn} \gamma_{mn}$ denotes the trace of $\gamma_{mn}$. Requiring that the perturbation \eqref{eqn: planar AdS param}
preserves the trace of the extrinsic curvature at $z=z_c$ yields
\begin{equation}\label{eqn: AdS planar bdry cond 2}
    \left.\partial_z \gamma\right|_\Gamma =0 \, .
\end{equation}
Using the parameterisation \eqref{eqn: planar AdS param}, the $zz$- and $zm$-components of the Einstein field equation respectively give
\begin{equation}\label{eqn: planar AdS G_zmu}
\begin{cases}
    \partial^m \partial^n \left(\gamma_{mn}-\eta_{mn} \gamma\right) + 2z^{-1}\partial_z \gamma = 0 \, , \\ 
    \partial^n \partial_z\left( \gamma_{mn}-\eta_{mn} \gamma\right) = 0 \, .
\end{cases}
\end{equation}
These equations imply that $\gamma=-\frac{1}{4}\partial^m v_m (z^2-z_c^2)+\chi$ and $
\partial^n (\gamma_{mn}-\eta_{mn}\gamma) = v_m$ for some scalar $\chi$ and vector $v_m$, which are both independent of $z$. Acting on \eqref{eqn: AdS planar bdry cond 1} with $\partial^m$, we find that $v_m = -\frac{2}{3}\partial_m \chi$. Imposing the boundary condition \eqref{eqn: AdS planar bdry cond 2}, we obtain $\partial^m v_m = - \frac{2}{3}\partial^2 \chi = 0$. 
In terms of $\gamma_{mn}$, these two equations imply that
\begin{equation} \label{eq: for appendix}
    \partial^n \gamma_{mn} = \frac{1}{3}\partial_m \gamma \, , \qquad \partial^m \partial^n \gamma_{mn} = \frac{1}{3}\partial^2 \gamma = 0 \, ,
\end{equation}
where $\gamma$ is now independent of $z$. Hence, by solving \eqref{eqn: AdS planar bdry cond 1}-
\eqref{eqn: planar AdS G_zmu}, one can decompose $h_{mn}$ as 
\begin{equation}\label{eqn: AdS planar metric param 2}
    h_{mn}(z,x^m) = \frac{\ell^2}{z^2}\left(\tilde{\gamma}_{mn}(z,x^m) + \frac{1}{3}\eta_{mn} \gamma(x^m)\right) \, ,
\end{equation}
where $\tilde{\gamma}_{mn}$ is a transverse-traceless tensor, and $\gamma$ is a $z$-independent scalar obeying $\partial^2 \gamma = 0$. The conformal boundary conditions for \eqref{eqn: AdS planar metric param 2} imply that
\begin{equation}
    \left.\tilde{\gamma}_{mn}\right|_\Gamma = 0 \, .
\end{equation}

Consider now the $mn$-components of the Einstein field equation.
By using \eqref{eqn: AdS planar metric param 2}, we obtain
\begin{equation}
    \partial^2 \tilde{\gamma}_{mn} + z^2 \partial_z \left(z^{-2}\partial_z \tilde{\gamma}_{mn}\right) = -\frac{1}{3}\partial_m \partial_n \gamma \, .
\end{equation}
We observe that, upon redefining $\tilde{\gamma}_{mn} \to \tilde{\gamma}_{mn} + \frac{1}{6}(z^2-z_c^2)\partial_m \partial_n \gamma$, the right hand side vanishes. This means that the general metric perturbation that obeys the conformal boundary conditions at $z=z_c$ is given by
\begin{equation}\label{eqn: AdS planar final exp}
    h_{mn}(z,x^m) = \frac{\ell^2}{z^2}\left(\tilde{\gamma}_{mn}(z,x^m) + \frac{1}{3}\left(\eta_{mn}+ \frac{1}{2}(z^2-z_c^2)\partial_m \partial_n  \right)\gamma(x^m)\right)  \, , \qquad  \left.\tilde{\gamma}_{mn}\right|_\Gamma = 0\, ,
\end{equation}
where the transverse-traceless tensor $\tilde{\gamma}_{mn}$ and the scalar $\gamma$ obey
\begin{equation}
\begin{cases}
    \partial^2 \tilde{\gamma}_{mn} + z^2 \partial_z \left(z^{-2}\partial_z \tilde{\gamma}_{mn}\right) = 0  \, , \\
    \partial^2 \gamma = 0 \, .
    \end{cases}
\end{equation}
The part of $h_{mn}$ that depends on $\gamma$ is a physical diffeomorphism, gauge equivalent to that analysed in the previous sub-subsection. Our analysis indicates that the bulk perturbation governed by the transverse-traceless tensor $\tilde{\gamma}_{mn}$ leads to a vanishing Weyl factor --- the Weyl mode is completely controlled by $\gamma$, and it is fixed uniquely by initial data localised at the boundary.

The equation for $\gamma$ can be solved straightforwardly. The general solution is given by
\begin{equation}
    \gamma(x^m) = \text{Re} \, \int_{\mathbb{R}^2}\frac{d^2 \bold{k}}{(2\pi)^2} \gamma(\bold{k}) e^{-i\omega t + i \bold{k} \cdot \bold{x}}  \, , \qquad \omega = |\bold{k}| \, ,
\end{equation}
where $\bold{k}$ is a two-dimensional vector and $\gamma(\bold{k})$ is an arbitrary function of $\bold{k}$. 

The solution for $\tilde{\gamma}_{mn}$ is also straightforward, but a bit longer, so we refer the reader to appendix \ref{app: planar perturbation} for details. The general solution for $\tilde{\gamma}_{mn}$ can be written as 
\begin{equation}
    \tilde{\gamma}_{mn}(z,x^m) = \text{Re} \int_0^\infty\frac{d q}{2\pi}\int_{\mathbb{R}^2}\frac{d^2 \bold{k}}{(2\pi)^2} \, \sum_{j=1}^2 \beta^{(j)}_{mn}(q,\bold{k}) \left(\frac{1+i q z}{1+i q z_c}e^{-iq(z-z_c)} - \frac{1-i q z}{1-i q z_c}e^{iq(z-z_c)}\right)e^{-i\omega t + i \bold{k} \cdot \bold{x}} \, , \label{eq: bulk modes planar}
\end{equation}
where $\beta^{(j)}_{mn}$ represent the two transverse-traceless polarisations of the metric perturbation and $\omega = \sqrt{|\bold{k}|^2+q^2}$. As $\omega$ is always real, the solutions are generally well behaved in time, mimicking the Dirichlet results in \cite{Andrade:2015gja}.

By using \eqref{cby}, the linearised conformal Brown-York stress-energy tensor associated with \eqref{eqn: AdS planar final exp} is given by
\begin{equation}
    T_{mn} = \left.\frac{\ell^2}{16 \pi G_N z_c^2}\left(\partial_z \tilde{\gamma}_{mn}+\frac{z_c}{3}\partial_m \partial_n \gamma\right)\right|_\Gamma \, .
\end{equation}
Once again, we can confirm that $T_{mn}$ is traceless and conserved. 

{\textbf{Asymptotic boundary limit.}} Let us consider the behavior of \eqref{eqn: AdS planar final exp} in the limit where the timelike boundary approaches the AdS$_4$ boundary.   Taking a perturbation which has the characteristic length scale of the order of $\ell$, namely $|\bold{k}|\ell\sim q\ell\sim 1$, along with $\tfrac{z_c}{\ell} \to 0$ and $\tfrac{z}{\ell} \to 0$ while keeping $\tfrac{z}{z_c}$ fixed, we obtain
\begin{equation}
    h_{mn} = \text{Re} \int_0^\infty\frac{d q}{2\pi}\int_{\mathbb{R}^2}\frac{d^2 \bold{k}}{(2\pi)^2} \, \frac{\ell^2}{z^2}\left[\frac{\eta_{mn}\gamma(\bold{k})}{3} -  \frac{z^2-z_c^2}{6}k_m k_n \gamma(\bold{k})- \sum_{j=1}^2 \frac{2i}{3}\left(z^3-z_c^3\right)q^3 \beta_{mn}^{(j)}(q,\bold{k})+ \cdots \right]e^{-i\omega t + i \bold{k}\cdot \bold{x}}\,,
\end{equation}
up to terms of order $\tfrac{z^3}{\ell^3}$. Setting $z_c=0$ and comparing with the standard Fefferman-Graham expansion, we observe that the Weyl mode, $\gamma(\bold{k})$, perturbs  $g^{(0)}_{mn}$ but not $g^{(3)}_{mn}$. It can be associated with a Fefferman-Graham gauge preserving diffeomorphism in the asymptotic boundary limit.

\subsection{Non-linear mode on AdS$_4$ black brane}

So far we have considered linearised perturbations about planar AdS$_4$. Upon implementing conformal boundary conditions, we identified a novel mode which is locally a diffeomorphism. As a final remark, and to complement the previous discussion of this subsection, we would like to consider the analogous setup about a black brane in AdS$_4$. The corresponding metric can be expressed in the following form
\begin{equation}\label{bb}
\frac{ds^2}{\ell^2} = \frac{dz^2}{z^2 f(z)} + \frac{1}{z^2}\left( - f(z) dt^2 + dx^2 + dy^2\right) \quad , \quad f(z) = 1 - \frac{2G_NMz^3}{\ell^4}    \,.
\end{equation}
Unlike the case of the planar AdS$_4$ background, the trace of the extrinsic curvature, $K$, on a surface of constant $z=z_c$ is now sensitive to the value of $z_c$. 

We would like to identify a hypersurface in the black brane geometry which obeys the conformal boundary conditions with constant $K$, and an induced metric conformally equivalent to $\eta_{mn}$
\begin{equation} \label{eq: flat bb bdy metric}
ds^2 |_\Gamma =  e^{{2\delta \bomega(t)}}  \frac{\ell^2 }{z_c^2} \left(- dt^2 +dx^2+dy^2\right)~.
\end{equation}
We will take $\delta \bomega(t)$ to be purely time dependent, but we will allow it to be non-linear. It is clear that $z_c$ can always be absorbed by a simple change in $\delta \bomega(t)$, so it should be viewed as a bookkeeping device.\footnote{Adding space dependence to $\delta\bomega$ is an interesting extension of this problem. It will require a more thorough analysis, beyond our scope, as the boundary mode $\delta\bomega$ would no longer be locally expressible as a diffeormophism.}

We now obtain a non-linear equation about the AdS$_4$ black brane background, analogous to (\ref{eqn: brane dynamics near AdS bdry}). 
Requiring that the timelike boundary is embedded in the AdS$_4$ black brane geometry (\ref{bb}), subject to our conformal boundary conditions, we are led to the following differential equation for $\delta\bomega$,
\begin{equation} \label{flat black brane dynamics}
    \ell^2 \partial_t^2 \delta \bomega = -\frac{3\ell^2}{z_c^2}e^{2\delta \bomega} + \frac{3G_N M z_c}{\ell^2} e^{-\delta \bomega} - 2 (\ell\partial_t \delta \bomega)^2+e^{\frac{\delta \bomega}{2}}\frac{K\ell^3}{z_c^2}\sqrt{e^{3\delta \bomega}-\frac{2G_N M z_c^3}{\ell^4}+\frac{z_c^2}{\ell^2}e^{\delta \bomega}(\ell\partial_t\delta \bomega)^2} \,.
\end{equation}

Note that upon setting $M=0$ and requiring $\delta \bomega(t)$ to be constant, this equation simply becomes $K\ell = 3$, as in the planar AdS$_4$ case. If $M$ is non-zero and we set $\delta \bomega$ to zero, the timelike boundary is at a constant $z=z_c$ slice of \eqref{bb}, and the value of $z_c$ is fixed as
\begin{equation}
    z_c^3 = \frac{\ell^4}{G_N M}\left(1-\frac{1}{9}\left(K^2\ell^2-K\ell\sqrt{K^2\ell^2-9}\right)\right) = \frac{\ell^4}{G_N M}\sqrt{\frac{2(K\ell-3)}{3}} + \mathcal{O}(K\ell-3) \, .
\end{equation}
Upon linearising \eqref{flat black brane dynamics} at finite $M$, we obtain the simpler equation
\begin{equation}\label{omegabb}
    \partial_t^2 \delta \bomega (t) = \Omega^2 \delta \bomega (t)~, \quad \text{with} \quad \Omega \equiv \frac{3 G_N M z_c^2}{\ell^2\sqrt{\ell^4-2 G_N  M z_c^3}} \,,
\end{equation}
which has solutions of the form $\delta \bomega (t) = e^{\pm \Omega \, t}$. The second order nature of (\ref{omegabb}) indicates that, as in our previous discussion, we must fix the dynamical data $\mathcal{C}_{\partial \Sigma} = \{\delta\bomega, \partial_t\delta\bomega \}|_{\partial \Sigma}$ to have a complete specification of the initial data.

Upon setting $M= 0$ in (\ref{omegabb}), such that $\Omega=0$,  we retrieve the simpler equation $\partial_t^2 \delta\bomega(t) =0$, in line with the analysis of subsection \ref{sec: diff planar}.  We also note that $\Omega$ is strictly real valued and diverges as the back brane horizon approaches the boundary, whereby $2 G_N  M z_c^3=\ell^4$. Such a divergence indicates that in the near-horizon region, where the metric is approximated to leading order by the Rindler spacetime, one must rescale the time coordinate to render it non-degenerate. Upon rescaling to the Rindler time coordinate, the rescaled expression for $\Omega$ is no longer divergent and our analysis naturally ties to the Rindler analysis of subsection 5.3 in \cite{Anninos:2024wpy}. What we are learning here, is that the Rindler boundary mode indeed non-linearises and persists away from the strict Rindler limit. 

The only non-vanishing components of the conformal Brown-York stress tensor (\ref{cby}) satisfy $T_{tt}=2T_{xx}=2T_{yy}$, with %
\begin{equation}
    T_{tt} = \frac{1}{12\pi \ell G_N} \left( K \ell \, e^{2 \delta\bomega }-3 e^{\delta\bomega /2} \sqrt{e^{3 \delta\bomega }+ e^{\delta\bomega } \left(\ell \partial_t\delta\bomega\right)^2 -\frac{2 G_N M}{\ell}} \right) \,,
\end{equation}
where we have set $\tfrac{z_c}{\ell}=1$ for notational simplicity. Upon imposing \eqref{flat black brane dynamics}, one can readily confirm that in addition to being traceless, $T_{mn}$ is also conserved with respect to boundary metric \eqref{eq: flat bb bdy metric}. The conformal energy follows immediately, by integrating the $tt$-component of the stress-tensor over a spatial slice. Interestingly, we see that for a finite conformal boundary in AdS$_4$, one can no longer decouple $\delta\bomega$ from the remaining dynamics. 

It might be tempting to view $\delta\bomega$ as a type of soft hair, along the lines of \cite{Hawking:2016msc}, which can dress the bare black brane geometry. However, in the case at hand, hair can carry physical energy. In this sense, the situation is similar to dressing a BTZ black hole with a Brown-Henneaux diffeomorphism \cite{Brown:1986nw}. A similar analysis holds for the black hole in global AdS$_4$, as detailed in appendix \ref{app: bh}.

\section{Sourcing the Weyl mode} 
\label{sec: source}

In this subsection we consider Euclidean AdS$_4$ subject to conformal boundary conditions. In particular, we are interested in the role of $K$ as a source for boundary operators. In ordinary AdS$_4$/CFT$_3$, the Euclidean problem produces a functional, $\mathcal{Z}$, whose arguments are the boundary values of the bulk fields. For the gravitational sector, one has a functional $\mathcal{Z}[[g_{i j}(\bold{x})]]$ of the conformal structure,  $[g_{i j}(\bold{x})]$, of a given boundary metric with coordinates $\bold{x}\in\mathbb{R}^3$. To compute the leading order contribution to $\mathcal{Z}$ one resorts to a saddle-point approximation whereby one seeks solutions to the Euclidean equations of motion that are smooth in the interior and asymptote to the specified boundary data. In the case at hand, the gravitational path integral will  produce a functional $\mathcal{Z}[[g_{ij}(\bold{x})],K(\bold{x})]$ of the conformal boundary data, as $K$ is no longer fixed to the value $K \ell = 3$. 

For an asymptotically AdS$_4$ spacetime, the configuration space near the AdS$_4$ boundary is organised in the form of a Fefferman-Graham expansion \cite{fefferman1985conformal}
\begin{equation}
\frac{ds^2}{\ell^2} = \frac{dz^2}{z^2} + \frac{1}{z^2} \left( g^{(0)}_{i j}+  z^2 g^{(2)}_{ij} + z^3 g_{ij}^{(3)} + \ldots\right) dx^i dx^j~,
\end{equation}
where $i,j$ denote indices tangential to the boundary, 
\begin{equation}
g^{(2)}_{ij} = \frac{1}{4} R[g^{(0)}_{ij}] g^{(0)}_{ij} - R_{ij}[g^{(0)}_{ij}]  \,,
\end{equation}
is minus the Schouten tensor, and $g_{ij}^{(3)}$ is traceless and transverse with respect to $g^{(0)}_{ij}$. The above series generally converges at some small but finite value of $z$ away from $z=0$. There exist residual diffeomorphisms that preserve the above asymptotic gauge. These involve a Weyl transformation of $g^{(0)}_{ij}$ accompanied by a particular transformation of the remaining coordinates. As such, the boundary data is that of a conformal metric.\footnote{In odd dimensions one must modify the discussion slightly due to anomalous transformations.} It is worth re-emphasising that for all asymptotic configurations, $K$ evaluated at the asymptotic boundary takes the same fixed value, namely $K = \tfrac{3}{\ell}$. Consequently, to impose boundary conditions for more general choices of $K$, we have to  move away from the asymptotic AdS$_4$ boundary. For instance, at a slice of constant but small $z=z_c$, we can compute the leading correction to be
\begin{equation}
K \ell = 3 + \frac{z_c^2}{4} R[g^{(0)}_{ij}] + \ldots ~,
\end{equation}
to leading order in the small $z_c$ expansion. The above expression for $K$ is now sensitive to the full boundary metric $g^{(0)}_{ij}$, and not just its conformal structure.

\subsection{Perturbed partition function}

We would like to deform $K(\bold{x})$ in an infinitesimal way, perturbing the boundary slightly away from its value $K\ell=3$ about the empty Euclidean AdS$_4$ solution. This could be taken, for instance, to be either the hyperbolic metric with flat slicing 
\begin{equation}\label{planar}
\frac{ds^2}{\ell^2} = \frac{1}{z^2} \left( dz^2 + {d\bold{x}^2} \right) ~, 
\end{equation}
where $\bold{x}\in \mathbb{R}^3$, or the hyperbolic metric with spherical slices
\begin{equation}
\frac{ds^2}{\ell^2} = \frac{dz^2}{z^2} + \frac{\ell^2}{z^2} \left( \frac{1}{4}  - \frac{z^2}{2\ell^2} + \frac{z^4}{4\ell^4} \right) d\Omega_3~,
\end{equation}
where $d\Omega_3$ is the round metric on the three-sphere, and $z\in \mathbb{R}^+$. For the sake of simplicity, in what follows we focus on (\ref{planar}).

At the level of the linearised equations of motion, we seek solutions that obey the conformal boundary conditions at $z=z_c$ that impose an induced metric conformally equivalent to the boundary metric $\delta_{ij}$, whilst the trace of the extrinsic curvature is $K = \tfrac{3}{\ell} + \,\delta K(\bold{x})$. 

In the planar case, for non-constant $\delta K(\bold{x})$, the linearised gravity solutions subject to our prescribed conformal boundary conditions enforce that the metric perturbation is locally pure gauge, that is,
\begin{equation}\label{eqn: hmunu planar ads}
    h_{\mu\nu} = \nabla_\mu \xi_\nu + \nabla_\nu \xi_\mu \, ,
\end{equation}
where $\nabla_\mu$ is the covariant derivative on the planar metric (\ref{planar}). By imposing that the induced metric at $z_c$ is conformally flat, we find
\begin{equation}
    \xi^i(z_c,\bold{x}) = 0 \, .
\end{equation}
The trace of the extrinsic curvature, to the linear order, fixes 
\begin{equation}
    \xi^z(z_c,\bold{x}) = \zeta(\bold{x}) \, , \qquad \delta K(\bold{x}) = \frac{z_c}{\ell}\partial_i \partial^i \zeta(\bold{x}) \, .
\end{equation}
We note that $\delta K(\bold{x})$ vanishes at $z_c = 0$, consistent with the fact that all Fefferman-Graham configurations have $K = \tfrac{3}{\ell}$ at the asymptotic boundary. The general diffeomorphism $\xi^\mu(z,\bold{x})$ that solves these equations is given by
\begin{equation}
    \xi^z(z,\bold{x}) =\zeta(\bold{x}) + \tilde{\xi}^z (z,\bold{x}) \, , \qquad \xi^i(z,\bold{x}) = \tilde{\xi}^i(z,\bold{x}) \, ,
\end{equation}
where $\tilde{\xi}^z (z,\bold{x})$ and $\tilde{\xi}^i (z,\bold{x})$ are residual gauge degrees of freedom obeying $\tilde{\xi}^z(z_c,\bold{x}) = \tilde{\xi}^i (z_c,\bold{x}) = 0$. Fixing the Fefferman-Graham gauge $h_{zz} = h_{zi} = 0$ imposes conditions on the remaining terms, resulting in
\begin{equation}\label{fgslicing}
\xi^z(z,\bold{x})  = -z \int \frac{d^3\bold{y}}{z_c^3}\, z_c \frac{\ell\delta K (\bold{y})}{4\pi\left|\bold{x}-\bold{y}\right|} \, , \qquad \xi^i(z,\bold{x})  = \frac{z^2-z_c^2}{2}\delta^{ij}\partial_j\int \frac{d^3\bold{y}}{z_c^3}\, z_c\frac{\ell\delta K (\bold{y})}{4\pi\left|\bold{x}-\bold{y}\right|} \, .
\end{equation}
In the strict AdS$_4$ boundary limit, we can recognize the above as a Fefferman-Graham preserving transformation (see for example equations (2.6) and (2.7) in \cite{Anninos:2010zf}). As (\ref{fgslicing}) involves a non-vanishing $\xi^z$ component, the above is a physical diffeomorphism that cannot be gauged away. In terms of the metric perturbation \eqref{eqn: hmunu planar ads}, we find
\begin{equation}\label{hK}
    h_{ij}(z,\bold{x}) = \left(\frac{2\ell^2}{z^2}\delta_{ij}+\left(1-\frac{z_c^2}{z^2}\right)\ell^2\partial_i \partial_j\right)\int \frac{d^3\bold{y}}{z_c^3}\, z_c \frac{\ell\delta K (\bold{y})}{4\pi\left|\bold{x}-\bold{y}\right|} \, .
\end{equation}
In particular, the perturbed Weyl factor of the boundary metric is given by
\begin{equation} \label{eq: bdy euclidean}
   ds^2|_{\Gamma} = \frac{\ell^2}{z_c^2} e^{2\bomega(\bold{x})} \delta_{ij} dx^i dx^j \,, \, \quad e^{\bomega(\bold{x})} \approx 1+ \delta \bomega(\bold{x}) \= 1 + \frac{z_c \ell}{4\pi} \int \frac{d^3\bold{y}}{z_c^3}\,  \frac{\delta K (\bold{y})}{\left|\bold{x}-\bold{y}\right|}\,.
\end{equation}
We note that $\delta K(\bold{x})$ controls, non-locally, the linearised Weyl mode of the boundary metric. In the $z_c\to 0$ limit while keeping $\frac{\left|d\bold{x}\right|}{z_c}$ and $\ell \delta K(\bold{x})$ fixed, the response of the Weyl factor due to $\delta K(\bold{x})$ vanishes.

Given the solution satisfying the conformal boundary conditions, we can compute the on-shell Euclidean Einstein action for the solutions of interest as a function of $\delta K(\bold{x})$. The Euclidean action is given by, %
\begin{equation}
-I_{\text{cl}} = \frac{1}{16\pi G_N}\int_{z_c}^{z_{\text{IR}}} dz\int d^3 \bold{x} \sqrt{\det g_{\mu\nu}} \left(R-2\Lambda\right)+\frac{1}{24\pi G_N} \int_{\Gamma} d^3 \bold{x} \, \sqrt{\det g_{mn}} K(\bold{x})~,
\end{equation}
where we will take $z_{\text{IR}}\to+\infty$ at the end. Given \eqref{eq: bdy euclidean}, it is easy to see that the boundary term becomes,
\begin{equation}
-I_{\text{bdy}} = \frac{1}{24\pi G_N} \int d^3 \bold{x} \, \frac{\ell^3}{z_c^3}e^{3\bomega(\bold{x})} K(\bold{x})~.
\end{equation}
As we would like to compute the two-point function of the Weyl mode, it is sufficient to expand the action to quadratic order in the metric perturbation $h_{\mu\nu}$. Using the on-shell solution \eqref{hK} and after integrating by parts (assuming that $\delta \bomega$ decays fast enough as $|\bold{x}|\to\infty$), the on-shell action is given by
\begin{equation}\label{Icl}
  - I_{\text{cl}} = \frac{3 z_{\text{IR}}\ell^2}{64 \pi z_c G_N}\int \frac{d^3 \bold{x}}{z_c^3} \ell^2\delta K(\bold{x})^2 + \frac{\ell^2}{24\pi G_N} \int \frac{d^3 \bold{x}}{z_c^3} \frac{d^3 \bold{y}}{z_c^3} \frac{z_c\ell^2\delta K(\bold{x}) \delta K(\bold{y})}{4\pi|\bold{x}-\bold{y}|} + \mathcal{O}(z_{\text{IR}}^{-1}) \, .
\end{equation}
Note that the first term, which is local in $\delta K(\bold{x})$, is divergent as $z_{\text{IR}}\to \infty$. It would be interesting to perform a more refined regularisation procedure for this term, perhaps along the lines of \cite{Castro:2024cmf}.
\subsection{Weyl factor two-point function}

Expression (\ref{Icl}) gives us, in the semiclassical limit, the generating function of correlations for the operator sourced by $\delta K(\bold{x})$. We can compute the two-point function as a simple example. This is the second functional derivative of $-I_{\text{cl}}$ with respect to $\delta K(\bold{x})$, which is given by
\begin{equation}
    - \frac{\delta^2 I_{\text{cl}}}{\delta K(\bold{x}) \delta K(\bold{y})} =  \frac{\ell^2}{12 \pi G_N} \frac{1}{z_c^6}\frac{z_c \ell^2}{4\pi\left|\bold{x}-\bold{y}\right|} \, ,
\end{equation}
for separate points $\bold{x}$ and $\bold{y}$. 
This is the two point function of a three dimensional massless scalar field. This is in line with the massless dispersion relation (\ref{eqn: bdry dispersion}) for $\delta \bomega(x^m)$ that we uncovered in our Lorentzian mode analysis, and more specifically the relation (\ref{eqn: AdS planar linearised Weyl from xi^z}) in the Lorentzian planar AdS$_4$ analysis.

\textbf{Comparison to Euclidean flat space.} It is worth  comparing the above analysis to the case of linearised gravity with a vanishing cosmological constant around $\mathbb{R}^4$, whose metric we take to be
\begin{equation}
ds^2 = dz^2 + d\bold{x}^2~.
\end{equation}
We choose the boundary to  reside at $z=0$, and as in the previous analysis, we impose that the induced metric is conformally equivalent to the  flat metric on $\mathbb{R}^3$ and the trace of the extrinsic curvature is $\delta K (\bold{x})$. A similar analysis to that for the planar AdS$_4$ geometry now yields the following linearised physical diffeomorphism
\begin{equation}\label{eqn: comparison euclidean flat}
    \xi^z (z,\bold{x}) = -\int d^3 \bold{y} \, \frac{\delta K(\bold{y})}{4\pi \left|\bold{x}-\bold{y}\right|}  + \tilde{\xi}^z (z,\bold{x}) \, , \qquad \xi^i (z,\bold{x}) = \tilde{\xi}^i (z,\bold{x}) \, , 
\end{equation}
where $\tilde{\xi}^z (z,\bold{x})$ and $\tilde{\xi}^i (z,\bold{x})$ are residual gauge degrees of freedom obeying $\tilde{\xi}^z(0,\bold{x}) = \tilde{\xi}^i (0,\bold{x}) = 0$. The corresponding metric perturbation can be computed using $h_{\mu\nu}=\partial_\mu \xi_\nu + \partial_\nu \xi_\mu$. Interestingly, the Weyl factor of the boundary metric is unaffected by $\delta K(\bold{x})$. This can be seen by noting that, for a general background geometry, the Weyl factor transforms under an infinitesimal diffeomorphism as
\begin{equation}
    \delta \omega(z,\bold{x}) = \frac{1}{3}\mathcal{D}_i \xi^i(z,\bold{x}) + \frac{1}{3}K \xi^z(z,\bold{x}) \, ,
\end{equation}
where $\mathcal{D}^i$ is the covariant derivative of the boundary metric. In the present case, $\mathcal{D}_i = \partial_i$ and $K=0$ implying that $\xi^z$ does not enter the linearised Weyl factor. Since $\xi^i(0,\bold{x})$ is zero according to \eqref{eqn: comparison euclidean flat}, then $\delta \omega(0,\bold{x}) = 0$ identically. This is in line with the observation in \cite{Anninos:2023epi} that the physical diffeomorphism that alters the corner angle $\beta$ about the standard Minkowski corner does not perturb the boundary value of the Weyl factor to linear order.

\begin{center}
\pgfornament[height=5pt, color=black]{83}
\end{center}
\vspace{5pt}

To conclude, we offer some general remarks on the current state of affairs. We view the formulation of the general problem in terms of the conformal boundary condition fixing the pair $([g_{mn}],K)$ as carrying certain merit. In particular, it is a formulation that can be extended to arbitrary value for the cosmological constant \cite{Anninos:2023epi,Anninos:2024wpy,Liu:2024ymn,Banihashemi:2024yye}, and ties naturally to the existing structure of the Fefferman-Graham expansion. Moreover, in the limit $K\ell \to \infty$ with an $S^2\times \mathbb{R}$ boundary metric, we recover the Minkowskian results of \cite{Anninos:2023epi} directly from the global AdS$_4$ analysis. 

In all cases analysed so far, one finds that a novel dynamical degree of freedom, $\bomega(x^m)$, must be included to the gravitational phase space. As such, our setup may shed light on the question of how many underlying degrees of freedom there are in a finite portion of spacetime. In this regard, it is interesting to note that the sharp distinction between the ultraviolet divergences, that are often removed by local boundary counterterms,  and the remaining quantum field theoretic data encountered in the standard AdS/CFT dictionary is no longer manifest. This feature is already illustrated  in the introduction by considering the gravitational path integral $\log \mathcal{Z}[S^3,K]$ in (\ref{Zs3}). Perhaps, from a more schematic viewpoint, what we are seeing is the necessity to consider a more local version of the holographic renormalisation group flow whereby we allow the ultraviolet cutoff scale, $\boldsymbol{\Lambda}_{\text{u.v.}}(x^m)$, to depend on the boundary spacetime coordinates. The effective action for $\bomega(x^m)$ follows from a Legendre transform of the dual field theory with respect to $\boldsymbol{\Lambda}_{\text{u.v.}}(x^m)$. 

We note that if we impose, instead, Dirichlet conditions near the AdS$_4$ boundary, the corresponding boundary Brown-York stress-tensor, $T_{mn}(x^n)$, will  acquire a non-vanishing trace, ${T_m}^m(x^m)$. From the bulk point of view, this trace is accounted for by the trace of the extrinsic curvature which is no longer fixed, in contrast to the case of the Fefferman-Graham expansion, or the conformal boundary conditions. One may ask whether there exists a new boundary mode for the Dirichlet problem. Near the Minkowski corner, the answer is partially affirmative, in that one finds large physical diffeomorphisms \cite{An:2021fcq,Anninos:2022ujl}. However, they come at the cost of a non-unique development of the equations. 

One of the main unresolved questions, left unanswered in our work so far, is the fate of the linearised modes that exhibit exponential growth about global AdS$_4$. To make progress on this issue, a non-linear analysis may be necessary. The growing modes indicate, in the very least, that the global AdS$_4$ solution (along with its $SO(3,2)$ isometry group) is no longer the vacuum state --- conformal symmetries appear to be broken. As $K\ell \to 3$, the breaking occurs at increasingly large angular momentum. To control the behaviour of the growing modes, we might consider adding further boundary terms at $\Gamma$, which would alter the dynamics of $\bomega(x^m)$, or further spacetime dependence on $K(x^m)$. For instance, a boundary cosmological constant
\begin{equation}
I_{\text{bdy}} = \lambda \int_{\Gamma} d^3x \sqrt{-\det g_{mn}} \,,
\end{equation}
adds an interaction term to the theory governing $\bomega(x^m)$. Along these lines, one may wish to consider the possibility of path integrating over the space of conformal metrics, such that the boundary theory is a type of conformal gravity, echoing ideas expressed in \cite{Banihashemi:2024yye}.

From a holographic point of view, what is missing is a clear interpretation for the meaning of $K(x^m)$ in the dual theory. Considerations of the Bekenstein-Hawking relation, as in (\ref{BH}),  appear to indicate that $K(x^m)$ is related to the number of underlying degrees of freedom in the finite-size theory. In parallel, $K(x^m)$ or its deviation away from the asymptotic value $K\ell=3$, appears to be related to the amount of conformal symmetry breaking. Relatedly, the conserved charges (\ref{eq: brown-york near boundary}) at the boundary $\Gamma$ diverge upon taking $K\ell\to 3$. In the standard AdS$_4$/CFT$_3$ literature, such divergences are related to the ultraviolet cutoff in the dual CFT$_3$. In our analysis, once the asymptotic AdS$_4$ boundary $\Gamma_\infty$ is brought in to some finite timelike boundary, $\Gamma$, the previously divergent terms are rendered finite and become sensitive to the dynamical mode $\bomega(x^m)$. The novel terms describe part of the physical content in the theory.

To make contact with concrete examples of AdS$_4$/CFT$_3$, it is worth generalising the conformal boundary conditions of \cite{An:2021fcq} to theories of supergravity. In particular, we may wish to understand the analogue condition for the whole gravitational supermultiplet. A natural proposal would be to consider the supersymmetric variation of the boundary data $([g_{mn}],K)$ and fix the whole collection at $\Gamma$. In particular, we may wish to fix the superconformal structure at $\Gamma$ along with a supersymmetric generalisation of $K(x^m)$. This, among other things such as the generalisation to other spacetime dimensions, and developing a more general string theoretic picture \cite{Silverstein:2022dfj,Ahmadain:2024uyo}, will be left for future considerations.

From a more general perspective, a sharper understanding of Euclidean gravity on a manifold with a finite size boundary may help clarify certain puzzling aspects of Euclidean quantum gravity. An interesting example that arises in this context comes from Euclidean gravity with $\Lambda>0$. Here, the Euclidean saddle corresponds to a round four-sphere. One can carve out a variety of compact three-manifolds from an $S^4$. The gravitational path integral over $S^4$ \cite{Gibbons:1976ue,Anninos:2020hfj}, though rich in structure exhibits some unusual features including (potentially) the presence of a phase \cite{Polchinski:1988ua}. On its own, the $S^4$ path integral has no external parameters that can be varied. In contrast, if we excise a solid cylinder, we obtain a portion of $S^4$ with an $S^2 \times S^1$ boundary, as  explored in \cite{Wang:2001gt,Anninos:2011zn,Anninos:2017hhn,Anninos:2018svg,Blacker:2023oan,Banihashemi:2022jys,Coleman:2021nor,Svesko:2022txo,Anninos:2024wpy,Batra:2024kjl,Silverstein:2024xnr}. It is of interest, then, to understand how features of the sphere path-integral are encoded in the excised sphere path integral. The presence of the phase, for example, may now depend on what boundary conditions are placed on the $S^2 \times S^1$ (see \cite{Maldacena:2024spf} for a related discussion). Moreover, the structure of the $S^2 \times S^1$ boundary may lend itself to an interpretation closer to that of a trace, making it easier to assess the underlying unitarity of the theory (or lack thereof). If we wish to fuse two portions of an excised sphere back together \cite{Anninos:2024wpy,Anninos:2022hqo},  we may wish to consider placing conformal boundary conditions at each boundary.










\chapter{Discussion}
%

In this final chapter, we summarise our results and discuss some future directions along with more speculative remarks.

\section{Quantum features of de Sitter}
\label{sec:dSholography}

De Sitter entropy \cite{Gibbons:1977mu} is perhaps among the most curious yet elusive quantities in theoretical physics. Its cousin, black hole entropy \cite{Bekenstein:1973ur,Hawking:1975vcx}, led to the radical idea that a black hole is an ordinary thermal system living in one less dimension, from which modern concepts in quantum gravity, such as holography \cite{tHooft:1993dmi,Susskind:1994vu}, have grown. A sharper understanding of de Sitter entropy may open a new window into the quantum nature of our Universe.

\paragraph{Thermally stable dS in a box.} A natural hypothesis is that de Sitter entropy is the thermal entropy of some system. In chapter \ref{chap: dS} we pursued this idea by enclosing the dS horizon within a timelike boundary obeying CBCs, see \cite{Banihashemi:2022jys,Banihashemi:2022htw,Coleman:2021nor,Batra:2024kjl,Silverstein:2024xnr} for a similar setup with DBCs. Evaluating the Euclidean gravitational path integral in the conformal canonical ensemble, we found that for boundaries with sufficiently large trace of the extrinsic curvature the horizon is thermally stable, with entropy given by the standard area formula,\footnote{In the context of the saddle point approximation of the Euclidean gravitational path integral, there is a notion of \textit{Euclidean stability}, which concerns the existence of negative eigenmodes of the Lichnerowicz operator about a Euclidean saddle. Such modes were studied for asymptotically flat black holes in \cite{Gross:1982cv}, and their relation to negative specific heat was established in \cite{Whiting:1988qr,Prestidge:1999uq,Reall:2001ag}. In the presence of a finite boundary, however, positive specific heat does not necessarily exclude Euclidean negative modes. Indeed, \cite{Liu:2024ymn} showed that a sufficiently large asymptotically flat black hole enclosed in a conformal box has a single negative mode despite having positive specific heat. This raises a question of relation among various notions of stability, especially in the presence of finite boundaries, which needs to be further understood.} 
\begin{equation}
\label{eq:stableCK}
C_K = -\tilde{\beta}^2 \partial_{\tilde{\beta}} \mathcal{S}\Big|_K > 0\,, \qquad \mathcal{S} = \frac{A}{4G_N}\,.
\end{equation}
This is in contrast with a Dirichlet boundary, for which the specific heat of the cosmic patch is always negative \cite{Banihashemi:2022jys,Draper:2022ofa}. It provides, for the first time, a consistent thermodynamic setting in which the microscopic origin of the dS horizon entropy can be probed.

\paragraph{Gluing mechanics.} In the absence of a boundary, the natural object of the $\Lambda>0$ Euclidean path integral in the semi-classical limit is the $S^4$ partition function. On its own the $S^4$ saddle appears featureless, carrying no external parameters that can be varied. Excising the sphere along a boundary obeying CBCs supplies such parameters, and one may then ask whether two excised spheres can be fused back into the full sphere. At the level of saddles we showed in chapter \ref{chap: dS} that this occurs precisely when the two pieces share the conformal structure of the induced metric, encoded here in the inverse conformal temperature $\tilde\beta$, and carry trace of the extrinsic curvature $K$ of opposite sign,
\begin{equation}
    \mathcal{Z}_{S^4}= \mathcal{Z}(\tilde \beta,K) \mathcal{Z}(\tilde \beta ,-K) \, .
\end{equation}
It would be interesting to assess this gluing beyond the saddle-point approximation, see \cite{Anninos:2025fer,Anninos:2026hia} for works along this line.

\section{Conformal boundaries near extremal black holes}
\label{sec:extremal}

The considerations of this thesis were largely confined to pure Einstein gravity. A richer structure emerges in four-dimensional Einstein-Maxwell theory, admitting near-extremal Reissner-Nordstr\"om black hole solutions carrying arbitrarily small temperature with macroscopic entropy and developing an infinitely long AdS$_2\times S^2$ throat near the horizon. Their low temperature thermodynamics has drawn renewed attention through the holographic dual \cite{Maldacena:2016upp} between AdS$_2$ Jackiw-Teitelboim (JT) gravity \cite{Jackiw:1984je,Teitelboim:1983ux} with DBCs and the Sachdev-Ye-Kitaev (SYK) model \cite{Sachdev:1992fk,Kitaev:2015}, in which the Schwarzian mode governs the leading departure from extremality. It is natural to ask how this structure is altered under CBCs. We summarise below the main results of \cite{Galante:2025tnt}, which does not form part of the thesis.

\paragraph{Low temperature phases with CBCs.} Placing the boundary in the near-horizon region of a near-extremal black hole involves the double limit
\begin{equation}
\label{eq:doublescale}
    \tilde \beta \to \infty\,, \qquad KQ \to 1\,,
\end{equation}
with extremal value $\tilde\beta_{\text{ex}} = 2\pi/\sqrt{K^2Q^2-1}$ and $\delta\tilde\beta \equiv \tilde\beta - \tilde\beta_{\text{ex}}$. The two dimensionless parameters $\tilde\beta$ and $KQ$ control the deviation from extremality, so the phase structure depends on the order in which the limits in \eqref{eq:doublescale} are taken. Denoting the extremal entropy by $\mathcal{S}_{\text{ex}} = \tfrac{\pi Q^2}{G_N}$, the deviation $\mathcal{S}-\mathcal{S}_{\text{ex}}$ ranges from the familiar linear-in-temperature behaviour to a quadratic $\tilde{\beta}^{-2}$ correction, and includes a genuinely new regime with the fractional scaling $(KQ-1)^{5/4}\,|\delta\tilde\beta|^{1/2}$ that is absent under DBCs. It would be interesting to find a microscopic theory, perhaps a generalisation of the SYK model, which reproduces this behaviour.

\paragraph{Dilaton-gravity theory with CBCs.} Reducing the static, spherically symmetric sector of four-dimensional Einstein--Maxwell theory on the transverse two-sphere yields a two-dimensional dilaton-gravity theory with action
\begin{equation}
    I_E= I_\text{top} - \frac{1}{16 \pi G_2}\int d^2x \sqrt{g}\, \left(\Phi R + U(\Phi)\right) + I_\text{bdry}\,.
\end{equation}
Fixing the total charge $Q$ through gauge-field boundary conditions compatible with CBCs, the reduced higher-dimensional CBCs become a novel one-parameter family of two-dimensional boundary conditions
\begin{equation}
\label{eq:2dCBC}
    \Phi^{-\alpha}\sqrt{h} = \text{fixed}\,, \qquad \Phi^{-\alpha}\left(\Phi K + \alpha\, n^\mu \nabla_\mu \Phi\right) = \text{fixed}\,,
\end{equation}
with $\alpha=3/4$, which one may equally treat as a general dilaton-gravity theory with $U(\Phi)$ and $\alpha$ arbitrary. The central lesson is that the linear dilaton potential of JT gravity no longer suffices once the boundary obeys \eqref{eq:2dCBC}. Expanding $\Phi=\Phi_0+\phi$ and truncating $U(\Phi)$ at linear order, the bulk JT contribution cancels the boundary term on-shell and leaves the extremal thermodynamics up to a constant shift of the conformal energy. The first genuine deviation therefore requires the quadratic term in $U(\Phi)$, unlike the DBC analysis where the linear term already generates the linear-in-temperature correction.

\paragraph{Double Schwarzian theories.} The role of the Schwarzian is sharpest for spherically reduced AdS$_3$, whose two-dimensional description is exactly JT gravity with negative cosmological constant. Placing the boundary near the conformal boundary of AdS$_3$, so that $K \ell_\text{AdS} = 2 + \epsilon^2 \kappa_2$ with $\epsilon\ll1$, and working partially on-shell, the one-dimensional effective action for the boundary reparametrisation $t(u)$ and renormalised dilaton $\Phi_r(u)$, with $\Gamma(u)\equiv\int^u \Phi_r$, is a difference of two Schwarzians,
\begin{equation}\label{eqn: double Schwarzian}
    I_\text{bdry}= - \frac{1}{8 \pi G_2}\int_{\partial \mathfrak{M}} du\, \left(S(t(u),u) - \frac{\kappa_2}{2}\, (\partial_u\Gamma)^2 - S(\Gamma(u),u)\right)\,.
\end{equation}
The first term is the Schwarzian of the standard DBC analysis, while the second, together with the $\kappa_2$ kinetic term, is the imprint of CBCs. Varying with respect to $\Phi_r\equiv e^{\omega}$ yields the one-dimensional Liouville equation $\omega'' + \kappa_2 e^{2\omega}=0$ for the boundary Weyl factor. The same relative minus sign arises when the Schwarzian is coupled to one-dimensional quantum gravity \cite{Anninos:2021ydw}. Since \eqref{eqn: double Schwarzian} is only partially on-shell, it would be interesting to compute the one-loop determinant and to understand how the boundary and bulk modes couple at higher orders in $\epsilon$.

\section{Stretched horizon limit and membrane paradigm}\label{sec:stretchedhorizon}

Taking the boundary to hover just above the horizon is a reminiscence of the membrane paradigm \cite{Damour:1978cg,Price:1986yy}, in which the black hole horizon is replaced by a fictitious viscous fluid membrane. That effective membrane behaves like a peculiar fluid, its bulk viscosity being negative, $\zeta_{\text{m.p.}} = -\frac{1}{16\pi}$, signaling ill-defined behaviour. It is therefore worthwhile to sharpen the notion of a stretched horizon living parametrically close to the de Sitter horizon, using CBCs as a mathematically precise implementation of the membrane programme. We study this briefly in chapter \ref{chap: dS} and at length in \cite{Anninos:2025zgr}, where we analyse linearised gravitational perturbations of the static patch subject to CBCs in the limit of large $|K\ell_{\text{dS}}|$.

\paragraph{Ill-posedness and the EFT perspective.} In the stretched horizon limit the growing modes of the Minkowski analysis of \cite{Anninos:2023epi} survive only in a parametrically large angular momentum sector set by $K\ell$. For $\mathfrak{L} \equiv \sqrt{l(l+1)} \gg |K\ell| \gg 1$, the allowed frequencies take the form
\begin{equation}\label{eq:largeLmodes}
\omega\ell = \frac{\mathfrak{L}}{|K\ell|} + \frac{\nu}{2^{1/3}}\left(\frac{\mathfrak{L}}{|K\ell|}\right)^{1/3} + \mathcal{O}(\mathfrak{L}^{-1/3})\,,
\end{equation}
where $\nu$ solves an Airy-type quantisation condition and, in the scalar sector, admits a pair of complex solutions with positive imaginary part, so that the modes grow at a rate scaling as $\mathfrak{L}^{1/3}$. It is precisely this behaviour that was used to establish the failure of continuous dependence on the initial data for general relativity subject to CBCs \cite{Liu:2025xij}. We find, moreover, that these modes decay exponentially away from the boundary, with a proper width scaling as $\tfrac{D}{\ell} \sim \mathfrak{L}^{-2/3}|K\ell|^{-1/3}$. Interestingly, the large dimensionless parameter $K\ell$ naturally creates a separation of scale. Moreover viewed as the leading term in a derivative expansion, general relativity cannot be trusted for such sharply localised large angular momentum modes, and whether they persist or are tamed once higher-derivative corrections are incorporated remains open, calling for a systematic treatment of CBCs within an effective field theory framework.

\paragraph{Soft modes and the disparity of horizons.} Below $|K\ell|$ the modes \eqref{eq:largeLmodes} are no longer present. In their place one finds soft modes, whose frequency to leading order is $\omega\ell = +i$ for any angular momentum, or $\omega \approx 2\pi i\, T_{\text{dS}}$ in terms of the de Sitter temperature measured by an inertial observer.\footnote{It is worth noting that such frequency appears also in the context of quantum chaos black hole \cite{Knysh:2024asf}.} Their boundary stress tensor is subleading relative to that of the other modes. These soft modes also diagnose the type of horizon the boundary encloses. To leading order the dynamics is a universal Rindler geometry, insensitive to the details of the horizon, and the disparity between horizons appears only at sub-leading order. For the cosmic patch of de Sitter and for a Schwarzschild black hole of horizon radius $r_{\text{h}}$, these read
\begin{eqnarray}\label{eq:softdisparity}
\frac{\omega^{(\text{dS})}\,\mathfrak{r}}{\sqrt{f(\mathfrak{r})}} &= \frac{2\pi i}{\tilde{\beta}}\left(1 - \frac{l(l+1)-2}{2K^2\ell^2} + \mathcal{O}((K\ell)^{-4})\right)\,, \nonumber \\
\frac{\omega^{(\text{BH})}\,\mathfrak{r}}{\sqrt{f(\mathfrak{r})}} &= \frac{2\pi i}{\tilde{\beta}}\left(1 - \frac{l(l+1)+1}{2K^2 r_{\text{h}}^2} + \mathcal{O}((K r_{\text{h}})^{-4})\right)\,,
\end{eqnarray}
where $\tilde{\beta}$ is the inverse conformal temperature at the boundary. The leading large-$l$ terms agree, whereas the order-one terms differ. It would be interesting to connect these subtle linearised differences to the sharper non-linear signatures that distinguish the two types of horizons, such as the Gao--Wald effect \cite{Gao:2000ga}.

\paragraph{Non-linear conformal fluids.} Encasing the de Sitter horizon in the stretched horizon limit, the shear and sound modes yield the transport coefficients of the dual fluid,
\begin{equation}\label{eq:transport}
\zeta = 0\,, \qquad \frac{\eta}{s} = \frac{1}{4\pi}\,,
\end{equation}
with $s \equiv \frac{1}{4G_N}$ is the entropy density. The shear viscosity-to-entropy density ratio obeys the celebrated universal formulae for strongly coupled fluid in the context of AdS/CFT \cite{Kovtun:2004de}. The vanishing bulk viscosity contrasts sharply with the membrane paradigm \cite{Price:1986yy} and points to conformal behaviour, consistent with the tracelessness of the conformal Brown-York stress tensor. For solutions slowly varying in units of $K$, this sector completes non-linearly into a compressible Navier-Stokes equation on the boundary, which reproduces the known results in the non-relativistic limit \cite{Bredberg:2011xw,Anninos:2011zn}. It would be interesting to couple the fluid to the remaining modes, whose frequencies fall outside the slowly-varying regime, or connect the results here to one in the fluid/gravity correspondence literature \cite{Bhattacharyya:2007vjd,Baier:2007ix}, see also \cite{Rangamani:2009xk}.

\paragraph{Emergence of boundary locality.} Several structures in the stretched horizon limit appear to enjoy a form of boundary locality. The fluid dynamics organise into a $(2+1)$-dimensional local structure, while the soft modes, with frequency independent of the angular momentum, hint at ultralocal behaviour at the boundary. Sharpening this observation would be worthwhile.

\section{AdS$_3$ with finite conformal boundary}
\label{sec:AdS3Liouville}

A natural first step towards a holographic description of gravity with a finite boundary obeying CBCs is to place that boundary parametrically close to the asymptotic boundary of AdS$_{d+1}$, reached as $K\ell \to d$, as shown in chapter \ref{chap: AdS}. Under DBCs the analogous setup is described holographically by a $T\bar{T}$-deformed CFT \cite{Smirnov:2016lqw, Cavaglia:2016oda,McGough:2016lol,Hartman:2018tkw}. Under CBCs, evidence for the holographic dual already appears in the (semi-)classical analysis of AdS$_{d+1}$ with CBCs in this limit. In chapter \ref{sec: bulk modes}, around global AdS$_4$ one finds an area worth of boundary Weyl modes with a three-dimensional gapless dispersion relation, while the spherical reduction of AdS$_3$ yields a boundary mode obeying the one-dimensional classical timelike Liouville equation \cite{Galante:2025tnt}.

\paragraph{Deformed timelike Liouville theory.} These classical analyses have been made precise in a proposed holographic dual of AdS$_3$ with CBCs \cite{Allameh:2025gsa}, which we now review. AdS$_3$ with CBCs of fixed trace of the extrinsic curvature $K$ is dual to the matter CFT$_2$ of standard AdS$_3$/CFT$_2$, with Brown--Henneaux central charge $c_m = 3\ell/2G_N$, coupled to timelike Liouville theory with $c_L = -c_m$, and deformed by an exactly marginal $\tilde T\bar{\tilde T}e^{-2\Phi}$-operator. Defining the theory at finite $c_m$ is generically hard. It simplifies when $c_m$ is large and positive, the classical limit, in which the timelike Liouville sector is governed by the action
\begin{equation}\label{eq:tLaction}
S_{tL} = \frac{1}{4\pi \mathfrak{b}^2}\int \text{d}^2x \sqrt{\tilde{g}}\left[-(\tilde{\nabla}\Phi)^2 - \tilde{R}\,\Phi + 4\pi \mathfrak{b}^2 \mu\, e^{2\Phi}\right]\,, \qquad \mathfrak{b}^2 \approx \frac{6}{c_m}\,.
\end{equation}
The seed theory couples timelike Liouville to the matter CFT$_2$, $S = S_m + S_{tL}$, whilst the deformed theory is defined by the flow
\begin{equation}\label{eq:deformflow}
    \frac{\partial S}{\partial \lambda} = \int \text{d}^2x \sqrt{\tilde g}\, \tilde T \bar{\tilde T}\, e^{-2\Phi} + \frac{c_m}{48 \pi \lambda} \int \text{d}^2x \sqrt{\tilde g}\, \tilde R \,,
\end{equation}
where $\tilde T \bar{\tilde T}\equiv \tfrac{1}{8}\big(\tilde T^{\mu\nu}\tilde T_{\mu\nu}-(\tilde T^\mu{}_\mu)^2\big)$ and $\tilde T_{\mu\nu}$ is the stress tensor of the total system with the cosmological constant contribution subtracted. The holographic dictionary is furnished by the parameter identifications
\begin{equation}\label{eq:dictionary}
    \mu = \frac{K\ell-2}{16 \pi G_N \ell} \, , \qquad \lambda = 16 \pi G_N\ell \, , \qquad c_L=-c_m = -\frac{3\ell}{2G_N} \, .
\end{equation}
Under this prescription the sphere and torus partition functions, evaluated on the homogeneous saddles, match between the bulk and boundary theories at finite deformation parameter $\lambda$ \cite{Allameh:2025gsa}.

\paragraph{Inhomogeneous solutions.} The torus matching assumes the dominant saddles are homogeneous. In \cite{Hamdan:2026olq} we put the proposal to a further test with an inhomogeneous torus boundary, uncovering Euclidean AdS$_3$ with \textit{undulating} boundaries whose radius oscillates along either cycle of the torus. Evaluating the semiclassical gravitational path integral on these saddles, to first order in the $K\ell \to 2$ limit, we find agreement for solutions that oscillate only mildly about the homogeneous configuration and disagreement for those with large oscillations. Understanding the origin of this mismatch would sharpen the proposed duality.

\paragraph{Beyond the semiclassical limit.} The check of the duality is so far semiclassical, namely $c_m \to \infty$. At finite $c_m$ the marginality of the dressed operator is generically spoiled beyond tree level. It would be interesting to understand the duality at finite $c_m$, at least at the one-loop level corresponding to the first $\tfrac{1}{c_m}$ correction.

\section{All-$\Lambda$ quasi-local holography}
\label{sec:allLambda}

The thermodynamics of pure Einstein gravity with an arbitrary cosmological
constant $\Lambda$, subject to CBCs, reveals an interesting thermodynamic structure in the
saddle-point approximation of the semi-classical limit. The relevant thermal
solutions are finite patches of, for instance, the Schwarzschild black hole
for $\Lambda=0$, Schwarzschild--anti-de~Sitter for $\Lambda<0$, and
Schwarzschild--de~Sitter for $\Lambda>0$.

\paragraph{Universal high-temperature entropy.}
In all of the above cases the entropy is given by the area of the horizon
enclosed by the geometry. For a $(d+1)$-dimensional spacetime, its
high-temperature expansion takes the form
\begin{equation}
\label{eq:universalS}
\mathcal{S}
= \frac{N_{\mathrm{d.o.f.}}(K)}{\tilde{\beta}^{\,d-1}} + \dots\,,
\end{equation}
where $\tilde{\beta}$ is the inverse conformal temperature and the dots denote
subleading terms. The leading coefficient $N_{\mathrm{d.o.f.}}$ is a function of $K$, given by\footnote{The flat case $\Lambda=0$
is recovered as the $|\Lambda|/K^{2}\to0$ limit, for either sign of $\Lambda$, in which case
$N_{\mathrm{d.o.f.}}\to \Omega_{d-1}(2\pi)^{d-1}/(4\,G_{d+1}\,K^{d-1})$.}
\begin{equation}
\label{eq:Ndof}
N_{\mathrm{d.o.f.}}(K)
= \frac{\Omega_{d-1}}{4G_N}
\left[
\frac{2\pi(d-1)}{d\Lambda}
\left(
\sqrt{K^{2}+\frac{2d}{d-1}\,\Lambda}-K
\right)
\right]^{d-1}\,,
\end{equation}
where $\Omega_{d-1}$ the volume of the unit $(d-1)$-sphere.
For any value of $\Lambda$, this function is monotonic in $K$, decreasing as
$K$ increases. The same leading coefficient appears also in Einstein-Maxwell-$\Lambda$
theory \cite{Banihashemi:2025qqi}, while it is modified, together with
the boundary condition itself, in Einstein-Gauss-Bonnet gravity theory \cite{Galante:2025emz}. See \cite{Banihashemi:2024yye,Banihashemi:2025qqi} also for the study of the first sub-leading coefficients.

The high-temperature behaviour of \eqref{eq:universalS} matches the
expectation from the thermal effective action of a local CFT$_{d}$ \cite{Benjamin:2023qsc,Allameh:2024qqp}, in
contrast to the Dirichlet problem. In
particular, for $\Lambda<0$ equation~\eqref{eq:universalS} reduces to the
standard result for an AdS black hole, conjectured to be dual to a CFT at
finite temperature in the context of AdS/CFT. Here the $\tilde{\beta}^{-(d-1)}$
scaling manifestly reflects the holographic nature of the black hole, namely the
entropy is extensive in the boundary spatial dimensions rather than in the
bulk ones.

Equation~\eqref{eq:universalS} suggests that a similar non-local reorganization of
degrees of freedom into one less dimension may be a universal feature of spacetime, for any value of the cosmological constant. We
hope this observation is a first step toward a unified, all-$\Lambda$
formulation of the holography of a finite region of spacetime.

\begin{appendices}

\chapter{Geometry of the boundary}\label{app:hyper basic} 
In this appendix, we clarify basic definitions and collect useful formulae for geometrical structure of a smooth codimension-one hypersurface $\Sigma$ in a Riemannian manifold $\mathcal{M}$ equipped with a metric $g_{\mu\nu}$. Here, greek indices $\mu,\nu,=,1,\dots,d$ refer to the indices on $\mathcal{M}$ whose coordinates are $x^\mu$, while $i,j=1,\dots,d-1$ refer to those on $\Sigma$ whose coordinates are $y^i$. The covariant derivative compatible with $g_{\mu\nu}$ is denoted as $\nabla_\mu$ with $\Gamma^\rho_{\mu\nu}$ referred to its Christoffel symbols.

\textbf{Embedding.} Locally, a hypersurface can be defined as a regular level set of a smooth function $F(x^\mu)$,
\begin{equation}
    \Sigma = \left\{x^\mu \in \mathcal{M} \, |\, F(x^\mu)=0\right\}\, , \qquad d F|_\Sigma \neq 0 \, ,
\end{equation}
which naturally leads to an embedding map, $X:\Sigma \to \mathcal{M}$. Written the embedding map as $x^\mu = X^\mu(y^i)$, the tangent vectos on $\Sigma$ is given by
\begin{equation}
    e^\mu_i \equiv \frac{\partial X^\mu}{\partial y^i} \, ,
\end{equation}
which satisfies $e_i^\mu \partial_\mu F =0$. Define $N\equiv \sqrt{g^{\rho\sigma}\partial_\rho F \partial_\sigma F}$, the unit normal one-form $n_\mu$ and its dual vector $n^\mu$ are given by
\begin{equation}\label{eqn: nmu appen}
    n_\mu \equiv \left.\frac{\partial_\mu F}{N}\right. \, , \qquad n^\mu \equiv g^{\mu\nu}n_\nu \, ,
\end{equation}
which satisfies $n^\mu n_\mu =1 $ and $n_\mu e^\mu_i=0$. 

\textbf{Induced metric.} Given $n^\mu$, one can define an orthogonal project onto the tangent directions on $\Sigma$,
\begin{equation}
    P^\mu{}_\nu \equiv \delta^\mu_\nu - n^\mu n_\nu \, ,
\end{equation}
which satisfies $P^\mu{}_\nu P^\nu{}_\rho = P^\mu{}_\rho$ and $P^\mu{}_\nu n^\nu = 0$. For any tensor $T_{\mu_1 \mu_2 \mu_3 \dots}$ on $\mathcal{M}$, its pullback is denoted as $T_{i_1i_2\dots}\equiv (X^*T)_{i_1i_2\dots}= e_{i_1}^{\mu_1}e_{i_2}^{\mu_2}\dots T_{\mu_1 \mu_2 \mu_3 \dots}$. Consequently, one defines the induced metric of $\Sigma$ as the pullback of the bulk metric,
\begin{equation}
    h_{ij} \equiv (X^* g)_{ij}= e^\mu_i e^\nu_j g_{\mu\nu} \,  , \qquad ds^2|_\Gamma = h_{ij}dy^i dy^j\, .
\end{equation}
The relation between the projector $P_{\mu\nu}$ and the induced metric $h_{ij}$ is 
\begin{equation}
    P^{\mu\nu} = e_i^\mu e_j^\nu h^{ij} \, ,
\end{equation}
where $\mu,\nu,\dots$ are raised using $g^{\mu\nu}$ while $i,j,\dots$ are raised using $h^{ij}$.

\textbf{Extrinsic curvature.} The extrinsic curvature tensor, or sometimes referred to as second fundamental form, is defined as
\begin{equation}
    K_{\mu\nu}\equiv P^\rho_\mu P^\sigma_\nu \nabla_\rho n_\sigma \, , \qquad K_{ij} \equiv (X^* K)_{ij} = e_i^\mu e_j^\nu \nabla_\mu n_\nu \, .
\end{equation}
There exists several useful formulae for computing $K_{ij}$. First is the level set representation, which is obtained by plugging in \eqref{eqn: nmu appen} to the definition,
\begin{equation}
    K_{ij} = \left.\frac{1}{N}e_i^\mu e^\nu_j\left(\partial_\mu \partial_\nu F - \Gamma^\rho_{\mu\nu}\partial_\rho F\right)\right. \, .
\end{equation}
Another useful representation of is through
the embedding $X^\mu$. By differentiating the identity $n_\mu e^\mu_i=0$ along the tangential direction, one obtains
\begin{equation}
    K_{ij} = -n_\mu \left(\partial_i \partial_j X^\mu + \Gamma^\mu_{\rho\sigma}\partial_i X^\rho \partial_j X^\sigma\right) \, .
\end{equation}
A final representation we consider is $K_{ij}$ as a normal Lie-derivative of the induced metric along the normal direction. Specifically, by extending $n_\mu$ in the neighbourhood of $\Sigma$, one can write
\begin{equation}
    K_{ij}=\frac{1}{2} \left(X^* \mathcal{L}_n g\right)_{ij} \, ,
\end{equation}
where $\mathcal{L}_n$ denotes the Lie-derivative along the vector $n^\mu$. 

\chapter{Useful formulae for Gaussian normal coordinates}\label{app:normal}

In this appendix, we provide some results on geometric quantities using Gaussian normal coordinates
\begin{equation}
	ds^2 \= dx_\perp^2 + \bar{g}_{mn} dx^m dx^n \, ,
\end{equation}
 that  are useful to derive some of the expressions in section \ref{sec:framework}. For instance, the extrinsic curvature in Gaussian normal coordinates is just given by,
\begin{equation}
	K_{m n} \= \frac{1}{2}\partial_\perp \bar{g}_{mn} \, .
\end{equation}
Christoffel symbols with indices in the perpendicular direction are simply given by, 
\begin{equation}
	\Gamma^\perp_{mn} \= - K_{mn} \, , \qquad \qquad \Gamma^m_{\perp n} \= K^m{}_n \, , \qquad \qquad \Gamma^\perp_{\perp \mu} \= \Gamma^\mu_{\perp \perp} \= \Gamma^\perp_{\perp \perp} \= 0 \,.
\end{equation}
This is useful to decompose spacetime quantities into tangential and perpendicular components. For instance, let $V^\mu$ be a vector on the bulk manifold. A divergence of $V^\mu$ with respect to the bulk covariant derivative can be rewritten as
\begin{eqnarray}
	\nabla_\mu V^\mu = \bar{g}^{mn} \nabla_m V_n + \nabla_\perp V_\perp 
	= \bar{g}^{mn} \left(\mathcal{D}_m V_n - \Gamma^\perp_{mn} V_\perp \right) + \partial_\perp V_\perp 
	= \mathcal{D}_m V^m + K V_\perp + \partial_\perp V_\perp \, .
\end{eqnarray}
We can also derive the formula for $\delta K$ in \eqref{eqn: bdry cond 2} in the main text. The variation of the extrinsic curvature is given by \cite{compendium},
\begin{equation}\label{eqn: appendix Kmn}
	\delta K_{mn} \= -\frac{1}{2} \mathcal{D}_m h_{\perp n} -\frac{1}{2} \mathcal{D}_n h_{\perp m} - \frac{1}{2} K_{mn} h_{\perp \perp}  + \frac{1}{2}\partial_\perp h_{mn} \, . 
\end{equation}
Using this, we find that
\begin{equation}
\begin{split} 
	\delta K & \= \delta K_{mn} \bar{g}^{mn} - K_{mn} h^{mn} = -\mathcal{D}^m h_{\perp m} - \frac{1}{2} K h_{\perp \perp} + \frac{\bar{g}^{mn}}{2}\partial_\perp h_{mn} - K^{mn}h_{mn}   \\
	& \= -\mathcal{D}^m h_{\perp m} - \frac{1}{2} K h_{\perp \perp} + \frac{1}{2}\partial_\perp h^m{}_m \, ,
	\end{split}
\end{equation}
where in the last line we used that $K^{mn} = - \frac{1}{2}\partial_\perp \bar{g}^{mn}$.
 
It is also useful to see how various quantities transform under small bulk diffeomorphisms in Gaussian normal coordinates. Specifically, under $x^\mu \rightarrow x^\mu + \epsilon \, \xi^\mu(x)$,
\begin{eqnarray}
	\begin{cases}
		h_{mn} &\rightarrow \, h_{mn} + \mathcal{D}_m \xi_n + \mathcal{D}_n \xi_m + 2K_{mn} \xi_\perp \, , \\
		h_{\perp m} &\rightarrow \, h_{\perp m} + \partial_\perp \xi_m - 2 K_m{}^n\xi_n + \mathcal{D}_m \xi_\perp \, , \\
		h_{\perp \perp} &\rightarrow \, h_{\perp \perp} + 2\partial_\perp \xi_\perp \, .
	\end{cases}
\end{eqnarray}
Using \eqref{eqn: appendix Kmn}, one finds that 
\begin{equation}
	K_{mn} \,\rightarrow\, K_{mn} + \xi^p\mathcal{D}_p K_{mn} + K_{mp}\mathcal{D}_n \xi^p + K_{np} \mathcal{D}_m \xi^p - \mathcal{D}_{(m} \mathcal{D}_{n)} \xi_\perp + \left(\partial_\perp K_{mn}\right) \xi_\perp \, .
\end{equation}

\chapter{Derivation of the gauge-fixed solution for the flat boundary} \label{app:derivation}

In this appendix, we provide a detailed derivation of the solution \eqref{eqn: gauge-fixed sol kx!=0}. Recall that the gauge constraints $T_\mu(h_{\mu\nu})=0$ after imposing the boundary conditions \eqref{bdy_cond222} lead to the set of algebraic equations \eqref{eqn: gauge cond kx!=0}, 
\begin{equation}
\begin{dcases}
               \alpha_{mn} \= 0 \,, & \left( k^m k^n + k_x^2 \eta^{mn} \right)\beta_{mn} \= 0 \,, \\
               2ik^n \beta_{x n} - k_x \alpha_{x x} \= 0  \,, &  k_x  \beta^m{}_m -2ik^n \alpha_{x n} \= 0 \,, \\
               u^{(1)m} k_x \beta_{x m}  \= 0 \,,          &   i u^{(1)m} k^n \beta_{mn} - u^{(1)m} k_x \alpha_{x m}  \= 0 \,, \\
               u^{(2)m} k_x \beta_{x m}  \= 0 \,,          &   i u^{(2)m} k^n \beta_{mn} - u^{(2)m} k_x \alpha_{x m}  \= 0 \,.
        \end{dcases}
\end{equation}
By choosing the gauge $\alpha_{x\mu}=0$, these equations reduce to
\begin{equation}\label{appendix: dum1}
\begin{dcases}
               \alpha_{\mu \nu} \= 0 \,, &  \eta^{mn}\beta_{mn}   \= 0 \,, \\
               2ik^n \beta_{x n} \= 0  \,, & k^m k^n\beta_{mn}  \= 0 \,, \\
               u^{(1)m} k_x \beta_{x m}  \= 0 \,,          &   i u^{(1)m} k^n \beta_{mn}   \= 0 \,, \\
               u^{(2)m} k_x \beta_{x m}  \= 0 \,,          &   i u^{(2)m} k^n \beta_{mn}   \= 0 \,.
        \end{dcases}
\end{equation}
Since $k^m$, $u^{(1)m}$, and $u^{(2)m}$ span the tangent vector space on the timelike boundary, we can rewrite the boundary metric as
\begin{equation}\label{appendix: decompose eta}
	\eta^{mn} \= -\frac{k^m k^n}{k_x^2} + u^{(1)m}u^{(1)n} + u^{(2)m}u^{(2)n} \, .
\end{equation}
Hence, the condition $\eta^{mn}\beta_{mn}=0$ can be rewritten as
\begin{equation}
	-\frac{1}{k_x^2} k^m k^n\beta_{mn} + u^{(1)m}u^{(1)n}\beta_{mn} + u^{(2)m}u^{(2)n}\beta_{mn} \= 0 \, .
\end{equation}
Combining with the condition $k^mk^n \beta_{mn}=0$, we find
\begin{equation}\label{appendix: dum2}
	u^{(1)m}u^{(1)n}\beta_{mn} \= -  u^{(2)m}u^{(2)n}\beta_{mn} \, .
\end{equation}
 
Now we use \eqref{appendix: decompose eta} to express $\beta_{xm}$ as
\begin{equation}
	\beta_{xm} \= \delta^n_m \beta_{xn} \= - k_m \left(\frac{k^n \beta_{xn}}{k_x^2}\right) + u^{(1)}_m \left(u^{(1)n}\beta_{xn}\right) + u^{(2)}_m \left(u^{(2)n}\beta_{xn}\right) \, .
\end{equation}
The last three equations on the left column of \eqref{appendix: dum1} then imply that $\beta_{xm} = 0$. Similarly, we can express $\beta_{mn}$ as
\begin{equation}
\begin{split}
	\beta_{mn} \=  & \delta_m^p \delta_n^q \beta_{pq} 
	 \=  k_mk_n\left( \frac{k^pk^q\beta_{pq}}{k_x^4}\right) - 2k_{(m} u^{(1)}_{n)} \left(\frac{k^pu^{(1)q}\beta_{pq}}{k_x^2}\right) - 2k_{(m} u^{(2)}_{n)} \left(\frac{k^pu^{(2)q}\beta_{pq}}{k_x^2}\right)  \\
	& + u^{(1)}_{m}u^{(1)}_{n} \left(u^{(1)p}u^{(1)q}\beta_{pq}\right) + u^{(2)}_{m}u^{(2)}_{n} \left(u^{(2)p}u^{(2)q}\beta_{pq}\right) + 2u^{(1)}_{(m}u^{(2)}_{n)} \left(u^{(1)p}u^{(2)q}\beta_{pq}\right) \, .
	\end{split}
\end{equation}
The first three terms are zero according to the last three equations on the right column of \eqref{appendix: dum1}. The fourth and fifth terms are related through \eqref{appendix: dum2}. Finally, defining $\beta^{(+)} \equiv u^{(1)m}u^{(1)n}\beta_{mn}$ and $\beta^{(\times)} \equiv u^{(1)m}u^{(2)n}\beta_{mn}$, the remaining terms are 
\begin{equation}
	\beta_{mn} \= \left(u^{(1)}_{m}u^{(1)}_{n} - u^{(2)}_{m}u^{(2)}_{n} \right) \beta^{(+)} + 2u^{(1)}_{(m}u^{(2)}_{n)} \beta^{(\times)} \, .
\end{equation}
Plugging this into \eqref{eqn: flat non-compact sol 1}, we arrive at \eqref{eqn: gauge-fixed sol kx!=0} in the main text,
\begin{eqnarray}
	h_{\mu \nu}dx^\mu dx^\nu \= \left[\beta^{(+)} \left(u^{(1)}_n u^{(1)}_m - u^{(2)}_n u^{(2)}_m\right) + 2 \beta^{(\times)} u^{(1)}_n u^{(2)}_m \right]  \sin(k_x x) e^{i k_n x^n} dx^n dx^m \,.
\end{eqnarray}

\chapter{Kodama-Ishibashi formalism}\label{sec: Kodama-Ishibashi}

In this appendix, we review the Kodama-Ishibashi formalism \cite{Kodama:2000fa,Kodama:2003jz}. This formalism will allow us to deal with gravitational polarisations directly through gauge invariant quantities.
 
Let us consider a four-dimensional spacetime of the following type:
\begin{equation}
	ds^2 \= g_{ab}dy^a dy^b+ r(y)^2 d\sigma^2 \, , \label{eqn: metric decomposition}
\end{equation}
where $r(y)$ is an arbitrary function of $y^a$, $g_{ab}dy^ady^b$ is a two-dimensional Lorentzian manfiold, called an orbit space, and $d\sigma^2 = \sigma_{ij} dx^i dx^j$ is a metric of a two-dimensional maximally symmetric space. 
 
Any metric perturbation can be uniquely decomposed into vector ($V$) and scalar ($S$) perturbations,
\begin{equation}
    h_{\mu \nu} \= h_{\mu \nu}^{(V)} + h_{\mu \nu}^{(S)} \, ,
\end{equation}
where
\begin{equation}
\begin{cases}
    h_{a b}^{(V)}  \= 0 \, , \\
    h_{a i}^{(V)}  \= rf^V_a \,\mathbb{V}_i \, , \\
    h_{ij}^{(V)}  \= 2r^2H^V \,\mathbb{V}_{ij} \, ,  
    \end{cases}
    \qquad \qquad 
\begin{cases}
    h_{a b}^{(S)} \= f_{ab}\,\mathbb{S} \, , \\ 
    h_{a i}^{(S)} \= rf^S_a \,\mathbb{S}_i  \, , \\
    h_{ij}^{(S)} \= 2r^2 H^S \,\mathbb{S}_{ij} + 2r^2\gamma \, \sigma_{ij} \mathbb{S}\, .
 \end{cases}
\end{equation}
The coefficients of the perturbations, $f_{ab}$, $f_{a}^{V/S}$, $H^{V/S}$, and $\gamma$, are functions of the orbit coordinate $y^a$. The vector and scalar harmonic tensors $\mathbb{V}_i$, $\mathbb{V}_{ij}$, $\mathbb{S}$, $\mathbb{S}_i$, and $\mathbb{S}_{ij}$ are defined as followed. Let $\tilde{\mathcal{D}}_i$, $\tilde{\mathcal{D}}^2$, and $\mathcal{K}$ be covariant derivative, Laplacian, and unit curvature with respect to the metric $\sigma_{ij}$. The scalar harmonic function $\mathbb{S}$ is an eigenfunction of the scalar Laplacian with eigenvalue $-k_S^2$,
\begin{equation}
	\left(\Tilde{\mathcal{D}}^2+k_S^2\right) \mathbb{S} \= 0 \, .
\end{equation}
Given $\mathbb{S}$, one can construct a gradient $\mathbb{S}_i$ and an associated symmetric traceless rank-2 tensor $\mathbb{S}_{ij}$ as 
\begin{equation}\label{eqn: Kodama S_i and S_ij def}
	\mathbb{S}_i \,\equiv\, -\frac{1}{k_S}\Tilde{\mathcal{D}}_i\mathbb{S}\,, \qquad\qquad  \mathbb{S}_{ij} \, \equiv \, \frac{1}{k_S^2} \Tilde{\mathcal{D}}_i \Tilde{\mathcal{D}}_j \mathbb{S} + \frac{1}{2}\sigma_{ij}\mathbb{S}  \, ,
\end{equation}
which satisfy the following properties
\begin{equation}
\begin{cases}
	\left(\Tilde{\mathcal{D}}^2+k_S^2-\mathcal{K}\right) \mathbb{S}_i \= 0 \, , \\
	\Tilde{\mathcal{D}}^i \mathbb{S}_i \= k_S \mathbb{S} \,, 
	\end{cases}
	\qquad \qquad
\begin{cases}
	\left(\Tilde{\mathcal{D}}^2+k_S^2-4\mathcal{K}\right) \mathbb{S}_{ij} \= 0 \, , \\
	\Tilde{\mathcal{D}}^j\mathbb{S}_{ij} \= \frac{k_S^2-2\mathcal{K}}{2k_S}\mathbb{S}_i\,, \\
	\mathbb{S}^i{}_i \= 0 \, .
	\end{cases}
\end{equation}
The vector harmonic tensor $\mathbb{V}_i$ is a divergenceless eigenvector of the vector Laplacian with eigenvalue $-k_V^2$,
\begin{equation}
	\Tilde{\mathcal{D}}^i \mathbb{V}_i \= 0 \,, \qquad \left(\Tilde{\mathcal{D}}^2+k_V^2\right) \mathbb{V}_i \= 0 \, .
\end{equation}
Given $\mathbb{V}_i$, one can construct an associated symmetric traceless rank-2 tensor $\mathbb{V}_{ij}$ as
\begin{equation}\label{eqn: Kodama V_ij def}
	\mathbb{V}_{ij} \, \equiv \, -\frac{1}{2k_V} \left(\Tilde{\mathcal{D}}_i\mathbb{V}_j + \Tilde{\mathcal{D}}_j\mathbb{V}_i\right) \,,
\end{equation}
which satisfies the following properties
\begin{equation}
\begin{cases}
	\left(\Tilde{\mathcal{D}}^2+k_V^2-3 \mathcal{K}\right) \mathbb{V}_{ij} \= 0\, , \\
	\Tilde{\mathcal{D}}^j\mathbb{V}_{ij} \= \frac{k_V^2-\mathcal{K}}{2k_V}\mathbb{V}_i\,, \\ 
	\mathbb{V}^i{}_i \= 0 \, .	
	\end{cases}
\end{equation}
Explicit expressions for $\mathbb{S}$ and $\mathbb{V}_i$ in the case of a two-dimensional sphere are provided below. Note also that the $k_S = 0$, $2 \mathcal{K}$, and $k_V = \mathcal{K}$ modes are special and must be analysed separately \cite{Kodama:2000fa,Kodama:2003jz}. 
 
A diffeomorphism can be decomposed in a similar way, i.e., $ \xi_\mu = \xi_\mu^{(V)} + \xi_\mu^{(S)}$, with
\begin{equation}
\begin{cases}
    \xi_a^{(V)} \= 0 \, , \\ 
     \xi_i^{(V)} \= r\mathcal{L}^V \, \mathbb{V}_i \,, 
    \end{cases}
    \qquad \qquad
    \begin{cases}
    \xi_a^{(S)} \= \mathcal{T}_a \, \mathbb{S} \, , \\ 
    \xi_i^{(S)} \= r \mathcal{L}^S \, \mathbb{S}_i \, , 
    \end{cases}
    \label{eqn: Kodama diff decomposition}
\end{equation}
where $\mathcal{T}_a$ and $\mathcal{L}^{V/S}$ are functions of the orbit coordinates $y^a$. In terms of these, gauge transformation of the metric peturbation is given by
\begin{equation}
	\begin{cases}
		\delta f_a^V \= r \mathcal{D}_a \left(\frac{\mathcal{L}^V}{r}\right) \, , \\
		\delta H^V \= -\frac{k_V}{r}\mathcal{L}^V\, ,
	\end{cases}
	\qquad \qquad
	\begin{cases}
		\delta f_{ab} \= \mathcal{D}_a \mathcal{T}_b + \mathcal{D}_b \mathcal{T}_a \, , \\
		\delta f_a^S \= r \mathcal{D}_a \left(\frac{\mathcal{L}^S}{r}\right) - \frac{k_S}{r}\mathcal{T}_a \, , \\ 
		\delta H^S \= -\frac{k_S}{r}\mathcal{L}^S \, , \\ 
		\delta \gamma \=  \frac{k_S}{2 r}\mathcal{L}^S + \frac{\mathcal{D}_a r}{r}\mathcal{T}^a \,,
	\end{cases}
\end{equation}
where $\mathcal{D}_a$ is the covariant derivative with respect to $g_{ab}$.
By looking at the gauge transformation, gauge invariant quantities can be constructed as follows,
\begin{equation}\label{eqn: KI gauge invariant 1}
	\begin{cases}
		F_a & \equiv\, f_a^V + \frac{r}{k_V} \mathcal{D}_a H^V \, , \\
		F &\equiv\, \gamma + \frac{H^S}{2} + \frac{\mathcal{D}^a r}{r} X_a \, , \\
		F_{ab} & \equiv\, f_{ab} + \mathcal{D}_a X_b + \mathcal{D}_b X_a \, ,
	\end{cases}
\end{equation}
where $X_a \,\equiv\, \frac{r}{k_S}\left(f^S_a + \frac{r}{k_S}\mathcal{D}_a H^S\right)$.
 
By virtue of the Einstein field equation, $F_a$, $F$, and $F_{ab}$  can be expressed in terms of the so-called master fields $\Phi^V$ and $\Phi^S$ as follows:
\begin{equation}\label{eqn: KI gauge invariant 2}
	\begin{cases}
		F^a &=\, \frac{1}{r}\epsilon^{ab}\mathcal{D}_b \left(r \Phi^V\right) \, , \\
		F &=\, \frac{1}{8 r^2 }\Big[2\left(k_S^2 - 2\right)+4 r \mathcal{D}^ar\mathcal{D}_a\Big]\left(r \Phi^S\right)\, , \\
    		F_{ab} &=\, \Big[\mathcal{D}_a\mathcal{D}_b - \frac{1}{2} g_{ab}\Box\Big]\left(r \Phi^S\right) \, .
	\end{cases}
\end{equation}
The master fields $\Phi^{V/S}$ represent two polarisations of the gravitational fluctuation whose dynamics are governed by master equations
\begin{equation}
    \left( g^{ab}\mathcal{D}_a \mathcal{D}_b- V_{V/S}(r)\right)\Phi^{V/S} \= 0 \, ,
\end{equation}
where the effective potentials are given by
\begin{equation}
	V_V(r) \= \frac{k_V^2 + \mathcal{K}}{r^2} \, , \qquad \qquad V_S(r) \= \frac{k_S^2}{r^2} \, ,
\end{equation}
for vector and scalar perturbations, respectively. In this analysis, we have excluded the presence of a black hole, for simplicity. For formulae with a black hole or a non-zero $\Lambda$, see \cite{Kodama:2000fa,Kodama:2003jz}.

\subsection*{Two-dimensional sphere}\label{sec: two-sphere}

For a two-dimensional sphere, the metric $\sigma_{ij}$ and the unit curvature $\mathcal{K}$ are 
\begin{equation}\label{eqn: def two-sphere metric}
	d\sigma^2 \= d\theta^2 + \sin^2\theta d\phi^2 \, , \qquad \qquad \mathcal{K} \= 1 \, .
\end{equation}
The scalar/vector harmonics $\mathbb{S}$ and $\mathbb{V}_i$ can be constructed as follows. For a given angular momentum $l \in \mathbb{N}_0$, the scalar harmonic $\mathbb{S}$ is given by a spherical harmonic function,
\begin{equation}
	\mathbb{S} \= Y_{lm}(\theta,\phi) \, , \qquad k_S^2 \= l(l+1) \, ,
\end{equation}
where $m$ is an integer taking value between $-l \leq m \leq l$. The vector harmonic $\mathbb{V}_i$ is given by
\begin{equation}
	\mathbb{V}_i \= \left(\star d \mathbb{S}\right)_i \, , \qquad k_V^2 \= l(l+1)-1,
\end{equation}
where $d$ and $\star$ are exterior derivative and hodge dual operator associated to the metric \eqref{eqn: def two-sphere metric}. Their associated vector and symmetric traceless tensor harmonics, $\mathbb{S}_i$, $\mathbb{S}_{ij}$, and $\mathbb{V}_{ij}$ follow \eqref{eqn: Kodama S_i and S_ij def} and \eqref{eqn: Kodama V_ij def}.
 
When $l\=0$, only $\mathbb{S}$ is defined, and it becomes a constant function. When $l\=1$, $\mathbb{S}_{ij}$ and $\mathbb{V}_{ij}$ are not defined. Also, $\mathbb{V}_i$ becomes a Killing vector of the two sphere.

\chapter{dS$_3$ Dirichlet thermodynamics} \label{sec:3d_Dirichlet}

In this appendix, we review the thermodynamics of the three-dimensional dS using Dirichlet boundary condition \cite{Coleman:2021nor}. Here, we use the standard coordinate for Euclidean Schwarzschild-de-Sitter space
\begin{equation}
	ds^2 \= \frac{f(r)}{f(\frakr_\text{tube})}d\tau^2 + \frac{dr^2}{f(r)} + r^2 d\phi^2 \, , \qquad f(r) \= \frac{\frakr_{\text{c}}^2-r^2}{\ell^2} \, ,
\end{equation}
where $\tau \,\sim\, \tau + \beta_D$ and $\phi \,\sim\, \phi + 2\pi$. The Euclidean time $\tau$ is measured with respect to the clock defined on the boundary $r\=\frakr_\text{tube}$. Similar to the analysis in the main text, $\frakr_c$ represents the physical size of the cosmological horizon. There is no black hole in this geometry. When $\frakr_c \neq 1$, there is a conical defect at the origin $r\=0$.

The boundary $r\=\frakr_\text{tube}$ has topology of $S^1 \times S^1$ and obeys Dirichlet boundary data, namely the induced metric at $r\=\frakr_\text{tube}$ is fixed to be
\begin{equation}
	\left.ds^2\right|_{r=\frakr_\text{tube}} \= d\tau^2 + \frakr_\text{tube}^2 d\phi^2 \, .
\end{equation} 
In this case, the boundary data consists of the Euclidean time period $\beta_D$ and the physical size of the boundary $\frakr_\text{tube}$. The Euclidean action $I_E$ is given by \eqref{euclidean_action} with $\alpha_{\text{b.c.}} = (D-1)$ and $D=3$.


Since the Dirichlet problem fixes $\beta_D$ and $\frakr_\text{tube}$, the partition function is a function of these, $\mathcal{Z}\=\mathcal{Z}(\beta_D,\frakr_\text{tube})$. The energy $E$ and entropy $\mathcal{S}$ follow from \eqref{eqn: defining thermo quant} with $\beta_D$ replacing $\tilde{\beta}$. The specific heat now becomes the specific heat at constant $\frakr$, $C_\frakr$, and is defined as in \eqref{eqn: defining thermo quant}, but with $\frakr_\text{tube}$ replacing $K$.

The solutions are divided into two classes as in the conformal problem, the pole patch and cosmic patch.

\textbf{Pole patch.} The first class of solutions is the pole patch solutions which restrict the spacetime to $r\,\in\,\left[0,\frakr_\text{tube}\right]$. In order to have a regular geometry, we must set $\frakr_c\=\ell$. Consequently, $\beta_D$ is a free parameter, and the Euclidean action is given by
\begin{equation}\label{eqn: Dirichlet euclidean action 1 3d}
	I_E^{(\text{pole})} \= -\frac{\beta_D \frakr_\text{tube}}{4G_N\ell}\sqrt{\frac{\ell^2}{\frakr_\text{tube}^2}-1} \, .
\end{equation}
It follows that the energy is given by
\begin{equation}
	\beta_D E \= {I_E^{(\text{pole})}} \, ,
\end{equation}
and the entropy and specific heat at constant $\frakr_\text{tube}$ are zero.

\textbf{Cosmic patch.} We now consider solutions with $r\,\in\, \left[\frakr_\text{tube},\frakr_{\text{c}}\right]$, so they contain the cosmological horizon. Regularity near the horizon $\frakr_c$ fixes
\begin{equation}
	\beta_D \= 2\pi\ell \sqrt{1- \frac{\frakr_\text{tube}^2}{\frakr_{\text{c}}^2}} \, .
\end{equation}
The Euclidean action for this case is given by
\begin{equation}\label{eqn: Dirichlet euclidean action 2 3d}
	I_E^{(\text{cosmic})} \= -\frac{\beta_D \frakr_\text{tube}}{4G_N \ell}\sqrt{\left(1-\frac{\frakr_\text{tube}^2}{\frakr_{\text{c}}^2}\right)^{-1}-1} \, .
\end{equation}
We note that, by setting $\frakr_{\text{c}} \=\ell$, \eqref{eqn: Dirichlet euclidean action 1 3d} and \eqref{eqn: Dirichlet euclidean action 2 3d} reproduce (B.8) and (B.16), respectively, in \cite{Coleman:2021nor}, but without adding the additional boundary subtraction term.

The energy of the cosmic patch is given by
\begin{equation}
	E \= \frac{1}{4G_N\ell}\sqrt{\frakr_{\text{c}}^2-\frakr_\text{tube}^2} \, .
\end{equation}
Defining a dimensionless energy $\mathcal{E} \,\equiv\, 2 \pi E \, \frakr_\text{tube} $, we find that
\begin{equation}
	\mathcal{E} + \frac{\pi \frakr_\text{tube}^2}{2 G_N \ell} \= \frac{\pi \frakr_\text{tube}^2}{2G_N \ell}\left(1+ \sqrt{-1+\frac{\frakr_{\text{c}}^2}{\frakr_\text{tube}^2}}\right) \, , 
\end{equation}
which agrees with the energy in \cite{Coleman:2021nor}, upon identifying
\begin{equation}
	\frac{3\ell}{2G_N} \to c  \, , \qquad\qquad \frac{\frakr_{\text{c}}^2}{\ell^2} \to \frac{12 \Delta}{c} - 1 \, , \qquad\qquad \frac{2G_N \ell}{\pi^2 \frakr_\text{tube}^2}  \to y \, .
\end{equation}

We can also compute the entropy and the specific heat with Dirichlet boundary conditions. For the entropy we obtain,
\begin{equation}
\mathcal{S} \= \frac{\pi \frakr_{\text{c}}}{2G_N} \,,
\end{equation}
which agrees with the Gibbons-Hawking entropy. In fact, this entropy (including a sub-leading logarithmic correction) can be matched to the entropy in a $T\bar{T} + \Lambda_2$-deformed CFT$_2$ \cite{Coleman:2021nor}. Finally, for the specific heat we obtain,
\begin{equation}
C_{\text{(Dirichlet)}} \= - \frac{\pi \frakr_{\text{c}}}{2G_N}\left(\frac{\frakr_{\text{c}}^2}{\frakr_\text{tube}^2}-1\right) \,,
\end{equation}
which is negative for all values of $\mathfrak{r}_{\text{tube}}$. Moreover, in the worldline limit, the specific heat diverges.

\chapter{Useful formulae for $D=4$}\label{sec: useful formulae for 4d thermo}

In this appendix, we provide useful formulae to study conformal thermodynamics of four-dimensional spacetime with $\Lambda \,>\, 0$. 

First, we recall the metric \eqref{eqn: euclidean sol 4D},
\begin{equation}
	ds^2 \= e^{2\omega} \left(\frac{f(r)}{f(\frakr)}d\tau^2 + \frac{dr^2}{f(r)} + r^2 d\theta^2 + r^2 \sin^2{\theta} d\phi^2 \right)\, , \qquad f(r) \= 1- \frac{2\mu}{r} - e^{2\omega} \frac{r^2}{\ell^2}\, .
\end{equation} 
Using a normal vector $\hat{n}\= \sqrt{f(r)}\partial_r$, one can express $e^{2\omega}$ in terms of $K\ell$ and $\mu/\frakr$. For $0\,<\,\mu/\frakr\,<\,1/3$, there exists a unique $e^\omega$ for any real $K \ell$ given by 
\begin{equation}\label{eqn: appendix dum1}
    e^{2\omega} \= \frac{3+\left(K^2\ell^2+9\right) \left(1-\frac{2\mu}{\frakr}\right) - K\ell \sqrt{\left(K^2\ell^2+9\right)\left(1-\frac{2\mu}{\frakr}\right)^2-1}}{2 \left(K^2\ell^2+9\right)\frakr^2/\ell^2 } \, , \qquad K \ell \in \mathbb{R} \, .
\end{equation}
For $1/3\,<\,\mu/\frakr\,<\,1/2$, there are two branches of $e^\omega$ which gives rise to the same $K \ell$ when $K \ell$ is positive. They are given by
\begin{equation}\label{eqn: appendix dum2}
	e^{2\omega_\pm} \= \frac{3+\left(K^2\ell^2+9\right) \left(1-\frac{2\mu}{\frakr}\right) \pm K\ell \sqrt{\left(K^2\ell^2+9\right)\left(1-\frac{2\mu}{\frakr}\right)^2-1}}{2 \left(K^2\ell^2+9\right)\frakr^2/\ell^2 } \, , \qquad K \ell \geq 0  \, .
\end{equation}
This also means that, for $1/3\,<\, \mu/\frakr \,<\, 1/2$, there is no $e^\omega$ that leads to negative $K \ell$.

We can write the radius of the boundary as $\frakr_\text{tube} \= e^\omega \frakr$. The cosmological horizon $\frakr_{\text{c}}$ and black hole horizon $\frakr_{\text{bh}}$ can also be written in terms of $e^\omega$ and $\mu$ by
\begin{equation}
    \frakr_{\text{c}} \= \frac{2}{\sqrt{3}} \cos\left(\frac{1}{3}\cos^{-1}\left(-3\sqrt{3}e^\omega \mu\right)\right) \, , \qquad \frakr_{\text{bh}} \= \frac{2}{\sqrt{3}}\sin\left(\frac{1}{3}\sin^{-1}\left(3\sqrt{3}e^\omega \mu\right)\right) \, .
\end{equation}
Given \eqref{eqn: appendix dum1} and \eqref{eqn: appendix dum2}, one can use these formulae to investigate the behaviour of $\frakr_\text{tube}$, $\frakr_{\text{c}}$, and $\frakr_{\text{bh}}$ while keeping $K \ell$ fixed and varying $\mu/\frakr$.

\chapter{Details of black hole patch computations in $D=4$}  \label{black hole patch app}

In this appendix we give some more details and explicit expressions of the computations done in section \ref{sec: bh patch}, for the black hole patch solutions in $D=4$. For convenience, in this section, we will always express $\frakr_{\text{c}}$ in terms of $\frakr_{\text{bh}}$.

Regularity of the geometry near the black hole horizon determines the conformal temperature of the black hole patch to be
\begin{equation}\label{eqn: bh ds4 beta}
	\tilde{\beta} \= \frac{4\pi \frakr_{\text{bh}}\ell\sqrt{\frakr_\text{tube}\ell^2 - \frakr_\text{tube}^3 - \frakr_{\text{bh}}\ell^2 + \frakr_{\text{bh}}^3}}{\frakr_\text{tube}^{3/2} \left(\ell^2-3 \frakr_{\text{bh}}^2\right)} \, ,
\end{equation}
which is greater than zero. The conformal temperature $\tilde{\beta}^{-1}$ has a lower bound $\tilde{\beta}^{-1}_{\text{min}} \= 2\pi$, which occurs in the Nariai limit, by setting $\frakr_\text{tube} \= \ell/\sqrt{3}$ and taking $\frakr_{\text{bh}} \rightarrow \ell/\sqrt{3}$ from below. 

Below this conformal temperature, the black hole patch solution does not exist. For larger conformal temperatures, $\tilde{\beta}^{-1} \, > \, \tilde{\beta}^{-1}_{\text{min}}$, there is a one-parameter family of black hole patches. To reach the high conformal temperature regime, $\tilde{\beta}\rightarrow 0$, one can take the near horizons limit, i.e. either $\frakr_\text{tube} \rightarrow \frakr_{\text{bh}}$ or $\frakr_\text{tube} \rightarrow \frakr_{\text{c}}$, or the small black hole limit $\frakr_{\text{bh}}/\ell \rightarrow 0$.

Requiring that the boundary has a constant trace of the extrinsic curvature $K$ fixes
\begin{equation}\label{eqn: bh ds4 K}
	K \ell \= \frac{4\frakr_\text{tube}\ell^2 - 6 \frakr_\text{tube}^3-3\frakr_{\text{bh}}\ell^2 + 3 \frakr_{\text{bh}}^3}{2\frakr_\text{tube}^{3/2}\sqrt{\frakr_\text{tube}\ell^2-\frakr_\text{tube}^3-\frakr_{\text{bh}}\ell^2+\frakr_{\text{bh}}^3}} \,.
\end{equation}
There is no upper or lower bound on $K\ell$. Similarly to the pole patch, the limit of $K \ell$ approaching negative infinity corresponds to pushing the boundary to be near the cosmological horizon. For $K\ell$ going to positive infinity, the boundary is pushed near the black hole horizon.

\textbf{Black hole patch thermodynamics.} The regulated action is given by 
\begin{equation}\label{eqn: onshell bh 4d}
	I_{E\text{, reg}}^{(\text{bh})} \= - \frac{\pi \frakr_{\text{bh}} \left(4\frakr_\text{tube}\ell^2 - 3 \frakr_{\text{bh}}\ell^2 - 3 \frakr_{\text{bh}}^3\right)}{3G_N\left(\ell^2-3\frakr_{\text{bh}}^2\right)} - I_E^{(\text{pole})} \, ,
\end{equation}
where we used \eqref{eqn: bh ds4 beta} and \eqref{eqn: bh ds4 K} to simplify the expression. The corresponding conformal energy and conformal entropy for the black hole patch are given by
\begin{equation}\label{eqn: conformal energy entropy bh 4d}
		E_\text{conf} \= - \frac{\frakr_\text{tube}^{3/2}\left(2\frakr_\text{tube}\ell^2-3\frakr_{\text{bh}} \ell^2+ 3 \frakr_{\text{bh}}^3\right)}{6 G_N\ell \sqrt{\frakr_\text{tube}\ell^2-\frakr_\text{tube}^3-\frakr_{\text{bh}}\ell^2+\frakr_{\text{bh}}^3}} - \frac{I_E^{(\text{pole})}}{\tilde{\beta}} \, , \qquad\qquad
		\mathcal{S}_\text{conf} \= \frac{\pi \frakr_{\text{bh}}^2}{G_N} \, .
\end{equation}
The conformal entropy agrees with the Bekenstein-Hawking entropy $A_\text{horizon}/4G_N$ where the horizon in the formula corresponds to the black hole horizon. 

Taking a small black hole limit, we find that 
\begin{equation}
	E_\text{conf} \rightarrow \frac{\frakr_{\text{bh}}}{2G_N} \frac{\frakr_\text{tube}}{\sqrt{1-\frac{\frakr_\text{tube}^2}{\ell^2}}} \, , \qquad \qquad \text{as} \quad \frakr_{\text{bh}}/ \ell \rightarrow 0  \, .
\end{equation}
Provided that $\frakr_{\text{bh}}/2G_N$ is the physical mass of the small black hole, we can see that $E_\text{conf}$ is indeed the energy as measured by the conformal clock defined on the boundary.

The specific heat at constant $K$ of the black hole patch is given by
\begin{equation}\label{eqn: spec heat bh ds4}
	C_K = \frac{2\pi \frakr_{\text{bh}}^2 \left(-\ell^2+3 \frakr_{\text{bh}}^2\right)\left(9\frakr_{\text{bh}}^2\left(\frakr_{\text{bh}}^2-\ell^2\right)^2+16 \frakr_{\text{bh}} \left(\frakr_{\text{bh}}^2-\ell^2\right)\frakr_\text{tube}\ell^2+8\frakr_\text{tube}^2\ell^4 - 4 \frakr_\text{tube}^4\ell^2\right)}{G_N\left(\ell^2+3\frakr_{\text{bh}}^2\right)\left(9 \frakr_{\text{bh}}^2\left(\frakr_{\text{bh}}^2-\ell^2\right)^2+2\frakr_{\text{bh}}\frac{\left(-9\ell^4-10\frakr_{\text{bh}}^2\ell^2+15\frakr_{\text{bh}}^4\right)}{\left(\ell^2+3\frakr_{\text{bh}}^2\right)}\frakr_\text{tube}\ell^2+8\frakr_\text{tube}^2\ell^4-4\frakr_\text{tube}^4\ell^2\right)} \, .
\end{equation}
Note that
\begin{equation}
	C_K \,\rightarrow\, 
	\begin{cases}
		- \frac{2 \pi \frakr_{\text{bh}}^2}{G_N} \, , & \text{as} \quad \frakr_{\text{bh}}/\ell \rightarrow 0 \, , \\
		+ \frac{2 \pi \frakr_{\text{bh}}^2}{G_N} \, , & \text{as} \quad \frakr_{\text{bh}} / \frakr_\text{tube} \rightarrow 1 \, .
	\end{cases}
\end{equation}

The first limit corresponds to a small black hole and, in that case, the specific heat is negative. On the opposite limit, when the black hole size is comparable to the boundary size, then the specific heat is positive.
%
%


\chapter{Plots of the regulated action, $E_\text{conf}$, and $C_K$ in $D=4$}\label{sec: additional figures}

\begin{figure}[p!]
        \centering
        \subfigure[reg. action, $K\ell=-7.5$]{
                \includegraphics[height=4.5cm]{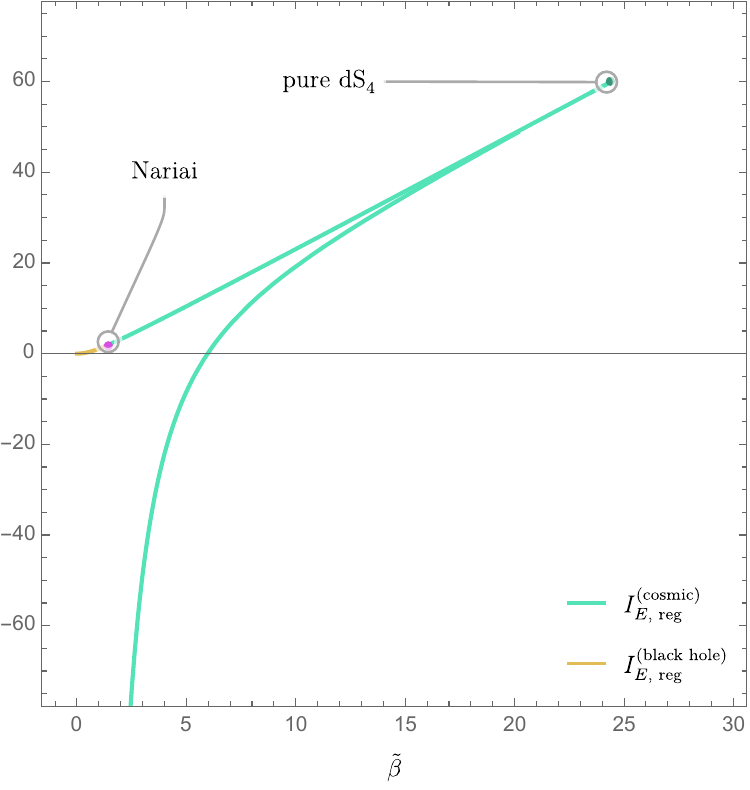}\label{fig: action -7.5}}  \quad\quad
        \subfigure[reg. action, $K\ell=-1.5$]{
                \includegraphics[height=4.5cm]{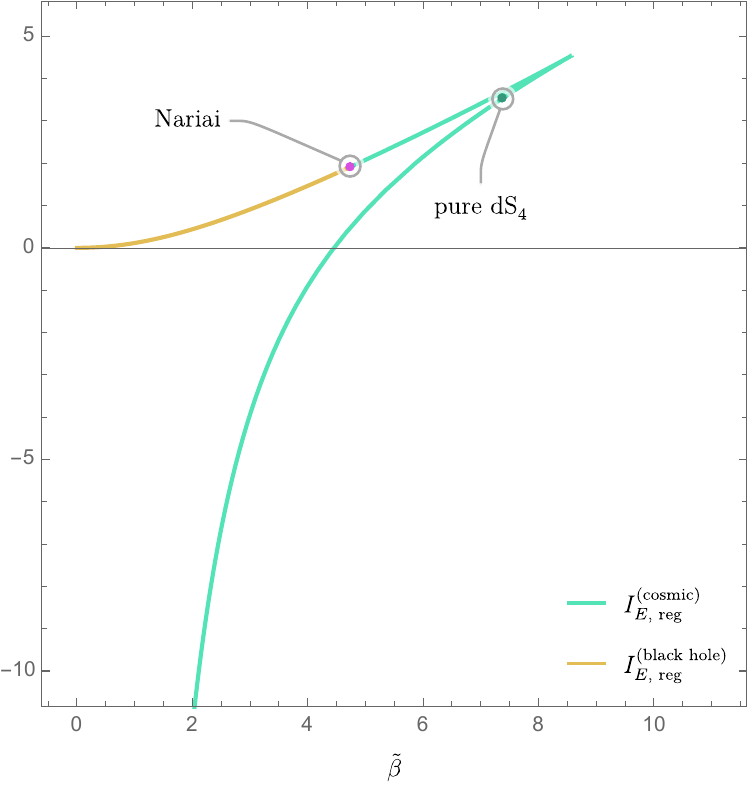}\label{fig: action -1.5}}  \quad\quad
        \subfigure[reg. action, $K\ell=6$]{
                \includegraphics[height=4.5cm]{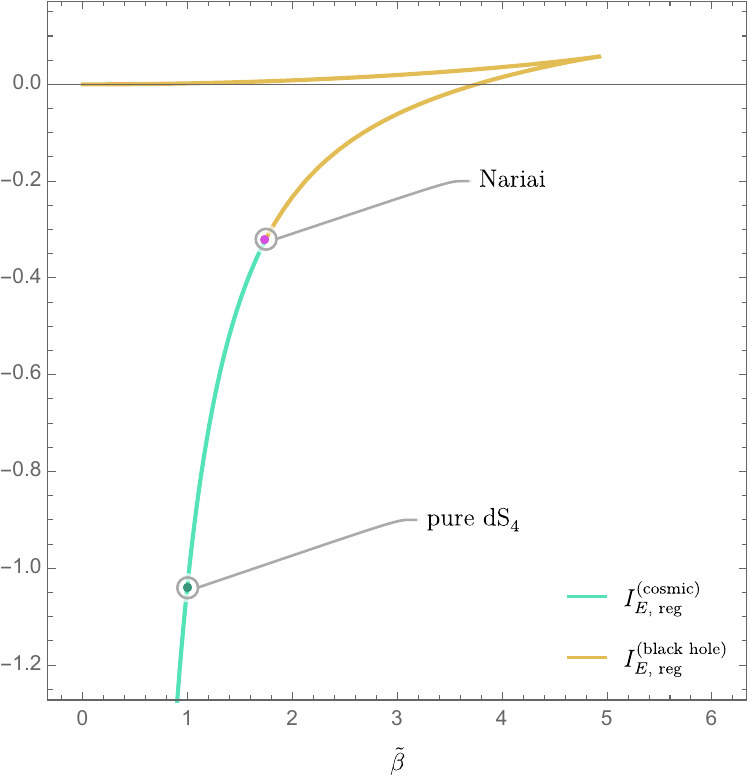}\label{fig: action 6}}  \quad\quad
        \subfigure[$E_\text{conf}$, $K\ell=-7.5$]{
                \includegraphics[height=4.5cm]{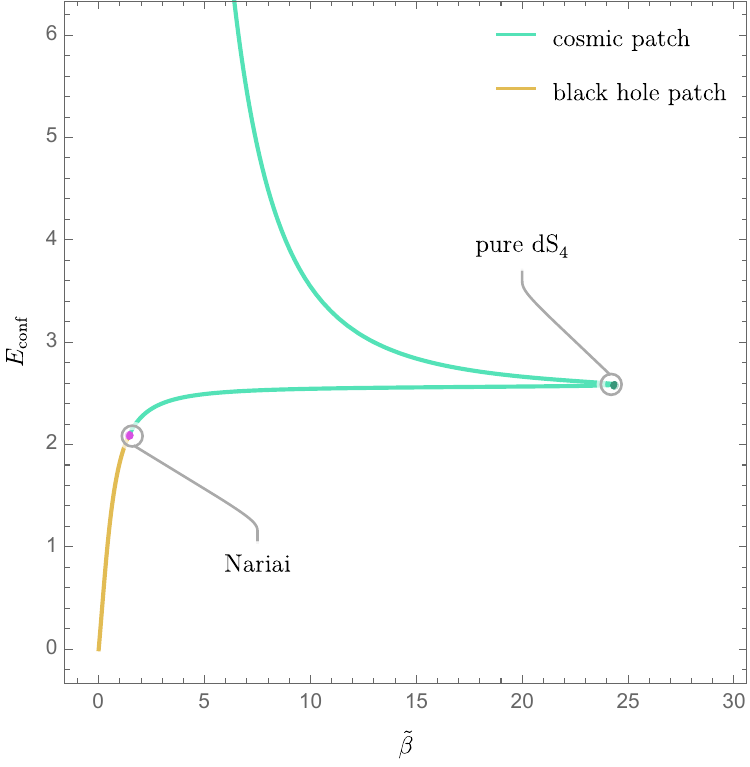}\label{fig: energy -7.5}}  \quad\quad
        \subfigure[$E_\text{conf}$, $K\ell=-1.5$]{
                \includegraphics[height=4.5cm]{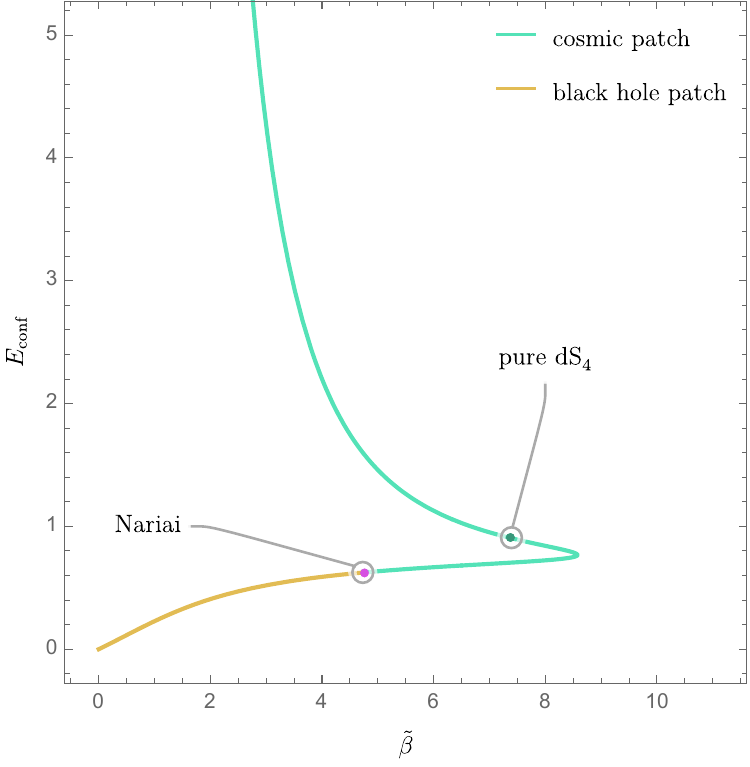}\label{fig: energy -1.5}}  \quad\quad
        \subfigure[$E_\text{conf}$, $K\ell=6$]{
                \includegraphics[height=4.5cm]{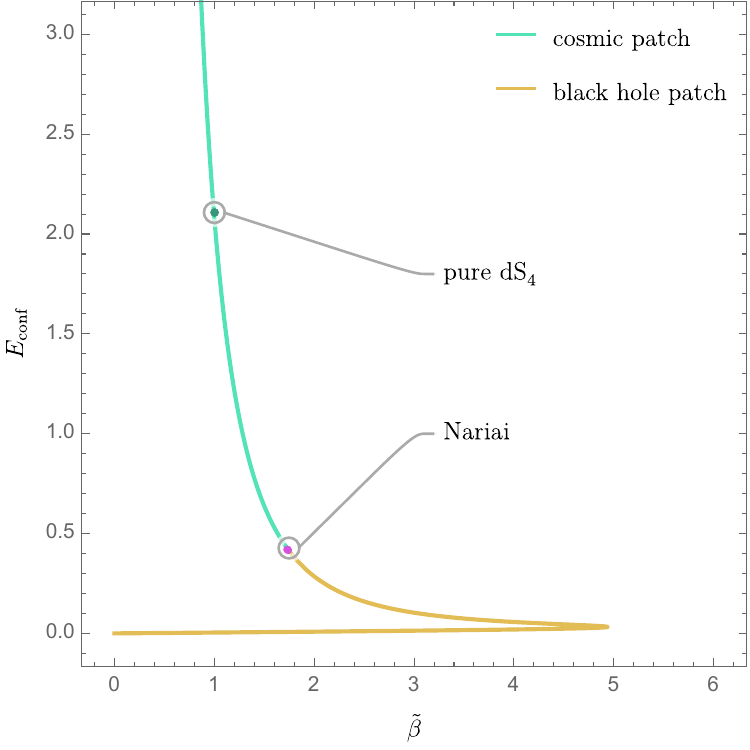}\label{fig: energy 6}}  \quad\quad
        \subfigure[$C_K$, $K\ell=-7.5$]{
                \includegraphics[height=4.5cm]{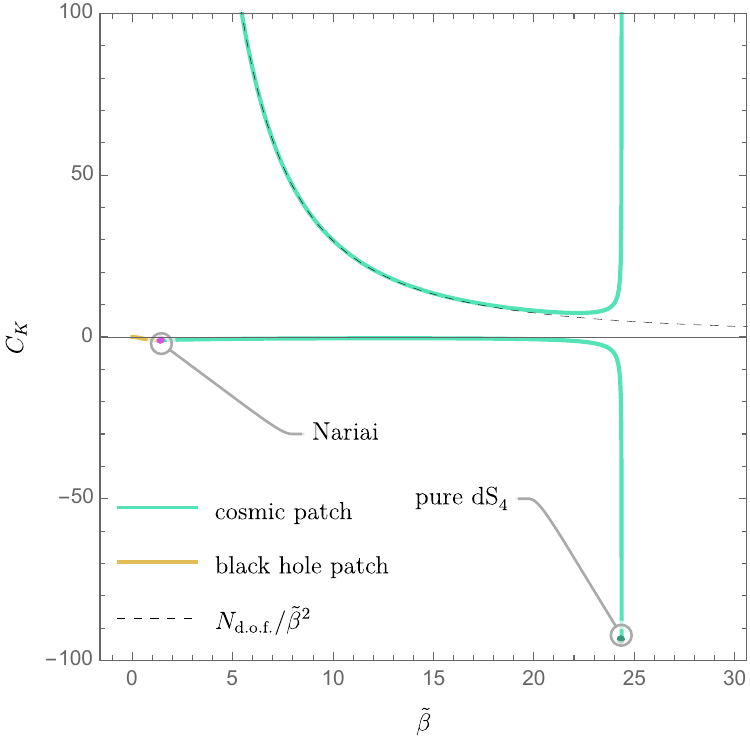}\label{fig: C -7.5}}  \quad\quad
        \subfigure[$C_K$, $K\ell=-1.5$]{
                \includegraphics[height=4.5cm]{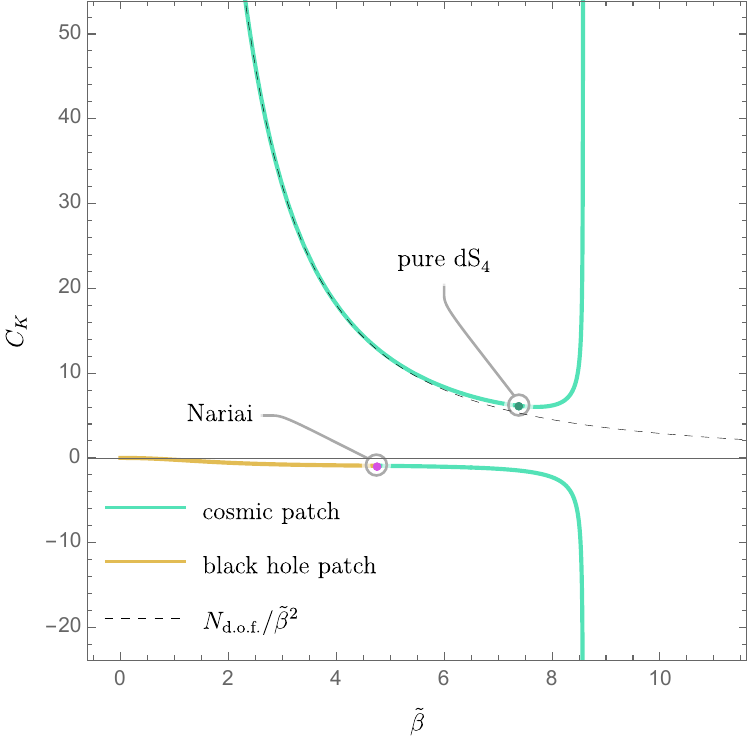}\label{fig: C -1.5}}  \quad\quad
        \subfigure[$C_K$, $K\ell=6$]{
                \includegraphics[height=4.5cm]{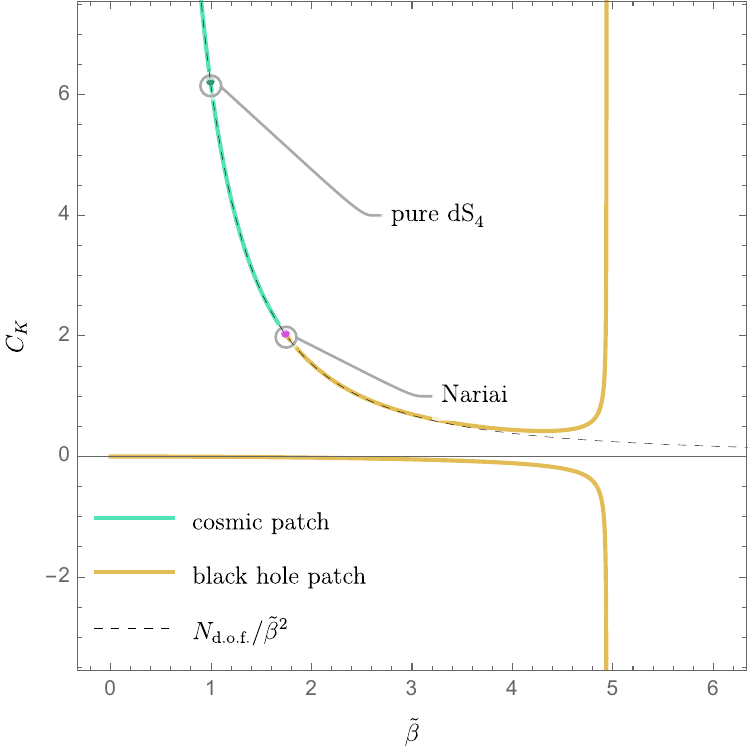}\label{fig: C 6}}      
                \caption{Plots of regulated action, $E_\text{conf}$, and $C_K$ as a function of $\tilde{\beta}$ at fixed $K\ell$. When evaluated on the cosmic (black hole) patch, the curve colour is green (yellow). The pure dS$_4$ and Nariai solutions are marked in dark green and purple dots, respectively. For the plots of $C_K$, we also display the conformal answer $N_\text{d.o.f.}/\tilde{\beta}^2$ as a black dashed curve.} \label{fig: appendix figures}
\end{figure}

In this appendix, we give numerical examples of the regulated action, conformal energy, and specific heat for dS$_4$ conformal thermodynamics as functions of $\tilde{\beta}$ for various values of $K\ell$. These are displayed in figure \ref{fig: appendix figures} for $K\ell \= -7.5$, $-1.5$, and $6.0$. These values of $K\ell$ are chosen to show pure dS$_4$ patches which are unstable, metastable, and stable, respectively. 

\chapter{$l\=0$ and $l\=1$ modes}\label{sec: l0/l1 modes}

In this appendix, we consider linearised dynamics of gravitational $l\=0$ and $l \=1$ modes. Following the analysis in \cite{Anninos:2023epi}, these modes are locally pure diffeomorphisms that become physical by the presence of the timelike boundary with fixed boundary data. In particular, we are interested in linearised perturbation $h_{\mu\nu}$ of the form,
\begin{equation}\label{eqn: physical diffeo perturbation}
    h_{\mu \nu} \= \nabla_\mu \xi_\nu + \nabla_\nu \xi_\mu \, ,
\end{equation}
for an arbitrary vector field $\xi^\mu$. The perturbation \eqref{eqn: physical diffeo perturbation} automatically satisfies the linearised Einstein field equation. The conditions that this perturbation preserves the conformal boundary data at $r\=\frakr$ lead to
\begin{equation}\label{eqn: bdry conds diffeo}
\begin{cases}
    \left.2\left(K_{mn} - \frac{K}{3}\bar{g}_{mn}\right)\left(\sqrt{1-\frac{r^2}{\ell^2}}\xi_r\right) + \left(\mathcal{D}_m \xi_n + \mathcal{D}_n \xi_m - \frac{2\bar{g}_{mn}}{3}\mathcal{D}^p \xi_p\right)\right|_{r=\frakr} &=\, 0 \, , \\
    \left.\left(\sqrt{1-\frac{r^2}{\ell^2}}\partial_r K - \mathcal{D}^m \mathcal{D}_m \right) \left(\sqrt{1-\frac{r^2}{\ell^2}}\xi_r\right) + \xi_m \mathcal{D}^m K \right|_{r=\frakr} &=\, 0 \, ,
\end{cases}
\end{equation}
where $K_{mn}$ and $\bar{g}_{mn}$ are the extrinsic curvature \eqref{eqn: lorentzian extrinsic} and induced metric \eqref{eqn: lorentzian induced metric} of $\Gamma$, respectively. The covariant derivative $\mathcal{D}_m$ is that associated to the induced metric $\bar{g}_{mn}$. 

At the linearised level, the perturbation \eqref{eqn: physical diffeo perturbation} is subject also to gauge redundancy in the form of the diffeomorphism
\begin{equation}\label{eqn: diffeo transf}
    x^\mu \,\to\, x^\mu + \epsilon\,\xi^{'\mu} \, , \qquad\qquad h_{\mu\nu} \,\to\, h_{\mu\nu} - \nabla_\mu \xi'_\nu - \nabla_\nu \xi'_\mu \, ,
\end{equation}
for an arbitrary vector field $\xi^{'\mu}$. Due to the presence of the boundary, the vector field $\xi^{'\mu}$ must preserve the boundary data and location of the boundary leading to \eqref{eqn: bdry conds diffeo} and $\left.\xi'^r\right|_{r=\frakr} \= 0$, respectively. This means that a large number of the perturbations \eqref{eqn: physical diffeo perturbation} obeying \eqref{eqn: bdry conds diffeo} can be gauged away by some suitable diffeomorphism \eqref{eqn: diffeo transf}. The exception is when the perturbation \eqref{eqn: physical diffeo perturbation} is constructed from the vector field $\xi^\mu$ which disturbs the location of the boundary, i.e.
\begin{equation}\label{eqn: disturbing location condition}
    \left.\xi^r\right|_{r=\frakr} \,\neq\, 0 \, .
\end{equation}
We therefore take \eqref{eqn: bdry conds diffeo} and \eqref{eqn: disturbing location condition} as boundary conditions for a physical metric perturbation \eqref{eqn: physical diffeo perturbation}.

\textbf{$l\=0$ modes.} Choosing the $h_{tr}\=0$ gauge, the general spherically symmetric ($l\=0$) vector field $\xi^\mu$ satisfying \eqref{eqn: bdry conds diffeo} and \eqref{eqn: disturbing location condition} is given by
\begin{equation}\label{eqn: l=0 sol}
    \xi_\mu dx^\mu \,\sim\, \frac{r}{\ell}\left(1-\frac{r^2}{\ell^2}\right)^{\omega^{(l=0)2}_\pm \ell^2/2}e^{-i \omega^{(l=0)}_\pm t} \left(dr - i \frac{1-r^2/\ell^2}{\omega^{(l=0)}_\pm r}\, dt\right)\, , 
\end{equation}
where the frequency $\omega^{(l=0)}_\pm \frakr \= \pm i \sqrt{2-\frac{\frakr^2}{\ell^2}}$ is purely imaginary. In the worldline limit, where $\frakr/\ell \,\to\, 0$, we match the result of $l\=0$ modes found in \cite{Anninos:2023epi}. In the strechted horizon limit, where $\frakr/\ell \,\to\, 1$, we find that these pair of modes coalesce to $\omega^{(l=0)}_\pm \ell \= \pm i$.

\textbf{$l\=1$ modes.} Taking again the $h_{tr}\=0$ gauge, the general $l\=1$ vector field $\xi^\mu$ satisfying \eqref{eqn: bdry conds diffeo} and \eqref{eqn: disturbing location condition} is given by
\begin{equation}\label{eqn: l=1 sol}
    \xi_\mu dx^\mu \,\sim\,\frac{e^{- i \omega^{(l=1)}_\pm t}}{\sqrt{1-r^2/\ell^2}} \left(\mathbb{S} \, dr + i {\omega^{(l=1)}_\pm r}\left(1-r^2/\ell^2\right) \,\mathbb{S} \, dt +\sqrt{2}\left(1-r^2/\ell^2\right) \,\mathbb{S}_i \,r \,d\Omega^i\right) \, ,
\end{equation}
where the frequency $\omega^{(l=1)}_\pm \ell \= \pm i$ is pure imaginary and $\frakr$-independent. Unlike the $l\=0$ modes, the metric perturbation constructed from \eqref{eqn: l=1 sol} is vanishing everywhere implying that \eqref{eqn: l=1 sol} is a Killing vector of the background dS$_4$. Near the worldline, where $\frakr/\ell \,\to\, 0$, these modes reproduce three translations and three Lorentz boosts of the flat spacetime. In the strectched horizon limit, where $\frakr/\ell \,\to\, 1$, we find that these modes become combinations of translations and Lorentz boosts in a local inertial frame near the boundary.

\chapter{Near-horizon diffeomorphisms}\label{diffeoapp}

In this appendix, we consider metric perturbations that are locally diffomorphisms for any $l$, and we consider a boundary that is located close to the cosmological horizon.

Let us first define a near-horizon parameter $\epsilon \,\equiv\,1-\tfrac{r}{\ell}$ and a near-horizon radial coordinate $\rho\,\equiv\,\tfrac{1}{{\epsilon}} \left({1-\tfrac{r}{\ell}}\right)$. It follows that in terms of $\rho$, the boundary is located at $\rho\=1$. 

Consider the following linearised diffeomorphism with complex frequency $\omega \ell \= \pm i$,
\begin{equation}\label{eqn: supertranslation}
    \xi_\mu dx^\mu \= e^{\pm t} \mathbb{S} \,dr - e^{\pm t} \left(1-\frac{r^2}{\ell^2}\right)  \left(\pm\mathbb{S}\, dt - \partial_i \mathbb{S}\, d\Omega^i\right)\, ,
\end{equation}
where $\mathbb{S}$ here is an arbitrary angle-dependent function obeying $\left|\frac{\partial_i \mathbb{S}}{\mathbb{S}}\right|\,\ll\,\frac{1}{\epsilon}$. By considering the associated linearised metric perturbation $h_{\mu\nu} \= \nabla_\mu \xi_\nu + \nabla_\nu \xi_\mu$, we find that
\begin{equation}
    \left.\delta K(h_{\mu\nu})\right|_{r=\frakr} \= \mathcal{O}(\sqrt{\epsilon})  \, , \qquad\qquad \left.\delta h^m{}_n\right|_{r=\frakr} \= \mathcal{O}(\epsilon)\delta^m_n \, .
\end{equation}
In terms of the near-horizon coordinate, we find that
\begin{equation}
    \left.\xi^\theta\right|_{\rho=1} \= \left.\xi^\phi\right|_{\rho=1} \= \mathcal{O}(\epsilon) \, , \qquad -\frac{\ell}{2}\left.\xi^\rho\right|_{\rho=1} \= \pm \left.\xi^t\right|_{\rho=1} \= e^{\pm t/\ell} \mathbb{S} + \mathcal{O}(\epsilon)\, .
\end{equation}
This means that, in the $\epsilon\,\to\,0$ limit, the diffeomorphism \eqref{eqn: supertranslation} preserves the conformal boundary data but not the location of the boundary. In particular, in the local inertial frame, these modes become angle-dependent radial/time translations. The frequencies of these modes coincide with the complex frequency modes, $\omega \ell = \pm i$, found in the cosmological horizon limit discussed in section \ref{sec: scalar pert}.

\chapter{Linearised Dirichlet problem} \label{app: Dirichlet problem}

In this appendix we comment on linearised dynamics about different backgrounds subject to Dirichlet boundary conditions. We discuss 
uniqueness and existence properties of the perturbed solution about Minkowski, Rindler, and planar AdS in four spacetime dimensions. In all cases, we work in a Fefferman-Graham-like gauge, $h_{z\mu}=0$, for all $z \geq z_c$.

\section{Uniqueness and existence for Minkowski, Rindler, and planar AdS$_4$}

\textbf{Minkowski corner.} Let us first consider the linearised problem about a Minkowski corner background. Specifically, we take the background metric to be
\begin{equation}
    ds^2 = dz^2 + \eta_{mn} dx^m dx^n \, ,
\end{equation}
and we place the timelike boundary at $z=z_c$. Working in a Fefferman-Graham-like gauge where
\begin{equation}
    h_{zz}=h_{zm}=0 \, , \qquad h_{mn} = \gamma_{mn} \, ,
\end{equation}
the $zz$- and $zm$-components of the Einstein field equations read
\begin{equation}\label{eqn: Mink corner Gzmu}
\begin{cases}
    \partial^m \partial^n \left(\gamma_{mn}-\eta_{mn} \gamma\right) = 0 \, , \\ 
    \partial^n \partial_z\left( \gamma_{mn}-\eta_{mn} \gamma\right) = 0 \, ,
\end{cases}
\end{equation}
where $\gamma \equiv \eta^{mn}\gamma_{mn}$ is the trace of $\gamma_{mn}$.
Now we impose the Dirichlet boundary conditions at $z=z_c$,
\begin{equation} \label{eq: dirichlet corner}
    \left.\gamma_{mn}\right|_\Gamma = \Gamma_{mn} \, ,
\end{equation}
where $\Gamma_{mn}$ is a symmetric tensor which only depends on $x^m$. By evaluating the first equation in \eqref{eqn: Mink corner Gzmu} at $z=z_c$ and imposing the Dirichlet boundary condition \eqref{eq: dirichlet corner}, we obtain an obstruction on generic Dirichlet data,
\begin{equation}
    \partial^m \partial^n \Gamma_{mn} - \partial^2 \Gamma^m{}_m=0 \, .
\end{equation}
The above implies that the linearised Ricci scalar for the induced metric must vanish in agreement with the non-linear analysis of \cite{An:2021fcq}.

\textbf{Rindler corner.} Next, we discuss perturbations about Rindler space,
\begin{equation}
    ds^2 = dz^2 -z^2dt^2 + \delta_{ab} dx^a dx^b \, ,
\end{equation}
and place the timelike boundary at $z=z_c$. We work again in the same Fefferman-Graham-like gauge, 
\begin{equation}\label{eqn: FGlikge gauge Rindler}
    h_{zz}=h_{zt}=h_{za}=0 \, , \qquad h_{ab}=\gamma_{ab} \, , \qquad h_{ta} = z v_a \, , \qquad h_{tt} = z^2s \, ,
\end{equation}
where $\gamma_{ab}$, $v_a$, and $s$ are $z$ and $x^m$-dependent variables. 
We consider imposing Dirichlet boundary conditions at $z=z_c$,
\begin{equation}\label{eqn: Dirichlet Rindler}
    \left.h_{mn}\right|_\Gamma = \Gamma_{mn}\,,
\end{equation}
where $\Gamma_{mn}$ only depends on $(t,x^a)$. 

Combining the linearised $tz$- and $zz$- Einstein equations with the Dirichlet boundary condition (\ref{eqn: Dirichlet Rindler}) one can show that
\begin{equation}\label{eqn: Rindler obst 2}
    \left(\partial_t^3-\partial_t\right)\Gamma^a{}_a = \left.\partial_a F^a\right|_\Gamma \, , 
\end{equation}
where
\begin{equation}
F^a \equiv z^2 \partial_z v^a - z v^a - z^2 \partial_t \partial^a s + 2 z \partial_t^2 v^a - z^2 \partial_t \partial^b \left(\gamma^a{}_b-{\delta^a}_b \gamma\right)~.
\end{equation}
Note that the first term in $F^a$ contains a $z$-derivative, so $F^a$ evaluated at $z=z_c$ cannot be purely determined from the Dirichlet data, $\Gamma_{mn}$. However, upon integrating over the non-compact spatial directions, the right hand side of \eqref{eqn: Rindler obst 2} becomes a boundary integral, i.e.,
\begin{equation}\label{eqn: Rindler obst 3}
    \int_{\mathbb{R}^2}d^2x \left(\partial_t^3-\partial_t\right)\Gamma^a{}_a = \lim_{r\to\infty}\int_{S^1} rd\theta \,  F_{r} \, ,
\end{equation}
where $r$ and $\theta$ denote polar coordinates on the spatial $\mathbb{R}^2$ of the Rindler spacetime. Assuming that $F_r$ decays as $\tfrac{1}{r^2}$ or faster as $r\to \infty$, the right hand side vanishes. We  note that although ${\Gamma^a}_a$ transforms under coordinate transformations at $\Gamma$ (decaying sufficiently fast at the boundary of $\mathbb{R}^2$), its integral over $\mathbb{R}^2$ does not. We thus obtain an obstruction on generic Dirichlet data.

A similar argument can be made also for the region near a horizon. In that case, the near horizon region is approximately the product of two-dimensional Rindler space and a two-sphere, and so the left hand side of \eqref{eqn: Rindler obst 3} is replaced by an integral over a two-sphere, leading to a similar obstruction to the one found in \cite{Bredberg:2011xw,Anninos:2011zn}.

Finally, let us consider the non-uniqueness  of  linearised gravity about the Rindler background  subject to  Dirichlet boundary conditions. We work again in the Fefferman-Graham-like gauge, \eqref{eqn: FGlikge gauge Rindler}. A solution that is only dependent on $z$ and $t$ is given by
\begin{equation}\label{eqn: Rindler zero mode}
    h_{\mu\nu}dx^\mu dx^\nu = z(z-z_c)f(t)dt^2 \, ,
\end{equation}
where $f(t)$ is an arbitrary function of time. This solution is locally diffeomorphic, namely one can write \eqref{eqn: Rindler zero mode} as $h_{\mu\nu}= \nabla_\mu \xi_\nu + \nabla_\nu \xi_\mu$, with
\begin{equation}
    \xi^\mu \partial_\mu = \frac{1}{2}\int_{\mathbb{R}} \frac{d\omega }{2\pi}\,\frac{f(\omega)e^{-i\omega t}}{1+\omega^2} \left(z_c\partial_z + \left(\frac{z+(z-z_c)\omega^2}{i\omega z}\right)\partial_t\right) \, , \qquad f(\omega) = \int_{\mathbb{R}}dt \, f(t) e^{i\omega t}\,.
\end{equation}
At the boundary $z=z_c$, \eqref{eqn: Rindler zero mode} obeys the Dirichlet boundary conditions \eqref{eqn: Dirichlet Rindler} for $\Gamma_{mn}=0$. As $\xi^z$ is non-vanishing at the boundary, this solution is physical and cannot be gauged away.

The non-uniqueness of the Rindler corner follows from the fact that \eqref{eqn: Rindler zero mode} contains an arbitrary function $f(t)$.  As it is an arbitrary function of time, one can always construct a solution that is localised in time away from the initial Cauchy surface at $t=0$ and $z>z_c$, so that the solution does not leave an imprint on the initial data. As such, specifying the initial conditions together with the Dirichlet boundary conditions on the Rindler corner background will never lead to a unique solution. The same argument also holds for the region $0<z<z_c$.\footnote{This property has also been observed by T. Zikopoulos in unpublished work.}

As an explicit example, take the bump function localised on $t\in \left(4,6\right)$ and vanishing elsewhere,
\begin{equation}
    f(t) = \begin{cases}
        \alpha\,e^{-(1-(5-t)^2)^{-1}} \,,& \quad t \in (4,6)\\
        0 \,,& \quad \text{else}
    \end{cases} \, , \qquad \alpha \in \mathbb{R} \, .
\end{equation}
Plugging this back in \eqref{eqn: Rindler zero mode}, it can be checked that the perturbation vanishes at the $t=0$ initial surface.

For the Euclidean Rindler problem, we instead consider linearised gravity about the background
\begin{equation}
ds^2 = z^2 d\tau^2 + dz^2 + dx^2 + dy^2~, \quad\quad 0<z<z_c\, , \quad \quad  \tau \sim \tau + 2\pi~.
\end{equation}
Due to the periodicity of $\tau$, the Euclidean counterpart of (\ref{eqn: Rindler zero mode}) is no longer regular at the Euclidean horizon. To avoid such a complication, it is useful to consider the kernel of the linearised Einstein field equation subject to the Dirichlet boundary conditions on this background without any gauge-fixing condition \cite{Witten:2018lgb}. Let us pick an arbitrary function $\zeta(\tau)$ satisfying the periodicity condition, $\zeta(\tau) = \zeta(\tau+2\pi)$. It can be shown that by perturbing the boundary so that the physical region is $0<z<z_c-\varepsilon\, z_c\partial_\tau\zeta(\tau)$, the induced metric at the boundary remains unchanged upon reparametrising $\tau$ coordinate by $\tau\to\tau+\varepsilon\,\zeta(\tau)$. As $\zeta(\tau)$ (and $\partial_\tau \zeta(\tau)$) is continuous across $\tau=2\pi$, the boundary remains smooth. As the origin, $z=0$, is unaffected by this perturbation, the space remains regular everywhere. Since there are infinitely many $\zeta(\tau)$ obeying the periodicity condition, it follows that the kernel of the linearised Einstein field equation, without fixing any gauge, is infinite-dimensional, and hence the Dirichlet problem is not elliptic.

In terms of the metric perturbation, this is equivalent to considering the following diffeomorphism,
\begin{equation}
    \xi^\mu\partial_\mu = \left(\tilde{\xi}^\tau(\tau,z) + \zeta(\tau)\right)\partial_\tau + \left(\tilde{\xi}^z(\tau,z)-z_c \partial_\tau \zeta(\tau)\right)\partial_z \, ,
\end{equation}
where $\tilde{\xi}^\tau(\tau,z)$ and $\tilde{\xi}^z(\tau,z)$ are arbitrary functions satisfying the periodicity condition and that become zero at $z=z_c$. Then the metric perturbation $h_{\mu\nu}=\nabla_\mu \xi_\nu + \nabla_\nu \xi_\mu$ becomes
\begin{equation}
    h_{\mu\nu}dx^\mu dx^\nu = 2z\left( \tilde{\xi}^z +z\partial_\tau\tilde{\xi}^\tau+(z-z_c)\partial_\tau \zeta \right)d\tau^2 + 2 \partial_z \tilde{\xi}^z dz^2+2\left(\partial_\tau \tilde{\xi}^z+z^2\partial_z \tilde{\xi}^\tau-z_c \partial_\tau^2\zeta\right)d\tau dz\,.
\end{equation}
It is straightforward to show that $h_{\mu\nu}$ at $z=z_c$ satisfies the Dirichlet boundary conditions, $\left.h_{ij}\right|_\Gamma=0$. %
At $z=0$, one can always choose $\tilde{\xi}^\tau(\tau,z)$ and $\tilde{\xi}^z(\tau,z)$ so that the perturbation is regular. As $\xi^z|_\Gamma$ is non-vanishing, the perturbation is physical and cannot be gauged away.

\textbf{Planar AdS$_4$.} We now consider perturbations about planar AdS$_4$ in the Fefferman-Graham gauge,
\begin{equation}
    h_{zz}=h_{zm}=0\, , \qquad h_{mn}=\frac{\ell^2}{z^2}\gamma_{mn} \, .
\end{equation}
We will show, shortly, that one can always go to the above gauge at the linearised level. Then, one finds that the $zz$- and $zm$-components of the Einstein field equations, \eqref{eqn: planar AdS G_zmu}, are given by
\begin{equation}\label{eqn: dummy1}
\begin{cases}
    \partial^m \partial^n \left(\gamma_{mn}-\eta_{mn} \gamma\right) + 2z^{-1}\partial_z \gamma = 0 \, , \\ 
    \partial^n \partial_z\left( \gamma_{mn}-\eta_{mn} \gamma\right) = 0 \, ,
\end{cases}
\end{equation}
where $\gamma=\eta^{mn}\gamma_{mn}$ denotes the trace part of $\gamma_{mn}$. Now we consider imposing the Dirichlet boundary conditions at $z=z_c$, 
\begin{equation}\label{eqn: dummy2}
    \left.\gamma_{mn}\right|_\Gamma = \Gamma_{mn}\,,
\end{equation}
where $\Gamma_{mn}$ is a symmetric tensor which depend on $x^m$. Combining \eqref{eqn: dummy1} and \eqref{eqn: dummy2}, we find that
\begin{equation}\label{eqn: dummy3}
\begin{cases}
    \partial^m \gamma_{mn} = \partial^m \Gamma_{mn} + \frac{1}{4}\partial_n \left(\partial^p \partial^q \Gamma_{pq}-\partial^2 \Gamma^p{}_p\right)\left(z_c^2-z^2\right) \, , \\
    \gamma = \Gamma^m{}_m + \frac{1}{4}\left(\partial^p \partial^q \Gamma_{pq} - \partial^2 \Gamma^p{}_p\right)\left(z_c^2-z^2\right) \, , 
\end{cases}
\end{equation}
Thus, at the linearised level and unlike the case for the Minkowski \cite{An:2021fcq} or the Rindler corner, there is no obstruction to the linearised Dirichlet problem about planar AdS$_4$. This is also true for the global AdS$_4$ case \cite{Andrade:2015fna} at the linearised level.

To discuss uniqueness, it is sufficient to set $\Gamma_{mn}=0$. The equations \eqref{eqn: dummy2} and \eqref{eqn: dummy3} then imply that $\gamma_{mn}$ is a transverse-traceless tensor subject to the boundary conditions $\left.\gamma_{mn}\right|_\Gamma=0$. This is precisely $\tilde{\gamma}_{mn}$ discussed in section \ref{sec: bulk planar AdS}. Upon imposing the $mn$-components of the Einstein field equations, this perturbation leads to a unique solution. In summary, this analysis shows that, by imposing Dirichlet boundary conditions \eqref{eqn: dummy2}, the linearised initial boundary value problem about planar AdS$_4$ is well-posed.

\textbf{Existence for linearised conformal boundary conditions.} Let us briefly comment on the existence property of conformal boundary conditions at the linearised level about planar AdS$_4$. We will allow for both a small perturbation in the conformal structure and in the trace of the extrinsic curvature, such that 
\begin{equation}\label{eqn: dummy4}
    \left.\gamma_{mn}-\frac{1}{3}\eta_{mn}\gamma\right|_\Gamma = \Gamma_{mn} \, , \qquad \left.z\partial_z \gamma\right|_\Gamma = 2\delta K \ell\, ,
\end{equation}
where $\Gamma_{mn}$ is a symmetric-traceless tensor, and $\delta K$ is a scalar. Both depend only on $x^m$. One can show that the equations \eqref{eqn: dummy1} together with \eqref{eqn: dummy4} imply that
\begin{equation}\label{eqn: dummy5}
\begin{cases}
    \partial^m \gamma_{mn} = \frac{1}{3}\partial_n \varphi + \partial^m \Gamma_{mn} + \frac{\partial_n \delta K\ell}{z_c^2}\left(z^2-z_c^2\right) \, , \\
    \gamma = \varphi + \frac{\delta K \ell}{z_c^2}\left(z^2-z_c^2\right) \, , \\
    \partial^2 \varphi = \frac{6\delta K \ell}{z_c^2}+ \frac{3}{2}\partial^m \partial^n \Gamma_{mn} \, .
\end{cases}
\end{equation}
The first two equations fix completely $\partial^m \gamma_{mn}$ and $\gamma$ in terms of the boundary data $\Gamma_{mn}$ and $K$ and a $z$-independent scalar $\varphi$, which satisfies the last equation. 
These equations do not impose any constraints on the allowed boundary data $(\Gamma_{mn},K)$. Instead, we have a new degree of freedom, $\varphi$, whose dynamics is localised at the boundary. Note that by setting both $\Gamma_{mn}=0$ and $\delta K =0$, \eqref{eqn: dummy5} becomes  \eqref{eq: for appendix}.

\section{Permissibility of Fefferman-Graham gauge} \label{app: FG}

As a last comment, we would like to show that it is always possible to use allowed diffeomorphisms in planar AdS$_4$ to impose the Fefferman-Graham gauge. This follows from the setup in section \ref{sec: diff planar}. Consider a general linearised solution that is composed of both a bulk perturbation $\gamma_{\mu\nu}$, and a linearised diffeomorphism. Then, using \eqref{eqn: planar AdS xi^m} and \eqref{eqn: AdS planar sol xi^z}, the $h_{z\mu}$ components of the linearised metric, see \eqref{eqn: AdS planar metric diffeo}, become
\begin{equation}
    h_{zz} = \frac{2\ell^2}{z^2}\left(-\frac{\zeta}{z}+\partial_z\tilde{\xi}^z - \frac{\tilde{\xi}^z}{z} + \frac{\gamma_{zz}}{2}\right) \, , \qquad h_{zm} = \frac{\ell^2}{z^2}\left(\partial_m \zeta + \partial_m \tilde{\xi}^z + \eta_{mn} \partial_z \tilde{\xi}^n+\gamma_{zm}\right) \, .
\end{equation}
At this level, $\gamma_{zz}$ and $\gamma_{zm}$ can be arbitrary functions of $z$ and $x^m$. We observe that by choosing $\tilde{\xi}^\mu$ to be
\begin{equation}
\begin{cases}
    \tilde{\xi}^m(z,x^m) = \frac{z_c^2-z^2}{2z_c}\eta^{mn}\partial_n \zeta(x^m)-\int^z_{z_c}dz'\,\eta^{mn}\gamma_{zn}(z',x^m)+\int^z_{z_c}dz' z' \int^{z'}_{z_c} \frac{dz''}{2z''}\,\partial^m \gamma_{zz}(z'',x^m) \, , \\
    \tilde{\xi}^z(z,x^m) = \left(\frac{z}{z_c}-1\right)\zeta(x^m)-z \int_{z_c}^z \frac{dz'}{2z'}\,\gamma_{zz}(z',x^m) \, ,
    \end{cases}
\end{equation}
the metric components $h_{zz}$ and $h_{zm}$ vanish everywhere on $z\geq z_c$, including at the timelike boundary.

\chapter{Near AdS$_4$ timelike boundary} \label{app: near infinity}

In this appendix we collect some useful formulae regarding the intrinsic/extrinsic geometry of a timelike boundary located near the conformal boundary of an asymptotically AdS$_4$ spacetime. 

Recall first that near the infinity, one can write the metric as
\begin{equation}
    \frac{ds^2}{\ell^2} = d\rho^2 + \frac{e^{2\rho}}{\ell^2}\left(g^{(0)}_{mn} + \ell^2e^{-2\rho}g^{(2)}_{mn} + \ell^3 e^{-3\rho} g^{(3)}_{mn}  + \mathcal{O}(e^{-4\rho}) \right)dx^m dx^n \, ,
\end{equation}
where $g^{(2)}_{mn}$ is minus the Schouten tensor of $g_{mn}^{(0)}$, see (\ref{schouten}), and $g_{mn}^{(3)}$ is a transverse-traceless tensor, also with respect to $g_{mn}^{(0)}$. Let us consider placing a timelike boundary $\Gamma$ at $\rho = \rho_c \gg 1$. The induced metric then reads
\begin{equation}\label{gmn FG gauge}
    \left.g_{mn}\right|_{\Gamma}=e^{2\rho_c}\left(g^{(0)}_{mn} + \ell^2e^{-2\rho_c}g^{(2)}_{mn} + \ell^3e^{-3\rho_c} g^{(3)}_{mn} + %
    + \mathcal{O}(e^{-4\rho_c}) \right) \, ,
\end{equation}
with the inverse metric given by
\begin{eqnarray}
    \left.g^{mn}\right|_{\Gamma}= e^{-2\rho_c}\left(g^{(0)mn}-\ell^2 e^{-2\rho_c}g^{(2)mn}-\ell^3e^{-3\rho_c}g^{(3)mn} + %
      \mathcal{O}(e^{-4\rho_c}\right) \, .
\end{eqnarray}

Using $K_{mn} = \frac{1}{2\ell}\partial_\rho g_{mn}$, the extrinsic curvature is given by
\begin{equation}\label{Krho}
    \left.K_{mn}\right|_{\Gamma} = \frac{ e^{2\rho_c}}{\ell}\left(g^{(0)}_{mn} - \frac{1}{2}\ell^3e^{-3\rho_c}g_{mn}^{(3)} %
    + \mathcal{O}(e^{-4\rho_c})\right) \, .
\end{equation}
It follows that the trace of the extrinsic curvature is given by
\begin{equation}\label{Krhoc}
    \left.K\ell\right|_{\Gamma} = 3 -\ell^2 e^{-2\rho_c}\text{tr}(g^{(2)}_{mn}) +%
    \mathcal{O}(e^{-4\rho_c}) \, ,
\end{equation}
where we have used the fact that $\text{tr}g^{(3)}=0$. One straightforwardly obtains $\text{tr} \, g_{mn}^{(2)} = -R[g^{(0)}_{mn}]/4$. To compute the conformal Brown-York tensor (\ref{cby}), let us choose the representative of the conformal class of the induced metric to be
\begin{equation}\label{bar g}
    \bar{g}_{mn} = e^{-2\rho_c}\left.g_{mn}\right|_{\Gamma} = g^{(0)}_{mn} +\ell^2 e^{-2\rho_c}g^{(2)}_{mn} +\ell^3 e^{-3\rho_c} g^{(3)}_{mn} %
    + \mathcal{O}(e^{-4\rho_c}) \, ,
\end{equation}
so that the Weyl factor is given by $e^\bomega = e^{\rho_c}$. Plugging \eqref{gmn FG gauge}, \eqref{Krho}, \eqref{Krhoc}, and $e^\bomega = e^{\rho_c}$ into (\ref{cby}), the conformal Brown-York tensor can be written in the large $\rho_c$ expansion as
\begin{equation}\label{TmnFG}
    T_{mn} = e^{\rho_c} T^{(-1)}_{mn} + T^{(0)}_{mn} %
    + \mathcal{O}(e^{-\rho_c}) \, ,
\end{equation}
where the term is explicitly given by
\begin{equation}\label{TmnFG2}
    \begin{cases}
        T^{(-1)}_{mn} = \frac{\ell}{8 \pi G_N}\left(g^{(2)}_{mn} - \frac{1}{3}\text{tr}(g^{(2)}_{mn} )g^{(0)}_{mn}\right)\,, \\
        T^{(0)}_{mn}  = \frac{3\ell^2}{16 \pi G_N}g^{(3)}_{mn}\, , %
    \end{cases}
\end{equation}
By construction, $T_{mn}$ is traceless with respect to the induced metric $\left.g_{mn}\right|_{\Gamma}$. %
Provided that $\text{tr}(g^{(2)}_{mn})$ is non-vanishing, we can invert \eqref{Krhoc} to write $e^{\rho_c}$ in a small $(K\ell -3)$ expansion,
\begin{equation}\label{eqn: rho_c as Kell-3}
    \frac{e^{\rho_c}}{\ell} = \sqrt{\frac{-\text{tr}(g^{(2)}_{mn})}{K \ell-3}} %
    + \mathcal{O}(1) \,.
\end{equation}
This equation allows us to rewrite the large $\rho_c$ expansion in terms of the small $(K\ell-3)$ expansion. As an example, the induced metric \eqref{gmn FG gauge} to leading order can be written as
\begin{equation}
    \left.g_{mn}\right|_{\Gamma} = \frac{-\ell^2\text{tr}(g^{(2)}_{mn})}{K\ell-3}g^{(0)}_{mn} + \mathcal{O}(K\ell-3)^{-1/2}\,.
\end{equation}
Finally, provided $\text{tr}(g^{(2)}_{mn})$ is non-vanishing and given (\ref{eqn: rho_c as Kell-3}), we can express the Brown-York stress energy tensor in terms of the conformal boundary data to get the result shown in \eqref{eq: brown-york near boundary}.

\chapter{Spherically symmetric diffeomophisms on the AdS$_4$ black hole} \label{app: bh}

In this appendix we consider the physical diffeomorphism with $l=0$ about the AdS$_4$ black hole background.

\section{Modes with $l=0$ in the Fefferman-Graham gauge} \label{app: FG l=0}

Here we extend the result in section \ref{l0modes} to the Fefferman-Graham gauge. To do so, we first recall that the most general spherically symmetric solution in the Fefferman-Graham gauge can be written as
\begin{equation}\label{eqn: spherically sym FG gauge}
    ds^2 = \frac{\ell^2dz^2}{z^2}-F(t,z)dt^2 + H(t,z)\ell^2 d\Omega^2_2\,,
\end{equation}
where $F(t,z)$ and $H(t,z)$ are arbitrary functions of $t$ and $z$. For the AdS$_4$ black hole solution, these functions are given by
\begin{equation}\label{eqn: AdS bh FG gauge}
    \begin{cases}
        F=\bar{F}(z)\equiv\frac{\ell^2}{z^2}+\frac{1}{2}-\frac{4G_NMz}{3\ell^2}+\frac{z^2}{16\ell^2}+...\,, \\
        H=\bar{H}(z)\equiv\frac{\ell^2}{z^2}-\frac{1}{2}+\frac{2G_NMz}{3\ell^2}+\frac{z^2}{16\ell^2}+... \, .
    \end{cases}
\end{equation}
The Schwarzchild gauge can be obtained by defining $r=\ell\sqrt{\bar{H}(z)}$. The timelike boundary is then located at $\r=\ell\sqrt{\bar{H}(z_c)}$. The trace of the extrinsic curvature of \eqref{eqn: spherically sym FG gauge} at $z=z_c$ is given by
\begin{equation}\label{eqn: K of AdS bh}
    K \ell = \left.-\frac{z \partial_z F}{2F}-\frac{z \partial_z H}{H}\right|_\Gamma \,.
\end{equation}
This equation implies that $z_c$ is implicitly a function of $K\ell$.

Now we consider metric \eqref{eqn: spherically sym FG gauge} which differs from the AdS$_4$ black hole solution by a linearised diffeomorphism. We require that this metric, at $z=z_c$, leads to a trace of the extrinsic curvature and conformal structure of the induced metric that match those of the AdS$_4$ black hole solution, namely
\begin{equation}
    \left.ds^2\right|_\Gamma = (1+2\delta\bomega)\left(-\bar{F}(z_c) dt^2 + \bar{H}(z_c)\ell^2d\Omega^2\right) \, , \qquad \left.K \ell\right|_\Gamma = \left.-\frac{z\partial_z \bar{F}}{2\bar{F}}-\frac{z\partial_z \bar{H}}{\bar{H}} \right|_\Gamma\,,
\end{equation}
where $\delta \bomega$ is a $z$-independent function describing the perturbed Weyl factor. Taking both $\tfrac{z}{\ell}\to 0$ and $\tfrac{z_c}{\ell}\to 0$ while keeping $\tfrac{z}{z_c}$ fixed, this metric admits an expansion in $\tfrac{z}{\ell}$ as
\begin{equation}\label{eqn: AdS BH FG gauge perturbed metric}
\begin{cases}
    F(t,z) &= \frac{\ell^2}{z^2}\left(1+2\delta \bomega\right) + \frac{1}{2}\left(1+\left(\frac{2z_c^2}{z^2}-1\right)2\delta\bomega\right) - \frac{4G_N M z}{3\ell^2}\left(1+\left(\frac{3z_c^3}{z^3}-1\right)\delta \bomega\right)+\mathcal{O}\left(\frac{z^2}{\ell^2}\right) \, ,  \\
    H(t,z) &= \frac{\ell^2}{z^2}\left(1+2\delta \bomega\right) - \frac{1}{2}\left(1+\frac{z_c^2}{z^2}2\delta \bomega\right) + \frac{2G_NMz}{3\ell^2}\left(1+\left(\frac{3z_c^3}{z^3}-1\right)\delta \bomega\right)+\mathcal{O}\left(\frac{z^2}{\ell^2}\right) \, .
    \end{cases}
\end{equation}
The perturbed Weyl factor $\delta \bomega$ satisfies the equation
\begin{equation}
    \ell^2\partial_t^2\delta\bomega =\left(1+\mathcal{O}\left(\frac{z_c^2}{\ell^2}\right)\right)\delta \bomega \, .
\end{equation}
In these equations, $z_c$ can be written perturbatively in terms of $K\ell-3$ as
\begin{equation}
    \frac{z_c}{\ell} = \sqrt{2(K\ell-3)} + \mathcal{O}(K\ell-3)^{3/2} \, .
\end{equation}
Setting $K\ell=3$ ($z_c=0$) and comparing with the standard Fefferman-Graham expansion, we observe that the Weyl mode, $\delta \bomega$, contributes to the $g^{(0)}_{mn}$ and $g^{(3)}_{mn}$ as
\begin{equation} \label{eq: lin g0 g3}
    g^{(0)}_{mn}= \bar{g}^{(0)}_{mn}\left(1+2\delta \bomega\right) \, , \qquad g^{(3)}_{mn}= \bar{g}^{(3)}_{mn}\left(1-\delta \bomega\right) \, ,
\end{equation}
where $\bar{g}^{(n)}_{mn}$ are those of the AdS$_4$ black hole solution. This is indeed the behavior of $g^{(0)}_{mn}$ and $g^{(3)}_{mn}$ obtained under a Fefferman-Graham gauge preserving diffeomorphism (see for instance (2.6) and (2.7) of \cite{Anninos:2010zf}).

\section{Dressing the global AdS$_4$ black hole} \label{app: bh}

We now perform a similar analysis to the one in section \ref{sec: non-linear diffeos}, but now for the global AdS$_4$ black hole. The background metric corresponds to 
\begin{equation}
ds^2 = -f(r) dt^2 + \frac{dr^2}{f(r)} + r^2 d\Omega_2^2 ~, \quad \text{with} \quad f(r) = 1- \frac{2 G_N M}{r} + \frac{r^2}{\ell^2} \,.
\end{equation}

The case of a boundary at a constant $r$ slice has been analysed above in section \ref{sec: conf by}. Here we would like to identify a hypersurface in the black hole geometry which obeys the conformal boundary conditions with constant $K$, and induced metric
\begin{equation} \label{eq: bh bdy metric}
ds^2 |_\Gamma =  e^{{2\delta \bomega(u)}} \left(- du^2 + \r^2 \, d\Omega_2^2 \right)~.
\end{equation}
Note that for convenience, we are rescaling the time coordinate with respect to \eqref{indmetric} and that $\r$ is fixed and part of the conformal boundary data. Requiring that $K$ is constant provides a non-linear differential equation for $\delta \bomega(u)$,
\begin{equation} \label{eq: brane dynamics bh}
    \r^2 \partial^2_u \delta \bomega(u) = K \ell e^{\frac{\delta\bomega}{2}} \sqrt{-\frac{2 \r G_N M}{\ell^2}+ \frac{\r^2}{\ell^2}  e^{\delta\bomega} \left((\r\partial_u \delta\bomega)^2+1\right)+\frac{\r ^4}{\ell^4} e^{3 \delta\bomega}}+ \frac{3 G_N M e^{-\delta\bomega}}{\r}-\frac{3 \r ^2 e^{2 \delta\bomega}}{\ell^2} - 2-2 (\r \partial_u \delta\bomega )^2 \,.
\end{equation}
Upon setting $M=0$ (and suitably rescaling time), this equation reduces to \eqref{eqn: brane eqn}. If $\delta\bomega$ is time independent, then this equation gives the trace of the extrinsic curvature of a constant $r=\r \, e^{\delta\bomega}$ slice of the black hole geometry,
\begin{equation}
    K \ell = \frac{e^{-\frac{3}{2} \delta\bomega} \left(-3 G_N \ell^2 M +2 \ell^2 \r  e^{\delta\bomega}+3 \r ^3 e^{3 \delta\bomega}\right)}{\r ^{3/2} \sqrt{-2 G_N \ell^2 M+\ell^2 \r  e^{\delta\bomega}+\r ^3 e^{3 \delta\bomega}}} \,.
\end{equation}
When $\r = \ell$ this is the $K\ell$ that appears in \eqref{omegaM}. Another interesting limit is to keep $M$ finite and linearise $\delta\bomega (u)$. In that case we obtain,
\begin{equation}\label{omegabh}
  \r^2  \partial_u^2 \delta \bomega (u) = \Omega^2 \delta \bomega (u)~, \quad \text{with} \quad \Omega \equiv \frac{\left(9 G_N^2 M^2 \ell^2 -8 G_N M \ell^2 \r +2 \ell^2 \r ^2+\r ^4\right)}{ \r^4 +\ell^2 \r^2 -2 G_N M \ell^2 \r} \,,
\end{equation}
which has solutions of the form $\delta \bomega (u) = e^{\pm \Omega  u}$. The second order nature of (\ref{omegabh}) indicates that, as in our previous discussions, the must fix the dynamical data $\mathcal{C}_{\partial \Sigma} = \{\delta\bomega, \partial_u\delta\bomega \}|_\Gamma$ to have a complete specification of the initial data. By setting $M=0$, we retrieve the exponentially growing modes of section \ref{l0modes} (appropriately rescaled by $f(\r)$ with $M=0$).

Next we can compute the conformal Brown-York stress tensor (\ref{cby}), that reads
\begin{equation} \label{eq: cby bh}
T_{uu} = \frac{2}{\ell^2}T_{\theta \theta} = \frac{2 }{\ell^2\sin^2{\theta}}T_{\phi \phi} = \frac{K \ell e^{2 \delta\bomega}-3 e^{\frac{\delta\bomega}{2}} \sqrt{-\frac{2 G_N \ell^2 M}{\r^3} + \frac{\ell^2}{\r^2}  e^{\delta\bomega} \left((\r \partial_u \delta\bomega)^2+1\right)+ e^{3 \delta\bomega}}}{12 \pi  G_N \ell}\,.
\end{equation}
Upon imposing \eqref{eq: brane dynamics bh}, this stress tensor is not only traceless, but also conserved, with respect to the metric \eqref{eq: bh bdy metric}.

\noindent \textbf{Asymptotic boundary limit.} We would like to analyse this setup as the timelike boundary $\Gamma$ gets near the conformal boundary of the AdS$_4$ black hole. In that limit, as in empty AdS, one can check that $K \ell \approx 3$, so one can consider the solution in an expansion for small $\delta K \ell \equiv K \ell - 3$. Note from \eqref{eq: brane dynamics bh} in the static limit, that $e^{\delta \bomega(u)}$ should diverge as $(K \ell -3)^{-1/2}$. 

In order to compare with the standard Fefferman-Graham expansion (where the conformal factor is constant and large), it is convenient to isolate this divergence in the conformal factor of the metric, such that,
\begin{equation}
    ds^2|_{\Gamma} = \frac{1}{K \ell -3} \left( e^{{2\delta \bomega(u)}} \left(- du^2 + \r^2 \, d\Omega_2^2 \right) \right) \,.
\end{equation}
On top of rescaling $\delta \bomega$ in \eqref{eq: cby bh}, this transformation adds an overall scaling to the stress tensor, $T_{mn} \to \frac{T_{mn}}{\sqrt{K \ell -3}}$. After this, $\delta \bomega$ needs to be finite in the $\delta K \ell \to 0$ limit, and we obtain that
\begin{eqnarray}
    e^{\delta\bomega(u)} = e^{\delta\bomega_0 (u)} + e^{\delta\bomega_1 (u)} \sqrt{\delta K \ell} + %
    \mathcal{O} \left(\delta K \ell \right) \,. 
\end{eqnarray}
Solving \eqref{eq: brane dynamics bh} order-by-order in $\delta K \ell$ gives dynamical equations for all the $\delta \bomega_n$. For brevity, here we report the first two equations,
\begin{equation}
    \begin{cases}
       \r^2 \partial_u^2 \delta\bomega_0 = e^{2 \delta \bomega_0} \frac{\r^2}{\ell^2}-\frac{1}{2}-\frac{1}{2} (\r \partial_u \delta\bomega_0)^2 \, ,\\
     \r^2 \partial_u^2  \delta\bomega_1 = \delta\bomega_1 \left(3 e^{2 \delta\bomega_0}\frac{\r^2}{\ell^2}-\frac{1}{2}-\frac{1}{2}(\r\partial_u \delta\bomega_0)^2\right)+ \r^2 (\partial_u \delta\bomega_0) (\partial_u \delta\bomega_1) \,. 
    \end{cases} \label{eq: bomega expansions}
\end{equation}
As seen in previous examples, these equations will be completely determined after we supplement initial data for $\delta \bomega$ and $\partial_u \delta \bomega$ at the corner $\partial \Sigma$.

We can also expand the Brown-York stress tensor \eqref{eq: cby bh} close to the boundary, that after the rescaling becomes
\begin{equation}
 T_{uu} = \frac{T_{uu}^{(-1)}}{\sqrt{\delta K \ell}} + T_{uu}^{(0)} %
 + \mathcal{O} \left( (\delta K \ell)^{1/2} \right) \,.
\end{equation}
Upon imposing \eqref{eq: bomega expansions}, we find that
\begin{equation} \label{eq: expansion tmn}
    \begin{cases}
        T_{uu}^{(-1)} = \frac{e^{2 \delta \bomega _0}}{12 \pi  \ell G_N} - \frac{\ell}{8 \pi  \r^2 G_N} \left(1+ (\r\partial \delta \bomega _0)^2\right) \,, \\
        T_{uu}^{(0)} = \frac{e^{-\delta \bomega _0} \ell M}{4 \pi  \r^3} + \frac{e^{-\delta \bomega _0} \left(2 e^{2 \delta \bomega _0} \delta \bomega _1+3 \ell^2 \partial_u \delta \bomega _0 \left(\partial_u \delta \bomega _0 \delta \bomega _1-\partial_u \delta \bomega _1\right)\right)}{12 \pi  \ell G_N} \,.
    \end{cases}
\end{equation}
The leading contribution is divergent in the strict $K\to \tfrac{3}{\ell}$ limit. This expression for the stress tensor near the boundary is compatible with the one in the main text, see \eqref{eq: brown-york near boundary}, upon identifying $g_{mn}^{(0)}dx^m dx^n$ with $e^{2\delta\bomega} \left(-du^2 + \r^2 d\Omega_2^2\right)$. Recall that $e^{\delta \bomega} = e^{\delta \bomega_0} + e^{\delta\bomega_1} \delta K \ell + \cdots$. Then the leading contribution to the stress tensor will come exclusively from $\delta \bomega_0$ and perfectly matches the divergent term in \eqref{eq: brown-york near boundary}.

The next term $T_{uu}^{(1)}$ receives contributions from two different sources. The first term is proportional to the mass of the black hole $M$. The scaling with $e^{-\delta\bomega_0}$ is compatible with the expected scaling for $g^{(3)}_{mn}$, see also the linearised analysis in \eqref{eq: lin g0 g3}. The second term is exactly the first subleading contribution that comes from the leading term in \eqref{eq: brown-york near boundary} upon expanding $\delta\bomega$.

One can continue this analysis order-by-order systematically, finding that when the timelike boundary is sufficiently away from the conformal boundary of AdS$_4$, $M$ and $\delta \bomega$ no longer decouple.

\chapter{Locally diffeomorphic modes for any $l$} \label{sec: phys_diffeos_l}

In this appendix we show that there are no physical diffeomorphisms for $l\geq 2$ satisfying conformal boundary conditions (\ref{gbar_global}) and (\ref{eq: K function of r}) about global AdS$_4$ (\ref{globalAdS4}). Consider a general-$l$ metric perturbation in the scalar sector, which is locally a diffeomorphism, $ h_{\mu\nu} = \nabla_\mu \xi_\nu + \nabla_\nu \xi_\mu $. Then we can parameterise the vector field $\xi^\mu$,
\begin{equation}
    \xi^r = r \zeta_r \,\mathbb{S}_l \, , \qquad \xi^t = r^{-2} \zeta_t \, \mathbb{S}_l \, , \qquad \xi^\theta = r^{-2} \zeta_\sigma \, \partial_\theta \mathbb{S}_l \, , \qquad \xi^\phi = r^{-2} \zeta_\sigma \, \frac{\partial_\phi \mathbb{S}_l}{\sin^2\theta} \, ,
\end{equation}
for some $(t,r)$-dependent functions $\zeta_r$, $\zeta_t$, and $\zeta_\sigma$. %

Now we consider inserting a timelike boundary at $r=\r$. For a generic $\r$, the induced metric and the trace of the extrinsic curvature at the boundary are given by
\begin{eqnarray}
    \left.ds^2\right|_{\Gamma} &=& -\left(1+\frac{\r^2}{\ell^2} + \varepsilon\,2\left(\frac{\r^2}{\ell^2}\zeta_r + \left(\frac{1}{\ell^2}+\frac{1}{\r^2}\right)\partial_t \zeta_t\right)\,\mathbb{S}_l\right)dt^2 + 2 \left(\partial_t \zeta_\sigma-\left(\frac{1}{\ell^2}+\frac{1}{\r^2}\right)\zeta_t \right)dt d\mathbb{S}_l \nonumber \\
    & & + \left(\r^2 + \varepsilon\, 2 \r^2\zeta_r\,\mathbb{S}_l\right)d\Omega^2 + \varepsilon\,2\zeta_\sigma \tilde{\nabla}_i\tilde{\nabla}_j \mathbb{S}_l\, d\Omega^i d\Omega^j \,\Bigg|_\Gamma\, , \label{eq: diffeo_any_d_metric}\\
    \left.K\ell\right|_{\Gamma} &=& \frac{2\ell^2+3\r^2}{\ell \r\sqrt{1+\frac{\r^2}{\ell^2}}} + \varepsilon\,\frac{\ell \, \mathbb{S}_l}{\r \left(1+\frac{\r^2}{\ell^2}\right)^{3/2}}\left(\r^2 \partial_t^2 \zeta_r + \left(\frac{\r^2}{\ell^2}(l^2+l-1)+(l^2+l-2)\right)\zeta_r\right)\Bigg|_\Gamma \, .
\end{eqnarray}
By requiring the perturbed induced metric \eqref{eq: diffeo_any_d_metric} to be conformal to \eqref{gbar_global}, we obtain that all three $\zeta_\sigma, \zeta_t,$ and $\zeta_r$ should identically vanish. So, in general, there are no physical diffeomorphisms in the scalar sector for $l\geq 2$. One can show, also, that diffeomorphisms with $l\ge 2$, built from vectorial spherical harmonics $\mathbb{V}_i$ are never physical.

Nonetheless, a notion of approximate diffeomorphisms emerges when we consider $\r \gg \ell$. In that case, by keeping $\zeta_r$, $\zeta_t$, and $\zeta_\sigma$ finite, the induced metric and the trace of the extrinsic curvature to first non-trivial order in the perturbative parameter are given by
\begin{eqnarray}\label{eqn: large r global ads induced metric}
    \left.ds^2\right|_\Gamma &=& \left.\left(1+ \varepsilon\, 2 \zeta_r \, \mathbb{S}_l\right)\left(- \frac{\r^2}{\ell^2}dt^2 + \r^2 d\Omega^2\right)\right|_\Gamma + \mathcal{O}\left(1 \right) \, , \\ \left.K\ell\right|_\Gamma &=& 3 + \left.\frac{\ell^2}{\r^2}\left(\frac{1}{2}+ \varepsilon \, \Big(\ell^2 \partial_t^2 \zeta_r + \left(l^2+l-1\right)\zeta_r\Big)\,\mathbb{S}_l\right)\right|_\Gamma + \mathcal{O}\left({\r^{-4}}\right) \, .\label{eqn: large r global ads trace of K}
\end{eqnarray}
It follows that the induced metric is automatically conformally equivalent to the cylinder in this limit. Note that $\zeta_t$ and $\zeta_\sigma$ do not appear in this analysis. Requiring that $\xi^\mu$ preserves the trace of the extrinsic curvature, we obtain a differential equation for $\zeta_r$ at the boundary,
\begin{equation}
  \left.  \partial_t^2 \zeta_r + \left(\frac{l(l+1)-1}{\ell^2}\right) \zeta_r \right|_\Gamma = 0 \, .
\end{equation}
Thus, we can write the general diffeomorphism that preserves conformal boundary conditions as
\begin{equation}
    \xi^r(t,\r,\theta,\phi) = \sum_{l\in\mathbb{N}^0}\r \left(e^{- i \omega^{(l)} t}f_{+,l}(\r) + e^{i \omega^{(l)} t}f_{-,l}(\r)\right) \mathbb{S}_l \, ,
\end{equation}
where 
\begin{equation}\label{masslessdisp}
    \omega^{(l)} \ell = \sqrt{l(l+1)-1} \,,
\end{equation} 
and $f_{+,l}(\r)$ and $f_{-,l}(\r)$ are constants of integration. Since the orthogonal component $\xi^r$ is non-vanishing on the boundary, this solution is physical and cannot be gauged away. Note that these frequencies coincide with the lowest frequencies found in section \ref{sec: bulk modes}.

From \eqref{eqn: large r global ads induced metric}, we can also compute the perturbation of the Weyl factor at the boundary,
\begin{equation}\label{eqn: any l diff sol}
    \delta \boldsymbol{\omega} |_{r=\r} = \varepsilon \,  \sum_{l\in\mathbb{N}} \left(e^{- i \omega^{(l)} t}f_{+,l}(\r) + e^{i \omega^{(l)} t}f_{-,l}(\r)\right) \mathbb{S}_l \, .
\end{equation}
For $l\geq 1$, these solutions do not exhibit exponential growth at late times. Only the $l=0$ sector leads to an exponentially growing solution.

We can also compute the conformal Brown-York stress-energy tensor for these near-diffeomorphic modes, for which we obtain,
\begin{equation} \label{conformal BY near diffeos}
    T_{mn} = \left.\frac{\ell}{8\pi G_N}\left(\frac{\bar{R}}{3} \bar{g}_{mn}-\bar{R}_{mn}+\left(\bar{\mathcal{D}}_m \bar{\mathcal{D}}_n+\frac{\bar{R} }{2}\bar{g}_{mn} -\bar{R}_{mn}\right) \delta \bomega\right)\right|_\Gamma \,,
\end{equation}
where $\bar{\mathcal{D}}_m$ and $\bar{R}_{mn}$ are the covariant derivative and Ricci tensor with respect to the metric
\begin{equation}
ds^2=-\frac{\r^2}{\ell^2}dt^2+\r^2 d\Omega_2^2~.
\end{equation}
One can readily confirm that $T_{mn}$ is traceless and covariantly conserved upon imposing \eqref{eqn: any l diff sol}.

Computing the subleading correction reveals that the large $\r$ approximation should breakdown when $\omega^{(l)}  \sim \mathcal{O}(\r)$. In terms of $K\ell$, we have $\tfrac{\ell}{\r} \sim \sqrt{K\ell-3}$, such that the above analysis breaks down when $\omega^{(l)} \ell \sim (\sqrt{K\ell-3})^{-1/2}$.



\chapter{Complex frequencies at large $l$} \label{app: large l}

In section \ref{sec: bulk modes}, we found that the behavior of the complex frequencies depended on the position of the boundary given by $K \ell$. When $K\ell \to \infty$, the complex frequencies at large $l$ asymptote the flat space behavior $\omega \r \approx \pm l \pm i 0.34 l^{1/3}$ \cite{Anninos:2023epi,Liu:2024ymn}. When $K\ell \to 3$, we used the WKB approximation to show that there were no complex frequencies. In this appendix, we provide numerical evidence of the intermediate regime.

For any fixed $K\ell$ and $l$, we numerically search for the complex frequencies. Then we vary $l$ for large values in between $l=30$ and $l=120$ and we keep track of the values of the complex frequencies. We find reasonable to assume the following ansatz for the frequencies at large $l$,
\begin{equation}
    \omega \ell (l) \approx \pm \, \omega_{\text{Re}} \, l^{\alpha_\text{Re}} \pm  i \, \omega_{\text{Im}} \, l^{\alpha_{\text{Im}}} \,. 
\end{equation}

\begin{figure}[h!]
        \centering
         \subfigure[$\alpha_{\text{Re}}$]{
                \includegraphics[scale=0.55]{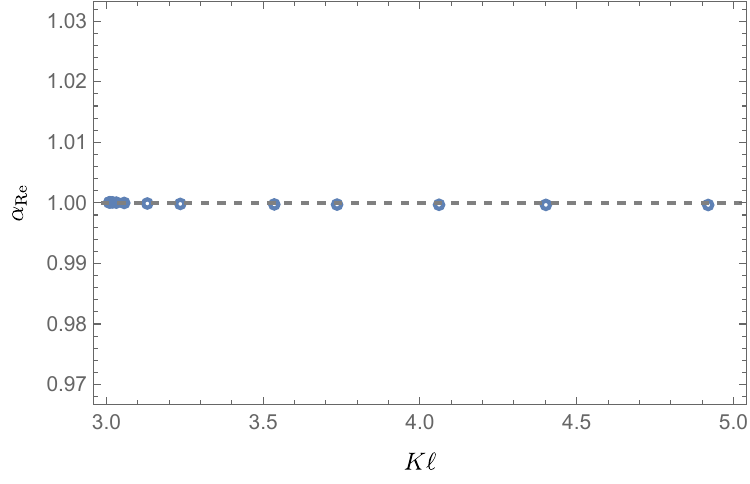}\label{fig: appA2}}  \quad
                 \subfigure[$\omega_{\text{Re}}$]{
                \includegraphics[scale= 0.55]{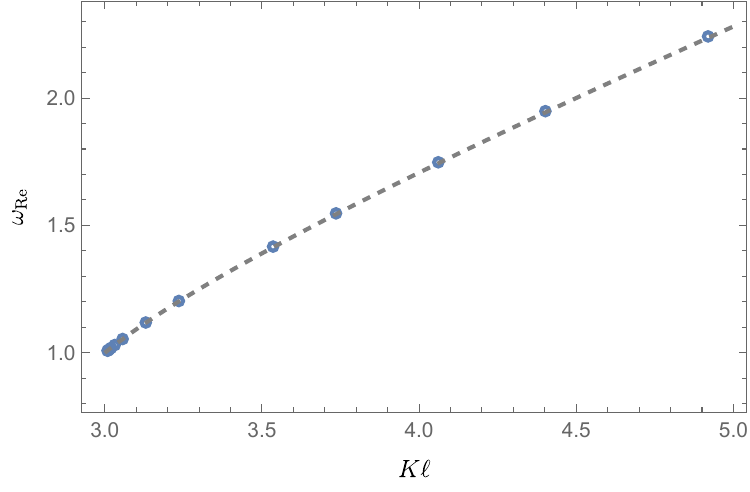} \label{fig: appB2}}                           
                \caption{Behavior of the real part of the complex frequencies at large $l$. In (a), in dashed gray, we show $\alpha_{\text{Re}}=1$. In (b), we show in dashed gray the analytic result \eqref{eq: app re}, which perfectly matches the numerical results for all the values of $K\ell$ analysed.} \label{fig: Re largel}
\end{figure}

The real part behaves in a universal manner, finding that $\alpha_{\text{Re}} \approx 1$ for any $K \ell$. We provide a numerical plot for both $\alpha_{\text{Re}}$ and $\omega_{\text{Re}}$ in figure \ref{fig: Re largel}. It is interesting to note that $\alpha_{\text{Re}}$ and $\omega_{\text{Re}}$ can be obtained analytically by doing a large $l$ expansion of \eqref{eqn: deltaK scalar}. The hypergeometric function can be approximated using a steepest descend method \cite{paris1,paris2}, after which we obtain $\alpha_{\text{Re}}=1$ and,
\begin{equation} \label{eq: app re}
    \omega_{\text{Re}} = \sqrt{\frac{K\ell}{8} \left(\sqrt{K^2\ell^2-8}+K\ell\right)-\frac{1}{2}} = \begin{cases}
        1 + (K\ell - 3) + \mathcal{O} (K\ell-3)^2 \, , \, \text{for} \, K\ell \to 3 \,, \\
        \frac{K\ell}{2}-\frac{1}{K\ell}+\mathcal{O}\left(\frac{1}{K\ell}\right)^3 \, ,  \, \text{for} \, K\ell \to \infty \,.
        \end{cases}
    \end{equation}

In the limit of $K\ell \to \infty$ this agrees with the flat space result. Moreover, it agrees with the numerical results for all values of $K \ell$ analysed, as seen in figure \ref{fig: appB2}.

\begin{figure}[h!]
        \centering
         \subfigure[$\alpha_{\text{Im}}$]{
                \includegraphics[scale=0.55]{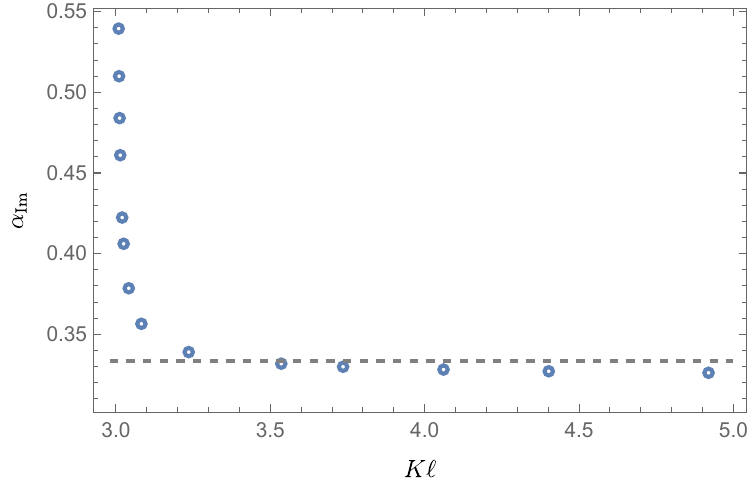}\label{fig: appA}}  \quad
                 \subfigure[$\omega_{\text{Im}}$]{
                \includegraphics[scale= 0.55]{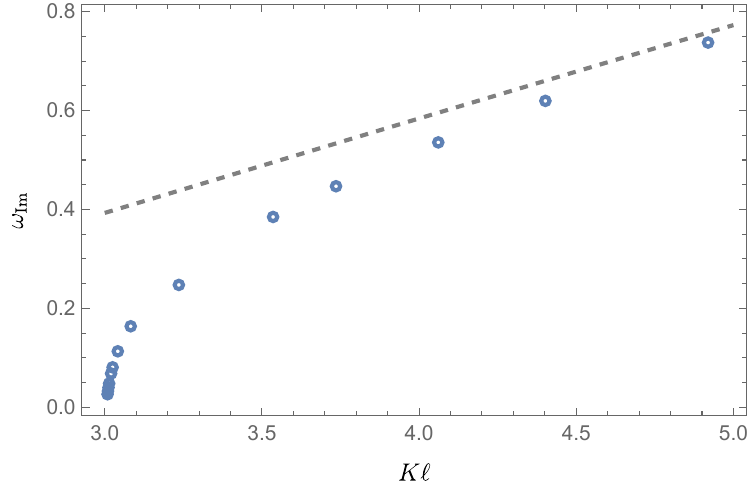} \label{fig: appB}}                           
                \caption{Behavior of the imaginary part of the complex frequencies at large $l$. In dashed gray, we show the flat space asymptotic value at large $l$. In (a), the small discrepancy at large $K\ell$ can be atributed to numerical error, while in (b), the coefficient becomes linear at large $K\ell$ due to the fact that frequencies scale with $\r$ instead of $\ell$ in the flat space limit.} \label{fig: Im largel}
\end{figure}

In contrast, the imaginary part behaviour appears to strongly depend on the position of the boundary. In figure \ref{fig: Im largel}, we plot both $\omega_{\text{Im}}$ and $\alpha_{\text{Im}}$ for a range of $K \ell$. Note that while $\alpha_{\text{Im}}$ grows as $K\ell \to 3$, $\omega_{\text{Im}}$ decays. One can check that, at least for the values of $K\ell$ and $l$ analysed, the full imaginary part $\omega_{\text{Im}} \, l^{\alpha_{\text{Im}}}$ does not seem to grow larger than the real part as the boundary moves towards the boundary of AdS$_4$, consistent with the WKB analysis of section \ref{sec: bulk modes}. It would be desirable to get an analytical understanding of the behavior of these frequencies for any value of $K \ell$, at large $l$. Also notice that these frequencies do not appear in flat AdS$_4$. In that case,  the reason might be that $K \ell$ is fixed to $K \ell =3$. It would be interesting to see if the complex frequencies appear as perturbations about the black brane geometry, where $K \ell$ is allowed to change. We leave these for future work.

\chapter{Bulk perturbations in planar AdS$_4$} \label{app: planar perturbation}

In this appendix we give an explicit derivation of the expression for the bulk metric perturbation \eqref{eq: bulk modes planar} about planar AdS$_4$. To find a solution for $\tilde{\gamma}_{mn}$, it is useful to consider the mode expansion
\begin{equation}
    \tilde{\gamma}_{mn}(z,x^m) = \sum_{j=1}^2 \beta^{(j)}_{mn}\varphi(z)e^{ik_mx^m} \, ,
\end{equation}
where $\beta^{(j)}_{mn}$ represent the two transverse-traceless polarisations of gravity, and $\varphi(z)$ is a solution to
\begin{equation}\label{eqn: planar AdS eqn for phi}
    q^2 \varphi + z^2 \partial_z \left(z^{-2}\partial_z \varphi\right) = 0 \, , \qquad q \equiv \sqrt{-k^m k_m} \, .
\end{equation}
The conformal boundary condition simplifies to 
\begin{equation}\label{eqn: planar AdS bdry cond for phi}
    \varphi(z_c)=0 \, .
\end{equation}
The asymptotic behaviour of $\tilde{\gamma}_{mn}$ or equivalently $\varphi$ as $z\to\infty$ depends on the value of $q$. Let us consider separately the case where $q =0$, $q$ is a positive imaginary number, and $q$ is a positive real number.

\textbf{Polynomial solutions.} Let us first consider the case where $q=0$. This is equivalent to making $k^m$ a null vector. The solution to \eqref{eqn: planar AdS eqn for phi} is given by
\begin{equation}
    \varphi(z) = c_1 z^3 + c_2 \, ,
\end{equation}
where $c_1$ and $c_2$ are arbitrary constants. Requiring that $h_{mn}$ is finite as $z\to \infty$ fixes $c_1 = 0$. The boundary condition \eqref{eqn: planar AdS bdry cond for phi} then sets $c_2 = 0$. Hence, there is no $\tilde{\gamma}_{mn}$ with $q =0$ that obeys the conformal boundary conditions.

\textbf{Exponentially growing/decaying solutions.} Consider the case where $\alpha\equiv iq \in \mathbb{R}_{>0}$. This is equivalent to make $k^m$ a spacelike vector. The solution to \eqref{eqn: planar AdS eqn for phi} is given by
\begin{equation}
    \varphi(z) = c_1 (1+\alpha z)e^{-\alpha z} + c_2 (1-\alpha z)e^{\alpha z} \, ,
\end{equation}
where $c_1$ and $c_2$ are arbitrary constants. Requiring that $h_{mn}$ is finite as $z\to\infty$ fixes $c_2 = 0$. The boundary condition \eqref{eqn: planar AdS bdry cond for phi} sets also $c_1=0$. This is similar to the case where $q=0$, namely there is no $\tilde{\gamma}_{mn}$ with purely imaginary $q$ that obeys the conformal boundary conditions.

\textbf{Oscillating solutions.} Consider the case where $q \in \mathbb{R}_{>0}$. This is equivalent to making $k^m$ a timelike vector. The solution to \eqref{eqn: planar AdS eqn for phi} is given by
\begin{equation}
    \varphi(z) = c_1 (1+i q z)e^{-i q z} + c_2 (1-iq z)e^{i q z} \, ,
\end{equation}
where $c_1$ and $c_2$ are real constants. As this solution is oscillating in the $z$-direction, the regularity of $h_{mn}$ as $z\to \infty$ requires certain normalisability condition on $\varphi$ after integrating the solution over $q$. Imposing the boundary condition \eqref{eqn: planar AdS bdry cond for phi}, we find
\begin{equation}
    \frac{c_1}{c_2} = -\frac{1-i q z_c}{1+i q z_c}e^{2iqz_c} \, .
\end{equation}
Therefore, the general $\tilde{\gamma}_{mn}$ that obeys the conformal boundary conditions is given by
\begin{equation}
   \tilde{\gamma}_{mn}(z,x^m) =  \text{Re} \, \int_0^\infty\frac{d q}{2\pi}\int_{\mathbb{R}^2}\frac{d^2 \bold{k}}{(2\pi)^2} \, \sum_{j=1}^2 \beta^{(j)}_{mn}(q,\bold{k}) \left(\frac{1+i q z}{1+i q z_c}e^{-iq(z-z_c)} - \frac{1-i q z}{1-i q z_c}e^{iq(z-z_c)}\right)e^{-i\omega t + i \bold{k} \bold{x}} \, ,
\end{equation}
where $\omega = \sqrt{|\bold{k}|^2+q^2}$.












\end{appendices}

\bibliographystyle{JHEP}
\bibliography{bibliography}

\end{document}